\documentclass{LMCS}

\def\dOi{11(3:6)2015}
\lmcsheading%
{\dOi}
{1--72}
{}
{}
{Dec.~17, 2014}
{Sep.~\phantom03, 2015}
{}

\ACMCCS{[{\bf Theory of computation}]: Logic---Linear logic\,/\,Proof
  theory; Semantics and reasoning---Program semantics---Categorical semantics}

\usepackage{hyperref}

\usepackage{amsfonts}
\usepackage{amsmath}
\usepackage{amsthm}
\usepackage{txfonts} \usepackage{pxfonts}
\usepackage{diagrams}
\usepackage{exscale} 
\usepackage{tensor}
\usepackage{framed}
\usepackage{bussproofs}

\usepackage{graphics} 
\usepackage{color} 
\usepackage{subfigure} 

\usepackage{changepage} 
\newenvironment{clmenv}[1][10mm]{\begin{adjustwidth}{#1}{}}{\end{adjustwidth}} 

\usepackage{amsmath, amssymb, latexsym, amsthm,url}
\usepackage{nccmath}
\usepackage{xxcolor}
\usepackage{tikz}
\usetikzlibrary{positioning}
\usepackage{mathtools}
\usepackage{pxfonts}
\scrollmode 
\newcommand{\tens}{\mathbin{\otimes}}

\usetikzlibrary{shapes}
\usetikzlibrary{positioning,arrows}

\theoremstyle{plain}\newtheorem*{cLm}{Claim}

\newcommand{\mll}{MLL$^{-}$} \newcommand{\mall}{MALL$^{-}$}

\newcommand{\mllmix}{MLL$^{-}$+Mix} \newcommand{\GRel}{\mathbf{GRel}}
\newcommand{\mllmixmaybe}{MLL$^{-}$(+Mix)}
\newcommand{\otensor}{\otimes}

\newcommand{\C}{\mathbb{C}}
\newcommand{\D}{\mathbb{D}}
\newcommand{\GC}{\mathbf{G}\mathbb{C}}
\newcommand{\G}[1]{\mathbf{G}#1}
\newcommand{\gc}[1]{\mathbf{G}_{#1}\mathbb{C}}
\newcommand{\F}{\mathbb{F}}
\newcommand{\FDVec}{\mathbf{FDVec}_{\mathbb{F}}}
\newcommand{\GFDVec}{\mathbf{GFDVec}_{\mathbb{F}}}
\newcommand{\after}{\circ}

\newcommand{\Rel}{\mathbf{Rel}}
\newcommand{\I}{\mathbf{I}}

\newcommand{\p}{} 
\newcommand{\N}{\mathbb{N}}
\newcommand{\R}{\mathbb{R}}
\newcommand{\K}{\mathcal{K}_{\mathbf{I}}}
\newcommand{\Mix}{\mathbf{Mix}}
\newcommand{\mix}{\mathbf{mix}}
\newcommand{\MIX}{\mathbf{MIX}}
\newcommand{\bbrk}[1]{[\![#1]\!]}
\newcommand{\GiC}{\mathbf{G}_{1}\mathbb{C}}
\newcommand{\GsC}{\mathbf{G}_{E}\mathbb{C}}
\newcommand{\Parf}{\mathbf{Par}}
\newcommand{\HypCoh}{\mathbf{HypCoh}}
\newcommand{\huh}{U} 

\newcommand{\Perm}{\mathop{\mathsf{Perm}}\nolimits}
\newcommand{\PPerm}{\mathop{\mathsf{PPerm}}\nolimits}
\newcommand{\cycle}{\mathbf{cycle}}

\newcommand{\tbigotimes}{\textstyle{\bigotimes}}

\newcommand{\tbiginvamp}{\displaystyle{\invamp}}

\newcommand{\binvamp}{\displaystyle{\invamp}}
\newcommand{\bforall}{\displaystyle{\forall}}
\newcommand{\botimes}{\textstyle{\bigotimes}}

\newcommand{\fctl}[3]{\mbox{$#1\colon #2\rTo #3$}\/}
\newcommand{\fromto}[2]{\mbox{$#1\rTo #2$\/}}

\newcommand{\cfct}[5]{\mbox{$ #1\rTo^{#2} #3\rTo^{#4} #5$\/}}

\newcommand{\inj}{\mathop{\mathsf{in}}\nolimits}

\newcommand{\hugha}[1]{} 
\newcommand{\hughb}[1]{} 

\newcommand{\hugh}[1]{#1}
\newcommand{\hughc}[1]{#1}
\newcommand{\hughd}[1]{#1}
\newcommand{\hughe}[1]{#1}
\newcommand{\hughf}[1]{#1}
\newcommand{\hughg}[1]{#1}
\newcommand{\hughh}[1]{#1}

\begin{document}
\title[Constructing Fully Complete Models of Multiplicative Linear Logic]{Constructing Fully Complete Models of \\ Multiplicative Linear Logic\rsuper*}

\author[A.~Schalk]{Andrea Schalk\rsuper a}	
\address{{\lsuper a}School of Computer Science, University of Manchester, Oxford Road, Manchester M13 9PL, UK}	
\email{andrea.schalk@manchester.ac.uk}  

\author[H.~P.~Steele]{Hugh P. Steele\rsuper b}	
\address{{\lsuper b}Universit\'e Paris 13, Sorbonne Paris Cit\'e, LIPN, CNRS, F-93430, Villetaneuse, France}	
\email{hugh.steele@lipn.univ-paris13.fr}  




\titlecomment{{\lsuper*}A preliminary version of this article appeared in the proceedings of LICS 2012~\cite{SS12}.}


\begin{abstract}
The multiplicative fragment of Linear Logic is the formal system in this family with the best understood proof theory, and the categorical models which best capture this theory are the fully complete ones. We demonstrate how the Hyland-Tan double glueing construction produces such categories, either with or without units, when applied to any of a large family of degenerate models. This process explains as special cases a number of such models from the literature. In order to achieve this result, we develop a tensor calculus for compact closed categories with finite biproducts. We show how the combinatorial properties required for a fully complete model are obtained by this glueing construction adding to the structure already available from the original category.
\end{abstract}

\maketitle

\section{Introduction}

Linear Logic~\cite{Gir87} is a well-known formal system that has
attracted interest from computer science as well as logicians. It has
a very well behaved proof theory, and categorical models for linear
logic also contain a model of the \hugh{(linear)} simply-typed $\lambda$-calculus. {\em
  Fully complete}~\cite{AJ94} models are those that equate precisely
those proofs considered equivalent by the proof theory, and which
exclusively contain morphisms that are the interpretation of some
proof.

\p The best understood fragment of linear logic is that of {\em
  unit-free Multiplicative Linear Logic},~\mll. To model that one
requires a $*$-autonomous category~\cite{Bar79}, but not all such
categories satisfy the desired full completeness property. \hugh{For example, compact closed categories~\cite{KL80}, which are thought of as \emph{degenerate} models of \mll, do not satisfy full completeness for the logic. The Chu construction~\cite{Chu79} creates $*$-autonomous categories which generally are not fully complete. Pre-existing studies of
`good' models in this stronger sense are~\cite{HO93, Loa94a, DHPP99}.}

\p In~\cite{Tan97} the {\em double glueing}\/ construction is
introduced (see also~\cite{HS03} for a general account), and it is
suggested that fully complete models may be obtained when this is
applied to three particular compact closed categories. The three
categories in question are the category $\mathbf{Rel}$ of sets and
relations, the category $\mathbf{FDVec}_{\mathbb{F}}$ of finite
dimensional vector spaces over an arbitrary field $\mathbb{F}$ of
characteristic 0, and the category of Conway games and history-free
strategies. However, the proof of the second---arguably the most
interesting case---is not completed in the cited work; and the
restriction regarding the characteristic of the field turns out to be
unnecessary. Furthermore no two of the three proofs lend themselves
to a common unification.

\p In this paper, a greatly expanded version of the extended abstract~\cite{SS12}, we provide an entirely new approach to proving \hughh{full completeness which can be applied to a large variety of models provided by {\em tensor-generated compact closed categories with finite biproducts} to which the double glueing construction has been applied.} 
In the process we develop a `tensor calculus' for such categories,
and discuss its combinatorial consequences. The full completeness
proofs consist of algorithms which calculate the required
proof-theoretic structure for a given natural transformation. As a
consequence using the tensor calculus has a very algorithmic flavour, and there is certainly interesting future work to be done to connect this with other such work, for example in game semantics. \hughh{Both $\mathbf{Rel}$ and $\mathbf{FDVec}_{\mathbb{F}}$ belong to this collection of compact closed categories, and so the result is indeed a generalisation of Tan's work.}

\p For the sake of self-containment, we start in Section~\ref{BackgroundSection} by offering some background information relevant to the theorem being proved. This includes a short introduction to unit-free multiplicative linear logic and its proof theory, as well as its categorical models and the double glueing construction which can be placed over them. We then provide a formal description of the `tensor calculus' and its validity for the resulting categories within Section~\ref{ArrowDecomp}. It is within Section~\ref{SectionGCFC} that we provide the proof of our main result, Theorem~\ref{fcomp}, which
says that all \hughg{compact closed categories with finite
biproducts satisfying an extremely weak version of full completeness give rise to fully complete \mll\ models via double
glueing}. Using a mixture of results from earlier sections and
from~\cite{Tan97}, we finish by demonstrating in Section~\ref{SectionMLLMixFC} that the same double glueing construction under a \emph{slack focused orthogonality}~\cite{HS03} can produce fully complete models of \mllmix, the multiplicative fragment of linear logic with the `Mix' rule (Theorem~\ref{mixfcomp}).

\p The results in this paper are taken from the thesis of the second
author~\cite{Ste13}. The thesis, to which we occasionally refer, offers further background and discusses
some issues in more detail than is possible here.

\section{Background} \label{BackgroundSection}

In this section we give a short account of the well-known proof theory of \mll\ \cite{DR89} and how this system can be modelled categorically~\cite{BFSS90,Blu93}. We repeat the
formal notion of full completeness for such models, and indicate what can be said about the models provided by compact closed categories. An introduction to double glueing and orthogonalities is also contained within this background survey.

\subsection{Cut-Free \mll~Proof Nets}

The linear logic fragment \mll\ possesses a beautiful proof theory
revolving around the notion of proof nets~\hughd{\cite{Mel06,Str06}}. Proof nets provide a
method of equating two distinct derivations which differ only due to
`bureaucracy'~\cite{Gir87}. Since derivations in \mll\ can be
normalised confluently, we are interested primarily in cut-free
\mll~proof nets. It is sufficient to consider right-sided sequents of
formulae built from literals (which in a derivation are created as
pairs, one positive, one negative) using the multiplicative
conjunction $\otensor$ and disjunction~$\invamp$. Given a deduction
in the system we can construct a graph by using the parse trees of
the final sequent, connecting those literals that are created
together in said proof via edges known as \emph{(axiom) links}. This
is the {\em proof net}\/ that corresponds to the derivation, and we
wish to equate those deductions that have the same proof net.

\subsubsection{Correctness Criteria}

While every proof net represents a derivation, it is possible to
create graphs which resemble proof nets but do not correspond to
correct derivations. Given a parse tree for a sequent constructed
from literals and the \mll\ connectives $\otensor$ and $\invamp$ we
use the term {\em proof structure} for a graph resulting from
connecting matching literals.

\hughh{In this paper, proof structures are written only as the sequent together with its set of axiom links connecting appropriate literals above it. An example is given below.}
\begin{center} \vspace{5mm}
\begin{tikzpicture}
  [auto, node
  distance=6mm, skip loop/.style={to path={-- ++(0,#1) -| (\tikztotarget)}}] \tikzstyle{every node} = [text depth=-5pt,text height=0.5ex]
   \node (1) {$((\,\alpha\phantom{''}$}; \node (a) [right of=1] {$\tens$} ;
  \node (2) [right of=a] {$\alpha^{\bot})$}; \node (b) [right of=2] {$\tens$} ;
  \node (3)  [right of=b, xshift=-1mm] {$\phantom{'}\alpha\phantom{'})$}; \node (c)  [right of=3] {$\invamp$};
  \node (4)  [right of=c] {$(\alpha^{\bot}$}; \node(d) [right of=4] {$\tens$} ;
  \node (5) [right of=d, xshift=-1mm] {$\phantom{'}\alpha\phantom{'})$}; \node (e) [right of=5] {$\invamp$} ;
  \node (6) [right of=e] {$((\,\alpha^{\bot}$}; \node (f) [right of=6] {$\invamp$} ;
  \node (7)  [right of=f,xshift=-1mm] {$\phantom{'}\alpha)\phantom{'}$}; \node (g)  [right of=7] {$\tens$};
  \node (8)  [right of=g] {$\alpha^{\bot})$};
\begin{scope}
\path   (1)  edge [black, skip loop =6mm, shorten >=3mm, shorten <=3mm]  (2);
\path (3)  edge  [black, skip loop =6mm, shorten >=3mm, shorten <=3mm] (4);
\path   (5)  edge [black, skip loop =6mm, shorten >=3mm, shorten <=3mm]  (6);
\path (7)  edge  [black, skip loop =6mm, shorten >=3mm, shorten <=3mm] (8);
\end{scope}
\end{tikzpicture} \vspace{5mm}
\end{center}

\p It is possible to check whether a given proof structure is, in fact,
a proof net~\cite{DR89} using certain \emph{correctness criteria}. A
\emph{switching} of a proof structure is a subgraph created by
removing exactly one of the two argument edges of each
$\invamp$-vertex. A proof structure is a proof net if and only if every one of its
switchings is both {\em acyclic}\/ and {\em connected}.

\subsubsection{MDNF~Proof Structures} \label{MDNFProofStructures}

Multiplicative linear logic does not possess all the distributivity laws associated with Boolean logic. However, there are \hugh{\emph{weak distributivity} laws}\footnote{\hugh{These are sometimes known as linear distributivity laws.}} which have the effect of ``weakening'' formulae into a state closer to (and to closure into) a disjunctive form.

\begin{prooftree}
\AxiomC{$A \,\otimes\, (B \,\invamp\, C)$}
\LeftLabel{\scriptsize($w^{LL}$)}
\UnaryInfC{$(A \,\otimes\, B) \,\invamp\, C$}
\end{prooftree}
\begin{prooftree}
\AxiomC{$A \,\otimes\, (B \,\invamp\, C)$}
\LeftLabel{\scriptsize($w^{LR}$)}
\UnaryInfC{$(A \,\otimes\, C) \,\invamp\, B$}
\end{prooftree}
\begin{prooftree}
\AxiomC{$(A \,\invamp\, B) \,\otimes\, C)$}
\LeftLabel{\scriptsize($w^{RL}$)}
\UnaryInfC{$B \,\invamp\, (A \,\otimes\, C)$}
\end{prooftree}
\begin{prooftree}
\AxiomC{$(A \,\invamp\, B) \,\otimes\, C)$}
\LeftLabel{\scriptsize($w^{RR}$)}
\UnaryInfC{$A \,\invamp\, (B \,\otimes\, C)$}
\end{prooftree}

\hugh{From now on, in the appropriate contexts, we use the following notation: $[M,N] \,=\, \{n \,\in\, \mathbb{N} \,:\, M\,\leq\,n\,\leq\,N\}$, and $[N] \,=\, [1,N]$.}
%

\begin{defi}
An \mll~formula $A$ is in \emph{multiplicative disjunctive normal form} (or~\emph{MDNF}) if \hughd{$A = \binvamp_{m=1}^{M}(\botimes_{l=1}^{L_{m}}(\alpha_{f(m,l)})^{\phi(m,l)})$ for literals $\{\alpha_{i}:i \in [N]\}$ for some $N \in \N$, $M,L_{1},\ldots,L_{M} \in \N$, and functions $f: \sum_{m=1}^{M} \{(m,l):l \in [L_{m}]\} \longrightarrow [N]$ and $\phi: \sum_{m=1}^{M} \{(m,l):l \in [L_{m}]\} \longrightarrow \{\hughf{\epsilon},\bot\}$ indicating the literal and polarity of said literal respectively, \hughf{where $\epsilon$ denotes an empty superscript, and therefore positivity}}. Each subformula \hugh{$\botimes_{l=1}^{L_{m}}(\alpha_{f(m,l)})^{\phi(m,l)}$} of $A$ for a given $m$ is called a \emph{block} of the formula. An \mll~sequent is considered to be in \emph{MDNF} if all its constituent formulae are in MDNF; and its blocks are the blocks of its formulae.
\end{defi}

The correctness criteria for proof structures over \mll~sequents which are in MDNF are even further simplified. 

\begin{fact} \label{FactMDNFCorrectness} \hugh{\cite{Ste13}}
  An MDNF proof structure is a proof net for \mll\ if and only if its maximal
  $\invamp$-free subgraph is a tree.
\end{fact}

\hughh{The simplicity of these MDNF proof structures and their correctness criteria is very useful to us when proving full completeness in a category. As is seen in Sections~\ref{SectionGCFC} and~\ref{SectionMLLMixFC}, seemingly weaker full completeness theorems concerning only sequents of this form can be shown to be equivalent to theorems dealing with all sequents.}

\p \hughh{Each block of literals in a MDNF~structure written as a sequent with axiom links can be seen as one large vertex without affecting the acyclicity and connectedness of graph (the switchings of their parse trees are indeed still trees), and the instances of `$\invamp$' can be ignored by Fact~\ref{FactMDNFCorrectness}. From this perspective, we can check this graph for acyclicity and connectedness very swiftly. For example, the MDNF proof structure below is clearly incorrect due to the cycle between the two blocks.}

\begin{center} \vspace{5mm}
\begin{tikzpicture}
  [auto, node
  distance=5mm, skip loop/.style={to path={-- ++(0,#1) -| (\tikztotarget)}}] \tikzstyle{every node} = [text depth=-5pt,text height=0.5ex]
   \node (1) {$(\phantom{'}\alpha\phantom{'}$}; \node (a) [right of=1] {$\tens$} ;
  \node (2) [right of=a] {$\alpha^{\bot})$}; \node (b) [right of=2] {$\invamp$} ;
  \node (3)  [right of=b] {$(\phantom{'}\alpha\phantom{'}$}; \node (c)  [right of=3] {$\tens$};
 \node (4)  [right of=c] {$\alpha^{\bot})$};
\begin{scope}
\path   (2)  edge [black, skip loop =6mm, shorten >=3mm, shorten <=3mm]  (3);
\path (1)  edge  [black, skip loop =7mm, shorten >=3mm, shorten <=3mm] (4);
\end{scope}
\end{tikzpicture} \vspace{2mm}
\end{center}

\subsubsection{The `Mix' Rule}

Although not a formal part of linear logic, the `Mix' rule is routinely seen as a part of the multiplicative structure. This is partly due to its insistence to being represented in many standard models, not least the category of coherence spaces~\cite{Gir87}. The proof theory of \mll\ is however not unduly made too difficult by the addition of this derivation rule. Thanks to \cite{FR94} we know that the only true difference which can occur between correct proof structures for \mll\ and \mllmix\ regards the disconnectedness of switchings.


\begin{fact} \hugh{\cite{FR94}}
An \mll~proof structure describes a proof net for \mllmix\ if and only if all its switchings are acyclic.
\end{fact}

\p It is also possible to create a `Mix' version of the MDNF criterion of the previous section:
\begin{fact} \hugh{\cite{FR94,Ste13}}
An MDNF proof structure is a proof net for \mllmix\ if and only if its maximal
  $\invamp$-free subgraph is a forest.
\end{fact}

\subsection{Sound Categorical Models} \label{CatModels}

Sound categorical models of \mll\ are found in the form of
$*$-autonomous categories~\cite{Bar79,See89}---these are symmetric monoidal
categories with a well-behaved self-duality. The underlying idea is
quite simple: Each symbol in the logic is interpreted by a functor
(of the appropriate arity) on the category; the monoidal structure
$\otensor$ gives conjunction, the duality $(-)^{\bot}$ allows negation,
and to model disjunction these can be combined to define a De Morgan
dual $-\,\invamp\,- = ((-)^{\bot}\otensor (-)^{\bot})^{\bot}$. We use the
latter functor also to interpret the commas separating formulae in a
sequent.

\p Hence every sequent in \mll\ determines a functor
\[\fromto{\C^{N} \times (\C^{op})^{N}}{\C}\] where $N \in \N^{+}$. \hugh{That is to say, each sequent is described by a multivariant endofunctor with $N$ co- and contravariant arguments respectively.} If
we look at these functors then we see that they are built by
\begin{itemize} 
\item applying the duality functor $(-)^{\bot}$ to each
  copy of~$\C^{op}$,
\item creating as many copies of the arguments as
  required, then reordering them appropriately,
\item applying the functors $\otensor$ and $\invamp$ to
  get a result in~$\C$.
\end{itemize}
We refer to the functors that can be built in this way as {\em \mll~functors}. Similarly, functors that correspond to MDNF~sequents are
referred to as {\em MDNF~functors}.
  
\p Assume we have a right-sided sequent interpreted by the \mll\
functor~$F$. The formula representing truth is interpreted by the functor whose
value is the constant $\I$, the unit for the monoidal structure. We
refer to this functor as $\K$, and allow ourselves to adjust its
source as needed. Every proof of the given sequent is interpreted by
a {\em dinatural transformation} from $\K$ (with the same source
as~$F$) to~$F$, which is a family of arrows $(\tau_{\mathbf{R}} \in
\C[\I,F(\mathbf{R},\mathbf{R})])_{\mathbf{R} \in \C^{N}}$ for which
the diagram below commutes for every $\fctl{\mathbf{f} =
  (f_{1},\ldots,f_{N})}{\mathbf{R}}{\mathbf{S}}$, where
$\mathbf{R},\mathbf{S} \in \C^{N}$.
\begin{displaymath}
\begin{diagram}[labelstyle=\scriptstyle,objectstyle=\textstyle,h=5ex]
  && F(\mathbf{R},\mathbf{R}) &&\\
  & \ruTo^{\tau_{\mathbf{R}}}&& \rdTo^{F(\mathbf{f},1_{\mathbf{R}})}	&\\
  \I &&&& F(\mathbf{S},\mathbf{R}) \\
  & \rdTo_{\tau_{\mathbf{S}}} && \ruTo_{F(1_{\mathbf{S}},\mathbf{f})} &\\
  & & F(\mathbf{S},\mathbf{S}) & &
\end{diagram}
\end{displaymath}

\begin{defi}
An {\em \mll} (respectively {\em MDNF}) {\em transformation}\/ is a dinatural transformation to an \mll\ (respectively MDNF)
functor from a constant functor $\K$ of appropriate source.
\end{defi}

It is possible to build \mllmixmaybe\ transformations corresponding to correct one-sided sequent derivations of the logic in an\label{SoundnessTechnique} inductive manner~\cite{Ste13} in any $*$-autonomous category. Furthermore, it can be shown that any two derivations which reduce to the same cut-free proof net are represented by the same \mll\ transformation in the category~\cite{LS06}. This suggests that $*$-autonomous categories are a sensible collection of models through which one can investigate \mll.

\p Dinatural transformations do not naturally compose with one another. However, they are capable of being composed with transformations natural in all components. The diagram below demonstrates the dinatural behaviour a composition of natural~$\mu$ and dinatural~$\tau$.

\label{DiNatsComposeStatement}
\begin{displaymath}
\begin{diagram}[labelstyle=\scriptstyle,objectstyle=\textstyle,h=5ex]
  && F(\mathbf{R},\mathbf{R}) & \rTo^{\mu_{(\mathbf{R},\mathbf{R})}} & G(\mathbf{R},\mathbf{R}) && \\
  & \ruTo^{\tau_{\mathbf{R}}}&& \rdTo^{F(\mathbf{f},1_{\mathbf{R}})}	&& \rdTo^{G(\mathbf{f},1_{\mathbf{R}})} & \\
  \I &&&& F(\mathbf{S},\mathbf{R}) & \rTo^{\mu_{(\mathbf{S},\mathbf{R})}} & G(\mathbf{S},\mathbf{R}) \\
  & \rdTo_{\tau_{\mathbf{S}}} && \ruTo_{F(1_{\mathbf{S}},\mathbf{f})} && \ruTo_{G(1_{\mathbf{S}},\mathbf{f})} & \\
  & & F(\mathbf{S},\mathbf{S}) & \rTo_{\mu_{(\mathbf{S},\mathbf{S})}} & G(\mathbf{S},\mathbf{S}) &&
\end{diagram}
\end{displaymath}
The equivalent result for precompositions is demonstrated in a dual manner.

\p One can now show that two proofs of the same sequent are interpreted
by the same dinatural transformation if and only if they have the
same proof net~\cite{LS06}. In other words, this categorical interpretation of
proofs fits very well with the existing proof theory for~\mll.

\p With $*$-autonomous categories being sound categorical models of \mll, it is expected that they should all contain natural transformations which model the \hugh{weak distributivity} laws discussed in Section~\ref{MDNFProofStructures}.
\begin{eqnarray*}
w^{LL}	&	:	& -_{1} \otimes (-_{2} \invamp -_{3}) \longrightarrow (-_{1} \otimes -_{2}) \invamp -_{3} \\
w^{LR} 	&	:	&	-_{1} \otimes (-_{2} \invamp -_{3}) \longrightarrow (-_{1} \otimes -_{3}) \invamp -_{2} \\
w^{RL} 	&	:	&	(-_{1} \invamp -_{2}) \otimes -_{3} \longrightarrow -_{2} \invamp (-_{1} \otimes -_{3}) \\
w^{RR} 	&	:	&	(-_{1} \invamp -_{2}) \otimes -_{3} \longrightarrow -_{1} \invamp (-_{2} \otimes -_{3})
\end{eqnarray*}

It is shown how one can construct each of them in~\cite{Ste13}.

\p The `Mix' Rule is modelled in a $*$-autonomous category if and only if there is a natural transformation $\fctl{\Mix}{(-)_{1} \otimes (-)_{2}}{(-)_{1} \invamp (-)_{2}}$. This is equivalent to there existing a `Mix' morphism $\fctl{\mix}{\bot}{\I}$~\cite{Tan97}.

\subsection{Full Completeness}

Full Completeness was first defined in~\cite{AJ94}, and it is meant
to describe the tightest possible connection between the logic and
its model. Not only are the interpretations of two proofs the same if
and only if they are equivalent in the proof theory, but the model
does not contain any representations of `non-proofs'.\footnote{The name `full completeness' is derived from its original non-dinatural interpretation sense, in that the property establishes the existence of a full functor between the model category and a free $*$-autonomous category. With this analogy in mind, due to every pair of derivations sharing the same cut-free proof net corresponding to the same \mllmixmaybe\ transformation, the satisfaction of the above definition may be more accurately described as full \emph{and faithful} completeness. However, for the sake of convenience and easy comprehension, we keep to the originally given name.}

\p In the case of the work in this paper, we are considering what this means from a dinatural interpretation of proofs. This provides us with the following definition, which originates from the notion as it is set out in~\cite{BS96}, and is further used in such works as~\cite{Loa94b,Tan97,Hag00}:

\begin{defi}
  A $*$-autonomous category $\C$ satisfies \emph{\mllmixmaybe\ full
    completeness} if every \mllmixmaybe\ transformation in the
  category is the interpretation of a cut-free proof net.
\end{defi}


\subsection{Compact Closed Categories with Finite Biproducts} \label{SectionCompCCs}
 
Compact closed categories~\cite{KL80} are particular degenerate $*$-autonomous
categories which possess a parameterised adjunction $B \,\otimes\, - \,\vdash\, - \,\otimes\,B^{*}$ (we denote the negation function $(-)^{*}$ for historical reasons in these categories). The existence of this \hugh{adjunction} induces an invertible `Mix' natural transformation $\fctl{\Mix}{(-)_{1} \,\otimes\, (-)_{2}}{(-)_{1} \,\invamp\, (-)_{2}}$ between the two functors modelling the binary connectives. Nonetheless it turns out that they can form the basis for constructing fully complete models, as is seen in Sections~\ref{SectionGCFC} and~\ref{SectionMLLMixFC}.

\p \hughc{The adjunction associated with compact closed categories creates a bijective correspondence $\fctl{v}{\left[ (-)_{2} \otimes (-)_{1},(-)_{3} \right]}{\left[(-)_{1},(-)_{3} \otimes (-)_{2}^{\,\,*}\right]}$. This, along with the coherence properties of symmetric monoidal categories and the functoriality of tensor, ensure that the \mll~transformations can be reproduced in the following manner:}
\begin{enumerate} \label{CompCCDinatModelling}
\item Take the \hughc{right natural isomorphism} on each of the basis \mll~functors,
\begin{displaymath}
\fctl{\rho_{i}}{(-)_{i} \,\otimes\, \I}{(-)_{i}},
\end{displaymath}
with $\fctl{(-)_{i}}{\C^{N}}{\C}$ the projection of the $i^{th}$ component of the product category $\C^{N}$ for each $i\,\in\,[N]$.
\item Use the parameterised adjunction associated with these categories with the \hughc{right isomorphism} to produce dinatural transformations \hughc{$d_{i} \,=\, v_{\I,(-)_{i},(-)_{i}}(\rho_{(-)_{i}}) \,:\,\K \longrightarrow (-)_{i} \otimes (-)_{i}^{\,\,*}$ for each argument}\footnote{\hughc{This dinatural transformation is indeed the collection of arrows $(\fctl{d_{X}}{\I}{X \otimes X^{*}})_{X \in \C}$ first discussed in~\cite{KL80}}}.
\item If the sequent whose derivation is being modelled contains $n_{i}$ positive/negative occurrences of the literal being modelled with the $i^{th}$ entries of the functors for each $i \in [N]$, then create the \mll~transformation
\begin{displaymath}
\I \rTo{\lambda^{-1}} \bigotimes_{i=1}^{N}\bigotimes_{j=1}^{n_{i}} \I \rTo{\hughc{\bigotimes_{i=1}^{N}\bigotimes_{j=1}^{n_{i}}d_{i}}} \bigotimes_{i=1}^{N}\bigotimes_{j=1}^{n_{i}}(-_{i} \otimes (-_{i})^{*})
\end{displaymath}
\item Use the associativity and symmetry isomorphisms $\alpha$ and $\sigma$ to rearrange brackets and the ordering of the literals so that those literals are found in the same order and the brackets in the same place as the functor requires. If the $i^{th}$ and $j^{th}$ literals are supposed to be connected by an axiom link, then it must be ensured that the two literals found in those positions are two that were created simultaneously by the exact same `axiom link' dinatural transformation. \hughd{The resulting dinatural transformation can be thought of as a \emph{fixed-point-free involution}, relating literals to their negations.}
\item Use the natural transformation $\mathbf{Mix}$ liberally to change all $\otimes$ functors into $\invamp$ functors where necessary to shape the target functor to create the \mll~functor in the category.
\end{enumerate}

\p\noindent Unfortunately, this advantage comes with a natural drawback: the unrestricted freedom to generate representations of axiom links within the \mll\ transformations means that unsequentialisable proof structures can be modelled just as easily as genuine proof nets. As such, \mll~full completeness cannot be satisfied by any of these categories.

\p In this paper we are more interested in the case where there is even more structure than that given above,
namely where the compact closed category $\C$ has finite biproducts.
Recall~\cite{Hou08} that $\C$ has finite products if and only if it
has finite sums if and only if it has finite biproducts. In this case
the scalars (that is the homset of endomorphisms on the tensor unit $\C[\I,\I]$) form a commutative unital semiring, and due to the
biproducts distributing over the tensor product $\C$ is enriched over
$\mathbf{CSMod}_{\C[\I,\I]}$---the category of commutative
semimodules over the semiring of scalars~\cite{Heu08}.

\p \hugh{Given a set of dinatural transformations of the same type $\{\tau_{i} \,:\, i\,\in\, I\}$ for some index set $I$, we define a linear combination $\sum_{i}s_{i}\cdot\tau_{i}$ of them as the collection of arrows $\left( \sum_{i}s_{i}\cdot (\tau_{i})_{\mathbf{A}}\right) _{\mathbf{A} \in \C}$.} It is easy to show that in such a category every linear combination
of dinatural transformations is another such, giving \hugh{a second} obstacle
to full completeness. Nonetheless, from the above limitations, there is a clear concept of
a compact closed category with finite biproducts being `as fully complete as it can hope to be'.

\begin{defi} \label{FeebleFCDefn}
  A compact closed category $\C$ with finite biproducts satisfies
  \emph{feeble full completeness} if every \mll\ transformation
  for the category is a linear combination of interpretations of
  proof structures over the same sequent.
\end{defi}

\p It is known from~\cite{CHS01} (proof reproduced in~\cite{Ste13}) that finite biproductal compact
closed categories whose tensor unit acts as a separator satisfy
feeble full completeness. These models are called
\emph{tensor-generated}, and there are many of them. Examples include
both $\mathbf{Rel}$ and $\mathbf{FDVec}_{\mathbb{F}}$ for any field
$\mathbb{F}$, and extend beyond these to include categories such as
that of finite-dimensional Hilbert spaces. Every compact closed
category with finite biproducts clearly has a non-trivial full
subcategory which is tensor-generated, namely the category generated
by the object $\I$ and the tensor, biproduct and duality functors.

\subsection{The Double Glueing Construction} \label{SectionDGlueing}

Double Glueing constructions~\cite{HS03} operate upon categories. They can be thought of as adding structure to objects in the form of two arrows, which then has to be preserved by the morphisms of the newly created category. This leaves a trivial forgetful functor~$\fctl{U}{\D}{\C}$ from the double-glued category to the original. When the added structure arrows are monomorphisms, this has the effect of generating homsets between two objects in the new category which are fundamentally subsets of the homsets of their underlying objects.~\cite{Ste13} That is,
\begin{displaymath}
\left\{Uf \,:\, f \,\in\,\D[(A,\alpha,\xi),(B,\beta,\zeta)]\right\} \,\subseteq\, \C[A,B].
\end{displaymath}

\p The most commonly seen double glueing construction used on top of models of linear logic is the so-called Hyland-Tan construction~\cite{Tan97}, inspired by Loader's linear logic predicates~\cite{Loa94b}. This double glueing is based upon structure arrows which take the form of injections into homsets in~$\mathbf{Set}$, and so are clear monomorphisms, allowing us to use the above fact. We define this construction formally below.

\begin{defi} \label{GCDefn}
Given a $*$-autonomous category $\C$ \hugh{with tensor unit $\I$, and letting $\bot \,=\, \I^{\bot}$}, the category $\GC$ is the category described with the following object set and homsets:
\begin{itemize}
\item Objects --- $Obj(\GC) = \{(A,U,X):\,A \in Obj(\C),\,U \subseteq \C[\I,A],\,X \subseteq \C[A,\bot]\}$
\item Arrows --- Arrows in $\GC[(A,U,X),(B,V,Y)]$ are described by single morphisms $f \in \C[A,B]$ such that:
	\begin{eqnarray*}
	f \after U \subseteq V	&	\mbox{and}	&	Y \after f \subseteq X
	\end{eqnarray*}
\end{itemize}
\end{defi}

In general, when discussing a $\GC$-object $(A,U,X)$, we refer to $U$ and $X$ as the object's sets of \emph{values} and \emph{covalues} respectively. We write
\begin{align*}
(A,U,X)_{Val}	&=	U \\
(A,U,X)_{CoVal}	&=	X
\end{align*}

As with any double glueing, the Hyland-Tan construction preserves $*$-autonomy, and it also removes the degeneracy of compact closed categories. Tensor products and negations of arrows are immediately inherited from the underlying category --- indeed, the functor $\fctl{\huh}{\GC}{\C}$ is $*$-autonomous. The object assignments are as follows:
\begin{eqnarray*}
(A,U,X)^{\bot}					&	=	&	(A^{\bot},X^{\bot} \after \chi_{\I},U^{\bot}) \\
(A,U,X) \otimes (B,V,Y)	&	=	&	(A \otimes B,(U \otimes V) \after \lambda_{\I},Z) \\
\I											&	=	&	(\I,\{1_{\I}\},\C[\I,\bot])
\end{eqnarray*}
where $Z = \left\{z \in \C[A \otimes B,\bot]:\, \begin{array}{l} \forall v \in V,\, A \rTo{\rho_{A}} A \otimes \I \rTo{1_{A} \otimes v} A \otimes B \rTo{z} \bot \in X, \\
\forall u \in U,\,B \rTo{\lambda_{B}} \I \otimes B \rTo{u \otimes 1_{B}} A \otimes B \rTo{z} \bot \in Y \end{array} \right\}$.

\subsubsection{Focused Orthogonalities} \label{SectionFocusedOrth}

Many examples of double-glued structures which have been investigated, particularly ones in which their glued nature is hidden, have restrictions on the objects which they are allowed to contain. For example, the category of Chu Spaces~\cite{Chu79} can be thought of as a full subcategory of a generalised version of the category $\Rel$ under the influence of a double-glueing \cite{Hug04}. Of course, the constraints must be defined sensibly in order to preserve the closure of the operations defined on such categories. These constraints are called \emph{orthogonalities}, and they can come in a variety of forms. In this paper, we wish to look at \emph{focused}, \emph{slack} orthogonalities.

\p Given any subset $E \,\subseteq\, \C[\I,\bot]$, we can construct an orthogonality in which two morphisms $f \in \C[\I,A]$ and $g \in \C[A,\bot]$ are considered orthogonal if and only if they compose to form an arrow in the chosen set. This is called the orthogonality \emph{focused on} $E$, and spawns a \emph{focused glueing} $\gc{E}$ whose values and covalues must be mutually orthogonal.

\begin{defi}
Given a $*$-autonomous category $\C$ and a set $E \subseteq \C[\I,\bot]$, the category $\gc{E}$ is defined as the smallest full subcategory of $\GC$ containing every object $(A,U,X)$ such that
\begin{displaymath}
\forall\,u \in U,\,x \in X,\,\,\,x \after u \in E.
\end{displaymath}
\end{defi}

\p Of course, the tensor unit $\I$ of $\GC$ is not contained in any of the orthogonally glued categories for which the focus is a proper subset of $\C[\I,\bot]$\footnote{The categories $\gc{E}$ and $\GC$ are clearly the same when $E = \C[\I,\bot]$}, meaning that the $*$-autonomous structure of $\GC$ is not preserved \textit{per se}. However, each one of these subcategories has another object within it which acts as legitimate unit for the tensor functor: the unit for $\gc{E}$ is $\I_{E} = (\I,\{1_{\I}\},E)$.

\p With this fact taken into account, and the fact that the categories are closed under the tensor product and negation functors, we know that all of these constructions preserve {$*$-autonomy} (with respect to the new unit definitions). Thus, just like the original $\mathbf{G}$-glueing, the glueings generate models of \mll.

\subsubsection{Properties}

The Hyland-Tan double glueing construction and its focused orthogonalities are particularly well-behaved, and in fact possess some properties which are immediately relevant to the coming results.

\pagebreak Given an arbitrary focused glueing $E \,\subseteq\, \C[\I,\bot]$, the value and covalue sets of an object are in fact precisely the homsets from $\I_{E}$ and to $\bot \,=\, \I^{\bot}$.

\begin{fact}{\cite{Ste13}} \label{CoValueArrowProp}
For every object $R \in Obj(\gc{E})$, $R_{Val} = \GC[\I_{E},R]$ and $R_{CoVal} = \GC[R,\bot_{E}]$, where $\bot_{E} \,=\, (\I_{E})^{\bot}$.
\end{fact}



We can also claim a large understanding of the $MLL^{-}$~transformations for these double-glued categories. Since homsets in these categories are in some way `stripped' versions of those from their underlying categories, it would be reasonable to assume that an analogous statement could be made about dinatural transformations. This is indeed correct, as can be deduced from the below proposition.

\begin{prop}{\cite{Tan97,Ste13}} \label{PropGPreservesDNTs}
Let $\tau:F \longrightarrow G$ be a dinatural transformation in $\gc{E}$. Then there is a dinatural transformation $\tilde{\tau}:\huh F \longrightarrow \huh G$ in $\C$ which defines $\tau$; that is, $\huh\tau_{\mathbf{R}} = \tilde{\tau}_{\mathbf{\huh\mathbf{R}}}$ for every $\mathbf{R} = (R_{1},\ldots,R_{N})$. 
\end{prop}
\proof
\hughc{Theorem~1.3.2 of~\cite{Tan97} provides the result for $E \,=\, \C[\I,\bot]$. As remarked on page~119 in~\cite{Ste13}, this proof only requires intermediary objects which are found in all categories of the given form, thus it extends to all subsets $E$.} \qed

The consequences of this proposition are marked. The transformations of feebly fully complete compact closed categories with biproducts take the form of linear combinations of fixed-point-free involutions~\cite{CHS01}. Therefore the same morphisms are used to described the transformations in the categories created by applying the glueing construction to them.

\p Finally, given a compact closed category $\C$ with finite biproducts, which is assured a morphism $\iota^{-1} = \rho_{\bot} \after v(\lambda_{\I}) \in \C[\bot,\I]$ and a separate zero morphism $0_{\bot,\I}$ in the homset $\C[\bot,\I]$, we note that every focused glueing $\mathbf{G}_{E}$ where $\{\iota\} \subsetneq \hugh{E} \subseteq \C[\I,\bot]$ stops this morphism being found in $\gc{E}[\bot_{E},\I_{E}]$, but $\gc{\{\iota\}}$ preserves this modelling of the mix rule.
\begin{fact}
For every compact closed category $\C$ with finite biproducts and a set $\hugh{E} \subseteq \C[\I,\bot]$ containing $\iota$, the category $\hugh{\gc{E}}$ models the mix rule if and only if $E = \{\iota\}$. (See~\cite{Tan97})
\end{fact}

For shorthand, we say $\gc{\{\iota\}} \,=\, \gc{1}$. The category is the subject of Section~\ref{SectionMLLMixFC}.

\section{Arrow Decomposition} \label{ArrowDecomp}

Every finite-dimensional vector space over a field $\F$ can be given a finite basis, meaning that all arrows in their category $\FDVec$ can take the shape of a matrix, or in fact a tensor if desired, over the underlying field. Although it is not possible to say that all arrows in a given compact closed category with finite biproducts can be reduced to an array-based form over a single input type, the multilinear representations still appear and can certainly be of use. \hughd{We introduce this generalisation of matrix representations of arrows\footnote{\hughf{A concept found in folklore, and briefly explained in~\cite{Hou08}.}} in the coming section, and show that \hughh{\mll~transformations and the calculations required in deducing sets of values and covalues in double-glued objects can take a simplified form when using this notation.}}

\begin{defi}
An \emph{array} with \emph{index set} $I$ over a set $X$ is a function $f:I \longrightarrow X$. An array can be considered to be \emph{$N$-dimensional} if its index set takes the form $\prod_{i=1}^{N}I_{i}$ for some $I_{1},\ldots,I_{N}$.
\end{defi}

\begin{defi}
Letting $n_{1},\ldots,n_{N} \in \N^{+}$, an \emph{$\prod_{k=1}^{N}n_{k}$-tensor} $t$ over a semiring $S$ is an $N$-dimensional array, each of whose components contain a value from $S$. The $(i_{1},\ldots,i_{N})^{th}$ entry is written $t_{i_{1}, \cdots, i_{N}}$.
\end{defi}

Indices of tensors are allowed to be separated by commas and semicolons to demarcate relevant groups of indices. Superscripts may also be used to facilitate writing, though once a notation style is chosen for a specific tensor it must be adhered to. The number of indices a tensor requires to be expressed (in this case $N$) is called its \emph{order}. Often tensors are written with general index variables to emphasise that they are indeed multidimensional arrays. Furthermore, long sets of indices can be replaced by bold `superindices'. For example, $t_{i_{1}, \cdots, i_{M}, j_{1},\cdots,j_{N}}$ can be rewritten $t_{\mathbf{i}\mathbf{j}}$ for shorthand, where $\mathbf{i} = (i_{1},\ldots,i_{M})$ and $\mathbf{j} = (j_{1},\ldots,j_{N})$. \hughf{\label{OverloadRef} Sometimes we overload the notation of the tensor so as to show the reader information concerning relative positions of indices with respect to other tensors or a set position.}

\p Calculations involving tensors over semirings are unsurprisingly done in an identical manner to those involving tensors over fields in standard multilinear algebra. All the differences which may occur concern how its entries sum and multiply together. The standard algebraic manipulations are found below.

\begin{itemize}
\item (Addition) --- Given two tensors $t$ and $u$ over the same index sets, their sum is clearly found in the same semimodule.
\begin{displaymath}
(t+u)_{i_{1}, \cdots, i_{M}} = t_{i_{1}, \cdots, i_{M}} + u_{i_{1}, \cdots, i_{M}}.
\end{displaymath}
\item (Composition) --- \hughf{Let $t$ and $u$ be $(M+N)$- and $(N+P)$-order tensors respectively with $N$ index positions ranging over the same index sets in both arrays.} The composition of the two is given by an $(M+P)$-order tensor for which each component is found to be as follows:
\begin{displaymath}
(tu)_{i_{1}, \cdots, i_{M};k_{1},\cdots,k_{P}} = \sum_{j_{1},\cdots,j_{N}}\left(t_{i_{1}, \cdots, i_{M};j_{1}, \cdots, j_{N}} \cdot u_{j_{1}, \cdots, j_{N};k_{1},\cdots,k_{P}}\right).
\end{displaymath}
\item (Product) --- The outer product of two tensors $t_{i_{1},\cdots,i_{M}}$ and $u_{j_{1},\cdots,j_{N}}$ is given by an $(M+N)$-tensor for which each entry is merely a product of entries from its factors.
\begin{displaymath}
(t \otimes u)_{i_{1},\cdots,i_{M};j_{1},\cdots,j_{N}} = t_{i_{1},\cdots,i_{M}} \cdot u_{j_{1},\cdots,j_{N}}.
\end{displaymath}
\end{itemize}

\noindent Examples of the first three manipulation techniques from above are seen ubiquitously in the forms of addition, multiplication and the trace operation of matrices. It should be noted that, if desired, one can consider tensor composition as the equivalent of finding the outer product of two tensors sharing indices and contracting the result.

\p There are \hugh{five types of tensor} which are seen continuously in various guises within this \hughf{paper}. Their definitions are given below.
\begin{defi}\label{TensorExamDefns} $\,$ \\
\begin{enumerate}
\item A \emph{zero tensor}, written $0_{\mathbf{i}}$ for some indices $\mathbf{i}$, is a tensor, all of whose entries contain the scalar $0$.
\item Similarly, a \emph{one tensor}, written $1_{\mathbf{i}}$, is a tensor whose entries all contain the scalar $1$.
\item A \emph{Kronecker delta}, written $\delta$, is an $(N \times N)$-tensor for any $N \in \mathbb{N}$ such that
\begin{displaymath}
\delta_{ij} = \left\{ \begin{array}{ll}
1 &	\mbox{ if }i=j \\
0 & \mbox{ otherwise}
\end{array} .\right.
\end{displaymath}
It is the tensor representation of the identity matrix. \\ \hugh{We also write $\tensor*{\delta}{*^{j_{1}}_{i_{1}}^{\;\cdots\;}_{\;\cdots}^{j_{N}}_{i_{N}}} \,=\, \bigotimes_{k=1}^{N}\delta_{i_{k}j_{k}}$ as shorthand}\footnote{\hughf{Note that these tensors are an example of where it is of use to overload the index notation as discussed on page~\pageref{OverloadRef}. It facilitates the understanding of which indices are connected by such a relation without having to resort to a yet more cumbersome notation.}}.
\item \hugh{A \emph{(full)} $M$\emph{-permutation over }$[n]$ is an
  $n^{M}$-tensor $p_{i_{1} \cdots i_{M}}$ such that for all $k \in
  [M]$, and $x_{l} \in [n]$ for each $l \neq k$, there exists an
  $x_{k} \in [n]$ so that $p_{i_{1} \cdots i_{M}} \cdot \prod_{l \neq
    k}\delta_{i_{l}x_{l}} = \delta_{i_{k}x_{k}}$. We use the notation $\Perm(M,n)$ to
  denote the set of all {$M$-permutations} over $[n]$. Note that a $1$-permutation over $[n]$ is simply a tensor of the form $\delta_{ix}$ for some $x \in [n]$.}
  
  \p \hugh{Of particular use are \emph{cycle permutation tensors}. We define $\cycle(L,n,r)$, with $L,n,r \,\in\, \mathbb{N}^{+}$, to be the $n^{L}$-tensor with entries defined as follows:
  \begin{displaymath}
  \cycle(L,n,r)_{i_{1} \cdots i_{L}} = \left\{ \begin{array}{cc} 1	&	\mbox{ if } \textstyle\sum_{j=1}^{L}i_{j} \equiv r \mbox{ mod } n \\
  																																0	&	\mbox{ otherwise}\end{array} \right. .
  \end{displaymath}}
\item \hugh{Generalising the above, a \emph{partial} $M$\emph{-permutation over }$[n]$ is an
  $n^{M}$-tensor $p_{i_{1} \cdots i_{M}}$ such that for all $k \in
  [M]$, and $x_{l} \in [n]$ for each $l \neq k$, there exists an
  $x_{k} \in [n]$ so that $p_{i_{1} \cdots i_{M}} \cdot \prod_{l \neq
    k}\delta_{i_{l}x_{l}}$ is equal to $\delta_{i_{k}x_{k}}$ or $0_{i_{k}}$. We use the notation $\PPerm(M,n)$ to
  denote the set of all $M$-permutations over $[n]$.}
\end{enumerate}
\end{defi}

\noindent Because of the number of indices which may be given to each tensor, the calculus of tensors can become very cluttered. The use of summation symbols adds to the excessive number of symbols in many expressions, and they can be thought of in many cases as unnecessary. As such, at some points it is useful to use the Einstein summation convention: whenever an expression containing tensors has two index positions involved within it which are summed together over the same index $j$ say, the summation symbol may be omitted without worry. This means that a simple composition rule, for example, can be rewritten $(tu)_{\mathbf{ik}} = t_{\mathbf{ij}}u_{\mathbf{jk}}$.

\subsection{Tensor Representation} \label{SectionTensorRep}

Decomposition of arrows between objects in the form of tensor products of direct sums is possible in a symmetric monoidal category with finite biproducts, and can be a very useful tool. Suppose that $\C$ is such a category, and consider the arbitrary arrow
\begin{displaymath}
\fctl{f}{\bigotimes_{l=1}^{N} \bigoplus_{j_{l}=1}^{n_{l}}A_{l,j_{l}}}{\bigotimes_{k=1}^{M} \bigoplus_{i_{k}=1}^{m_{k}}B_{k,i_{k}}}.
\end{displaymath}
Due to the preservation of products by the tensor product\footnote{\hughd{Since there is a natural isomorphism $1_{C} \cong (-)^{**}$ in any compact closed category, the functor $\fctl{B \otimes -}{\C}{\C}$ in a compact closed category for any object $B$ can be seen as both a left and right adjoint, and as such preserves both products and coproducts~\cite{Mac97}.}}, we know that the following set of $m_{1}$ arrows describe $f$.
\begin{displaymath}
\left\{\left(\pi_{i_{1}} \otimes \bigotimes^{M}_{k=2}1_{\bigoplus^{m_{k}}_{i=1}B_{k,i}}\right) \after f : i_{1} \in [m_{1}]\right\}
\end{displaymath}
We can follow the same procedure of decomposing $f$ using the projections \linebreak $1_{\bigoplus^{m_{1}}_{i=1}B_{1,i}} \otimes \cdots \otimes \pi_{i_{k}} \otimes \cdots \otimes 1_{\bigoplus^{m_{M}}_{i=1}B_{1,i}}$ for all $k$ of the object. Furthermore, due to the bifunctoriality of the tensor product, each of the projections is independent of the others. The injections operate similarly; and as such, it is possible to separate $f$ yet further and describe it using a set of $\prod^{M}_{k=1}m_{k} \times \prod^{N}_{l=1}n_{l}$ arrows $\{f_{i_{1}, \ldots ,i_{M};j_{1}, \ldots ,j_{N}} : \forall k,l\,\,
i_{k} \in [m_{k}],\, j_{l} \in [n_{l}]\}$, where
\begin{displaymath}
  f_{i_{1}, \ldots ,i_{M};j_{1}, \ldots ,j_{N}} = {\bigotimes_{k=1}^{M} \pi_{i_{k}}} \after f \after {\bigotimes_{l=1}^{N} \inj_{j_{l}}}.
\end{displaymath}

This decomposition of $f$ is known as a \emph{tensor representation}. It is generally inconvenient to write down the resultant tensor as a single entity due to the number of dimensions required. This is, however, of little concern to us, because we are able to view the entries of the tensor on a case-by-case basis.

\p The greatest implication this has is to the freedom of expression one has when decomposing arrows concerning the tensor product: given two arrows in the prescribed form above, it is possible to gain a tensor representation of their tensor product. This comes about almost trivially, with each entry in the new tensor being created by producing the tensor product of one entry from each of the representations of the more primitive morphisms. Given arbitrary arrows $f \in \C[A,B]$ and $g \in \C[C,D]$, with $A = \bigotimes_{a=1}^{N_{A}} \bigoplus_{k_{a}=1}^{n_{A,a}}A_{a,j_{a}}$, $B = \bigotimes_{b=1}^{N_{B}} \bigoplus_{j_{b}=1}^{n_{B,b}}B_{l,j_{b}}$, $C~=~\bigotimes_{c=1}^{N_{C}} \bigoplus_{k_{c}=1}^{n_{C,c}}C_{c,i_{c}}$ and $D = \bigotimes_{d=1}^{N_{D}} \bigoplus_{j_{d}=1}^{n_{D,d}}D_{d,i_{d}}$, we find that
\begin{eqnarray*}
(f \otimes g)_{i_{1},\cdots,i_{N_{C}+N_{D}};j_{1},\cdots,j_{N_{A}+N_{B}}} & = & \bigotimes_{l=1}^{N_{C}+N_{D}}\pi_{i_{l}} \after (f \otimes g) \after \bigotimes_{l=1}^{N_{A}+N_{B}}\inj_{j_{l}} \\
& = & \left(\bigotimes_{l=1}^{N_{C}}\pi_{i_{l}} \after f \after \bigotimes_{l=1}^{N_{A}}\inj_{j_{l}}\right) \otimes \left(\bigotimes_{l=N_{C}+1}^{N_{C}+N_{D}}\pi_{i_{l}} \after g \after \bigotimes_{l=N_{A}+1}^{N_{A}+N_{B}}\inj_{j_{l}}\right) \\
& = & f_{i_{1},\cdots,i_{N_{C}};j_{1},\cdots,j_{N_{A}}} \otimes g_{i_{N_{C}+1},\cdots,i_{N_{C}+N_{D}};j_{N_{A}+1},\cdots,j_{N_{A}+N_{B}}}
\end{eqnarray*}

\noindent Rather unsurprisingly, addition and composition operate in an almost identical manner as they do in the matrix algebra. Identity and zero morphisms between objects of the form given above are represented by Kronecker deltas and zero tensors respectively. Letting $A = \bigotimes_{l=1}^{N_{A}} \bigoplus_{k_{l}=1}^{n_{A,l}}A_{l,j_{l}}$, $B = \bigotimes_{l=1}^{N_{B}} \bigoplus_{j_{l}=1}^{n_{B,l}}B_{l,i_{l}}$, and $f,g~\in~\C[A,B]$,

\begin{eqnarray*}
(0_{A,B})_{j_{1}, \ldots ,j_{N_{B}};k_{1}, \ldots ,k_{N_{A}}} 	& = & \bigotimes_{l=1}^{N_{B}}\pi_{j_{l}} \after 0_{A,B} \after \bigotimes_{l=1}^{N_{A}}\inj_{k_{l}} \\
								&	=	&	0_{\bigotimes_{l=1}^{N_{A}}A_{l,k_{l}},\bigotimes_{l=1}^{N_{B}}B_{l,j_{l}}}
\end{eqnarray*}

\begin{eqnarray*}
(f+g)_{i_{1},\ldots,i_{N_{B}};j_{1},\ldots,j_{N_{A}}}			&	=	&	\bigotimes_{l=1}^{N_{B}}\pi_{i_{l}} \after (f+g) \after \bigotimes_{l=1}^{N_{A}}\inj_{j_{l}} \\
								& = & \left(\bigotimes_{l=1}^{N_{B}}\pi_{i_{l}} \after f \after \bigotimes_{l=1}^{N_{A}}\inj_{j_{l}}\right) + \left(\bigotimes_{l=1}^{N_{B}}\pi_{i_{l}} \after g \after \bigotimes_{l=1}^{N_{A}}\inj_{j_{l}}\right) \\
								&	=	&	f_{i_{1},\ldots,i_{N_{B}};j_{1},\ldots,j_{N_{A}}} + g_{i_{1},\ldots,i_{N_{B}};j_{1},\ldots,j_{N_{A}}}
\end{eqnarray*}

\begin{eqnarray*}
(1_{A})_{i_{1}, \ldots , i_{N_{A}}; j_{1}, \ldots , j_{N_{A}}} & = & \bigotimes_{l=1}^{N_{A}}\pi_{i_{l}} \after 1_{A} \after \bigotimes_{l=1}^{N_{A}}\inj_{j_{l}} \\
		& = & \bigotimes_{l=1}^{N_{A}}\pi_{i_{l}} \after \bigotimes_{l=1}^{N_{A}}1_{\bigoplus_{k=1}^{n_{A,l}}A_{l,k}} \after \bigotimes_{l=1}^{N_{A}}\inj_{j_{l}} \\
		& = & \bigotimes_{l=1}^{N_{A}} \left(\pi_{i_{l}} \after 1_{\bigoplus_{k=1}^{n_{A,l}}A_{l,k}} \after \inj_{j_{l}}\right) \\
		& = & \bigotimes_{l=1}^{N_{A}} \left(\pi_{i_{l}} \after \sum_{k=1}^{n_{A,l}}(\inj_{k} \after \pi_{k}) \after \inj_{j_{l}}\right) \\
		& = & \sum_{k_{1},\ldots,k_{N_{A}}} \bigotimes_{l=1}^{N_{A}} (\pi_{i_{l}} \after \inj_{k_{l}} \after \pi_{k_{l}} \after \inj_{j_{l}}) \\
				& = & \tensor*{\delta}{*^{j_{1}}_{i_{1}}^{\;\cdots\;}_{\;\cdots}^{j_{N_{A}}}_{i_{N_{A}}}} = \left\{\begin{array}{ll}
      1_{\bigotimes_{l}^{N_{A}} A_{l,i_{l}}}	&	\hugh{\forall l \in [N_{A}]\, , \,i_{l}=j_{l}}	\\
      0_{\bigotimes_{l}^{N_{A}} A_{l,j_{l}},A_{l,i_{l}}}	&	otherwise \end{array}		
		\right.
\end{eqnarray*}

\noindent Then taking arbitrary $C = \bigotimes_{l=1}^{N_{C}} \bigoplus_{i_{l}=1}^{n_{C,l}}B_{l,i_{l}}$ and $\fctl{h}{B}{C}$,
\begin{eqnarray*}
(h \after f)_{i_{1},\ldots,i_{N_{C}};k_{1},\ldots,k_{N_{A}}}	& = &	(h \after 1_{B} \after f)_{i_{1},\ldots,i_{N_{C}};k_{1},\ldots,k_{N_{A}}} \\
									& = & \bigotimes_{l=1}^{N_{C}}\pi_{i_{l}} \after h \after \sum_{j_{1},\ldots,j_{N_{B}}} \left(\bigotimes^{N_{B}}_{l=1}\inj_{j_{l}} \after \bigotimes^{N_{B}}_{l=1}\pi_{j_{l}}\right) \after f \after \bigotimes_{l=1}^{N_{A}}\inj_{k_{l}} \\
									& = & \sum_{j_{1},\ldots,j_{N_{B}}} \left(\bigotimes_{l=1}^{N_{C}}\pi_{i_{l}} \after h \after  \bigotimes^{N_{B}}_{l=1}\inj_{j_{l}}\right) \after \left(\bigotimes^{N_{B}}_{l=1}\pi_{j_{l}} \after f \after \bigotimes_{l=1}^{N_{A}}\inj_{k_{l}}\right) \\
									& = & \sum_{j_{1},\ldots,j_{N_{B}}}h_{i_{1},\ldots,i_{N_{C}};j_{1},\ldots,j_{N_{B}}} \after f_{j_{1},\ldots,j_{N_{B}};k_{1},\ldots,k_{N_{A}}}
\end{eqnarray*}\smallskip

There is a specific type of interaction composition and tensor multiplicaton which ought to be noted, namely when a composition passing through an object $A \otimes B$ contains an arrow of the form $f \otimes 1_{B}$ or $1_{A} \otimes g$ for some arrows $f \neq 1_{A}$ or $g \neq 1_{B}$. Consider the composition $\cfct{C \otimes B}{f \otimes 1_{B}}{A \otimes B}{h}{D}$; the tensor representations of the composite morphisms are written, naturally, as $(f \otimes 1_{B})_{j_{1},\cdots,j_{N_{A}},j'_{1},\cdots,j'_{N_{B}};k_{1} \cdots k_{N_{C}},k'_{1},\cdots,k'_{N_{B}}}$ and $h_{i_{1},\cdots,i_{N_{D}};j_{1},\cdots,j_{N_{A}},j'_{1},\cdots,j'_{N_{B}}}$. However, $(f \otimes 1_{B})_{\mathbf{j},\mathbf{j}';\mathbf{k},\mathbf{k}'} = f_{\mathbf{j};\mathbf{k}} \otimes \delta_{\mathbf{j}'}^{\mathbf{k}'}$, and the only effect the Kronecker delta on the end representation of the composition is the change of index, which is superficial. In these types of situation, we consider ourselves at liberty to think instead of the composition of the two tensors $f_{\mathbf{j};\mathbf{k}}$ and $h_{\mathbf{i};\mathbf{j},\mathbf{j}'}$, summing over the indices of $\mathbf{j}$ only.

\p The monoidality of the category $\C$ allows for scalar multiplication of arrows to be modelled by the synonymous operation on tensors.
\begin{eqnarray*}
(s \cdot f)_{i_{1},\ldots,i_{N_{B}};j_{1},\ldots,j_{N_{A}}} & = & \bigotimes_{l=1}^{N_{B}}\pi_{i_{l}} \after ((\lambda_{B} \after (s \otimes 1_{B}) \after \lambda^{-1}_{B}) \after f) \after \bigotimes_{l=1}^{N_{A}}\inj_{j_{l}} \\
& = & ((\lambda_{\bigotimes_{l=1}^{N_{B}}B_{l,i_{l}}} \after (s \otimes 1_{\bigotimes_{l=1}^{N_{B}}B_{l,i_{l}}}) \after \lambda^{-1}_{\bigotimes_{l=1}^{N_{B}}B_{l,i_{l}}})) \left(\bigotimes_{l=1}^{N_{B}}\pi_{i_{l}} \after f \after \bigotimes_{l=1}^{N_{A}}\inj_{j_{l}}\right) \\
& = & s \cdot f_{i_{1},\ldots,i_{N_{B}};j_{1},\ldots,j_{N_{A}}}
\end{eqnarray*}\smallskip

When the source and target objects of an arrow are tensor products of direct sums of the tensor unit $\I$, the arrow's tensor representation acts as a standard tensor over \linebreak $\C[\I \otimes \cdots \otimes \I,\I \otimes \cdots \otimes \I]$. These homsets are trivially in bijective correspondence with the set of scalars $\C[\I,\I]$, and so we can work purely with tensors over this semiring.

\p Of course, in a compact closed category with finite biproducts, we are not restricted to building objects over the two functors $- \otimes -$ and $- \oplus -$: the contravariant endofunctor $(-)^{*}$ is also available. We now show how tensor representations can be given to morphisms between objects built with this functor.

\p Consider a morphism $\fctl{f}{A}{C \otimes B^{*}}$, where $A = \bigoplus^{a}_{k=1}A_{k}$, $B = \bigoplus^{b}_{j=1}B_{j}$, and \linebreak $C = \bigoplus^{c}_{i=1}C_{i}$. The extent to which this arrow can be decomposed using the tensor representation system given above is less than one would hope. The negation of $B$ means that it is not possible to break it down without extending the representation. The parameterised adjunction $- \otimes B \dashv - \otimes B^{*}$ described in Section~\ref{SectionCompCCs} makes this extension possible.

\p The isomorphism $v$ associated with the adjunction relates $f$ to an arrow \linebreak $\fctl{v^{-1}_{A,B,C}(f)}{B \otimes A}{C}$. This new arrow is certainly in exactly the shape seen of morphisms at the beginning of this section; $B$ is no longer shackled by the negation functor and its projections may be used in a decomposition in the same way as $A$ and $C$. The set of arrows $\{v^{-1}_{A,B,C}(f)_{ijk}:i \in [c], j \in [b], k \in [a]\}$, where
\begin{displaymath}
v^{-1}_{A,B,C}(f)_{ijk} = \pi_{k} \after v^{-1}_{A,B,C}(f) \after (\inj_{i} \otimes \inj_{j})
\end{displaymath}
determines $v^{-1}(f)$ uniquely; and since $v$ is an isomorphism the arrows provide a unique description of $f$ as well.

\p Using the isomorphism once more, this time on every arrow in the above set, we are able to understand $f$ by viewing a set of morphisms that have a superficially identical form to it: $\{v_{A_{i},B_{j},C_{k}}(v^{-1}_{A,B,C}(f)_{ijk}):i \in [c], j \in [b], k \in [a]\}$. The parameterised adjunction offers a simplification of the descriptions of each of these arrows.
\begin{eqnarray*}
v_{A_{i},B_{j},C_{j}}(v^{-1}_{A,B,C}(f)_{ijk})	&	=	&	v_{A_{i},B_{j},C_{j}}(\pi_{k} \after v^{-1}_{A,B,C}(f) \after (\inj_{i} \otimes \inj_{j})) \\
		& = & (\pi_{i} \otimes (\inj_{j})^{*}) \after f \after \inj_{i} 
\end{eqnarray*}

The principle above, where the duals of injection arrows into $B$ are used in place of projections, is entirely generalisable to multiple instances of dual objects in a tensor product. Furthermore, the inherent duality of compact closed categories allows morphisms from tensor products with dualised elements to be decomposed by substituting injection arrows for duals of projection arrows. Suppose that we have an arrow
\begin{displaymath}
  \fctl{g}{\bigotimes_{l=1}^{N_{A}}
    (\bigoplus_{k_{l}=1}^{n_{A,l}}A_{l,k_{l}})^{\phi(l)}}
  {\bigotimes_{l=1}^{N_{B}} (\bigoplus_{i_{l}=1}^{n_{B,l}}
    B_{l,j_{k}})^{\psi(l)}}  
\end{displaymath}
where the functions \hughf{$\fctl{\phi}{[N_{A}]}{\{\epsilon,*\}}$ and $\fctl{\psi}{[N_{B}]}{\{\epsilon,*\}}$ depict the polarity of a direct sum in the tensor product with respect to the negation functor, $\epsilon$ denoting an empty superscript once more.} We define the \emph{tensor representation} of such an arrow to be the set of morphisms \hughf{$\left(g_{\mathbf{j};\mathbf{k}}\right)_{\mathbf{j} \in \prod_{l=1}^{N_{B}}}[n_{B,l}],\mathbf{k} \in \prod_{l=1}^{N_{A}} [n_{A,l}]$}, where
\begin{displaymath}
  g_{j_{1}, \ldots ,j_{N_{B}};k_{1}, \ldots ,k_{N_{A}}} = {\bigotimes_{l=1}^{N_{B}} \nu_{l,j_{l}}} \after g \after {\bigotimes_{l=1}^{N_{A}} \mu_{l,k_{l}}},
\end{displaymath}
\begin{eqnarray*}
  \nu_{l,j_{l}} = \left\{ \begin{array}{ll} \pi_{j_{l}} & \psi(l)
      = \epsilon \\ (\inj_{j_{l}})^{*} & \psi(l)=* \end{array}\right. 
  &\mbox{and}& \mu_{l,k_{l}} = \left\{ \begin{array}{ll} \inj_{k_{l}} &
      \phi(l) = \epsilon \\ (\pi_{k_{l}})^{*} & \phi(l)=* .\end{array}
  \right.
\end{eqnarray*}
We can extend this using a simple induction so that the functions $\phi$ and $\psi$ \hughg{may adopt the new range of the set of strings using only the letter $*$, $F\{*\}$ as we shall name it at this point\footnote{\hughf{It is more traditionally written with a Kleene star, but the authors believe the resulting expression $\{*\}^{*}$ leads to confusing overloading.}}, so that we can apply the duality functor as many times as wished to a direct sum.} The tensor representation of the arrow $g$ remains the same, but the definitions of $\nu$ and $\mu$ are generalised accordingly.
\begin{eqnarray*}
  \nu_{l,j_{l}} = \left\{ \begin{array}{ll} 
   (\pi_{j_{l}})^{\psi(l)} & \psi(l) \in \hughf{F\{**\}} \\
   (\inj_{j_{l}})^{\psi(l)} & \psi(l) \notin \hughf{F\{**\}} \end{array}\right. 
  &\mbox{and}& \mu_{l,k_{l}} = \left\{ \begin{array}{ll} \inj_{k_{l}} &
      \phi(l) \in \hughf{F\{**\}} \\ (\pi_{k_{l}})^{*} & \phi(l) \notin \hughf{F\{**\}} .\end{array}
  \right.
\end{eqnarray*}

Of course, the negation of a number of positive and negative direct sums of objects together in a tensor product is also possible, and is in fact necessary if the $\invamp$ functor is to be defined. It would therefore be useful to produce tensor representations for arrows between objects containing such components. The principle being used in the earlier instances of negation may be translated directly to produce such entities. Given an arrow
\begin{displaymath}
\fctl{f}{A^{\alpha}}{D^{\delta} \otimes (B^{\beta} \otimes C^{\gamma})^{*}}
\end{displaymath}
where $A = \bigoplus_{l}A_{l}$, $B = \bigoplus_{j}B_{j}$, $C = \bigoplus_{k}C_{k}$ and $D = \bigoplus_{i}D_{i}$ are direct sums of objects acted on by the negation functor $\alpha$, $\beta$, $\gamma$ and $\delta$ times respectively. There is a bijective correspondence $v_{A^{\alpha},B^{\beta} \otimes C^{\gamma},D^{\delta}}^{-1}$ to $\C[A^{\alpha},D^{\delta} \otimes (B^{\beta} \otimes C^{\gamma})^{*}]$ to $\C[A^{\alpha} \otimes (B^{\beta} \otimes C^{\gamma}),D^{\delta}]$, and every morphism in this target set can be decomposed and be represented by a tensor containing arrows in the form shown below.
\begin{displaymath}
(v^{-1}_{A^{\alpha},B^{\beta} \otimes C^{\gamma},D^{\delta}}(f))_{i;l,j,k} = \nu_{i} \after v^{-1}_{A^{\alpha},B^{\beta} \otimes C^{\gamma},D^{\delta}}(f) \after (\mu_{l} \otimes (\mu_{j} \otimes \mu_{k})).
\end{displaymath}
Using the correspondence $v$ once again, we see that each morphism in the aforementioned tensor is related to an arrow which may be placed into a new representation: the relative of the above arrow is
\begin{eqnarray*}
f_{i,j,k;l} & = & v_{A_{l}^{\alpha},B_{j}^{\beta} \otimes C_{k}^{\gamma},D_{i}^{\delta}}((v^{-1}_{A^{\alpha},B^{\beta} \otimes C^{\gamma},D^{\delta}}(f))_{i;l,j,k}) \\
						& = & v_{A_{l}^{\alpha},B_{j}^{\beta} \otimes C_{k}^{\gamma},D_{i}^{\delta}}(\nu_{i} \after v^{-1}_{A^{\alpha},B^{\beta} \otimes C^{\gamma},D^{\delta}}(f) \after (\mu_{l} \otimes (\mu_{j} \otimes \mu_{k}))) \\
						& = & (\nu_{i} \otimes (\mu_{j} \otimes \mu_{k})^{*}) \after f \after \mu_{l}
\end{eqnarray*}

The arrows of this form bundled into an appropriate set are defined to be the tensor representation of the arrow, and it is once again clear that the concept is generalisable to morphisms whose sources and targets may contain any number of tensor products affected by the functor $(-)^{*}$. The self-duality of compact closed categories once more means that a similar line of reasoning can be afforded to dealing with negation in the source of an arrow by swapping the roles of injection and projection arrows. A logical extension to the argument can be constructed to deal with tensor products with the duality functor applied numerous times. Using the symmetry natural isomorphism $\sigma$ sensibly it is possible to express tensor representations for arrows between objects which are tensor products of tensor products of direct sums of primitive objects under any number of instances of the duality functor. Furthermore, due to induction, this style of reasoning is extendable to any arrow between objects built using indecomposable objects and the duality and tensor functors. We give new, recursive definitions for the arrow sets $\mu$ and $\nu$. These are, however, reliant on the concept of index sets in a category theoretical sense.

\begin{defi} \label{CatTensorIndexSetDefn}
An \emph{index set} of an instance of an object $A$ in $\C$, written $\iota(A)$, is defined inductively as follows:
\begin{itemize}
\item If $A = \bigoplus_{i=1}^{n}A_{i}$ for some instances of objects $A_{1},\ldots,A_{n}$ we wish to be considered indecomposable, then $\iota(A) = [n]$.
\item If $A = B^{*}$ for some object $B$, then $\iota(A) = \iota(B)$.
\item If $A = \bigotimes_{l=1}^{N}A_{l}$ for some objects $A_{1},\ldots,A_{N}$, then $\iota(A) = \prod_{l}^{N}\iota(A_{l})$.
\end{itemize}\medskip
\end{defi}

Equivalently, the index set of an object $A = \bigotimes_{l=1}^{N}\bigoplus_{i=1}^{n_{l}}A_{l,i}$ for which we consider its instances of $A_{l,i}$ indecomposable for each $l$ and $i$ is the set of $N$-tuples where for every $l \in [N]$, the $l^{th}$ component ranges over the number of objects in the $l^{th}$ direct sum.

\p We are now in a position to define $\mu$ and $\nu$.
\begin{defi}
The \emph{injection} and \emph{projection arrow functions} of an object $A$ built using the tensor product and negation functors over direct sums of indecomposable objects in a compact closed category $\C$ with finite biproducts, written $\mu_{A}$ and $\nu_{A}$ respectively, have domain $\iota(A)$ and are defined recursively as follows:
\begin{itemize}
\item If $A = \bigoplus_{i=1}^{n}A_{i}$ for indecomposable $A_{1},\ldots,A_{n}$, $\mu_{A}(i) = \inj_{i}$ and $\nu_{A}(i) = \pi_{i}$.
\item If $A = B^{*}$ for some $B$, then $\mu_{A}(\mathbf{i}) = (\nu_{B}(\mathbf{i}))^{*}$ and $\nu_{A}(\mathbf{i}) = (\mu_{B}(\mathbf{i}))^{*}$.
\item If $A = \bigotimes_{l=1}^{N}A_{l}$ for some $A_{1},\ldots,A_{N}$, then $\mu_{A}(\mathbf{i}_{1},\ldots,\mathbf{i}_{N}) = \bigotimes_{l=1}^{N}\mu_{A_{l}}(\mathbf{i}_{l})$ and \\  $\nu_{A}(\mathbf{i}_{1},\ldots,\mathbf{i}_{N}) = \bigotimes_{l=1}^{N}\nu_{A_{l}}(\mathbf{i}_{l})$, where each $\mathbf{i}_{l} \in \iota(A_{l})$.
\end{itemize}\medskip
\end{defi}

\noindent The definition above facilitates the description of the tensor representation of every arrow built in the manner that has been discussed earlier.

\begin{defi} \label{TensorRepDefn}
A \emph{tensor representation} of an arrow $f:A \longrightarrow B$, where $A$ and $B$ are objects built using tensor products and the negation functor over direct sums of instances of objects considered indecomposable, is defined as the set of morphisms $\{f_{\mathbf{i},\mathbf{j}}:\mathbf{i} \in \iota(B),\, \mathbf{j} \in \iota(A)\}$ where
\begin{displaymath}
f_{\mathbf{i},\mathbf{j}} = \nu_{\mathbf{i}} \after f \after \mu_{\mathbf{j}}
\end{displaymath}\vspace{-1 mm}
\end{defi}
Note how this definition of tensor representation is consistent with the earlier, more rudimentary forms which do not consider the possibility of negation existing beyond the direct sums. The entries in these tensors are still merely composites containing arrows; and addition of morphisms is a consequence of enrichment over $\mathbf{CMon}$, so the connection between tensor and arrow addition is maintained, including the zero morphisms. Similarly, scalar multiplication is unaffected. The extended definition of the identity arrow of any object $A$ is created recursively from prior knowledge of the standard diagonal matrix representation of the identity arrow for direct sums of objects and the preservation of identity morphisms by functors, and takes the form $\sum_{I \in \iota(A)}(\mu_{A}(I) \after \nu_{A}(I))$.

\begin{itemize}
\item If $A = \bigoplus_{i=1}^{n}A_{i}$ for indecomposable objects $A_{1},\ldots,A_{n}$, then the representation of the identity morphism $1_{A}$ is
\begin{displaymath}
(1_{A})_{i,j} = \sum_{k=1}^{n}(\inj_{k} \after \pi_{k}) = \sum_{I \in \iota(A)}(\mu_{A}(I) \after \nu_{A}(I)).
\end{displaymath}
\item If $A = B^{*}$ for some object $B$, then 
\begin{eqnarray*}
1_{A} = 1_{B^{*}} = (1_{B})^{*} & = & \left(\sum_{\mathbf{i} \in \iota(B)}(\mu_{B}(\mathbf{i}) \after \nu_{B}(\mathbf{i}))\right)^{*} = \sum_{\mathbf{i} \in \iota(A)}(\mu_{B}(\mathbf{i}) \after \nu_{B}(\mathbf{i}))^{*} \\
																& = & \sum_{\mathbf{i} \in \iota(A)}((\nu_{B}(\mathbf{i}))^{*} \after (\mu_{B}(\mathbf{i}))^{*}) = \sum_{\mathbf{i} \in \iota(A)}(\mu_{A}(\mathbf{i}) \after \nu_{A}(\mathbf{i})).
\end{eqnarray*}
\item If $A = \bigotimes_{l=1}^{N}A_{l}$ for some objects $A_{1},\ldots,A_{N}$, then
\begin{eqnarray*}
1_{A} = 1_{\bigotimes_{l=1}^{N}A_{1}} = \bigotimes_{l=1}^{N}1_{A_{l}} & = & \bigotimes_{l=1}^{N}\left(\sum_{\mathbf{i}_{l} \in \iota_{A_{l}}} (\mu_{A_{l}}(\mathbf{i}_{l}) \after \nu_{A_{l}}(\mathbf{i}_{l}))\right) \\
																& = & \sum_{(\mathbf{i}_{1},\cdots,\mathbf{i}_{N}) \in \prod_{l=1}^{N}\iota(A_{l})}\left(\bigotimes_{l=1}^{N}\mu_{A_{l}}(\mathbf{i}_{l}) \after \bigotimes_{l=1}^{N}\nu_{A_{l}}(I_{l})\right) \\
																& = & \sum_{\mathbf{i} \in \iota(A)}(\mu_{A}(\mathbf{i}) \after \nu_{A}(\mathbf{i})).
\end{eqnarray*}\medskip
\end{itemize}
Because of the lack of change required in the shape of this arrow from earlier versions, the manner in which composition operates is also preserved.

\p Unlike when we consider only the simpler form of arrow being placed into a multilinear representation, which lacks any use of the duality functor, it cannot be taken for granted that the tensor describing an arrow between two objects built over direct sums of $\I$, and tensor and negation functors can be viewed as being a standard tensor over $\C[\I,\I]$. This is because the entries take the form of arrows between tensor products over tensor products of both $\I$ and $\I^{*}$. \hughg{Fortunately, $\fctl{\lambda_{\I^{*}} \after v_{\I,\I,\I}(\rho_{\I})}{\I}{\I^{*}}$ is an isomorphism}; and we can compose suitable tensor products built from it, its inverse, dualities of both these arrows and the identity morphism $1_{I}$ to ``remove'' the dual instantiations of the unit, and so therefore allow us to view the tensors as being over \hugh{$\C[\I \otimes \cdots \otimes \I,\I \otimes \cdots \otimes \I] \cong \C[\I,\I]$.}

\subsection{Describing Transformations with Tensors} \label{SectionTransExtensors}

The representations described in this \hugh{section} offer a simplification to the form of morphisms between objects built from the three functors expected to exist in a compact closed category with finite biproducts. It therefore follows naturally that certain arrows that are part of an \mll~transformation can be viewed as tensors. In this section we consider an object $F(\mathbf{A},\mathbf{A})$ in a compact closed category $\C$ satisfying feeble full completeness for an arbitrary \mll~functor $F$ where $\mathbf{A} = (n_{1}\I,\ldots,n_{N}\I)$, \hugh{where $n\I \,=\, \bigoplus_{i=1}^{n}\I$,} and give the arrows into the object which are constituents of \mll~transformations.

\p It is known that every \mll~transformation in a compact closed category with finite biproducts satisfying feeble full completeness is a linear combination of fixed-point-free involutions~\cite{CHS01}, with an involution being the equivalent of an appropriate number of instances of the unit dinatural transformation $\eta$ joined by a tensor product being postcomposed with a series of symmetry natural isomorphisms.\footnote{Associativity natural isomorphisms are suppressed in this explanation, but this can be perceived as self-evident.} Each pair of objects created together by an instance of $\eta$ models a pair of literals joined by an axiom link in a cut-free proof net. As such, it makes sense to initially provide the tensor representation of $\eta_{n\I}$ for arbitrary $n \in \N$. The unit transformation is derived from the bijective correspondence connected to $\C$ being applied to the identity arrows in $\C$: for every $A \in \C$, $\eta_{A} = v_{\I,A,A}(\lambda_{A})$; and because of this the arrow $\fctl{\eta_{nI}}{\I}{n\I \otimes (n\I)^{*}}$ is easily shown to have an $n \times n$ identity matrix as its tensor representation.
\begin{eqnarray*}
(\eta_{n\I})_{ij} & = & \nu_{ij} \after \eta_{n\I} \after \mu_{*} \\
									& = & \lambda_{\I} \after (1_{\I} \otimes \chi^{-1}) \after (\pi_{i} \otimes (\inj_{j})^{*}) \after v(\lambda_{n\I}) \after 1_{I} \\
									& = &	\lambda_{\I} \after (1_{\I} \otimes \chi^{-1}) \after v(\delta_{ij}) \\
									& = & \left\{\begin{array}{ll}
									\chi^{-1} \after (\lambda_{\I} \after v(\lambda_{\I})) = \chi^{-1} \after \chi = 1 & i=j \\
									\chi^{-1} \after (\lambda_{\I} \after v(0_{\I \otimes \I,\I})) = \chi^{-1} \after 0_{\I,\I^{*}} = 0 & i \neq j
									\end{array} \right.
\end{eqnarray*}
We therefore are at liberty to express an axiom link as a Kronecker delta tensor, with one index referring to the object modelling the positive literal and the other the negative literal. This concept may be extended now by tensor multiplication to \mll~functors with more than one pair of literals to being joined. Letting $\alpha$ be modelled by $n\I$, the arrow modelling the proof structure
\begin{center} \vspace{5mm}
\begin{tikzpicture}
  [auto, node
  distance=6mm, skip loop/.style={to path={-- ++(0,#1) -| (\tikztotarget)}}] \tikzstyle{every node} = [text depth=-5pt,text height=0.5ex]
   \node (1) {$((\,\alpha\phantom{''}$}; \node (a) [right of=1] {$\tens$} ;
  \node (2) [right of=a] {$\alpha^{\bot})$}; \node (b) [right of=2] {$\tens$} ;
  \node (3)  [right of=b, xshift=-1mm] {$\phantom{'}\alpha\phantom{'})$}; \node (c)  [right of=3] {$\invamp$};
  \node (4)  [right of=c] {$(\alpha^{\bot}$}; \node(d) [right of=4] {$\tens$} ;
  \node (5) [right of=d, xshift=-1mm] {$\phantom{'}\alpha\phantom{'})$}; \node (e) [right of=5] {$\invamp$} ;
  \node (6) [right of=e] {$((\,\alpha^{\bot}$}; \node (f) [right of=6] {$\invamp$} ;
  \node (7)  [right of=f,xshift=-1mm] {$\phantom{'}\alpha)\phantom{'}$}; \node (g)  [right of=7] {$\tens$};
  \node (8)  [right of=g] {$\alpha^{\bot})$};
\begin{scope}
\path   (1)  edge [red, skip loop =6mm, shorten >=3mm, shorten <=3mm]  (2);
\path (3)  edge  [red, skip loop =6mm, shorten >=3mm, shorten <=3mm] (4);
\path   (5)  edge [red, skip loop =6mm, shorten >=3mm, shorten <=3mm]  (6);
\path (7)  edge  [red, skip loop =6mm, shorten >=3mm, shorten <=3mm] (8);
\end{scope}
\end{tikzpicture} \vspace{5mm}
\end{center}

\noindent is represented by the tensor $\tensor*{\delta}{*^{j_{1}}_{i_{1}}^{\;\cdots\;}_{\;\cdots}^{j_{M}}_{i_{M}}}$, where the indices $i_{l}$ and $j_{l}$ are connected to the objects representing the $l^{th}$ positive and negative literals respectively. In a more general form, if the literals which are connected via axiom links are less well-ordered so that the $l^{th}$ positive literal is connected to the $\sigma(l)^{th}$ negative literal for some permutation $\sigma \in S_{M}$, then the modelling tensor is $\tensor*{\delta}{*^{\sigma(j_{1})}_{i_{1}}^{\;\cdots\;}_{\;\cdots}^{\sigma(j_{M})}_{i_{M}}}$.
\newpage

\begin{exa}
The tensor $\tensor*{\delta}{*^{j_{4}}_{i_{1}}^{j_{2}}_{i_{2}}^{j_{1}}_{i_{3}}^{j_{3}}_{i_{4}}}$ represents the proof structure below.
\end{exa} 
\begin{center} \vspace{5mm}
\begin{tikzpicture}
  [auto, node
  distance=6mm, skip loop/.style={to path={-- ++(0,#1) -| (\tikztotarget)}}] \tikzstyle{every node} = [text depth=-5pt,text height=0.5ex]
   \node (1) {$((\,\alpha\phantom{''}$}; \node (a) [right of=1] {$\tens$} ;
  \node (2) [right of=a] {$\alpha^{\bot})$}; \node (b) [right of=2] {$\tens$} ;
  \node (3)  [right of=b, xshift=-1mm] {$\phantom{'}\alpha\phantom{'})$}; \node (c)  [right of=3] {$\invamp$};
  \node (4)  [right of=c] {$(\alpha^{\bot}$}; \node(d) [right of=4] {$\tens$} ;
  \node (5) [right of=d, xshift=-1mm] {$\phantom{'}\alpha\phantom{'})$}; \node (e) [right of=5] {$\invamp$} ;
  \node (6) [right of=e] {$((\,\alpha^{\bot}$}; \node (f) [right of=6] {$\invamp$} ;
  \node (7)  [right of=f,xshift=-1mm] {$\phantom{'}\alpha)\phantom{'}$}; \node (g)  [right of=7] {$\tens$};
  \node (8)  [right of=g] {$\alpha^{\bot})$};
\begin{scope}
\path   (1)  edge [blue, skip loop =7mm, shorten >=3mm, shorten <=3mm]  (8);
\path (2)  edge  [blue, skip loop =6mm, shorten >=3mm, shorten <=3mm] (5);
\path   (3)  edge [blue, skip loop =5mm, shorten >=3mm, shorten <=3mm]  (4);
\path (6)  edge  [blue, skip loop =6mm, shorten >=3mm, shorten <=3mm] (7);
\end{scope}
\end{tikzpicture} \vspace{5mm}
\end{center}

As discussed in Section~\ref{SectionCompCCs}, \mll~functors may model sequents which contain more than one instance of a single literal; and such entities can have more than one set of axiom links attached to them. Furthermore, regardless of the number of repeated literals, scalar multiplications on arrows, and so natural and dinatural transformations, are always possible. This gives rise to the possibility of transformations describing linear combinations of sets of axiom links on a single sequent. Such linear combinations are modelled by \mll~transformations which are linear combinations of transformations representing single sets of axiom links, whose constituent arrows are linear combinations as well. It follows, therefore, that the tensor representation of an arrow $\tau_{\mathbf{A}}$ for some \mll~transformation $\tau$ where $\mathbf{A} = (n_{1}\I,\ldots,n_{N}\I)$ is in the following form:
\begin{displaymath}
\tensor*{\tau}{*^{j_{1}}_{i_{1}}^{\;\cdots\;}_{\;\cdots}^{j_{M}}_{i_{M}}} = \sum_{\sigma \in S_{M}}s_{\sigma} \cdot \tensor*{\delta}{*^{\sigma(j_{1})}_{i_{1}}^{\;\cdots\;}_{\;\cdots}^{\sigma(j_{M})}_{i_{M}}}
\end{displaymath}
where $s_{\sigma} \in \C[\I,\I]$ for every $\sigma \in S_{M}$.

\begin{exa}
The tensor $\tensor*{\tau}{*^{j_{1}}_{i_{1}}^{j_{2}}_{i_{2}}^{j_{3}}_{i_{3}}^{j_{4}}_{i_{4}}} = \tensor*{\delta}{*^{j_{1}}_{i_{1}}^{j_{2}}_{i_{2}}^{j_{3}}_{i_{3}}^{j_{4}}_{i_{4}}} - \tensor*{\delta}{*^{j_{4}}_{i_{1}}^{j_{2}}_{i_{2}}^{j_{1}}_{i_{3}}^{j_{3}}_{i_{4}}}$ represents the linear combination of axiom links on the \mll~formula given below.
\end{exa}
\begin{center} \vspace{5mm}
\begin{tikzpicture}
  [auto, node
  distance=6mm, skip loop/.style={to path={-- ++(0,#1) -| (\tikztotarget)}}] \tikzstyle{every node} = [text depth=-5pt,text height=0.5ex]
   \node (1) {$((\,\alpha\phantom{''}$}; \node (a) [right of=1] {$\tens$} ;
  \node (2) [right of=a] {$\alpha^{\bot})$}; \node (b) [right of=2] {$\tens$} ;
  \node (3)  [right of=b, xshift=-1mm] {$\phantom{'}\alpha\phantom{'})$}; \node (c)  [right of=3] {$\invamp$};
  \node (4)  [right of=c] {$(\alpha^{\bot}$}; \node(d) [right of=4] {$\tens$} ;
  \node (5) [right of=d, xshift=-1mm] {$\phantom{'}\alpha\phantom{'})$}; \node (e) [right of=5] {$\invamp$} ;
  \node (6) [right of=e] {$((\,\alpha^{\bot}$}; \node (f) [right of=6] {$\invamp$} ;
  \node (7)  [right of=f,xshift=-1mm] {$\phantom{'}\alpha)\phantom{'}$}; \node (g)  [right of=7] {$\tens$};
  \node (8)  [right of=g] {$\alpha^{\bot})$};
  \node (z1) [left of=1] {};
  \node (t1) [above of= z1] {$\times$};
  \node (t2) [above of= t1] {$\times$};
  \node (s1) [left of= t1] {$-1$};
  \node (s2) [left of= t2] {$1$};
\begin{scope}
\path   (1)  edge [red, skip loop =11mm,shorten >=8mm, shorten <=8mm]  (2);
\path (3)  edge  [red, skip loop =11mm,shorten >=8mm, shorten <=8mm] (4);
\path   (5)  edge [red, skip loop =11mm,shorten >=8mm, shorten <=8mm]  (6);
\path (7)  edge  [red, skip loop =11mm,shorten >=8mm, shorten <=8mm] (8);
\path   (1)  edge [blue, skip loop =7mm, shorten >=3mm, shorten <=3mm]  (8);
\path (2)  edge  [blue, skip loop =6mm, shorten >=3mm, shorten <=3mm] (5);
\path   (3)  edge [blue, skip loop =5mm, shorten >=3mm, shorten <=3mm]  (4);
\path (6)  edge  [blue, skip loop =6mm, shorten >=3mm, shorten <=3mm] (7);
\end{scope}
\end{tikzpicture} \vspace{5mm}
\end{center}

\p Of course, not every proof structure (and none which satisfies the correctness criteria) is built over exactly one formula only containing tensors. It is therefore important to understand the effects of replacing some instances of the functor $\otimes$ with those of $\invamp$. Tensor representations of arrows between arbitrary objects are shown to exist in Section~\ref{SectionTensorRep}; this immediately implies that they exist for ones containing the par functor as well. After all, it is the de Morgan dual of the tensor product using $(-)^{*}$ as negation.

\p We can see how involutions are viewed tensorially when certain tensor products are changed by considering $\fctl{\Mix}{- \otimes -}{- \invamp -}$ described in Section~\ref{CatModels} and its representation. In compact closed categories, the `Mix' transformation is not just natural but isomorphic, and is in fact built from the same correspondence $v:\C[-_{A} \otimes -_{B},-_{C}] \cong \C[-_{A},-_{C} \otimes (-_{B})^{*}]$:
\begin{displaymath}
\Mix_{A,B} = v_{A \otimes B,A^{*} \otimes B^{*},I}(\lambda_{I} \after (v^{-1}(\lambda_{B^{*}}^{-1}) \otimes v^{-1}(\lambda_{A^{*}}^{-1})) \after \tilde{\sigma}),
\end{displaymath}
where $\tilde{\sigma}$ is the composition of $\sigma$ natural isomorphisms such that
\begin{displaymath}
\fctl{\tilde{\sigma}_{A,B,C,D}}{A \otimes B \otimes C \otimes D}{D \otimes B \otimes C \otimes A}.
\end{displaymath}
Via an inductive argument on the size of $A$ and $B$, we find that the  representation of $\mbox{Mix}_{A,B}$ reduces to an identity , as long as both $A$ and $B$ are unit-generated. The natural consequence of this is that one need not even think of the multiplicative functor being used once the tensor calculus is employed: we differentiate between arrows with the same tensors purely by looking at their signatures.

\p We can take this knowledge and provide an tensorial method to look at the the weakening natural transformations. Consider $w^{LL}~:~-_{1}~\otimes~(-_{2}~\invamp~-_{3})~\longrightarrow~(-_{1}~\otimes~-_{2})~\invamp~-_{3}$. If the three inputs to the natural transformation are unit-generated, then we know from above that the instantiations of the $-\invamp-$ functor may be viewed in exactly the same manner as the usages of the $-\otimes-$ functor. It becomes obvious that, just like the \mll\ transformations above, $w^{LL}$ is perceived as a tensor in the same way as the associativity isomorphism. That is, letting $I_{k}$ and $J_{k}$ be the superindices denoting the $k^{th}$ entries of the source and target functors respectively, we find that $(w^{LL})_{I_{1}I_{2}I_{3}J_{1}J_{2}J_{3}} = \tensor*{\delta}{*^{J_{1}}_{I_{1}}^{J_{2}}_{I_{2}}^{J_{3}}_{I_{3}}}$. The other weakening transformations are defined similarly.

\subsection{Describing Double-Glued Objects with Tensors} \label{SectionDGTensor}

\hughg{The sets of values and covalues of an object $R$ in $\GC$ contain morphisms from~$I$ and to~$I^{*}$ respectively. As such, if $UR$ is built solely from $I$, $(-)^{*}$ and $- \otimes -$ in the underlying category, then the elements of $R_{Val}$ and $R_{CoVal}$ can be represented by tensors over the semiring of scalars of $\C$.}

\p Given $R,S \,\in\, \GC$, $E \,\subseteq\, \C[\I,\I^{*}]$, we provide the shape of the values and covalues created from the $*$-autonomous structure. One can see particularly how the formulation of these sets is greatly simplified for the negation and tensor product.

\begin{itemize}
\item $\I \,=\, \left( \I,\{1_{\I}\}, \lambda_{\I^{*}} \after v_{\I,\I,\I}(\rho_{\I}) \after E \right)$
\item $R^{\bot}_{Val} \,=\, R_{CoVal}$; $R^{\bot}_{CoVal} \,=\, R_{Val}$.
\item $(R \,\otimes\, S)_{Val} \,=\, \left\{ r_{\mathbf{i}}s_{\mathbf{j}} \,:\, r_{\mathbf{i}} \in R_{Val},\, s_{\mathbf{j}} \in S_{Val} \right\}$
\item $(R \,\otimes\, S)_{CoVal} \,=\, \left\{ z_{\mathbf{ij}} \,:\, \forall r_{\mathbf{i}} \in R_{Val},\, z_{\mathbf{ij}}r_{\mathbf{i}} \,\in\, S_{CoVal},\,\, \forall s_{\mathbf{j}} \in S_{Val},\, z_{\mathbf{ij}}s_{\mathbf{j}} \,\in\, R_{CoVal} \right\}$

\end{itemize}

\section{\mll~Full Completeness for $\GC$} \label{SectionGCFC}

For the purposes of this section, we let $\C$ be an arbitrary compact closed category with biproducts satisfying feeble full completeness (Definition~\ref{FeebleFCDefn}). Feeble full completeness allows us to assume that every \mll~transformation (Section~\ref{CatModels}) in this category is described by a linear combination of fixed-point-free involutions, each involution representing an axiom link connecting the two literals being modelled. It is shown in \cite{Tan97,HS03} that $\GC$ must be $*$-autonomous, and so it is already known that all dinatural transformations modelling correct \mll~proof structures are found in the glued category. Fact~\ref{PropGPreservesDNTs} tells us that the dinatural transformations in the glued category are also described by linear combinations of fixed-point-free involutions. What still remains to be shown, however, is that the double glueing construction not only preserves the feeble full completeness of $\C$, but removes enough arrows to ensure that the only linear combinations of proof structures which are still modelled are not linear combinations at all, but denotations of single proof nets.

\p With such strong constraints on the arrows within transformations already placed, the strategy to do this becomes in essence remarkably simple. For an \mll~transformation $\tau:\K \longrightarrow UF$ in $\C$ to be found in $\GC$, it must be the case that $\tau_{U\mathbf{R}}$ is found as an arrow from $\I$ to $F(\mathbf{R},\mathbf{R})$ for every $\mathbf{R} \in (\GC)^{N}$, i.e. the set of values for the $\GC$-object $F(\mathbf{R},\mathbf{R})$. Alternatively, it may be said that $\tau$ does \emph{not} translate into $\GC$ if there is a tuple of $\GC$-objects $\mathbf{R}$ where $\tau_{U\mathbf{R}}$ does not belong to $F(\mathbf{R},\mathbf{R})_{Val}$. We provide tuples which expose how some arrows describing incorrect proof structures and impure linear combinations do not find themselves in all the sets of values needed to ensure they remain transformations in $\GC$.

Every compact closed category with finite biproducts has a full subcategory closed under all three of the characteristic functors which is generated solely by its tensor unit $\I$. By choosing `test objects' for the tuples from this subcategory, we ensure the proof is as general as possible. Furthermore, if the tuple $\mathbf{R}$ consists of test objects whose underlying $\C$-objects are of the form $n\I$ for some $n \in \N^{+}$, the shape of the object $F(\mathbf{R},\mathbf{R})$ must be akin to those discussed in Section~\ref{SectionTensorRep}, and tensor representations of arrows may be considered instead, simplifying the process noticeably.

\hughh{The structure of the coming proof can be viewed as follows:}
\begin{description}
\item[\rm(\S 4.1-4.5)] We first prove that the only MDNF~transformations in~$\GC$ model correct proof nets, a property which we call \emph{MDNF Full Completeness}. 
	\begin{enumerate}[label=(\autoref{SectionZeroTransProof})]
	\item[(\autoref{SectionMDNFObjects})] We introduce the families of test objects $\{A_{n}\,:\,n\,\in\,\mathbb{N}\}$ and $\{C_{n}\,:\,n\,\in\,\mathbb{N}\}$, and calculate the sets of values for the objects given by MDNF functors when instantiated using a chosen object from either one of these categories. These value sets are dependent on the families of full and partial permutations given in Definition~\ref{TensorExamDefns}.
	\item[(\autoref{SectionZeroTransProof})] It is shown that $\GC$ only contains MDNF~transformations modelling linear combinations of proof structures whose scalar multiples sum to $1$ (Proposition~\ref{OnlyOneLemma}). In particular, MDNF~transformations containing only zero morphisms (\emph{zero transformations}) are not found in the glued category.
	\item[(\autoref{SubsectionAcyclic})] A proof that every MDNF~transformation in~$\GC$ models linear combinations of acyclic proof structures is given.
		\begin{itemize}
		\item Consider a transformation $\fctl{\tau}{\K}{UF}$ in $\C$ modelling a linear combination containing a cyclic proof structure.
		\item Use Algorithm~\ref{AcyclicAlg} to produce partial permutations for all the blocks of a given cyclic structure.
		\item Proposition~\ref{AcyclicLemma}, together with technical lemmas~\ref{AcyclicClaim1} and~\ref{AcyclicClaim2}, are used to show that the generated partial permutations can be used to prove that  he tensor representation (Section~\ref{SectionTransExtensors}) of the transformation instantiated with a single object $n\I$ for large enough $n$ cannot be found in the instantiation of $F$ where all arguments are $A_{n}$. This disproves the possible existence of the equivalent transformation in~$\GC$.
		\end{itemize}
	\item[(\autoref{SubsectionConnectedness})] Following a similar strategy to the above, we deduce that every MDNF~transformation in~$\GC$ models linear combinations of correct proof structures (i.e. connectedness is proved).
		\begin{itemize}
		\item We consider a transformation $\fctl{\tau}{\K}{UF}$ in $\C$ modelling a linear combination of acyclic, disconnected proof structures, noting all the structures must be disconnected (Lemma~\ref{OneDiscThenAllClaim}).
		\item We create appropriate full permutations for all blocks except one using Algorithm~\ref{DisconnectTupleAlg} and Lemma~\ref{ConAlgCompleteLemma}.
		\item Proposition~\ref{ConnectedLemma} explains how these permutations when composed with a tensor representation of appropriate dimensions of the linear combination of disconnected proof structures produces a zero tensor. This proves that the representation could not exist in the instantiation of $F$ where all arguments are $C_{n}$, meaning that the transformation cannot exist in the double-glued category.
		\end{itemize}
	\item[(\autoref{SubsectionUniqueness})] It is found that only singular, unique MDNF~proof nets can be modelled in~$\GC$.
		\begin{itemize}
		\item The method is nearly identical to those seen above, considering a transformation in $\C$ to $F$ modelling a true linear combination of proof nets (that is, one containing at least two different proof nets whose scalars are non-zero).
		\item Algorithm~\ref{UniqueTupleAlg} generates partial permutations for all except one block containing exactly one literal (which is a \emph{leaf} in each MDNF~proof net).
		\item Proposition~\ref{UniqueLemma}, which lightly makes use of the full form of Proposition~\ref{OnlyOneLemma}, shows that the tensor representation of this linear combination cannot be found in the set of values of the functor when instantiated using an appropriate $A_{n}$ once again, and so MDNF~full completeness holds.
		\end{itemize}
	\end{enumerate}
\item[(\autoref{SectionMLLExtend})] Finally, we show that the full completeness of MDNF transformations extends to that for all \mll~transformations.
		\begin{itemize}
		\item Algorithms~\ref{MLLtoMDNFCycleAlg} and~\ref{MDNFAlg} take advantage of the natural isomorphisms and weak distributivity natural transformations in $*$-autonomous categories to create natural transformations which compose with general \mll~transformations to give ones for MDNF~functors. These transformations are used in Theorem~\ref{fcomp} that MDNF~full completeness implies \mll~full completeness, thus giving the desired result.
		\end{itemize}\medskip
\end{description}

\p\noindent  It should be noted that the lemmata used in this full completeness proof find themselves in an unusual order. In previous \mll~full completeness results, particularly those requiring glueing constructions \cite{Loa94b,Tan97,Hag00}, it is generally proved first that only \mll~proof structures are modelled in the category, and their correctness is shown afterwards. In the coming proof we do the exact reverse. The existence of cyclic proof structures in modelled linear combinations is disproved, and disconnectedness of modelled proof structures in the glued category is shown to be unallowed. Only then do we prove that `impure' linear combinations and scalar multiples of proof nets are not represented by transformations in $\GC$. Without assuming acyclicity the most natural approach to show the `purity' (or `uniqueness' as we refer to it from now on) of the allowed linear combinations required; the results of following this train of thought is seen in Section~\ref{SectionMLLMixFC}.

\p Most parts of the coming proof method require the use of an algorithm to produce a number of permutation tensors, and these are given in each of the corresponding subsections. The intuition behind how and why the permutations are of use are given alongside these algorithms before providing each formal, generalised proof.

\subsection{Test Objects and Permutations in MDNF Functors} \label{SectionMDNFObjects}

It turns out that only two fundamental types of test object are required for full completeness to be proved. We define them both for each $n \in \N^{+}$ as follows:
\begin{itemize}
\item $A_{n} := (n\I,\{\inj_{x} \in \C[\I,n\I]: x \in [n]\} \cup \{0_{I,n\I}\},\{\pi_{x} \in \C[n\I,\I]: x \in [n]\} \cup \{0_{n\I,\I}\})$
\item $C_{n} := (n\I,\{\inj_{x} \in \C[\I,n\I]: x \in [n]\},\{\pi_{x} \in \C[n\I,\I]: x \in [n]\})$
\end{itemize}

\hughh{As discussed in Section~\ref{SectionDGTensor}, we can represent these objects using sets of tensors instead of collections of arrows. Their new form is given below.}
\begin{itemize}
\item $A_{n} := (n\I,\{\delta_{ix}: x \in [n]\} \cup \{0_{i}\},\{\delta_{ix}: x \in [n]\} \cup \{0_{i}\})$
\item $C_{n} := (n\I,\{\delta_{ix}: x \in [n]\},\{\delta_{ix}: x \in [n]\})$
\end{itemize}

It is clear, particularly from the tensor representations, that $A_{n}$ and $C_{n}$ are self-dual --- that is, $A_{n} \cong A_{n}^{\bot}$ and $C_{n} \cong C_{n}^{\bot}$ --- for all positive natural numbers $n$. As a consequence the sets of tensor representations of values and covalues of the objects $F(\mathbf{R},\mathbf{R})$ and $F'(\mathbf{R},\mathbf{R})$ are the same if $F'$ is the \mll~functor representing the same sequent as $F$, only with all instances of negative literals converted to positive instances, and $\mathbf{R}$ is filled only with objects such as $A_{n}$ and $C_{n}$. Because of this, in these situations we are at liberty to pretend that all instances of literals are positive in the formulae when viewing arrows solely through their multilinear arrays.

\p Tensor powers of these objects (and therefore tensor products only containing positive and negative instances of them) reproduce the sets of higher-order permutations, both full and partial as defined in Definition~\ref{TensorExamDefns}. This can be demonstrated using a standard inductive argument. We use the notation $X^{\otimes N}$ for the $N^{th}$ tensor power of an object $X$.

\begin{lem} \label{ClaimTensorPowers}
For each $N \in \N^{+}$, $C_{n}^{\otimes N} = \left((n\I)^{\otimes N},\{\tensor*{\delta}{*^{x_{1}}_{i_{1}}^{\;\cdots\;}_{\;\cdots}^{x_{N}}_{i_{N}}} : (x_{1},\ldots,x_{N}) \in [n]^{N}\},\Perm(N,n)\right)$. Similarly, $A_{n}^{\otimes N} = \left((n\I)^{\otimes N},\{\tensor*{\delta}{*^{x_{1}}_{i_{1}}^{\;\cdots\;}_{\;\cdots}^{x_{N}}_{i_{N}}} : (x_{1},\ldots,x_{N}) \in [n]^{N}\} \cup \{0_{i_{1} \cdots i_{N}}\},\PPerm(N,n)\right)$.
\end{lem}
\proof
The claim is trivially true for $N = 1$ in both cases, and it is also clear that
\begin{displaymath}
\left|A_{n}^{\otimes N}\right| = \left|A_{n}\right|^{\otimes N} = (n\I)^{\otimes N} = \left|C_{n}\right|^{\otimes N} = \left|C_{n}^{\otimes N}\right|.
\end{displaymath}

\noindent To prove that the values and covalues are as desired for any positive $N$, we assume it is true for a natural number $M$ and show that it remains true for $M+1$. \hughg{Section~\ref{SectionDGTensor} contains the standard calculation rules that are used in this proof.} Starting with the values of the tensor power, we find
\begin{eqnarray*}
(A_{n}^{\otimes M+1})_{Val}	& = & \left(A_{n}^{\otimes M} \otimes A_{n}\right)_{Val} \\
																								& = & \left\{ u_{i_{1} \cdots i_{M}} \cdot v_{i_{M+1}}: u_{i_{1} \cdots i_{M}} \in (A_{n}^{\otimes M})_{Val}, \,v_{i_{M+1}} \in (A_{n})_{Val} \right\} \\
																								& = & \{ \tensor*{\delta}{*^{x_{1}}_{i_{1}}^{\;\cdots\;}_{\;\cdots}^{x_{M}}_{i_{M}}} \cdot \delta_{i_{M+1}x_{M+1}}: (x_{1},\ldots,x_{M}) \in [n]^{M},\, x_{M+1} \in [n]\} \\
																								&   & \cup \{ \tensor*{\delta}{*^{x_{1}}_{i_{1}}^{\;\cdots\;}_{\;\cdots}^{x_{M}}_{i_{M}}} \cdot 0_{i_{M+1}}: (x_{1},\ldots,x_{M}) \in [n]^{M}\} \\
																								&   & \cup \{0_{i_{1} \cdots i_{M}} \cdot \delta_{i_{M+1}x_{M+1}} : x_{M+1} \in [n]\} \cup \{0_{i_{1} \cdots i_{M}} \cdot 0_{i_{M+1}}\} \\
																								& = & \{ \tensor*{\delta}{*^{x_{1}}_{i_{1}}^{\;\cdots\;}_{\;\cdots}^{x_{M+1}}_{i_{M+1}}} : (x_{1},\ldots,x_{M+1}) \in [n]^{M+1}\} \cup \{0_{i_{1} \cdots i_{M+1}}\}.
\end{eqnarray*}

\noindent Finally, the covalues are evaluated to be as required.
\begin{eqnarray*}
(A_{n}^{\otimes M+1})_{CoVal}	& = & \left(A_{n}^{\otimes M} \otimes A_{n}\right)_{CoVal} \\
																									& = & \left\{z_{i_{1} \cdots i_{M+1}}: \forall\, u_{i_{1} \cdots i_{M}} \in (A_{n}^{\otimes M})_{Val},
																									\, z_{i_{1} \cdots i_{M} i_{M+1}} \cdot u_{i_{1} \cdots i_{M}} \in (A_{n})_{CoVal} , \right. \\
																									&   & \left. \; \forall\, v_{i_{M+1}} \in (A_{n})_{Val},\, z_{i_{1} \cdots i_{M} i_{M+1}} \cdot v_{i_{M+1}} \in (A_{n}^{\otimes M})_{CoVal}\right\} \\
																									& = & \left\{z_{i_{1} \cdots i_{M+1}}: \exists y \in [n] ,\, z_{i_{1} \cdots i_{M+1}} \cdot 0_{i_{1} \cdots i_{M}} \in \{\delta_{i_{M+1}y}, 0_{i_{M+1}}\},\right. \\
																									&   &	\;\; \forall \mathbf{x} \in [n]^{M}, \exists y \in [n] ,\, z_{i_{1} \cdots i_{M+1}} \cdot \tensor*{\delta}{*^{x_{1}}_{i_{1}}^{\;\cdots\;}_{\;\cdots}^{x_{M}}_{i_{M}}} \in \{\delta_{i_{M+1}y}, 0_{i_{M+1}}\}, \\
																									&   &	\;\; \exists \mathbf{y} \in [n]^{M} ,\, z_{i_{1} \cdots i_{M+1}} \cdot 0_{i_{M+1}} \in \{\tensor*{\delta}{*^{y_{1}}_{i_{1}}^{\;\cdots\;}_{\;\cdots}^{y_{M}}_{i_{M}}}, 0_{i_{1} \cdots i_{M}}\}, \\
																									&   &	\;\left.\;  \forall x_{M+1} \in [n], \exists \mathbf{y} \in [n]^{M} ,\, z_{i_{1} \cdots i_{M+1}} \cdot \delta_{i_{M+1}x_{M+1}} \in \PPerm(M,n)\right\} \\
																									& = & \left\{z_{i_{1} \cdots i_{M+1}}: \forall _{j=1}^{M+1} \exists y \in [n] ,\, z_{i_{1} \cdots i_{M+1}} \cdot \prod_{k \neq j} \delta_{i_{k}x_{k}} = \{\delta_{i_{j}y}, 0_{i_{j}}\}\right\} \\
																									& = & \PPerm(M+1,n)
\end{eqnarray*}

The inductive evaluations of the values and covalues for $C_{n}^{\otimes M+1}$ are simpler but essentially identical. \qed

If a functor $F$ is in multiplicative disjunctive normal form, then the object $F(\mathbf{R},\mathbf{R})$ takes the form
\vspace{1mm}
\begin{displaymath}
  \binvamp_{m=1}^{M}(\botimes_{l=1}^{L_{m}}R^{\phi(m,l)})
\end{displaymath}
\smallskip

\noindent for some $M$ and $L_{1},\ldots,L_{M}$, where $\mathbf{R} = (R,\ldots,R)$ for some object $R = (n\I,U,X)$, and $\phi: \bigcup_{m=1}^{M}\{(m,l):l \in [L_{m}]\} \longrightarrow \{0,1\}$ indicates the number of times the negation functor is applied to one of the instances of the object $R$. If $R$ is self-dual like each all objects of the form $A_{n}$ and $C_{n}$, and we continue to use tensor representations, then the function $\phi$ becomes irrelevant and may be ignored to all intents and purposes. The sets of values and covalues in tensor representation form of the object $F(\mathbf{R},\mathbf{R})$ are then the same as those of the object $\binvamp_{m=1}^{M}R^{\otimes L_{m}}$. Knowing the values and covalues of the tensor powers of the objects $A_{n}$ and $C_{n}$ for each $n$ allows us to calculate $F(\mathbf{R},\mathbf{R})_{Val}$ and $F(\mathbf{R},\mathbf{R})_{CoVal}$ concretely for $\mathbf{R} = \mathbf{A}_{n} = (A_{n},\ldots,A_{n})$ or $\mathbf{C}_{n} = (C_{n},\ldots,C_{n})$. As in the previous chapter, to aid clarity of arguments, the names of the indices used in these sets are based upon the polarity of the literal to which they are tied: `$i$'-indices represent positive literals, and `$j$'-indices negative ones. In a block $m$, we say that there are $P_{m}$ positive and $N_{m}$ negative literals; and in total there are $L_{m} = P_{m} + N_{m}$ literals.

\begin{eqnarray*}
  F(\mathbf{A}_{n},\mathbf{A}_{n})_{Val}	& = & \Big\{\tensor*{z}{*^{j_{(1,1)}}_{i_{(1,
          1)}}^{\;\cdots\;}_{\;\cdots}^{j_{(M, N_{M})}}_{i_{(M,
          P_{M})}}}
    : \forall k \in [M], 
    \bforall_{m \neq k} a^{m} \in \PPerm(L_m,n), \\ 
  & & \;\exists
  (x_{1},\ldots,x_{P_{k}},y_{1},\ldots,y_{N_{k}}) \in
  [n]^{L_{k}} \\ 
  & &  \;z^{\textbf{j}}_{\textbf{i}}\cdot\textstyle\prod_{m \neq
    k}\tensor*{(a^{m})}{*^{j_{(m,1)}}_{i_{(m,
        1)}}^{\;\cdots\;}_{\;\cdots}^{j_{(m, N_{m})}}_{i_{(m,
        P_{m})}}} 
  =  
  \tensor*{\delta}{*^{x_{1}}_{i_{(k,
        1)}}^{\;\cdots\;}_{\;\cdots}^{y_{N_k}}_{j_{(k, N_k)}}}
  \text{ or } 
  0_{i_{(k,1)} \cdots j_{(k,N_{k})}}\Big\} \\
 & & \\
	F(\mathbf{A}_{n},\mathbf{A}_{n})_{CoVal}	& = & \left\{\prod_{m=1}^{M}\tensor*{(a^{m})}{*^{j_{(m,1)}}_{i_{(m,
        1)}}^{\;\cdots\;}_{\;\cdots}^{j_{(m, N_{m})}}_{i_{(m,P_{m})}}} : a^{m} \in \PPerm(L_{m},n)\right\}
\end{eqnarray*}

\hughg{\begin{eqnarray*} \label{MDNFvaluesDesc}
  F(\mathbf{C}_{n},\mathbf{C}_{n})_{Val}	& = & \Big\{\tensor*{z}{*^{j_{(1,1)}}_{i_{(1,
          1)}}^{\;\cdots\;}_{\;\cdots}^{j_{(M, N_{M})}}_{i_{(M,
          P_{M})}}}
    : \forall k \in [M], 
    \bforall_{m \neq k} c^{m} \in \Perm(L_m,n), \\ 
  & & \;\exists
  (x_{1},\ldots,x_{P_{k}},y_{1},\ldots,y_{N_{k}}) \in
  [n]^{L_{k}} \\ 
  & &  \;z^{\textbf{j}}_{\textbf{i}}\cdot\textstyle\prod_{m \neq
    k}\tensor*{(c^{m})}{*^{j_{(m,1)}}_{i_{(m,
        1)}}^{\;\cdots\;}_{\;\cdots}^{j_{(m, N_{m})}}_{i_{(m,
        P_{m})}}} 
  =  
  \tensor*{\delta}{*^{x_{1}}_{i_{(k,
        1)}}^{\;\cdots\;}_{\;\cdots}^{y_{N_k}}_{j_{(k, N_k)}}}\Big\} \\
 & & \\
	F(\mathbf{C}_{n},\mathbf{C}_{n})_{CoVal}	& = & \left\{\prod_{m=1}^{M}\tensor*{(c^{m})}{*^{j_{(m,1)}}_{i_{(m,
        1)}}^{\;\cdots\;}_{\;\cdots}^{j_{(m, N_{m})}}_{i_{(m,P_{m})}}} : c^{m} \in \Perm(L_{m},n)\right\}
\end{eqnarray*}}

\noindent The derivation of these sets follows from a simple induction on the number of blocks of tensor powers which exist in an MDNF~formula, using the same principles as demonstrated in Lemma~\ref{ClaimTensorPowers} and Section~\ref{SectionDGTensor}.

\p If an MDNF~transformation $\tau$ in $\C$ representing a linear combination of sets of axiom links over an MDNF~sequent modelled by a functor $F$ is to exist in the glued category, then $\tau_{\mathbf{R}}$ must belong to both $F(\mathbf{A}_{n},\mathbf{A}_{n})$ and $F(\mathbf{C}_{n},\mathbf{C}_{n})$ when $\mathbf{R} = (n\I,\ldots,n\I) \in Obj(\C)^{N}$. The rest of this section is dedicated to showing how we can always find permutations (either full or partial) that expose the inability of the aforementioned arrow to belong to at least one of these sets if the transformation models an unwanted axiom link combination.

\subsection{Zero Transformations} \label{SectionZeroTransProof}

A zero transformation is an \mll~transformation found in every compact closed category with finite biproducts whose constituent arrows are all zero morphisms. They could be viewed from a certain perspective as the representation of the statement of a sequent being provable with no evidence, even if the sequent is in fact unprovable by standard linear logic reasoning. There is no place for such transformations in fully complete models of \mll, and so it is fortunate that their absence from the category $\GC$ can be deduced with minimal effort.

\p It is actually worthwhile proving a stronger result than merely the zero transformations being excluded from the glued category. We show that a linear combination of proof structures cannot possibly be modelled by an \mll~transformation in $\GC$ unless the sum of the scalars given to each of the proof structures totals exactly $1$. Although this is not a particularly strong result---in particular in categories such as $\mathbf{GRel}$, where the result is in fact equivalent to stating that zero transformations are forbidden---it is necessary for the final full completeness proof to be finished. The test object which provides the clearest proof of the desired result is $C_{1} = (\I,\{1\},\{1\})$. The object is in fact the tensor unit $\I_{1}$ of the sub-$*$-autonomous category $\GiC$ defined in Section~\ref{SectionFocusedOrth}, and the properties its position as a unit of a model of \mllmix\ bestows upon it are pivotal in the lemma below.

\begin{prop} \label{OnlyOneLemma} Every \mll~transformation in $\GC$
  models a linear combination of proof structures where the scalars applied to the constituent linkings sum to $1$.
\end{prop}
\proof
Using induction, we observe that, for every \mll~functor $F$, the arrows in the values and covalues of $F(\mathbf{C}_{1},\mathbf{C}_{1})$ have the tensor representation $1$. 
\begin{itemize}
\item The base case is trivially true: $C_{1} = C_{1}$, and $(C_{1})_{Val} = \{1\} = (C_{1})_{CoVal}$.
\item $R^{\bot} = ((UR)^{*},R_{CoVal},R_{Val}) = (R^{*},\{1\},\{1\})$.
\item $R \otimes S = (U(R \otimes S),R_{Val} \otimes S_{Val},\{z: 1 \cdot z \in \{1\},\,z \cdot 1 \in \{1\}\}) = (U(R \otimes S),\{1\},\{1\})$.
\item $R \invamp S = (R^{\bot} \otimes S^{\bot})^{\bot}$, and so this step may be deduced from the previous two.
\end{itemize}

A linear combination of proof structures is modelled by a tensor $\textstyle\sum_{\beta} s_{\beta}\cdot
    \tensor*{\delta}{*^{j_{\beta(1,1)}}_{i_{(1,
          1)}}^{\;\cdots\;}_{\;\cdots}^{j_{\beta(M, P_{M})}}_{i_{(M,
          P_M)}}}$, summing over bijections $\beta$ from the set of indices for positive indices to the set of indices for negative ones, with $s_{\beta} \in \C[\I,\I]$ for all $\beta$, when the underlying $\C$-objects of the inputs to $F$ are of the form $nI$ for some $n \in \N^{+}$. When $n=1$, as is the case when $C_{1}$ is the only $\GC$-object being used in $F$, the deltas become trivial (all the indices must be given the value $1$), meaning that the tensor becomes $\textstyle\sum_{\beta} s_{\beta}$; that is, the proof structures are modelled by the sum of the scalars associated with the proof structures.

 \p The only scalar found in $F(\mathbf{C}_{1},\mathbf{C}_{1})$ is $1$ regardless of the form of $F$, meaning morphisms for any sum of scalars not equalling $1$ do not lift to the homset $\GC[\I,F(\mathbf{C}_{1},\mathbf{C}_{1})]$. As such, because not all the arrows required to form them in the glued category can be seen, transformations describing linear combinations of proof structures whose scalars do not add up to $1$ in their semiring cannot translate into $\GC$ either. \qed

\begin{cor}
No zero transformation $0_{\K,F}:\K \longrightarrow F$ in $\C$ exists in the category $\GC$.
\end{cor}

\subsection{Acyclicity} \label{SubsectionAcyclic}

It is first shown that each proof structure that is part of a linear combination being modelled in $\GC$ satisfies the acyclicity criterion of Danos and Regnier \cite{DR89}. The concept behind how this is done can be understood by considering a few simple examples. The most basic example of a cyclic proof structure involves a single axiom link:
\begin{center} \vspace{5mm}
\begin{tikzpicture}
  [auto, node
  distance=5mm, skip loop/.style={to path={-- ++(0,#1) -| (\tikztotarget)}}] \tikzstyle{every node} = [text depth=-5pt,text height=0.5ex]
   \node (1) {$L$}; \node (a) [right of=1] {$\tens$} ;
  \node (2) [right of=a] {$L^{\bot}$};
\begin{scope}
\path   (1)  edge [black, skip loop =6mm, shorten >=3mm, shorten <=3mm]  (2);
\end{scope}
\end{tikzpicture} \vspace{5mm}
\end{center}

\p Modelling $L$ with $A_{n}$ for some $n > 1$, the values of the resultant object are square $(n \times n)$-matrices with at most one entry being non-zero, and if such an entry exists it must be the number $1$. However, the tensor representation of the proof structure desired is $\delta_{ij}$, which for this representation of $L$ has $n$ non-zero entries. There are too many entries containing non-zero positions. This fault still remains if we consider scalar multiples of the proof structure.\footnote{Multiplying the structure by zero naturally removes the problem. However, this produces the zero morphism; the existence of a zero transformation has already been contradicted in Proposition~\ref{OnlyOneLemma}.}

The same problem occurs with larger cycles containing more than one axiom link and more than one block.
\begin{center} \vspace{5mm}
\begin{tikzpicture}
  [auto, node
  distance=5mm, skip loop/.style={to path={-- ++(0,#1) -| (\tikztotarget)}}] \tikzstyle{every node} = [text depth=-5pt,text height=0.5ex]
   \node (1) {$(L$}; \node (a) [right of=1] {$\tens$} ;
  \node (2) [right of=a] {$L^{\bot})$}; \node (b) [right of=2] {$\invamp$} ;
  \node (3)  [right of=b] {$(L$}; \node (c)  [right of=3] {$\tens$};
  \node (4)  [right of=c] {$L^{\bot})$}; \node(d) [right of=4] {$\invamp$} ;
  \node (5) [right of=d] {$(L$}; \node (e) [right of=5] {$\tens$} ;
  \node (6) [right of=e] {$L^{\bot})$};
\begin{scope}
\path (1)  edge  [black, skip loop =7mm, shorten >=3mm, shorten <=3mm] (6);
\path   (3)  edge [black, skip loop =6mm, shorten >=3mm, shorten <=3mm]  (2);
\path (5)  edge  [black, skip loop =6mm, shorten >=3mm, shorten <=3mm] (4);
\end{scope}
\end{tikzpicture} \vspace{5mm}
\end{center}

The proof structure above is described by $\tensor*{\delta}{*^{j_{3}}_{i_{1}}^{j_{1}}_{i_{2}}^{j_{2}}_{i_{3}}}$ ($i_{k}$ and $j_{k}$ relating to the positive and negative literals in the $k^{th}$ block as is the norm). One of the criteria a tensor $\tensor*{z}{*^{j_{1}}_{i_{1}}^{j_{2}}_{i_{2}}^{j_{3}}_{i_{3}}}$ must satisfy if $\bbrk{L} = A_{n}$ for large enough $n$ in order for it to belong to the values of the corresponding object is for every pair of partial ($2$-)permutations over $[n]$, $a_{i_{1}j_{1}}$ and $b_{i_{2}j_{2}}$ say, to create an $(n \times n)$-matrix with at most one non-zero entry when composed with it. If we say that $a_{i_{1}j_{1}} = \tensor*{\delta}{*^{1}_{i_{1}}^{2}_{j_{1}}} + \tensor*{\delta}{*^{3}_{i_{1}}^{4}_{j_{1}}}$ and $b_{i_{2}j_{2}} = \tensor*{\delta}{*^{2}_{i_{2}}^{5}_{j_{2}}} + \tensor*{\delta}{*^{4}_{i_{2}}^{6}_{j_{2}}}$ --- which are indeed partial permutations --- we find
\begin{eqnarray*}
a_{i_{1}j_{1}} \cdot b_{i_{2}j_{2}} \cdot \tensor*{\delta}{*^{j_{3}}_{i_{1}}^{j_{1}}_{i_{2}}^{j_{2}}_{i_{3}}} & = & (\tensor*{\delta}{*^{1}_{i_{1}}^{2}_{j_{1}}} + \tensor*{\delta}{*^{3}_{i_{1}}^{4}_{j_{1}}}) \cdot (\tensor*{\delta}{*^{2}_{i_{2}}^{5}_{j_{2}}} + \tensor*{\delta}{*^{4}_{i_{2}}^{6}_{j_{2}}}) \cdot \tensor*{\delta}{*^{j_{3}}_{i_{1}}^{j_{1}}_{i_{2}}^{j_{2}}_{i_{3}}} \\
 & = & (\tensor*{\delta}{*^{1}_{i_{1}}^{2}_{j_{1}}} \cdot \tensor*{\delta}{*^{2}_{i_{2}}^{5}_{j_{2}}} \cdot \tensor*{\delta}{*^{j_{3}}_{i_{1}}^{j_{1}}_{i_{2}}^{j_{2}}_{i_{3}}}) + (\tensor*{\delta}{*^{1}_{i_{1}}^{2}_{j_{1}}} \cdot \tensor*{\delta}{*^{4}_{i_{2}}^{6}_{j_{2}}} \cdot \tensor*{\delta}{*^{j_{3}}_{i_{1}}^{j_{1}}_{i_{2}}^{j_{2}}_{i_{3}}}) \\
 &   & \; + \;(\tensor*{\delta}{*^{3}_{i_{1}}^{4}_{j_{1}}} \cdot \tensor*{\delta}{*^{2}_{i_{2}}^{5}_{j_{2}}} \cdot \tensor*{\delta}{*^{j_{3}}_{i_{1}}^{j_{1}}_{i_{2}}^{j_{2}}_{i_{3}}}) + (\tensor*{\delta}{*^{3}_{i_{1}}^{4}_{j_{1}}} \cdot \tensor*{\delta}{*^{4}_{i_{2}}^{6}_{j_{2}}} \cdot \tensor*{\delta}{*^{j_{3}}_{i_{1}}^{j_{1}}_{i_{2}}^{j_{2}}_{i_{3}}}) \\
 & = & \tensor*{\delta}{*^{j_{3}}_{1}^{2}_{2}^{5}_{i_{3}}} + \tensor*{\delta}{*^{j_{3}}_{1}^{2}_{4}^{6}_{i_{3}}} + \tensor*{\delta}{*^{j_{3}}_{3}^{4}_{2}^{5}_{i_{3}}} + \tensor*{\delta}{*^{j_{3}}_{3}^{4}_{4}^{6}_{i_{3}}} = \tensor*{\delta}{*^{5}_{i_{3}}^{1}_{j_{3}}} + \tensor*{\delta}{*^{6}_{i_{3}}^{3}_{j_{3}}}
\end{eqnarray*}
The resulting matrix clearly has two non-zero entries: when $i_{3} = 5$ and $j_{3} = 1$; and when $i_{3} = 6$ and $j_{3} = 3$. As such, this particular cyclic structure is shown to be represented neither in the set of values of $(A_{n} \otimes A_{n}^{\bot}) \invamp (A_{n} \otimes A_{n}^{\bot}) \invamp (A_{n} \otimes A_{n}^{\bot})$, nor the \mll~transformations of $\GC$.

\p The reason why this argument is possible is based on the fact that every block involved in a cycle has two literals incident to axiom links in the cycle. The two literals in the first block may be given tensors containing two non-zero positions which compose with the proof structure tensor because they are in the same block, and $(A_{n} \otimes A_{n}^{\bot})_{CoVal}$ contains all partial permutations; the same is true of those in the second block. If they were not, as would be the situation in the proof structure below, only one non-zero entry would be able to be found: tensors in $(A_{n} \invamp A_{n}^{\bot})_{CoVal}$ only have at most one value of $1$, with the rest being zeroes.

\begin{center} \vspace{5mm}
\begin{tikzpicture}
  [auto, node
  distance=5mm, skip loop/.style={to path={-- ++(0,#1) -| (\tikztotarget)}}] \tikzstyle{every node} = [text depth=-5pt,text height=0.5ex]
   \node (1) {$L$}; \node (a) [right of=1] {$\invamp$} ;
  \node (2) [right of=a] {$L^{\bot}$}; \node (b) [right of=2] {$\invamp$} ;
  \node (3)  [right of=b] {$(L$}; \node (c)  [right of=3] {$\tens$};
  \node (4)  [right of=c] {$L^{\bot})$}; \node(d) [right of=4] {$\invamp$} ;
  \node (5) [right of=d] {$(L$}; \node (e) [right of=5] {$\tens$} ;
  \node (6) [right of=e] {$L^{\bot})$};
\begin{scope}
\path (1)  edge  [black, skip loop =7mm, shorten >=3mm, shorten <=3mm] (6);
\path   (3)  edge [black, skip loop =6mm, shorten >=3mm, shorten <=3mm]  (2);
\path (5)  edge  [black, skip loop =6mm, shorten >=3mm, shorten <=3mm] (4);
\end{scope}
\end{tikzpicture} \vspace{2mm}
\end{center}

The objective of the coming proof is to generate partial permutations with two non-zero positions as above for all bar one of the blocks connected to the axiom links in a chosen cycle; and these tensors should compose with the Kronecker delta tensor representing an incorrect proof structure in a linear combination being considered to create a tensor with more than one entry not equalling zero. Of course, the scenarios offered so far have been curtailed in two ways: all of their axiom links are involved in the cycle (and there is only one cycle), and linear combinations of two or more distinct proof structures are absent. The algorithm below deals with both of these problems, and this is discussed in more detail after its description.

\begin{algo} \label{AcyclicAlg} Input: A cyclic MDNF~proof structure with linking $\lambda$ containing $M$ blocks, the $m^{th}$ of which containing $L_{m}$ literals, and one of its minimal cycles $\hat{\lambda}$. \\ Output: A number $n \in \N^{+}$;
  tensors $a^{1},\ldots,a^{M}$ such that $\hughf{(a^{m})}_{i_{1} \cdots
    i_{L_{m}}} \in \PPerm(L_{m},n)$ for each $m$.
\end{algo}
\begin{enumerate}
\item Let $i=0$, and note that, at this point, none of the links in $\lambda$ has been dealt
  with.

\item Find a link $l \in \lambda \backslash \hat{\lambda}$ which has
  not been dealt with yet.
  \begin{enumerate}
  \item If one should exist, then assign the number $i+1$ to both
    literals incident to~$l$. Increment $i$, and go to Step~2.
  \item If one does not exist, then move to Step~3.
  \end{enumerate}

\item Find a link $l \in \hat{\lambda}$ which has not been dealt with
  yet.
  \begin{enumerate}
  \item If one should exist, then assign both the numbers $i+1$ and
    $i+2$ to both literals incident to $l$. Increase the value of $i$
    by $2$, and restart Step~3.
  \item If one does not exist, then move to Step~4.
  \end{enumerate}
	
\item For each tensor product of literals which does not contain a
  literal incident to a link within $\hat{\lambda}$, place the values
  assigned to each literal into a tuple in the same order as their
  literals appear in the subformula. This tuple `belongs' to that
  subformula.

\item For each tensor product of literals which does contain a
  literal incident to at least one link within $\hat{\lambda}$,
  create two tuples as follows:
  \begin{enumerate}
  \item Place the lowest values assigned to each literal into a tuple
    in the same order as their literals appear in the subformula.
  \item Place the highest values assigned to each literal into a
    tuple in the same order as their literals appear in the
    subformula.\footnote{If a literal has been assigned only one
      number, then this number is indeed considered both the highest
      and lowest value.}
  \end{enumerate}
	
\item Set $n = i$; and for each $m \in [M]$, define an element
  $\hughf{(a^{m})}_{i_{1} \cdots i_{L_{m}}}$ of $\PPerm(L_{m},n)$ as follows:
  \begin{displaymath}
    \hughf{(a^{m})}_{i_{1} \cdots i_{L_{m}}} = \left\{ 
      \begin{array}{ll}
        1 & \text{if }(i_{1},\ldots,i_{L_{m}}) \text{ is a tuple
          for block } m \\ 
        0 & \text{otherwise.} 
      \end{array}
    \right.
  \end{displaymath}\smallskip
\end{enumerate}

\noindent The tensors created in this algorithm are indeed always partial permutations. The tensors associated with blocks which are not connected to the cycle $\hat{\lambda}$ only ever contain one non-zero entry, and that entry is $1$. In this type of situation it is abundantly clear that there are only $1$s and $0$s as entries, and there cannot be two $1$s in the same column. For those blocks connected to the cycle, there are two $1$s in their tensors, but since the two tuples which define their positions in the tensor differ in more than one component, they cannot exist in the same column, and so they can exist together in a partial permutation. An example of how the algorithm functions with a typical input is given below.

\begin{exa} \label{AcyclicExam1}
Consider the sum of the two linkings $\lambda_{1}$ and $\lambda_{2}$ seen in the following diagram:
\end{exa}
\begin{center} \vspace{5mm}
\begin{tikzpicture}
  [auto, node
  distance=5mm, skip loop/.style={to path={-- ++(0,#1) -| (\tikztotarget)}}] \tikzstyle{every node} = [text depth=-5pt,text height=0.5ex]
  \node (z1) { }; 
   \node (1) [right of=z1] {$L$}; \node (a) [right of=1] {$\invamp$} ;
  \node (2) [right of=a] {$L^{\bot}$}; \node (b) [right of=2] {$\invamp$} ;
  \node (3)  [right of=b] {$(L$}; \node (c)  [right of=3] {$\tens$};
  \node (4)  [right of=c] {$L^{\bot})$}; \node(d) [right of=4] {$\invamp$} ;
  \node (5) [right of=d] {$(L$}; \node (e) [right of=5] {$\tens$} ;
  \node (6) [right of=e] {$L^{\bot}$}; \node (f) [right of=6] {$\tens$} ;
  \node (7)  [right of=f] {$L)$}; \node (g)  [right of=7] {$\invamp$};
  \node (8)  [right of=g] {$L^{\bot}$};
  \node (t1) [above of= z1] {$\lambda_{2}$};
  \node (t2) [above of= t1] {$\lambda_{1}$};
\begin{scope}
\path   (1)  edge [red, skip loop =11mm,shorten >=8mm, shorten <=8mm]  (2);
\path (3)  edge  [red, skip loop =12mm,shorten >=8mm, shorten <=8mm] (6);
\path   (5)  edge [red, skip loop =11mm,shorten >=8mm, shorten <=8mm]  (4);
\path (7)  edge  [red, skip loop =11mm,shorten >=8mm, shorten <=8mm] (8);
\path   (1)  edge [blue, skip loop =7mm, shorten >=3mm, shorten <=3mm]  (6);
\path (2)  edge  [blue, skip loop =6mm, shorten >=3mm, shorten <=3mm] (3);
\path   (5)  edge [blue, skip loop =6mm, shorten >=3mm, shorten <=3mm]  (4);
\path (8)  edge  [blue, skip loop =6mm, shorten >=3mm, shorten <=3mm] (7);
\end{scope}
\end{tikzpicture} \vspace{5mm}
\end{center}

{\noindent\textit{The tensor representation of this linear combination of axiom links is  $\tensor*{\delta}{*^{j_{(2,1)}}_{i_{(1,1)}}^{j_{(4,1)}}_{i_{(3,1)}}^{j_{(3,1)}}_{i_{(4,1)}}^{j_{(5,1)}}_{i_{(4,2)}}} + \tensor*{\delta}{*^{j_{(4,1)}}_{i_{(1,1)}}^{j_{(2,1)}}_{i_{(3,1)}}^{j_{(3,1)}}_{i_{(4,1)}}^{j_{(5,1)}}_{i_{(4,2)}}}$. We apply Algorithm \ref{AcyclicAlg}, choosing $\lambda_{1}$ to be $\lambda$, and the subset of axiom links drawn below to be $\hat{\lambda}$ (with representation $\tensor*{\delta}{*^{j_{(4,1)}}_{i_{(3,1)}}^{j_{(3,1)}}_{i_{(4,1)}}}$)}} 
\begin{center} \vspace{2mm}
\begin{tikzpicture}
  [auto, node
  distance=5mm, skip loop/.style={to path={-- ++(0,#1) -| (\tikztotarget)}}] \tikzstyle{every node} = [text depth=-5pt,text height=0.5ex]
 \node (z1) { }; 
    \node (1) [right of=z1] {$L$}; \node (a) [right of=1] {$\invamp$} ;
   \node (2) [right of=a] {$L^{\bot}$}; \node (b) [right of=2] {$\invamp$} ;
   \node (3)  [right of=b] {$(L$}; \node (c)  [right of=3] {$\tens$};
   \node (4)  [right of=c] {$L^{\bot})$}; \node(d) [right of=4] {$\invamp$} ;
   \node (5) [right of=d] {$(L$}; \node (e) [right of=5] {$\tens$} ;
   \node (6) [right of=e] {$L^{\bot}$}; \node (f) [right of=6] {$\tens$} ;
   \node (7)  [right of=f] {$L)$}; \node (g)  [right of=7] {$\invamp$};
   \node (8)  [right of=g] {$L^{\bot}$};
  \node (t1) [above of= z1] { };
  \node (t2) [above of= t1] { };
\begin{scope}
\path (3)  edge  [magenta, skip loop =7mm, shorten >=3mm, shorten <=3mm] (6);
\path   (5)  edge [magenta, skip loop =6mm, shorten >=3mm, shorten <=3mm]  (4);
\end{scope}
\end{tikzpicture} \vspace{5mm}
\end{center}
	
	\begin{enumerate}[label={\cW4 \& }\arabic*.]
\item[2.] Step 2 from the algorithm is iterated twice. Starting with the left most axiom link from $\lambda \backslash \hat{\lambda}$, the literals, along with the numbers to them are as follows:
\begin{center} \vspace{0mm}
\begin{tikzpicture}
  [auto, node
  distance=5mm, skip loop/.style={to path={-- ++(0,#1) -| (\tikztotarget)}}] \tikzstyle{every node} = [text depth=-5pt,text height=0.5ex]
  \node (z1) { }; 
     \node (1) [right of=z1] {$L$}; \node (a) [right of=1] {$\invamp$} ;
    \node (2) [right of=a] {$L^{\bot}$}; \node (b) [right of=2] {$\invamp$} ;
    \node (3)  [right of=b] {$(L$}; \node (c)  [right of=3] {$\tens$};
    \node (4)  [right of=c] {$L^{\bot})$}; \node(d) [right of=4] {$\invamp$} ;
    \node (5) [right of=d] {$(L$}; \node (e) [right of=5] {$\tens$} ;
    \node (6) [right of=e] {$L^{\bot}$}; \node (f) [right of=6] {$\tens$} ;
    \node (7)  [right of=f] {$L)$}; \node (g)  [right of=7] {$\invamp$};
    \node (8)  [right of=g] {$L^{\bot}$};
  \node (t1) [above of= z1] { };
  \node (t2) [above of= t1] { };
  \node (11) [below of=1] {$1$};
  \node (12) [below of=2] {$1\,\,$};
  \node (17) [below of=7] {$2$};
  \node (18) [below of=8] {$2\,$};
\begin{scope}
\path   (1)  edge [red, skip loop =6mm, shorten >=3mm, shorten <=3mm]  (2);
\path (7)  edge  [red, skip loop =6mm, shorten >=3mm, shorten <=3mm] (8);
\end{scope}
\end{tikzpicture} \vspace{5mm}
\end{center}
	
\item[3.] Like Step 2, Step 3 must also be repeated. Starting once again from the left, the following number assignments are given:
\begin{center} \vspace{2mm}
\begin{tikzpicture}
  [auto, node
  distance=5mm, skip loop/.style={to path={-- ++(0,#1) -| (\tikztotarget)}}] \tikzstyle{every node} = [text depth=-5pt,text height=0.5ex]
  \node (z1) { }; 
     \node (1) [right of=z1] {$L$}; \node (a) [right of=1] {$\invamp$} ;
    \node (2) [right of=a] {$L^{\bot}$}; \node (b) [right of=2] {$\invamp$} ;
    \node (3)  [right of=b] {$(L$}; \node (c)  [right of=3] {$\tens$};
    \node (4)  [right of=c] {$L^{\bot})$}; \node(d) [right of=4] {$\invamp$} ;
    \node (5) [right of=d] {$(L$}; \node (e) [right of=5] {$\tens$} ;
    \node (6) [right of=e] {$L^{\bot}$}; \node (f) [right of=6] {$\tens$} ;
    \node (7)  [right of=f] {$L)$}; \node (g)  [right of=7] {$\invamp$};
    \node (8)  [right of=g] {$L^{\bot}$};
  \node (t1) [above of= z1] { };
  \node (t2) [above of= t1] { };
  \node (13) [below of=3] {3};
  \node (14) [below of=4] {5};
  \node (15) [below of=5] {5};
  \node (16) [below of=6] {3};
  \node (23) [below of=13] {4};
  \node (24) [below of=14] {6};
  \node (25) [below of=15] {6};
  \node (26) [below of=16] {4};
\begin{scope}
\path (3)  edge  [magenta, skip loop =7mm, shorten >=3mm, shorten <=3mm] (6);
\path   (5)  edge [magenta, skip loop =6mm, shorten >=3mm, shorten <=3mm]  (4);
\end{scope}
\end{tikzpicture} \vspace{5mm}
\end{center}
	
\item[4 \& 5.] Tuples for the blocks are now created by merging the numbers for each of their literals.

\begin{center} \vspace{5mm}
\begin{tikzpicture}
  [auto, node
  distance=5mm, skip loop/.style={to path={-- ++(0,#1) -| (\tikztotarget)}}] \tikzstyle{every node} = [text depth=-5pt,text height=0.5ex]
 \node (z1) { }; 
    \node (1) [right of=z1] {$L$}; \node (a) [right of=1] {$\invamp$} ;
   \node (2) [right of=a] {$L^{\bot}$}; \node (b) [right of=2] {$\invamp$} ;
   \node (3)  [right of=b] {$(L$}; \node (c)  [right of=3] {$\tens$};
   \node (4)  [right of=c] {$L^{\bot})$}; \node(d) [right of=4] {$\invamp$} ;
   \node (5) [right of=d] {$(L$}; \node (e) [right of=5] {$\tens$} ;
   \node (6) [right of=e] {$L^{\bot}$}; \node (f) [right of=6] {$\tens$} ;
   \node (7)  [right of=f] {$L)$}; \node (g)  [right of=7] {$\invamp$};
   \node (8)  [right of=g] {$L^{\bot}$};
  \node (t1) [above of= z1] {$\lambda_{2}$};
  \node (t2) [above of= t1] {$\lambda_{1}$};
  \node (11) [below of=1] {[1]};
  \node (12) [below of=2] {[1]};
  \node (13) [below of=3] {[3}; \node (1c) [below of=c] {,};
  \node (14) [below of=4] {5]};
  \node (15) [below of=5] {[5}; \node (1e) [below of=e] {,};
  \node (16) [below of=6] {3}; \node (1f) [below of=f] {,};
  \node (17) [below of=7] {2]};
  \node (18) [below of=8] {[2]};
  \node (23) [below of=13] {[4}; \node (2c) [below of=1c] {,};
  \node (24) [below of=14] {6]};
  \node (25) [below of=15] {[6}; \node (2e) [below of=1e] {,};
  \node (26) [below of=16] {4}; \node (2f) [below of=1f] {,};
  \node (27) [below of=17] {2]};
\begin{scope}
\path   (1)  edge [red, skip loop =11mm,shorten >=8mm, shorten <=8mm]  (2);
\path (3)  edge  [red, skip loop =12mm,shorten >=8mm, shorten <=8mm] (6);
\path   (5)  edge [red, skip loop =11mm,shorten >=8mm, shorten <=8mm]  (4);
\path (7)  edge  [red, skip loop =11mm,shorten >=8mm, shorten <=8mm] (8);
\path   (1)  edge [blue, skip loop =7mm, shorten >=3mm, shorten <=3mm]  (6);
\path (2)  edge  [blue, skip loop =6mm, shorten >=3mm, shorten <=3mm] (3);
\path   (5)  edge [blue, skip loop =6mm, shorten >=3mm, shorten <=3mm]  (4);
\path (8)  edge  [blue, skip loop =6mm, shorten >=3mm, shorten <=3mm] (7);
\end{scope}
\end{tikzpicture} \vspace{5mm}
\end{center}	

\item[6.] The tensors $a^{1},\ldots,a^{5}$ are then created from the tuples above. 
\begin{eqnarray*}
& & \hughf{(a^{1})}_{i_{(1,1)}}										= \tensor*{\delta}{*^{1}_{i_{(1,1)}}} \\
& & \hughf{(a^{2})}_{j_{(2,1)}}										= \tensor*{\delta}{*^{1}_{j_{(2,1)}}} \\
& & \hughf{(a^{3})}_{i_{(3,1)}j_{(3,1)}}					= \tensor*{\delta}{*^{3}_{i_{(3,1)}}^{5}_{j_{(3,1)}}} + \tensor*{\delta}{*^{4}_{i_{(3,1)}}^{6}_{j_{(3,1)}}} \\
& & \hughf{(a^{4})}_{i_{(4,1)}j_{(4,1)}i_{(4,2)}}	= \tensor*{\delta}{*^{5}_{i_{(4,1)}}^{3}_{j_{(4,1)}}^{2}_{i_{(4,2)}}} + \tensor*{\delta}{*^{6}_{i_{(4,1)}}^{4}_{j_{(4,1)}}^{2}_{i_{(4,2)}}} \\
& & \hughf{(a^{5})}_{j_{(5,1)}}										= \tensor*{\delta}{*^{2}_{j_{(5,1)}}}
\end{eqnarray*}\medskip
\end{enumerate}

\noindent In the tuples underneath each of the blocks in the MDNF~formula given in Part 4/5 of the above example, it can be seen that each of the numbers being assigned to the literals occur exactly twice. It can be easily verified that this occurs with any valid choice of input to the algorithm: before a number is assigned to a pair of literals in either of Steps 2 or 3, the counter used to provide that number is incremented so previously assigned numbers are never reused; and numbers are assigned to two literals at a time. Furthermore, two literals are given the same number (or numbers) of another literal if and only if they share an axiom link in the chosen linking $\lambda$. This can be realised by noting that Steps 2 and 3 are actually, in principle, numbering the axiom links of $\lambda$, and the literals, and only the literals, incident to an axiom link take the numbers associated with that link. This principle allows us to differentiate the chosen $\lambda$ from all other linkings, as can be seen in the coming claims. 

\begin{lem} \label{AcyclicClaim1}
Suppose that we have a linear combination of proof structures including a non-zero instance of a linking $\lambda$ which contains a minimal cycle $\hat{\lambda}$, and choose a block $k$ through which $\hat{\lambda}$ passes. Then a tensor modelling a linking $\lambda'$ when composed with all the tensors except $a^{k}$ created by Algorithm \ref{AcyclicAlg} (when using $\lambda$ and $\hat{\lambda}$ as inputs) will result in a zero tensor if $\lambda$ and $\lambda'$ do not have an identical set of axiom links not connected to block $k$.
\end{lem}
\proof
Suppose there is a link $l \in \lambda' \backslash \lambda$ not connected to block $k$. Then the tensor modelling $\lambda'$ must take the form $\bar{\lambda'}_{ij\mathbf{ij}} = \delta_{ij}\omega_{\mathbf{ij}}$ for some tensor $\omega$, with $i$ and $j$ being indices associated with the literals connected via $l$. For the blocks $m_{i}$ and $m_{j}$ containing the literals allocated the index $i$ and $j$, the algorithm creates partial permutations built from summations of one or two tensors of the form $\delta_{ix}\gamma_{\mathbf{i}'}$ and $\delta_{jy}\chi_{\mathbf{j}'}$, with $x,y \in \N^{+}$ and $\gamma$ and $\chi$ products of Kronecker deltas whose details are of little relevance.

\p Since the two literals connected to $l$ are not linked in $\lambda$, the numbers given to those literals in the tuples created by the algorithm are different, and therefore we know $x \neq y$ for each of the parts of the partial permutations $a^{m_{i}}$ and $a^{m_{j}}$ of the form $\delta_{ix}\gamma_{\mathbf{i}'}$ and $\delta_{jy}\chi_{\mathbf{j}'}$. We therefore find that $\delta_{ij}\delta_{ix}\delta_{jy} = \delta_{xy} = 0$, meaning $(\delta_{ij}\omega_{\mathbf{ij}})(\delta_{ix}\gamma_{\mathbf{i}'})(\delta_{jy}\chi_{\mathbf{j}'}) = (\delta_{ij}\delta_{ix}\delta_{jy})\omega_{\mathbf{ij}}\gamma_{\mathbf{i}'}\chi_{\mathbf{j}'} = 0$ in all entries, and therefore $a^{m_{i}}a^{m_{j}}\bar{\lambda'} = 0$ and $\bar{\lambda'} \cdot \prod_{m \neq k}a^{m} = 0$ for the appropriate indices. \qed

The claim above can be seen clearly in action in Example \ref{AcyclicExam1}. If we assume that block $k = 4$, which is certainly in the cycle $\hat{\lambda}$ in $\lambda = \lambda_{1}$, then we see that the link in $\lambda_{2}$ from the left-most negative literal to the positive literal directly to its right, represented by the tensor $\delta_{i_{(3,1)}j_{(2,1)}}$, is neither in $\lambda_{1}$ nor incident to block~$4$. The algorithm provides two partial permutations $\hughf{(a^{2})}_{j_{(2,1)}} = \tensor*{\delta}{*^{1}_{j_{(2,1)}}}$ and $\hughf{(a^{3})}_{i_{(3,1)},j_{(3,1)}} = \tensor*{\delta}{*^{3}_{i_{(3,1)}}^{5}_{j_{(3,1)}}} + \tensor*{\delta}{*^{4}_{i_{(3,1)}}^{6}_{j_{(3,1)}}}$ which are intended to be composed directly with $\delta_{i_{(3,1)}j_{(2,1)}}$, and they produce a zero tensor as desired.
\begin{displaymath}
\delta_{i_{(3,1)}j_{(2,1)}} \cdot \tensor*{\delta}{*^{1}_{j_{(2,1)}}} \cdot (\tensor*{\delta}{*^{3}_{i_{(3,1)}}^{5}_{j_{(3,1)}}} + \tensor*{\delta}{*^{4}_{i_{(3,1)}}^{6}_{j_{(3,1)}}}) = \tensor*{\delta}{*^{1}_{3}^{5}_{j_{(3,1)}}} + \tensor*{\delta}{*^{1}_{4}^{6}_{j_{(3,1)}}} = 0_{j_{(3,1)}}
\end{displaymath}
The effects of the tensor relating to $\lambda_{2}$ are therefore eradicated when composed with all of $a^{1}$, $a^{2}$, $a^{3}$ and $a^{5}$.

\p There are examples of linear combinations of proof structures and choice of block $k$, unlike the one given above, where there is a linking $\lambda'$ which shares all of its axiom links not incident with block $k$ with $\lambda$. Lemma~\ref{AcyclicClaim1} is rendered useless in these situations when attempting to differentiate all other sets of axiom links from $\lambda$. The example below typifies such a dilemma.

\begin{exa} \label{AcyclicExam2}
Consider the sum of the three linkings $\lambda_{1}$,$\lambda_{2}$ and $\lambda_{3}$ seen in the diagram below.
\end{exa}
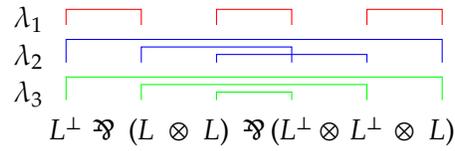
\begin{center} \vspace{5mm}
\begin{tikzpicture}
  [auto, node
  distance=5mm, skip loop/.style={to path={-- ++(0,#1) -| (\tikztotarget)}}] \tikzstyle{every node} = [text depth=-5pt,text height=0.5ex]
 \node (z1) { }; 
    \node (1) [right of=z1] {$L^{\bot}$}; \node (a) [right of=1] {$\invamp$} ;
   \node (2) [right of=a] {$(L$}; \node (b) [right of=2] {$\tens$} ;
   \node (3)  [right of=b] {$L)$}; \node (c)  [right of=3] {$\invamp$};
   \node (4)  [right of=c] {$(L^{\bot}$}; \node(d) [right of=4] {$\tens$} ;
   \node (5) [right of=d] {$L^{\bot}$}; \node (e) [right of=5] {$\tens$} ;
   \node (6) [right of=e] {$L)$};
  \node (t1) [above of= z1] {$\lambda_{3}$};
  \node (t2) [above of= t1] {$\lambda_{2}$};
  \node (t3) [above of= t2] {$\lambda_{1}$};
\begin{scope}
\path   (1)  edge [red, skip loop =16mm,shorten >=13mm, shorten <=13mm]  (2);
\path (3)  edge  [red, skip loop =16mm,shorten >=13mm, shorten <=13mm] (4);
\path   (5)  edge [red, skip loop =16mm,shorten >=13mm, shorten <=13mm]  (6);
\path   (1)  edge [blue, skip loop =12mm,shorten >=8mm, shorten <=8mm]  (6);
\path (2)  edge  [blue, skip loop =11mm,shorten >=8mm, shorten <=8mm] (4);
\path   (5)  edge [blue, skip loop =10mm,shorten >=8mm, shorten <=8mm]  (3);
\path (1)  edge  [green, skip loop =7mm, shorten >=3mm, shorten <=3mm] (6);
\path (2)  edge  [green, skip loop =6mm, shorten >=3mm, shorten <=3mm] (5);
\path (3)  edge  [green, skip loop =5mm, shorten >=3mm, shorten <=3mm] (4);
\end{scope}
\end{tikzpicture} \vspace{5mm}
\end{center}
\textit{Their corresponding tensor representation is $\tensor*{\delta}{*^{j_{(1,1)}}_{i_{(2,1)}}^{j_{(3,1)}}_{i_{(2,2)}}^{j_{(3,2)}}_{i_{(3,1)}}} + \tensor*{\delta}{*^{j_{(1,1)}}_{i_{(3,1)}}^{j_{(3,2)}}_{i_{(2,1)}}^{j_{(3,1)}}_{i_{(2,2)}}} + \tensor*{\delta}{*^{j_{(1,1)}}_{i_{(3,1)}}^{j_{(3,1)}}_{i_{(2,1)}}^{j_{(3,2)}}_{i_{(2,2)}}}$. Choosing $\lambda_{3}$ to be $\lambda$, and taking the subset of links given in the figure below to be $\hat{\lambda}$ (with representation $\tensor*{\delta}{*^{j_{(3,2)}}_{i_{(2,1)}}^{j_{(3,1)}}_{i_{(2,2)}}}$)}
\begin{center} \vspace{5mm}
\begin{tikzpicture}
  [auto, node
  distance=5mm, skip loop/.style={to path={-- ++(0,#1) -| (\tikztotarget)}}] \tikzstyle{every node} = [text depth=-5pt,text height=0.5ex]
 \node (z1) { }; 
    \node (1) [right of=z1] {$L^{\bot}$}; \node (a) [right of=1] {$\invamp$} ;
   \node (2) [right of=a] {$(L$}; \node (b) [right of=2] {$\tens$} ;
   \node (3)  [right of=b] {$L)$}; \node (c)  [right of=3] {$\invamp$};
   \node (4)  [right of=c] {$(L^{\bot}$}; \node(d) [right of=4] {$\tens$} ;
   \node (5) [right of=d] {$L^{\bot}$}; \node (e) [right of=5] {$\tens$} ;
   \node (6) [right of=e] {$L)$};
  \node (t1) [above of= z1] {$\hat{\lambda}$};
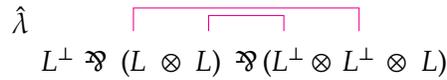
\begin{scope}
\path (2)  edge  [magenta, skip loop =7mm, shorten >=3mm, shorten <=3mm] (5);
\path (3)  edge  [magenta, skip loop =6mm, shorten >=3mm, shorten <=3mm] (4);
\end{scope}
\end{tikzpicture} \vspace{5mm}
\end{center}

\begin{enumerate}[label={\cW4 \& }\arabic*.]
\item[2.] Step 2 from the algorithm is only used once. We provide the single pair of literals connected to the sole link in $\lambda \backslash \hat{\lambda}$ with the number $1$.
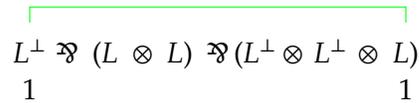
\begin{center} \vspace{5mm}
\begin{tikzpicture}
  [auto, node
  distance=5mm, skip loop/.style={to path={-- ++(0,#1) -| (\tikztotarget)}}] \tikzstyle{every node} = [text depth=-5pt,text height=0.5ex]
 \node (z1) { }; 
    \node (1) [right of=z1] {$L^{\bot}$}; \node (a) [right of=1] {$\invamp$} ;
   \node (2) [right of=a] {$(L$}; \node (b) [right of=2] {$\tens$} ;
   \node (3)  [right of=b] {$L)$}; \node (c)  [right of=3] {$\invamp$};
   \node (4)  [right of=c] {$(L^{\bot}$}; \node(d) [right of=4] {$\tens$} ;
   \node (5) [right of=d] {$L^{\bot}$}; \node (e) [right of=5] {$\tens$} ;
   \node (6) [right of=e] {$L)$};
  \node (11) [below of=1] {1};
  \node (16) [below of=6] {1};
\begin{scope}
\path (1)  edge  [green, skip loop =6mm, shorten >=3mm, shorten <=3mm] (6);
\end{scope}
\end{tikzpicture} \vspace{5mm}
\end{center}

\item[3.] Step 3 is repeated twice. Starting from the left, the assign numbers as follows:

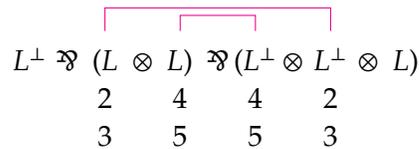
\begin{center} \vspace{5mm}
\begin{tikzpicture}
  [auto, node
  distance=5mm, skip loop/.style={to path={-- ++(0,#1) -| (\tikztotarget)}}] \tikzstyle{every node} = [text depth=-5pt,text height=0.5ex]
 \node (z1) { }; 
    \node (1) [right of=z1] {$L^{\bot}$}; \node (a) [right of=1] {$\invamp$} ;
   \node (2) [right of=a] {$(L$}; \node (b) [right of=2] {$\tens$} ;
   \node (3)  [right of=b] {$L)$}; \node (c)  [right of=3] {$\invamp$};
   \node (4)  [right of=c] {$(L^{\bot}$}; \node(d) [right of=4] {$\tens$} ;
   \node (5) [right of=d] {$L^{\bot}$}; \node (e) [right of=5] {$\tens$} ;
   \node (6) [right of=e] {$L)$};
  \node (t1) [above of= z1] { };
  \node (12) [below of=2] {2};
  \node (13) [below of=3] {4}; 
  \node (14) [below of=4] {4};
  \node (15) [below of=5] {2};
  \node (22) [below of=12] {3};
  \node (23) [below of=13] {5}; 
  \node (24) [below of=14] {5};
  \node (25) [below of=15] {3}; 
\begin{scope}
\path (2)  edge  [magenta, skip loop =7mm, shorten >=3mm, shorten <=3mm] (5);
\path (3)  edge  [magenta, skip loop =6mm, shorten >=3mm, shorten <=3mm] (4);
\end{scope}
\end{tikzpicture} \vspace{5mm}
\end{center}
\pagebreak
\item[4 \& 5.] The numbers created in the previous steps are now used to make tuples.

\begin{center} \vspace{5mm}
\begin{tikzpicture}
  [auto, node
  distance=5mm, skip loop/.style={to path={-- ++(0,#1) -| (\tikztotarget)}}] \tikzstyle{every node} = [text depth=-5pt,text height=0.5ex]
 \node (z1) { }; 
    \node (1) [right of=z1] {$L^{\bot}$}; \node (a) [right of=1] {$\invamp$} ;
   \node (2) [right of=a] {$(L$}; \node (b) [right of=2] {$\tens$} ;
   \node (3)  [right of=b] {$L)$}; \node (c)  [right of=3] {$\invamp$};
   \node (4)  [right of=c] {$(L^{\bot}$}; \node(d) [right of=4] {$\tens$} ;
   \node (5) [right of=d] {$L^{\bot}$}; \node (e) [right of=5] {$\tens$} ;
   \node (6) [right of=e] {$L)$};
  \node (t1) [above of= z1] {$\lambda_{3}$};
  \node (t2) [above of= t1] {$\lambda_{2}$};
  \node (t3) [above of= t2] {$\lambda_{1}$};
  \node (11) [below of=1] {[1]};
  \node (12) [below of=2] {[2}; \node (1b) [below of=b] {,};
  \node (13) [below of=3] {4]}; 
  \node (14) [below of=4] {[4}; \node (1d) [below of=d] {,};
  \node (15) [below of=5] {2}; 
  \node (16) [below of=6] {1]};
  \node (11) [below of=11] {[1]};
  \node (12) [below of=12] {[3}; \node (2b) [below of=1b] {,};
  \node (23) [below of=13] {5]}; 
  \node (24) [below of=14] {[5}; \node (2d) [below of=1d] {,};
  \node (25) [below of=15] {3};  \node (2e) [below of=1e] {,};
  \node (26) [below of=16] {1]};
\begin{scope}
\path   (1)  edge [red, skip loop =16mm,shorten >=13mm, shorten <=13mm]  (2);
\path (3)  edge  [red, skip loop =16mm,shorten >=13mm, shorten <=13mm] (4);
\path   (5)  edge [red, skip loop =16mm,shorten >=13mm, shorten <=13mm]  (6);
\path   (1)  edge [blue, skip loop =12mm,shorten >=8mm, shorten <=8mm]  (6);
\path (2)  edge  [blue, skip loop =11mm,shorten >=8mm, shorten <=8mm] (4);
\path   (5)  edge [blue, skip loop =10mm,shorten >=8mm, shorten <=8mm]  (3);
\path (1)  edge  [green, skip loop =7mm, shorten >=3mm, shorten <=3mm] (6);
\path (2)  edge  [green, skip loop =6mm, shorten >=3mm, shorten <=3mm] (5);
\path (3)  edge  [green, skip loop =5mm, shorten >=3mm, shorten <=3mm] (4);
\end{scope}
\end{tikzpicture} \vspace{5mm}
\end{center}

\item[6.] The tensors $a^{1},\ldots,a^{3}$ are then created from the tuples above.
\begin{eqnarray*}
& & \hughf{(a^{m})}_{j_{(1,1)}}										 =  \tensor*{\delta}{*^{1}_{j_{(1,1)}}} \\
& & \hughf{(a^{m})}_{i_{(2,1)}i_{(2,2)}}					 =  \tensor*{\delta}{*^{2}_{i_{(2,1)}}^{4}_{i_{(2,2)}}} + \tensor*{\delta}{*^{3}_{i_{(2,1)}}^{5}_{i_{(2,2)}}} \\
& & \hughf{(a^{m})}_{j_{(3,1)}j_{(3,2)}i_{(3,1)}}	 = 	\tensor*{\delta}{*^{4}_{j_{(3,1)}}^{2}_{j_{(3,2)}}^{1}_{i_{(3,1)}}} + \tensor*{\delta}{*^{5}_{j_{(3,1)}}^{3}_{j_{(3,2)}}^{1}_{i_{(3,1)}}}
\end{eqnarray*}
Then $\bar{\lambda_{2}}a^{1}a^{2} \,\neq\, 0 \,\neq\, \bar{\lambda_{3}}a^{1}a^{2}$.\medskip
\end{enumerate}

\noindent Fortunately, we are able to find another claim which provides enough information to differentiate any such rogue sets of axiom links from the chosen linking $\lambda$ sufficiently.

\begin{lem} \label{AcyclicClaim2}
Suppose that we have a linear combination of proof structures including a non-zero instance of a linking $\lambda$ which contains a minimal cycle $\hat{\lambda}$, and choose a block $k$ through which $\hat{\lambda}$ passes. Let $(\mathbf{x}^{1},\mathbf{y}^{1})$ and $(\mathbf{x}^{2},\mathbf{y}^{2})$ be the two tuples created for block $k$ by Algorithm \ref{AcyclicAlg} with $\lambda$ and $\hat{\lambda}$ over their associated \mll~formula as its inputs, reordered so the $l^{th}$ positions of $\mathbf{x}^{1}$ and $\mathbf{x}^{2}$ relate to the $l^{th}$ positive literal (given index $i_{(k,l)}$), and $\mathbf{y}^{1}$ and $\mathbf{y}^{2}$ to the $l^{th}$ negative literal with $j_{(k,l)}$. If a set of axiom links $\lambda'$ in the linear combination is similar enough to $\lambda$ that their axiom links not connected to block $k$ are the same, then the tensor $\bar{\lambda'}$ when composed with all the tensors created by Algorithm \ref{AcyclicAlg} except $a^{k}$, and either one of the tensors $\tensor*{\delta}{*^{\mathbf{x}^{1}}_{\mathbf{i}_{(k,-)}}^{\mathbf{y}^{1}}_{\mathbf{j}_{(k,-)}}}$ and $\tensor*{\delta}{*^{\mathbf{x}^{2}}_{\mathbf{i}_{(k,-)}}^{\mathbf{y}^{2}}_{\mathbf{j}_{(k,-)}}}$, the result is the scalar $1$ if $\lambda'=\lambda$, and $0$ otherwise.
\end{lem}
\proof
We start by showing that zero is created in the case that $\lambda \neq \lambda'$. In this situation, there is an axiom link in $\lambda'$ incident to block $k$ which is not found in $\lambda$. Without loss of generality, we assume that this link is connected to block $k$ by a positive literal given index $i_{(k,l_{i})}$ in the tensor representation; it is adjacent to a negative literal with index $j_{(m,l_{j})}$ for some $m$ and $l_{j}$. The tensor representation of the axiom link is therefore $\delta_{i_{(k,l_{i})}}^{j_{(m,l_{j})}}$, and the representation of $\lambda'$ has this as a factor.

\p The value (or values) found in the $(l_{j})^{th}$ position of the tuple (or tuples) associated with block $m$ in this case are not the same as $x^{p}_{l_{i}}$ for $p \,=\, 1 \mbox{ or } 2$, since the link is not found in $\lambda$: the tensor $\hughf{(a^{m})}_{\mathbf{i}_{(m,-)}\mathbf{j}_{(m,-)}}$ has form $\delta_{j_{(m,l_{j})}}^{y_{l_{j}}} \cdot \prod_{l \neq l_{j}}\delta_{j_{(m,l)}}^{y_{l}}$ or $\delta_{j_{(m,l_{j})}}^{y^{1}_{l_{j}}} \cdot \prod_{l \neq l_{j}}\delta_{j_{(m,l)}}^{y^{1}_{l}} + \delta_{j_{(m,l_{j})}}^{y^{2}_{l_{j}}} \cdot \prod_{l \neq l_{j}}\delta_{j_{(m,l)}}^{y^{2}_{l}}$, and we find that
\begin{displaymath}
\bar{\lambda'}_{\mathbf{i}}^{\mathbf{j}} \cdot \tensor*{\delta}{*^{\mathbf{x}^{1}}_{\mathbf{i}_{(k,-)}}^{\mathbf{y}^{1}}_{\mathbf{j}_{(k,-)}}} \cdot \hughf{(a^{m})}_{\mathbf{i}_{(m,-)}\mathbf{j}_{(m,-)}} = 0_{\mathbf{i}'}^{\mathbf{j}'}
\end{displaymath}
for appropriate superindices $\mathbf{i}'$ and $\mathbf{j}'$, from which this part of the claim is a trivial consequence. This is because, for both $r = 1$ and $2$,
\begin{displaymath}
\delta_{i_{(k,l_{i})}}^{j_{(m,l_{j})}} \cdot \delta_{j_{(m,l_{j})}}^{y_{l_{j}}} \cdot \delta_{i_{(k,l_{i})}}^{x^{r}_{i_{(k,l_{i})}}} = \delta^{y_{l_{j}}}_{x^{r}_{i_{(k,l_{i})}}} = 0.
\end{displaymath}

Now we consider the case when $\lambda' = \lambda$. Suppose that two literals, one of each polarity, given indices $i_{(m_{i},l_{i})}$ and $j_{(m_{j},l_{j})}$ are connected by an axiom link in $\lambda$. Then $\bar{\lambda}_{\mathbf{i}}^{\mathbf{j}}$ has $\delta_{i_{(m_{i},l_{i})}}^{j_{(m_{j},l_{j})}}$ as a factor.

\p If blocks $m_{i}$ and $m_{j}$ are not connected to $\hat{\lambda}$, then it must be the case that $\delta_{i_{(m_{i},l_{i})}}^{x_{(m_{i},l_{i})}}$ and $\delta_{i_{(m_{i},l_{i})}}^{x_{(m_{i},l_{i})}}$ are factors of $\hughf{(a^{m_{i}})}_{\mathbf{i}_{(m_{i},-)}}$ and $\hughf{(a^{m_{j}})}_{\mathbf{j}_{(m_{j},-)}}$ respectively for the $x_{(m_{i},l_{i})}$ and $x_{(m_{j},l_{j})}$ defined in the algorithm for these index position. We know that $x_{(m_{i},l_{i})} = x = x_{(m_{j},l_{j})}$ for some $x$ for both constant tensors due to the axiom link being in $\lambda$.
\begin{displaymath}
\delta_{i_{(m_{i},l_{i})}}^{j_{(m_{j},l_{j})}} \cdot \delta_{i_{(m_{i},l_{i})}}^{x} \cdot \delta_{j_{(m_{j},l_{j})}}^{x} = \delta_{1}^{1} = 1
\end{displaymath}
As such, $\bar{\lambda}_{\mathbf{i}}^{\mathbf{j}} \cdot \delta_{i_{(m_{i},l_{i})}}^{x_{(m_{i},l_{i})}} \cdot \delta_{j_{(m_{j},l_{j})}}^{x_{(m_{j},l_{j})}}$ produces Kronecker deltas representing an MDNF~proof structure identical to that of $\lambda$, but with the literals incident to the chosen axiom link removed (together with said axiom link). Following this line of argument to its logical conclusion, we find, when $\Gamma(\hat{\lambda})$ means all blocks not adjacent to $\hat{\lambda}$, that
\begin{displaymath}
\bar{\lambda}_{\mathbf{i}}^{\mathbf{j}} \cdot \prod_{m \notin \Gamma(\hat{\lambda})}\hughf{(a^{m})}_{\mathbf{i}_{(m,-)}\mathbf{j}_{(m,-)}} \cdot \prod_{q} \delta_{q}^{z(q)}
\end{displaymath}
reduces to the tensor representation of the axiom links solely in the neighbouring blocks of $\hat{\lambda}$, with the $q$ in the product being indices for each literal found in the blocks connected to the links in $\hat{\lambda}$ which are adjacent via axiom links to blocks not connected to the cycle.

We now consider tuples $\{b^{m}:m \in \Gamma(\hat{\lambda})\}$ for the remaining indices in the blocks connected to $\hat{\lambda}$. That is, for each of these $m$,
\begin{displaymath}
\hughf{(a^{m})}_{\mathbf{i}_{(m,-)}\mathbf{j}_{(m,-)}} = \hughf{(b^{m})}_{\mathbf{i}'_{(m,-)}\mathbf{j}'_{(m,-)}} \cdot \prod_{q_{l}} \delta_{q_{l}}^{z_{l}}
\end{displaymath}
for appropriate superindices.

\p Suppose that there is an axiom link connected to blocks connected to the cycle $\hat{\lambda}$ but not connected to block $k$. Then its representation takes the form $\delta_{i_{(m_{i},l_{i})}}^{j_{(m_{j},l_{j})}}$ for some $m_{i}$, $m_{j}$, $l_{i}$, $l_{j}$. It cannot be the case that $m_{i} = m_{j}$, since we have chosen for $\hat{lambda}$ to be a minimal cycle in $\lambda$ (as prescribed by Algorithm~\ref{AcyclicAlg}). If it were, it would actually be the entire cycle. We are therefore left with the only possibility that $m_{i} \,\neq\, m_{j}$.


If $m_{i} \neq m_{j}$, then there will be a factor of the original composition which is equivalent to the following:
\begin{displaymath}
\delta_{i_{(m,l_{i})}}^{j_{(m,l_{j})}} \cdot \hughf{(b^{m_{i}})}_{\mathbf{i}'_{(m_{i},-)}\mathbf{j}'_{(m_{i},-)}} \cdot \hughf{(b^{m_{j}})}_{\mathbf{i}''_{(m_{j},-)}\mathbf{j}''_{(m_{j},-)}}.
\end{displaymath}
The tensor $\delta_{i_{(m,l_{i})}}^{j_{(m,l_{j})}}$ is acting as the representation of an axiom link between the two literals represented by the indices $i_{(m,l_{i})}$ and $j_{(m,l_{j})}$.

\p The two tuples $b^{m_{i}}$ and $b^{m_{j}}$ are partial permutations containing exactly two non-zero positions. This is seen by how they are found by the blocks $a^{m_{i}}$ and $a^{m_{j}}$: the `$a$' blocks are partial permutations with exactly two non-zero positions, and since they can be formed from the two `$b$' tuples by use of a single Kronecker delta the same must be true of their factors. By assumption, and without loss of generality regarding which tuple contains the positive literal of the axiom link and which contains the negative, we know that the tuples $b^{m_{i}}$ and $b^{m_{j}}$ are tuples which take the forms below.
\begin{align*}
\hughf{(b^{m_{i}})}_{\mathbf{i}'_{(m_{i},-)}\mathbf{j}'_{(m_{j},-)}}	&=	\tensor*{\delta}{*^{\mathbf{p}^{1}}_{\mathbf{i}''_{(m_{i},-)}}^{\mathbf{q}^{1}}_{\mathbf{j}'_{(m_{i},-)}}}\tensor*{\delta}{*^{r_{1}}_{i_{(m_{i},l_{i})}}} \,+\, \tensor*{\delta}{*^{\mathbf{p}^{2}}_{\mathbf{i}''_{(m_{i},-)}}^{\mathbf{q}^{2}}_{\mathbf{j}'_{(m_{i},-)}}}\tensor*{\delta}{*^{r_{2}}_{i_{(m_{i},l_{i})}}} \\
\hughf{(b^{m_{j}})}_{\mathbf{i}'_{(m_{i},-)}\mathbf{j}'_{(m_{j},-)}}	&=	\tensor*{\delta}{*^{\mathbf{s}^{1}}_{\mathbf{i}'_{(m_{i},-)}}^{\mathbf{t}^{1}}_{\mathbf{j}''_{(m_{i},-)}}}\tensor*{\delta}{*^{r_{1}}_{j_{(m_{j},l_{j})}}} \,+\, \tensor*{\delta}{*^{\mathbf{s}^{2}}_{\mathbf{i}'_{(m_{i},-)}}^{\mathbf{t}^{2}}_{\mathbf{j}''_{(m_{i},-)}}}\tensor*{\delta}{*^{r_{2}}_{j_{(m_{j},l_{j})}}}
\end{align*}
for some appropriately sized constant superindices $\mathbf{p}^{1},\mathbf{p}^{2},\mathbf{q}^{1},\mathbf{q}^{2},\mathbf{s}^{1},\mathbf{s}^{2},\mathbf{t}^{1},\mathbf{t}^{2}$ and constants $r_{1},r_{2}$, whilst letting $\mathbf{i}''_{(m_{i},-)}$ and $\mathbf{j}'_{(m_{j},-)}$ be the free superindices which are $\mathbf{i}'_{(m_{i},-)}$ and $\mathbf{j}'_{(m_{j},-)}$ with $i_{(m_{i},l_{i})}$ and $j_{(m_{j},l_{j})}$ removed respectively. We know this because Algorithm~\ref{AcyclicAlg} ensures that the tuples created for two blocks of tensors will have entries which match each other if and only if there is an axiom link in $\lambda$ connecting the literals whose positions they are representing. Since the axiom links of $\lambda'$ and $\lambda$ which are not connected to block $k$ are the, this crosses over to this linking.

\p Once these representations have been created, it becomes clear that the tensor composition from before reduces to nothing more than another partial permutation with two non-zero entries.
\begin{align*}
\delta_{i_{(m,l_{i})}}^{j_{(m,l_{j})}} \cdot \hughf{(b^{m_{i}})}_{\mathbf{i}'_{(m_{i},-)}\mathbf{j}'_{(m_{i},-)}} \cdot \hughf{(b^{m_{j}})}_{\mathbf{i}''_{(m_{j},-)}\mathbf{j}''_{(m_{j},-)}} &= \delta_{i_{(m,l_{i})}}^{j_{(m,l_{j})}} \cdot \left( \tensor*{\delta}{*^{\mathbf{p}^{1}}_{\mathbf{i}''_{(m_{i},-)}}^{\mathbf{q}^{1}}_{\mathbf{j}'_{(m_{i},-)}}}\tensor*{\delta}{*^{r_{1}}_{i_{(m_{i},l_{i})}}} \,+\, \tensor*{\delta}{*^{\mathbf{p}^{2}}_{\mathbf{i}''_{(m_{i},-)}}^{\mathbf{q}^{2}}_{\mathbf{j}'_{(m_{i},-)}}}\tensor*{\delta}{*^{r_{2}}_{i_{(m_{i},l_{i})}}} \right) \\
	&\qquad \cdot \left( \tensor*{\delta}{*^{\mathbf{s}^{1}}_{\mathbf{i}'_{(m_{i},-)}}^{\mathbf{t}^{1}}_{\mathbf{j}''_{(m_{i},-)}}}\tensor*{\delta}{*^{r_{1}}_{j_{(m_{j},l_{j})}}} \,+\, \tensor*{\delta}{*^{\mathbf{s}^{2}}_{\mathbf{i}'_{(m_{i},-)}}^{\mathbf{t}^{2}}_{\mathbf{j}''_{(m_{i},-)}}}\tensor*{\delta}{*^{r_{2}}_{j_{(m_{j},l_{j})}}} \right) \\
	&=	\tensor*{\delta}{*^{\mathbf{p}^{1}}_{\mathbf{i}''_{(m_{i},-)}}^{\mathbf{q}^{1}}_{\mathbf{j}'_{(m_{i},-)}}}\tensor*{\delta}{*^{r_{1}}_{i_{(m_{i},l_{i})}}}\tensor*{\delta}{*^{\mathbf{s}^{1}}_{\mathbf{i}'_{(m_{i},-)}}^{\mathbf{t}^{1}}_{\mathbf{j}''_{(m_{i},-)}}}\tensor*{\delta}{*^{r_{1}}_{j_{(m_{j},l_{j})}}}\delta_{i_{(m,l_{i})}}^{j_{(m,l_{j})}} \\
	&\qquad +\, \tensor*{\delta}{*^{\mathbf{p}^{1}}_{\mathbf{i}''_{(m_{i},-)}}^{\mathbf{q}^{1}}_{\mathbf{j}'_{(m_{i},-)}}}\tensor*{\delta}{*^{r_{1}}_{i_{(m_{i},l_{i})}}}\tensor*{\delta}{*^{\mathbf{s}^{2}}_{\mathbf{i}'_{(m_{i},-)}}^{\mathbf{t}^{2}}_{\mathbf{j}''_{(m_{i},-)}}}\tensor*{\delta}{*^{r_{2}}_{j_{(m_{j},l_{j})}}}\delta_{i_{(m,l_{i})}}^{j_{(m,l_{j})}} \\
	&\qquad +\, \tensor*{\delta}{*^{\mathbf{p}^{2}}_{\mathbf{i}''_{(m_{i},-)}}^{\mathbf{q}^{2}}_{\mathbf{j}'_{(m_{i},-)}}}\tensor*{\delta}{*^{r_{2}}_{i_{(m_{i},l_{i})}}}\tensor*{\delta}{*^{\mathbf{s}^{1}}_{\mathbf{i}'_{(m_{i},-)}}^{\mathbf{t}^{1}}_{\mathbf{j}''_{(m_{i},-)}}}\tensor*{\delta}{*^{r_{1}}_{j_{(m_{j},l_{j})}}}\delta_{i_{(m,l_{i})}}^{j_{(m,l_{j})}} \\
	&\qquad +\, \tensor*{\delta}{*^{\mathbf{p}^{2}}_{\mathbf{i}''_{(m_{i},-)}}^{\mathbf{q}^{2}}_{\mathbf{j}'_{(m_{i},-)}}}\tensor*{\delta}{*^{r_{2}}_{i_{(m_{i},l_{i})}}}\tensor*{\delta}{*^{\mathbf{s}^{2}}_{\mathbf{i}'_{(m_{i},-)}}^{\mathbf{t}^{2}}_{\mathbf{j}''_{(m_{i},-)}}}\tensor*{\delta}{*^{r_{2}}_{j_{(m_{j},l_{j})}}}\delta_{i_{(m,l_{i})}}^{j_{(m,l_{j})}} \\
	&=	\tensor*{\delta}{*^{\mathbf{p}^{1}}_{\mathbf{i}''_{(m_{i},-)}}^{\mathbf{q}^{1}}_{\mathbf{j}'_{(m_{i},-)}}}\tensor*{\delta}{*^{\mathbf{s}^{1}}_{\mathbf{i}'_{(m_{i},-)}}^{\mathbf{t}^{1}}_{\mathbf{j}''_{(m_{i},-)}}}\delta_{r_{1}}^{r_{1}} \\
	&\qquad +\, \tensor*{\delta}{*^{\mathbf{p}^{1}}_{\mathbf{i}''_{(m_{i},-)}}^{\mathbf{q}^{1}}_{\mathbf{j}'_{(m_{i},-)}}}\tensor*{\delta}{*^{\mathbf{s}^{2}}_{\mathbf{i}'_{(m_{i},-)}}^{\mathbf{t}^{2}}_{\mathbf{j}''_{(m_{i},-)}}}\delta_{r_{1}}^{r_{2}} \\
	&\qquad +\, \tensor*{\delta}{*^{\mathbf{p}^{2}}_{\mathbf{i}''_{(m_{i},-)}}^{\mathbf{q}^{2}}_{\mathbf{j}'_{(m_{i},-)}}}\tensor*{\delta}{*^{\mathbf{s}^{1}}_{\mathbf{i}'_{(m_{i},-)}}^{\mathbf{t}^{1}}_{\mathbf{j}''_{(m_{i},-)}}}\delta_{r_{2}}^{r_{1}} \\
	&\qquad +\, \tensor*{\delta}{*^{\mathbf{p}^{2}}_{\mathbf{i}''_{(m_{i},-)}}^{\mathbf{q}^{2}}_{\mathbf{j}'_{(m_{i},-)}}}\tensor*{\delta}{*^{\mathbf{s}^{2}}_{\mathbf{i}'_{(m_{i},-)}}^{\mathbf{t}^{2}}_{\mathbf{j}''_{(m_{i},-)}}}\delta_{r_{2}}^{r_{2}}  \\
	&= \tensor*{\delta}{*^{\mathbf{p}^{1}}_{\mathbf{i}''_{(m_{i},-)}}^{\mathbf{q}^{1}}_{\mathbf{j}'_{(m_{i},-)}}}\tensor*{\delta}{*^{\mathbf{s}^{1}}_{\mathbf{i}'_{(m_{i},-)}}^{\mathbf{t}^{1}}_{\mathbf{j}''_{(m_{i},-)}}} \,+\, \tensor*{\delta}{*^{\mathbf{p}^{2}}_{\mathbf{i}''_{(m_{i},-)}}^{\mathbf{q}^{2}}_{\mathbf{j}'_{(m_{i},-)}}}\tensor*{\delta}{*^{\mathbf{s}^{2}}_{\mathbf{i}'_{(m_{i},-)}}^{\mathbf{t}^{2}}_{\mathbf{j}''_{(m_{i},-)}}}
\end{align*}
The representation could be said to be equivalent to the block of literals which would be created by merging to two blocks to which $b^{1}$ and $b^{2}$ are connected and removing the linked literals. The same idea applies to the two tuples that are created.

\p This is certainly a repeatable process for all axiom links not connected to block $k$. Following this procedure as far as possible, we find that the answer to the original composition becomes the same as
\begin{displaymath}
c_{\mathbf{i}_{(k,-)}\mathbf{j}_{(k,-)}} \cdot \tensor*{\delta}{*^{\mathbf{x}^{r}}_{\mathbf{i}_{(k,-)}}^{\mathbf{y}^{r}}_{\mathbf{j}_{(k,-)}}} \cdot \prod_{q}\delta_{q}^{z(q)}.
\end{displaymath}
The tensor $c$ is the partial permutation described by two tuples providing the locations of the non-zero entries of block $k$: the concatenation of all the first tuples of blocks which are not $k$, with tuple positions relating to indices describing literals not connected to block $k$ by an axiom link deleted; and the same with the second tuples. In Example~\ref{AcyclicExam2}, taking that $k=3$, it happens that $c_{\mathbf{i}_{(3,-)}\mathbf{j}_{(3,-)}} = \hughf{(a^{3})}_{j_{(3,1)}j_{(3,2)}j_{(3,1)}}$.

\p Due to the algorithm ensuring that certain tuple positions are kept
equal to one another, \[c_{\mathbf{i}''_{(k,-)}\mathbf{j}''_{(3,-)}}
\cdot \prod_{q}\delta_{q}^{z(q)} =
\tensor*{\delta}{*^{\mathbf{x}^{1}}_{\mathbf{i}_{(k,-)}}^{\mathbf{y}^{1}}_{\mathbf{j}_{(k,-)}}}
+
\tensor*{\delta}{*^{\mathbf{x}^{2}}_{\mathbf{i}_{(k,-)}}^{\mathbf{y}^{2}}_{\mathbf{j}_{(k,-)}}}\,.
\]
 It then follows trivially that 
\[(\tensor*{\delta}{*^{\mathbf{x}^{1}}_{\mathbf{i}_{(k,-)}}^{\mathbf{y}^{1}}_{\mathbf{j}_{(k,-)}}}
+
\tensor*{\delta}{*^{\mathbf{x}^{2}}_{\mathbf{i}_{(k,-)}}^{\mathbf{y}^{2}}_{\mathbf{j}_{(k,-)}}})
\cdot
\tensor*{\delta}{*^{\mathbf{x}^{r}}_{\mathbf{i}_{(k,-)}}^{\mathbf{y}^{r}}_{\mathbf{j}_{(k,-)}}}
=
\tensor*{\delta}{*^{\mathbf{x}^{1}}_{\mathbf{x}^{r}}^{\mathbf{y}^{1}}_{\mathbf{y}^{r}}}
+
\tensor*{\delta}{*^{\mathbf{x}^{2}}_{\mathbf{x}^{r}}^{\mathbf{y}^{2}}_{\mathbf{y}^{r}}}
= 1\,.\eqno{\qEd}
\] \medskip

\noindent Example \ref{AcyclicExam2} fits the scenario where Lemma~\ref{AcyclicClaim2} is used neatly. Assuming block $3$ is the chosen block $k$, it is clear that every axiom link in $\lambda_{2} = \lambda$ not incident to that block is found in $\lambda_{3}$ and vice versa. However, multiplying $\bar{\lambda_{3}}$ with $a^{1}$ and $a^{2}$ gives the following tensor:
\begin{eqnarray*}
\bar{\lambda}_{j_{(1,1)}i_{(2,1)}i_{(2,2)}j_{(3,1)}j_{(3,2)}i_{(3,1)}}\hughf{(a^{1})}_{j_{(1,1)}}\hughf{(a^{2})}_{i_{(2,1)}i_{(2,2)}} & = & \tensor*{\delta}{*^{j_{(1,1)}}_{i_{(3,1)}}^{j_{(3,1)}}_{i_{(2,1)}}^{j_{(3,2)}}_{i_{(2,2)}}} \tensor*{\delta}{*^{1}_{j_{(1,1)}}} (\tensor*{\delta}{*^{2}_{i_{(2,1)}}^{4}_{i_{(2,2)}}} + \tensor*{\delta}{*^{3}_{i_{(2,1)}}^{5}_{i_{(2,2)}}}) \\
& = & \tensor*{\delta}{*^{1}_{i_{(3,1)}}^{j_{(3,1)}}_{2}^{j_{(3,2)}}_{4}} + \tensor*{\delta}{*^{1}_{i_{(3,1)}}^{j_{(3,1)}}_{3}^{j_{(3,2)}}_{5}}
\end{eqnarray*} 

\p\noindent Multiplying the resulting tensor with either $\tensor*{\delta}{*^{4}_{j_{(3,1)}}^{2}_{j_{(3,2)}}^{1}_{i_{(3,1)}}}$ or $\tensor*{\delta}{*^{5}_{j_{(3,1)}}^{3}_{j_{(3,2)}}^{1}_{i_{(3,1)}}}$ will immediately result in producing a $0$ as an output.
\begin{displaymath}
(\tensor*{\delta}{*^{1}_{i_{(3,1)}}^{j_{(3,1)}}_{2}^{j_{(3,2)}}_{4}} + \tensor*{\delta}{*^{1}_{i_{(3,1)}}^{j_{(3,1)}}_{3}^{j_{(3,2)}}_{5}}) \cdot  \tensor*{\delta}{*^{4}_{j_{(3,1)}}^{2}_{j_{(3,2)}}^{1}_{i_{(3,1)}}} =
\tensor*{\delta}{*^{1}_{1}^{4}_{2}^{2}_{4}} + \tensor*{\delta}{*^{1}_{1}^{4}_{3}^{2}_{5}} = 0
\end{displaymath}
\begin{displaymath}
(\tensor*{\delta}{*^{1}_{i_{(3,1)}}^{j_{(3,1)}}_{2}^{j_{(3,2)}}_{4}} + \tensor*{\delta}{*^{1}_{i_{(3,1)}}^{j_{(3,1)}}_{3}^{j_{(3,2)}}_{5}}) \cdot  \tensor*{\delta}{*^{5}_{j_{(3,1)}}^{3}_{j_{(3,2)}}^{1}_{i_{(3,1)}}} =
\tensor*{\delta}{*^{1}_{1}^{5}_{2}^{3}_{4}} + \tensor*{\delta}{*^{1}_{1}^{5}_{3}^{3}_{5}} = 0
\end{displaymath}
Note in particular how composing with $\tensor*{\delta}{*^{4}_{j_{(3,1)}}^{2}_{j_{(3,2)}}^{1}_{i_{(3,1)}}}$ produces two tensors being summed together which either have the right numbers to use in the entry positions but in the wrong order as in the first (due to the axiom links connecting with the block in a different manner), or the wrong numbers altogether like in the second (due to using the tensor created from the information from the upper tuple rather than the lower). The dual of this remark can be found by considering composition with $\tensor*{\delta}{*^{4}_{j_{(3,1)}}^{2}_{j_{(3,2)}}^{1}_{i_{(3,1)}}}$.

\p Armed with the two claims created from the earlier simple principle, we are in a position to express why MDNF~transformations from $\C$ only translate to $\GC$ if the linear combination of proof structures they model do not contain a cyclic proof structure.

\begin{prop} \label{AcyclicLemma} Every MDNF transformation in $\GC$
  models a linear combination of acyclic proof structures.
\end{prop}
\proof
  Suppose that the functor $\bbrk{F}$ is in MDNF, and therefore for
  any $n \in \N^{+}$
  \begin{displaymath}
    \bbrk{F}(\mathbf{A}_{n},\mathbf{A}_{n}) =
    \tbiginvamp_{m=1}^{M}(\tbigotimes_{l=1}^{L_{m}}A_{n}^{\phi(m,l)}). 
  \end{displaymath}
  The tensor representations of the values of the above object are
  given in Section~\ref{SectionMDNFObjects}

  Let $\tau = (\tau_{\mathbf{R}} \in
  \C[\I,|\bbrk{F}|(\mathbf{R},\mathbf{R})])_{\mathbf{R} \in \C^{N}}$
  be an MDNF transformation in $\C$, modelling a linear combination
  of proof structures, one of which is cyclic. For the object
  $\mathbf{R} = (n\I,\ldots,n\I)$, the component $\tau_{\mathbf{R}}$
  is represented by the tensor
  \begin{displaymath}
    \tensor*{\tau}{*^{j_{(1,1)}}_{i_{(1,1)}}^{\;\cdots\;}_{\;\cdots}^{j_{(M,
          N_M)}}_{i_{(M,P_{M})}}}
    = \textstyle\sum_{\beta} s_{\beta}\cdot
    \tensor*{\delta}{*^{j_{\beta(1,1)}}_{i_{(1,
          1)}}^{\;\cdots\;}_{\;\cdots}^{j_{\beta(M, P_{M})}}_{i_{(M,
          P_M)}}}.
  \end{displaymath}
  One of the bijections, $\zeta$ say, is such that $s_{\zeta} \neq
  0$, and there is a cycle between the blocks.

  We show that $\tau^{\mathbf{j}}_{\mathbf{i}} \notin
  \bbrk{F}(\mathbf{A}_{n},\mathbf{A}_{n})_{Val} =
  \GC[\I,\bbrk{F}(\mathbf{A}_{n},\mathbf{A}_{n})]$ for some choice of
  $n \in \N^{+}$. We choose $n$ and define a set of partial
  permutations $\{a^{m}: m \in [M] \}$, one for each block of tensor
  products of literals, using Algorithm~\ref{AcyclicAlg}; and we
  nominate any one of the blocks that is part of the chosen cycle,
  calling it $k$. We know from Lemmas~\ref{AcyclicClaim1} and~\ref{AcyclicClaim2} that there are two distinct entry tuples
  $(\mathbf{x}^{1},\mathbf{y}^{1}) = (x^{1}_{(k,1)},
  \ldots,x^{1}_{(k,P_{k})}, y^{1}_{(k,1)}, \ldots,
  y^{1}_{(k,N_{k})})$ and $(\mathbf{x}^{2},\mathbf{y}^{2}) =
  (x^{2}_{(k,1)}, \ldots, x^{2}_{(k,P_{k})}, y^{2}_{(k,1)}, \ldots,
  y^{2}_{(k,N_{k})})$ for $a^{k}$ where
  $\hughf{(a^{k})}_{\mathbf{x}^{2},\mathbf{y}^{2}} = 1 =
  \hughf{(a^{k})}_{\mathbf{x}^{2},\mathbf{y}^{2}}$ that, for $l \in \{1,2\}$, 
  \begin{displaymath}
    \tensor*{\delta}{*^{\mathbf{x}^{l}}_{\mathbf{i}_{(k,-)}}^{\mathbf{y}^{l}}_{\mathbf{j}_{(k,-)}}} \cdot \tensor*{\delta}{*^{x^{l}_{(k, 1)}}_{i_{(k,
          1)}}^{\;\cdots\;}_{\;\cdots}^{x^{l}_{(k, 
          P_{k})}}_{i_{(k,P_{k})}}^{y^{l}_{(k, 1)}}_{j_{(k,
          1)}}^{\;\cdots\;}_{\;\cdots}^{y^{l}_{(k, 
          N_{k})}}_{j_{(k,N_{k})}}}
    \cdot
    \tau^{\mathbf{j}}_{\mathbf{i}}\cdot\textstyle\prod_{m \not=
      k}\hughf{(a^{m})_{i_{(m,1)}\,\cdots\,i_{m,P_{m}}\,j_{(m,1)}\,\cdots\,j_{(m,N_{m})}}} = s_{\zeta} \neq 0.
  \end{displaymath}
  The tensor $\tau^{\mathbf{j}}_{\mathbf{i}}$, when multiplied with
    these partial permutations, produces a tensor with at least two
    non-zero entries, meaning it fails the criterion desired of it to
    belong to $\bbrk{F}(\mathbf{A}_{n},\mathbf{A}_{n})_{Val}$. As such
    $\tau_{\mathbf{R}}$ is not an arrow in
    $\GC[\I,\bbrk{F}(\mathbf{A}_{n},\mathbf{A}_{n})]$, and consequently
    $\tau$ cannot be seen as an \mll\ transformation in $\GC$. \qed
  
  \subsection{Connectedness} \label{SubsectionConnectedness}
  
  The previous subsection holds the proof that MDNF~transformations in $\GC$ only describe linear combinations of acyclic proof structures. This section is devoted to proving the proof structures are connected as well. Once again we can see the intuition behind the coming proof using a couple of small examples.
  
  \p Consider the simplest \hughf{disconnected} proof structure.
  \begin{center} \vspace{5mm}
  \begin{tikzpicture}
    [auto, node
    distance=5mm, skip loop/.style={to path={-- ++(0,#1) -| (\tikztotarget)}}] \tikzstyle{every node} = [text depth=-5pt,text height=0.5ex]
     \node (1) {$L$}; \node (a) [right of=1] {$\invamp$} ;
    \node (2) [right of=a] {$L^{\bot}$}; \node (b) [right of=2] {$\invamp$} ;
    \node (3)  [right of=b] {$L$}; \node (c)  [right of=3] {$\invamp$};
   \node (4)  [right of=c] {$L^{\bot}$};
  \begin{scope}
  \path   (1)  edge [black, skip loop =6mm, shorten >=3mm, shorten <=3mm]  (2);
  \path (3)  edge  [black, skip loop =6mm, shorten >=3mm, shorten <=3mm] (4);
  \end{scope}
  \end{tikzpicture} \vspace{5mm}
  \end{center}
If we say $\bbrk{L} = C_{n}$ for some $n > 1$, it is quickly observed that the values of \linebreak $\bbrk{L \invamp L^{\bot} \invamp L \invamp L^{\bot}}$ are described by the $4$-permutations over $[n]$. These are all tensors of the form $\tensor*{z}{*^{j_{1}}_{i_{1}}^{j_{2}}_{i_{2}}}$ ($i_{l}$ and $j_{l}$ being linked to the $l^{th}$ positive and negative literals respectively) which, when composed with any three constant tensors, produce another constant tensor.
  
  \p The proof structure is described by $\tensor*{\delta}{*^{j_{1}}_{i_{1}}^{j_{2}}_{i_{2}}}$. It is already known that this is not a full $4$-permutation, and therefore does not belong to the set of values. However, we require a more generalisable perspective in order to the learn from the example. If we compose $\delta_{i_{1}}^{j_{1}}$ with $\delta_{i_{1}}^{1}$ and $\delta_{j_{1}}^{2}$, which are both $1$-permutations over $[n]$, we see that $\delta_{i_{1}}^{j_{1}} \cdot \delta_{i_{1}}^{1} \cdot \delta_{j_{1}}^{2} = \delta_{1}^{2} = 0$, meaning $\tensor*{\delta}{*^{j_{1}}_{i_{1}}^{j_{2}}_{i_{2}}} \cdot \delta_{i_{1}}^{1} \cdot \delta_{j_{1}}^{2} = 0 \cdot \delta_{i_{2}}^{j_{2}} = 0_{i_{2}j_{2}}$. Remembering the description of values of MDNF~objects built solely from $C_{n}$ in Section~\ref{SectionMDNFObjects}, it becomes clear that $\tensor*{\delta}{*^{j_{1}}_{i_{1}}^{j_{2}}_{i_{2}}}$ cannot belong in $(C_{n} \invamp C_{n}^{\bot} \invamp C_{n} \invamp C_{n}^{\bot})_{Val}$: the zero tensor does not belong to $(C^{\bot}_{n})_{Val}$, and composing any tensor from the non-empty $(C_{n})_{CoVal}$ with $0_{i_{2}j_{2}}$ produces $0_{j_{2}}$.
  
  \p In the above example it was possible to find ($1$-)permutations for both blocks in the left-hand component which, when composed with the Kronecker delta modelling the axiom link incident to their associated literals, produce the scalar $0$. The generalisability of this idea becomes further evident when a more complicated proof structure is considered. 
  
  \begin{center} \vspace{5mm}
\begin{tikzpicture}
  [auto, node
  distance=5mm, skip loop/.style={to path={-- ++(0,#1) -| (\tikztotarget)}}] \tikzstyle{every node} = [text depth=-5pt,text height=0.5ex]
 \node (z1) { }; 
    \node (1) [right of=z1] {$L^{\bot}$}; \node (a) [right of=1] {$\invamp$} ;
   \node (2) [right of=a] {$(L$}; \node (b) [right of=2] {$\tens$} ;
   \node (3)  [right of=b] {$L)$}; \node (c)  [right of=3] {$\invamp$};
   \node (4)  [right of=c] {$(L^{\bot}$}; \node(d) [right of=4] {$\tens$} ;
   \node (5) [right of=d] {$L^{\bot}$}; \node (e) [right of=5] {$\tens$} ;
   \node (6) [right of=e] {$L)$};  \node (f) [right of=6] {$\invamp$} ;
   \node (7)  [right of=f] {$L$}; \node (g)  [right of=7] {$\invamp$};
   \node (8) [right of=g] {$L$};  \node (h) [right of=8] {$\invamp$} ;
   \node (9)  [right of=h] {$L$}; \node (i)  [right of=9] {$\invamp$};
   \node (10)  [right of=i] {$L^{\bot}$};
\begin{scope}
\path (1)  edge  [black, skip loop =6mm, shorten >=3mm, shorten <=3mm] (2);
\path (3)  edge  [black, skip loop =6mm, shorten >=3mm, shorten <=3mm] (4);
\path (5)  edge  [black, skip loop =7mm, shorten >=3mm, shorten <=3mm] (8);
\path (6)  edge  [black, skip loop =6mm, shorten >=3mm, shorten <=3mm] (7);
\path (9)  edge  [black, skip loop =6mm, shorten >=3mm, shorten <=3mm] (10);
\end{scope}
\end{tikzpicture} \vspace{5mm}
\end{center}  
  The above structure contains far more axiom links, and some blocks have more than one literal within them. Its links are described by $\bar{\lambda} = \tensor*{\delta}{*^{j_{(1,1)}}_{i_{(2,1)}}^{j_{(3,1)}}_{i_{(2,2)}}^{j_{(3,2)}}_{i_{(5,1)}}^{j_{(3,3)}}_{i_{(4,1)}}^{j_{(7,1)}}_{i_{(6,1)}}}$, with the left-most component's links given by $\bar{\lambda^{l}} = \tensor*{\delta}{*^{j_{(1,1)}}_{i_{(2,1)}}^{j_{(3,1)}}_{i_{(2,2)}}^{j_{(3,2)}}_{i_{(5,1)}}^{j_{(3,3)}}_{i_{(4,1)}}}$.
  
  \p The first five blocks constitute the left-most component, and five permutations which can annihilate the above tensors are found below.
  \vspace{3mm}
  \begin{eqnarray*}
  \hughf{(c^{1})}_{j_{(1,1)}}						& = &	\delta_{j_{(1,1)}}^{1} \\
  \hughf{(c^{2})}_{i_{(2,1)}i_{(2,2)}}	& = &	\cycle(2,n,5)_{i_{(2,1)}i_{(2,2)}} \\
  \hughf{(c^{3})}_{j_{(3,1)}j_{(3,2)}j_{(3,3)}}	& = &	\cycle(3,n,11)_{j_{(3,1)}j_{(3,2)}j_{(3,3)}} \\
  \hughf{(c^{4})}_{i_{(4,1)}}						& = &	\delta_{i_{(4,1)}}^{2} \\
  \hughf{(c^{5})}_{i_{(5,1)}}						& = &	\delta_{i_{(5,1)}}^{3}
  \end{eqnarray*}
  \vspace{3mm}
  \begin{eqnarray*}
  & & (c^{1})_{j_{(1,1)}}(c^{2})_{i_{(2,1)}i_{(2,2)}}(c^{3})_{j_{(3,1)}j_{(3,2)}j_{(3,3)}}(c^{4})_{i_{(4,1)}}(c^{5})_{i_{(5,1)}} \tensor*{(\bar{\lambda^{l}})}{*^{j_{(1,1)}}_{i_{(2,1)}}^{j_{(3,1)}}_{i_{(2,2)}}^{j_{(3,2)}}_{i_{(4,1)}}^{j_{(3,3)}}_{i_{(5,1)}}}	\\ 
  & & \quad \quad \quad \quad =	\delta_{j_{(1,1)}}^{1}(c^{2})_{i_{(2,1)}}i_{(2,2)}(c^{3})_{j_{(3,1)}j_{(3,2)}j_{(3,3)}}\delta_{i_{(4,1)}}^{2}\delta_{i_{(5,1)}}^{3} \tensor*{\delta}{*^{j_{(1,1)}}_{i_{(2,1)}}^{j_{(3,1)}}_{i_{(2,2)}}^{j_{(3,2)}}_{i_{(5,1)}}^{j_{(3,3)}}_{i_{(4,1)}}} \\
  & & \quad \quad \quad \quad = \cycle(2,n,5)_{i_{(2,1)}i_{(2,2)}}\cycle(3,n,11)_{j_{(3,1)}j_{(3,2)}j_{(3,3)}} \tensor*{\delta}{*^{1}_{i_{(2,1)}}^{j_{(3,1)}}_{i_{(2,2)}}^{j_{(3,2)}}_{3}^{j_{(3,3)}}_{2}} \\
  & & \quad \quad \quad \quad = (\cycle(2,n,5)_{i_{(2,1)}i_{(2,2)}} \tensor*{\delta}{*^{1}_{i_{(2,1)}}}) \cdot (\cycle(3,n,11)_{j_{(3,1)}j_{(3,2)}j_{(3,3)}} \tensor*{\delta}{*^{j_{(3,1)}}_{i_{(2,2)}}^{j_{(3,3)}}_{2}}) \cdot \tensor*{\delta}{*^{j_{(3,2)}}_{3}} \\
  & & \quad \quad \quad \quad = (\tensor*{\delta}{*^{4}_{i_{(2,2)}}} \cdot \cycle(2,n,9)_{i_{(2,2)}j_{(3,2)}}) \cdot  \tensor*{\delta}{*^{j_{(3,2)}}_{3}} \\
  & & \quad \quad \quad \quad = \tensor*{\delta}{*^{5}_{j_{(3,2)}}^{j_{(3,2)}}_{3}} = \tensor*{\delta}{*^{5}_{3}} = 0
  \end{eqnarray*}
  It is therefore true that $(c^{1})_{j_{(1,1)}}(c^{2})_{i_{(2,1)}i_{(2,2)}}(c^{3})_{j_{(3,1)}j_{(3,2)}j_{(3,3)}}(c^{4})_{i_{(4,1)}}(c^{5})_{i_{(5,1)}}(c^{6})_{i_{(6,1)}}(\bar{\lambda})^{\mathbf{j}}_{\mathbf{i}} = 0_{j_{(7,1)}}$ 
  for all $c^{6} \in \Perm(1,n)$, and so $\bar{\lambda}$ is not a value, meaning that $\lambda$ is not modelled in $\GC$ by an \mll~transformation.
  
  \p The lesson to be learned from the arithmetic above is that although cycle permutations are being used for the permutations $c^{1},\ldots,c^{5}$ above\footnote{It should be remembered that $\cycle(1,n,x)_{i} = \delta_{i}^{x}$.}, this is only due to their ease in comprehension. What is most important is that certain positions in the permutations have value $1$. The proof structure figure below shows a choice of positions in the permutations corresponding to each block in the proof structure from above which ensure that their composition with the tensor $\bar{\lambda}$ yields a zero tensor.
  \begin{center} \vspace{5mm}
\begin{tikzpicture}
  [auto, node
  distance=5mm, skip loop/.style={to path={-- ++(0,#1) -| (\tikztotarget)}}] \tikzstyle{every node} = [text depth=-5pt,text height=0.5ex]
 \node (z1) { }; 
    \node (1) [right of=z1] {$L^{\bot}$}; \node (a) [right of=1] {$\invamp$} ;
   \node (2) [right of=a] {$(L$}; \node (b) [right of=2] {$\tens$} ;
   \node (3)  [right of=b] {$L)$}; \node (c)  [right of=3] {$\invamp$};
   \node (4)  [right of=c] {$(L^{\bot}$}; \node(d) [right of=4] {$\tens$} ;
   \node (5) [right of=d] {$L^{\bot}$}; \node (e) [right of=5] {$\tens$} ;
   \node (6) [right of=e] {$L)$};  \node (f) [right of=6] {$\invamp$} ;
   \node (7)  [right of=f] {$L$}; \node (g)  [right of=7] {$\invamp$};
   \node (8) [right of=g] {$L$};  \node (h) [right of=8] {$\invamp$} ;
   \node (9)  [right of=h] {$L$}; \node (i)  [right of=9] {$\invamp$};
   \node (10)  [right of=i] {$L^{\bot}$};
  \node (11) [below of=1] {[1]};
  \node (12) [below of=2] {[1}; \node (1b) [below of=b] {,};
  \node (13) [below of=3] {4]}; 
  \node (14) [below of=4] {[4}; \node (1d) [below of=d] {,};
  \node (15) [below of=5] {5};  \node (1e) [below of=e] {,};
  \node (16) [below of=6] {2]}; 
  \node (17) [below of=7] {[2]};
  \node (18) [below of=8] {[3]};
\begin{scope}
\path (1)  edge  [black, skip loop =6mm, shorten >=3mm, shorten <=3mm] (2);
\path (3)  edge  [black, skip loop =6mm, shorten >=3mm, shorten <=3mm] (4);
\path (5)  edge  [black, skip loop =7mm, shorten >=3mm, shorten <=3mm] (8);
\path (6)  edge  [black, skip loop =6mm, shorten >=3mm, shorten <=3mm] (7);
\path (9)  edge  [black, skip loop =6mm, shorten >=3mm, shorten <=3mm] (10);
\end{scope}
\end{tikzpicture} \vspace{5mm}
  \end{center}
  Notice that each pair of literals in the proof structure are only given the same number for their entry positions if they are connected by an axiom link, and that all bar one of the linked pairs share a number, one of the rogue pair being the leaf which is the $5^{th}$ block. Because of this, it is possible to follow a chain of compositions of permutations with $\bar{\lambda}$ in an order such that the permutation of a block $m$ is not considered until those of all the blocks whose unique path to the $5^{th}$ block in the block graph passes through $m$ have been (leaving $c^{5}$ to last). This ensures that the product of all the permutations corresponding to blocks greater than or equal to $m$ in the block graph of the structure with respect to the partial tree ordering induced by the vertex representing the $5^{th}$ block with the primitive Kronecker deltas of $\bar{\lambda}$ having at least one index in common with one of the permutations reduces to a Kronecker delta $\tensor*{\delta}{*^{x}_{i}}$, where $i$ is the sole remaining index relating to the literal connected via an axiom link to a literal not in the set of blocks greater than or equal to $m$, and $x$ is the number in the tuple placed underneath the same literal. The inevitable consequence is that the final Kronecker delta associated with the axiom link connected to the right-most leaf of the component having its indices substituted for two distinct numbers, which is equal to zero.
  
  \p Of course, we are not in a position to assume we are only dealing with singular proof structures and their scalar multiples --- linear combinations of proof structures have not yet been discounted. Lemma \ref{AcyclicLemma} allows us to assume that all proof structures in a linear combination being considered are acyclic. This has a rather useful consequence, namely that if any one proof structure in a linear combination of them is disconnected, then all the others are as well.
  
  \begin{lem} \label{OneDiscThenAllClaim}
  If a sequent $S$ can be bestowed with a valid set of axiom links which induce an acyclic yet disconnected proof structure, then all possible valid linkings are also disconnected.
  \end{lem}
  \proof
  Given a set of
  proof structures over $S$, we know that all the structures have the same number
  of edges. The number of axiom links in a single structure is equal to the number of pairs of literals which exist in $S$, and so independent of the position of the links; and since the proof structures are built over the same parse forest (namely the one described by $S$), the number of edges which are not axiom links are equal as well. The number of $\invamp$ connectives is trivially only dependent on $S$, and so the number of $\invamp$-vertices in each structure does not vary. The number of edges in a switching of a proof structure is equal to the number in the entire proof structure minus the number of $\invamp$-vertices in it, meaning that every switching of every proof structure over $S$ has the same number of edges.
  
  \p The number of vertices in a proof structure over $S$ (and therefore in each of its switchings), $n$ say, is equal to the total number of literals and connectives in the sequent; for a graph over this number of vertices to be connected the number of edges must be greater than $n-1$. If there is an acyclic and disconnected proof structure, its switchings must have strictly fewer than $n-1$ edges: acyclicity provides an upper bound of $n-1$, and the disconnectedness discounts the possibility of the number being exactly $n-1$. Therefore \emph{all} switchings of all other proof structures have strictly fewer than $n-1$ edges, which implies that the structures are all disconnected. \qed
  
  This fact leaves us with a slightly less daunting task. We merely need to place more restrictions on where entries must contain the value $1$ for each of the permutations being composed with the tensor describing the incorrect \mll~proof structure. Different proof structures clearly lead to different components, and this could be viewed as creating a moving target. However, the claim above ensures that no problems are caused. We choose one leaf in the example to be the block not to be given a permutation to compose with the MDNF~tensor, and let ourselves be prepared to add restrictions to permutations for any of the other blocks if they are in a different component from that leaf for any one of the proof structures being modelled in the linear combination.
  
  \p The first task is to create a set of permutations over some $[n]$ for an unacceptable linear combination of proof structures over a sequent $F$ which makes it possible to derive a disproof of its associated tensor's existence in the set of values $\bbrk{F}(\mathbf{C}_{n},\mathbf{C}_{n})$.
  
  \begin{algo} \label{DisconnectTupleAlg} Input: A linear combination of
    \hughf{acyclic, disconnected} MDNF proof structures of the same sequent
    containing $M$ blocks. \\ Output: A number $n \in \N^{+}$; tensors
    $b^{1},\ldots,b^{M}$ such that $\hughf{(b^{m})}_{i_{1} \cdots i_{L_{m}}} \in
    \PPerm(L_{m},n)$ for each $m$.
  \end{algo}
  \begin{enumerate}
  \item Let $\{\lambda_{1},\ldots,\lambda_{K}\}$ be the set of axiom
    links in a linear combination of acyclic yet disconnected MDNF
    proof structures which are multiplied by a non-zero scalar. For
    each of the blocks containing exactly one literal (\emph{leaves})
    assign distinct values $1,\ldots,B$. We assign variables
    $v_{1},\ldots,v_{M}$ to each of the $M$ blocks, with $v_{m} =
    L_{m}$ as the initial setting. We call these variables
    \emph{valencies}. Let $i = B+1$ and $k=1$.
  
  \item Take $\lambda_{k}$ and the $\invamp$-free subgraph of the parse
    tree, and consider the component of the graph containing the lowest
    numbered component which does not contain block $1$. Choose the
    leaf with the highest number in that component and change its
    valency to $0$.
  
  \item Choose the first block $m$ in the component such that $v_{m} =
    1$.
    \begin{enumerate}
    \item If one does exist and it is a leaf, let $v_{m} = 0$. If it is
      not adjacent to another leaf in the graph, then assign the number
      given to that leaf to the literal to which it is connected and decrement the valency of the block of the connected literal.
      Restart Step~3.
    \item If one does exist but it is not a leaf, then mark the sole
      literal not yet allocated a number as an `exit' literal and go to
      Step~4.
    \item If there are no more blocks of valency $1$ and $k<K$, create
      a new \\ $L_{m}$-tuple $\mathbf{x}$ for each block $m$ in the
      component, and for every $i \in [L_{m}]$ let $x_{i}$ be the label given to the $i^{th}$ literal of the
      block. If the $i^{th}$ literal is the exit of the block, then
      mark $x_{i}$ as an exit entry. Delete duplicate tuples associated
      with each block, and remove all the labels and marks from all the
      literals except the leaves. Increment $k$ and return to Step~2.
    \item Otherwise, terminate the algorithm after declaring that $n =
      2^{i}-1$ and stating that for each $m \in [2,M]$,
      \begin{displaymath}
        \hughf{(b^{m})}_{i_{1} \cdots i_{L_{m}}} = \left\{ 
          \begin{array}{ll}
            1 & \text{if }(i_{1},\ldots,i_{L_{m}}) \text{ is a tuple for } m \\
            0 & \text{otherwise.} 
          \end{array}
        \right.
      \end{displaymath}
    \end{enumerate}
  \item Check to see whether the numbers given to each of the non-exit
    literals correspond exactly to those of a tuple $\mathbf{x} =
    (x_{1},\ldots,x_{L_{m}})$ already associated with the block
    (meaning that if $u_{i}$ is the number given to the $i^{th}$
    literal in the block, that $u_{i} = x_{i}$ for every $i$ for which
    $u_{i}$ is defined).
    \begin{enumerate}
    \item If so, then assign the final unused number in $\mathbf{x}$ to
      the exit literal.
    \item If not, assign $i$ to the exit literal and increment~$i$.
    \end{enumerate}
    Assign the number to the literal with which the exit literal shares
    an axiom link, unless that literal is a leaf and already has been
    assigned a number. Decrement the valencies of both block $m$ and
    the block with which the exit literal of block $m$ is linked by an
    axiom link in $\lambda_{k}$. Go to Step~3.
  \end{enumerate}
  
  \begin{exa} \label{DisconnectExam1}
  Consider the sum of the three linkings $\lambda_{1}$, $\lambda_{2}$ and $\lambda_{2}$ provided below in red, blue and green respectively.
  
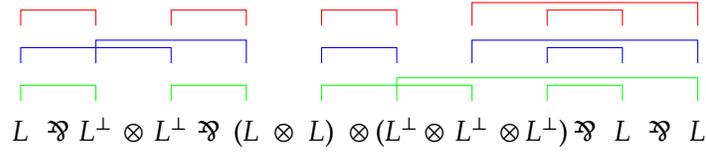
\begin{center} \vspace{5mm}
  \begin{tikzpicture}
    [auto, node
    distance=5mm, skip loop/.style={to path={-- ++(0,#1) -| (\tikztotarget)}}] \tikzstyle{every node} = [text depth=-5pt,text height=0.5ex]
   \node (z1) { }; 
      \node (1) [right of=z1] {$L$}; \node (a) [right of=1] {$\invamp$} ;
     \node (2) [right of=a] {$L^{\bot}$}; \node (b) [right of=2] {$\tens$} ;
     \node (3)  [right of=b] {$L^{\bot}$}; \node (c)  [right of=3] {$\invamp$};
     \node (4)  [right of=c] {$(L$}; \node(d) [right of=4] {$\tens$} ;
     \node (5) [right of=d] {$L)$}; \node (e) [right of=5] {$\tens$} ;
     \node (6) [right of=e] {$(L^{\bot}$};  \node (f) [right of=6] {$\tens$} ;
     \node (7)  [right of=f] {$L^{\bot}$}; \node (g)  [right of=7] {$\tens$};
     \node (8) [right of=g] {$L^{\bot})$};  \node (h) [right of=8] {$\invamp$} ;
     \node (9)  [right of=h] {$L$}; \node (i)  [right of=9] {$\invamp$};
     \node (10)  [right of=i] {$L$};
  \begin{scope}
  \path (1)  edge  [red, skip loop =16mm,shorten >=13mm, shorten <=13mm] (2);
  \path (3)  edge  [red, skip loop =16mm,shorten >=13mm, shorten <=13mm] (4);
  \path (5)  edge  [red, skip loop =16mm,shorten >=13mm, shorten <=13mm] (6);
  \path (7)  edge  [red, skip loop =17mm,shorten >=13mm, shorten <=13mm] (10);
  \path (8)  edge  [red, skip loop =16mm,shorten >=13mm, shorten <=13mm] (9);
  \path (1)  edge  [blue, skip loop =11mm,shorten >=8mm, shorten <=8mm] (3);
  \path (2)  edge  [blue, skip loop =12mm,shorten >=8mm, shorten <=8mm] (4);
  \path (5)  edge  [blue, skip loop =11mm,shorten >=8mm, shorten <=8mm] (6);
  \path (7)  edge  [blue, skip loop =12mm,shorten >=8mm, shorten <=8mm] (10);
  \path (8)  edge  [blue, skip loop =11mm,shorten >=8mm, shorten <=8mm] (9);
  \path (1)  edge  [green, skip loop =6mm, shorten >=3mm, shorten <=3mm] (2);
  \path (3)  edge  [green, skip loop =6mm, shorten >=3mm, shorten <=3mm] (4);
  \path (5)  edge  [green, skip loop =6mm, shorten >=3mm, shorten <=3mm] (7);
  \path (6)  edge  [green, skip loop =7mm, shorten >=3mm, shorten <=3mm] (10);
  \path (8)  edge  [green, skip loop =6mm, shorten >=3mm, shorten <=3mm] (9);
  \end{scope}
  \end{tikzpicture} \vspace{5mm}
  \end{center}
  The linear combination's tensor representation is 
\[\tensor*{\bar{\lambda}}{*^{j_{(2,1)}}_{i_{(1,1)}}^{j_{(3,1)}}_{i_{(4,1)}}^{j_{(5,1)}}_{i_{(4,2)}}^{j_{(5,2)}}_{i_{(6,1)}}^{j_{(5,3)}}_{i_{(7,1)}}}
=
\tensor*{\delta}{*^{j_{(2,1)}}_{i_{(1,1)}}^{j_{(3,1)}}_{i_{(4,1)}}^{j_{(5,1)}}_{i_{(4,2)}}^{j_{(5,3)}}_{i_{(6,1)}}^{j_{(5,2)}}_{i_{(7,1)}}}
+
\tensor*{\delta}{*^{j_{(3,1)}}_{i_{(1,1)}}^{j_{(2,1)}}_{i_{(4,1)}}^{j_{(5,1)}}_{i_{(4,2)}}^{j_{(5,3)}}_{i_{(6,1)}}^{j_{(5,2)}}_{i_{(7,1)}}}
+
\tensor*{\delta}{*^{j_{(2,1)}}_{i_{(1,1)}}^{j_{(3,1)}}_{i_{(4,1)}}^{j_{(5,2)}}_{i_{(4,2)}}^{j_{(5,3)}}_{i_{(6,1)}}^{j_{(5,1)}}_{i_{(7,1)}}}\,.
\]
 We apply Algorithm~\ref{DisconnectTupleAlg}.
  \end{exa}
  \begin{enumerate}[label={\cW4 \& }\arabic*.]
  \item[1.] We first attach values to the leaves in each of the proof structures described. The value $i$ reaches $6$.
  
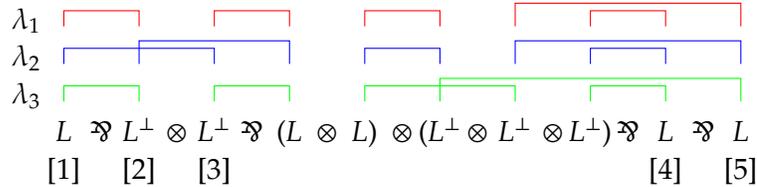
\begin{center} \vspace{5mm}
  \begin{tikzpicture}
    [auto, node
    distance=5mm, skip loop/.style={to path={-- ++(0,#1) -| (\tikztotarget)}}] \tikzstyle{every node} = [text depth=-5pt,text height=0.5ex]
   \node (z1) { }; 
      \node (1) [right of=z1] {$L$}; \node (a) [right of=1] {$\invamp$} ;
     \node (2) [right of=a] {$L^{\bot}$}; \node (b) [right of=2] {$\tens$} ;
     \node (3)  [right of=b] {$L^{\bot}$}; \node (c)  [right of=3] {$\invamp$};
     \node (4)  [right of=c] {$(L$}; \node(d) [right of=4] {$\tens$} ;
     \node (5) [right of=d] {$L)$}; \node (e) [right of=5] {$\tens$} ;
     \node (6) [right of=e] {$(L^{\bot}$};  \node (f) [right of=6] {$\tens$} ;
     \node (7)  [right of=f] {$L^{\bot}$}; \node (g)  [right of=7] {$\tens$};
     \node (8) [right of=g] {$L^{\bot})$};  \node (h) [right of=8] {$\invamp$} ;
     \node (9)  [right of=h] {$L$}; \node (i)  [right of=9] {$\invamp$};
     \node (10)  [right of=i] {$L$};
    \node (t1) [above of= z1] {$\lambda_{3}$};
    \node (t2) [above of= t1] {$\lambda_{2}$};
    \node (t3) [above of= t2] {$\lambda_{1}$};
    \node (11) [below of=1] {[1]};
    \node (12) [below of=2] {[2]}; 
    \node (13) [below of=3] {[3]}; 
    \node (14) [below of=4] { }; 
    \node (15) [below of=5] { }; 
    \node (16) [below of=6] { };
    \node (17) [below of=7] { };
    \node (18) [below of=8] { };
    \node (19) [below of=9] {[4]};
    \node (20) [below of=10] {[5]};
  \begin{scope}
  \path (1)  edge  [red, skip loop =16mm,shorten >=13mm, shorten <=13mm] (2);
  \path (3)  edge  [red, skip loop =16mm,shorten >=13mm, shorten <=13mm] (4);
  \path (5)  edge  [red, skip loop =16mm,shorten >=13mm, shorten <=13mm] (6);
  \path (7)  edge  [red, skip loop =17mm,shorten >=13mm, shorten <=13mm] (10);
  \path (8)  edge  [red, skip loop =16mm,shorten >=13mm, shorten <=13mm] (9);
  \path (1)  edge  [blue, skip loop =11mm,shorten >=8mm, shorten <=8mm] (3);
  \path (2)  edge  [blue, skip loop =12mm,shorten >=8mm, shorten <=8mm] (4);
  \path (5)  edge  [blue, skip loop =11mm,shorten >=8mm, shorten <=8mm] (6);
  \path (7)  edge  [blue, skip loop =12mm,shorten >=8mm, shorten <=8mm] (10);
  \path (8)  edge  [blue, skip loop =11mm,shorten >=8mm, shorten <=8mm] (9);
  \path (1)  edge  [green, skip loop =6mm, shorten >=3mm, shorten <=3mm] (2);
  \path (3)  edge  [green, skip loop =6mm, shorten >=3mm, shorten <=3mm] (4);
  \path (5)  edge  [green, skip loop =6mm, shorten >=3mm, shorten <=3mm] (7);
  \path (6)  edge  [green, skip loop =7mm, shorten >=3mm, shorten <=3mm] (10);
  \path (8)  edge  [green, skip loop =6mm, shorten >=3mm, shorten <=3mm] (9);
  \end{scope}
  \end{tikzpicture} \vspace{5mm}
  \end{center}
  We make sure all the other components have valency the same as the number of literals within them: $v_{3} = 1$, $v_{4} = 4$, $v_{5} = 3$, $v_{6} = 1$ and $v_{7} = 1$.
  
  \item[2.] We proceed with $\lambda_{1}$. With this set of axiom links the first block not in the same component as the first is the third. The component containing block $3$ contains the $4^{th}$, $5^{th}$, $6^{th}$ and $7^{th}$ blocks as well. We change the valency of block $7$ to $0$.
  
  \item[3 \& 4.] We must repeat the processes contained in Steps~3 and~4 four times (once for each axiom link in $\lambda_{1}$ connected to the component containing block $3$).
  	\begin{itemize}
  	\item Block $3$ is a leaf and connects to the first literal of block $4$. That literal is therefore given the same number allocated to block $3$ ($3$), and the second literal is marked as an exit. Block $4$ now has valency $1$, and block $3$ has valency $0$.
  	\item Block $4$ is now the first block to have valency $1$. Its second literal, is given $i=6$ as its tuple entry, as is the first literal of the fifth block, which is adjacent. The valencies of blocks $4$ and $5$ become $0$ and $2$. The counter $i$ is incremented to $7$.
  	\item Block $6$, a leaf, is now the first block with valency $1$. It has already been given value $4$, so we give this to its adjacent literal --- the third literal in the fifth block. We decrement the block valencies.
  	\item Block $5$ is now the only one with valency~$1$ in the component. Its only empty position --- the second --- is allocated the value $i = 7$. However, it is connected to the last leaf, and so we do not give its adjacent literal the same number (it already has been given the entry $5$).
  	\end{itemize}
  	We are left with the following tuples, with the numbers in squares emphasising exit literals for that particular linking:
  
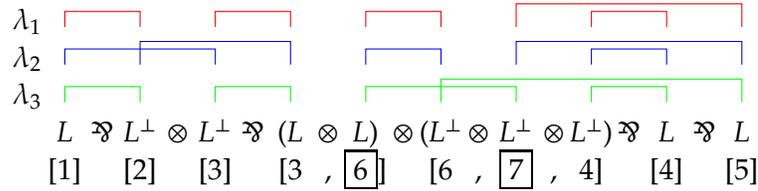
\begin{center} \vspace{5mm}
  \begin{tikzpicture}
    [auto, node
    distance=5mm, skip loop/.style={to path={-- ++(0,#1) -| (\tikztotarget)}}] \tikzstyle{every node} = [text depth=-5pt,text height=0.5ex]
   \node (z1) { }; 
      \node (1) [right of=z1] {$L$}; \node (a) [right of=1] {$\invamp$} ;
     \node (2) [right of=a] {$L^{\bot}$}; \node (b) [right of=2] {$\tens$} ;
     \node (3)  [right of=b] {$L^{\bot}$}; \node (c)  [right of=3] {$\invamp$};
     \node (4)  [right of=c] {$(L$}; \node(d) [right of=4] {$\tens$} ;
     \node (5) [right of=d] {$L)$}; \node (e) [right of=5] {$\tens$} ;
     \node (6) [right of=e] {$(L^{\bot}$};  \node (f) [right of=6] {$\tens$} ;
     \node (7)  [right of=f] {$L^{\bot}$}; \node (g)  [right of=7] {$\tens$};
     \node (8) [right of=g] {$L^{\bot})$};  \node (h) [right of=8] {$\invamp$} ;
     \node (9)  [right of=h] {$L$}; \node (i)  [right of=9] {$\invamp$};
     \node (10)  [right of=i] {$L$};
    \node (t1) [above of= z1] {$\lambda_{3}$};
    \node (t2) [above of= t1] {$\lambda_{2}$};
    \node (t3) [above of= t2] {$\lambda_{1}$};
    \node (11) [below of=1] {[1]}; 
    \node (12) [below of=2] {[2]}; 
    \node (13) [below of=3] {[3]}; 
    \node (14) [below of=4] {[3};  \node (1d) [below of=d] {,};
    \node (15) [below of=5] {\fbox{6}]}; 
    \node (16) [below of=6] {[6}; \node (1f) [below of=f] {,};
    \node (17) [below of=7] {\fbox{7}}; \node (1g) [below of=g] {,};
    \node (18) [below of=8] {4]}; 
    \node (19) [below of=9] {[4]}; 
    \node (20) [below of=10] {[5]};
  \begin{scope}
  \path (1)  edge  [red, skip loop =16mm,shorten >=13mm, shorten <=13mm] (2);
  \path (3)  edge  [red, skip loop =16mm,shorten >=13mm, shorten <=13mm] (4);
  \path (5)  edge  [red, skip loop =16mm,shorten >=13mm, shorten <=13mm] (6);
  \path (7)  edge  [red, skip loop =17mm,shorten >=13mm, shorten <=13mm] (10);
  \path (8)  edge  [red, skip loop =16mm,shorten >=13mm, shorten <=13mm] (9);
  \path (1)  edge  [blue, skip loop =11mm,shorten >=8mm, shorten <=8mm] (3);
  \path (2)  edge  [blue, skip loop =12mm,shorten >=8mm, shorten <=8mm] (4);
  \path (5)  edge  [blue, skip loop =11mm,shorten >=8mm, shorten <=8mm] (6);
  \path (7)  edge  [blue, skip loop =12mm,shorten >=8mm, shorten <=8mm] (10);
  \path (8)  edge  [blue, skip loop =11mm,shorten >=8mm, shorten <=8mm] (9);
  \path (1)  edge  [green, skip loop =6mm, shorten >=3mm, shorten <=3mm] (2);
  \path (3)  edge  [green, skip loop =6mm, shorten >=3mm, shorten <=3mm] (4);
  \path (5)  edge  [green, skip loop =6mm, shorten >=3mm, shorten <=3mm] (7);
  \path (6)  edge  [green, skip loop =7mm, shorten >=3mm, shorten <=3mm] (10);
  \path (8)  edge  [green, skip loop =6mm, shorten >=3mm, shorten <=3mm] (9);
  \end{scope}
  \end{tikzpicture} \vspace{5mm}
  \end{center}
  
  \item[2.] The process is then repeated with $\lambda_{2}$ and $\lambda_{3}$. Particular points to note are the following:
  	\begin{itemize}
  	\item The component containing block $1$ can change (note that blocks $1$ and $2$ are not respected with respect to $\lambda_{2}$, but the two are adjacent with when the linkings $\lambda_{1}$ and $\lambda_{3}$ are used).
  	\item In the third iteration for $\lambda_{3}$, block $4$ is reached at the same time in the same way as in $\lambda_{1}$. As such, the number $6$ is reused for the tuple entry for its second literal.
  	\end{itemize}
  We now have three sets of tuples for each block (possibly repeated):
  
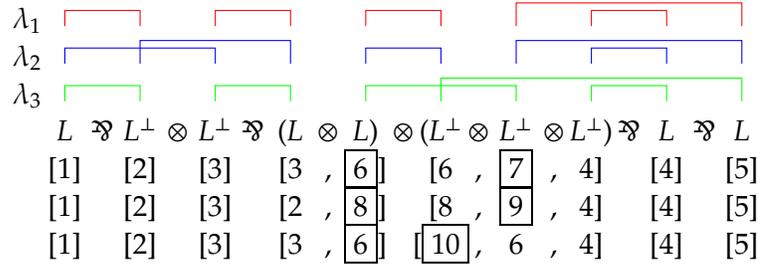
\begin{center} \vspace{5mm}
  \begin{tikzpicture}
    [auto, node
    distance=5mm, skip loop/.style={to path={-- ++(0,#1) -| (\tikztotarget)}}] \tikzstyle{every node} = [text depth=-5pt,text height=0.5ex]
   \node (z1) { }; 
      \node (1) [right of=z1] {$L$}; \node (a) [right of=1] {$\invamp$} ;
     \node (2) [right of=a] {$L^{\bot}$}; \node (b) [right of=2] {$\tens$} ;
     \node (3)  [right of=b] {$L^{\bot}$}; \node (c)  [right of=3] {$\invamp$};
     \node (4)  [right of=c] {$(L$}; \node(d) [right of=4] {$\tens$} ;
     \node (5) [right of=d] {$L)$}; \node (e) [right of=5] {$\tens$} ;
     \node (6) [right of=e] {$(L^{\bot}$};  \node (f) [right of=6] {$\tens$} ;
     \node (7)  [right of=f] {$L^{\bot}$}; \node (g)  [right of=7] {$\tens$};
     \node (8) [right of=g] {$L^{\bot})$};  \node (h) [right of=8] {$\invamp$} ;
     \node (9)  [right of=h] {$L$}; \node (i)  [right of=9] {$\invamp$};
     \node (10)  [right of=i] {$L$};
    \node (t1) [above of= z1] {$\lambda_{3}$};
    \node (t2) [above of= t1] {$\lambda_{2}$};
    \node (t3) [above of= t2] {$\lambda_{1}$};
    \node (11) [below of=1] {[1]}; 
    \node (12) [below of=2] {[2]}; 
    \node (13) [below of=3] {[3]}; 
    \node (14) [below of=4] {[3};  \node (1d) [below of=d] {,};
    \node (15) [below of=5] {\fbox{6}]}; 
    \node (16) [below of=6] {[6}; \node (1f) [below of=f] {,};
    \node (17) [below of=7] {\fbox{7}}; \node (1g) [below of=g] {,};
    \node (18) [below of=8] {4]}; 
    \node (19) [below of=9] {[4]}; 
    \node (20) [below of=10] {[5]};
    \node (21) [below of=11] {[1]}; 
    \node (22) [below of=12] {[2]}; 
    \node (23) [below of=13] {[3]}; 
    \node (24) [below of=14] {[2};  \node (2d) [below of=1d] {,};
    \node (25) [below of=15] {\fbox{8}]}; 
    \node (26) [below of=16] {[8}; \node (2f) [below of=1f] {,};
    \node (27) [below of=17] {\fbox{9}}; \node (2g) [below of=1g] {,};
    \node (28) [below of=18] {4]}; 
    \node (29) [below of=19] {[4]}; 
    \node (30) [below of=20] {[5]};  
    \node (31) [below of=21] {[1]}; 
    \node (32) [below of=22] {[2]}; 
    \node (33) [below of=23] {[3]}; 
    \node (34) [below of=24] {[3};  \node (3d) [below of=2d] {,};
    \node (35) [below of=25] {\fbox{6}]}; 
    \node (36) [below of=26] {[\fbox{10}}; \node (3f) [below of=2f] {,};
    \node (37) [below of=27] {6}; \node (3g) [below of=2g] {,};
    \node (38) [below of=28] {4]}; 
    \node (39) [below of=29] {[4]}; 
    \node (40) [below of=30] {[5]};
  \begin{scope}
  \path (1)  edge  [red, skip loop =16mm,shorten >=13mm, shorten <=13mm] (2);
  \path (3)  edge  [red, skip loop =16mm,shorten >=13mm, shorten <=13mm] (4);
  \path (5)  edge  [red, skip loop =16mm,shorten >=13mm, shorten <=13mm] (6);
  \path (7)  edge  [red, skip loop =17mm,shorten >=13mm, shorten <=13mm] (10);
  \path (8)  edge  [red, skip loop =16mm,shorten >=13mm, shorten <=13mm] (9);
  \path (1)  edge  [blue, skip loop =11mm,shorten >=8mm, shorten <=8mm] (3);
  \path (2)  edge  [blue, skip loop =12mm,shorten >=8mm, shorten <=8mm] (4);
  \path (5)  edge  [blue, skip loop =11mm,shorten >=8mm, shorten <=8mm] (6);
  \path (7)  edge  [blue, skip loop =12mm,shorten >=8mm, shorten <=8mm] (10);
  \path (8)  edge  [blue, skip loop =11mm,shorten >=8mm, shorten <=8mm] (9);
  \path (1)  edge  [green, skip loop =6mm, shorten >=3mm, shorten <=3mm] (2);
  \path (3)  edge  [green, skip loop =6mm, shorten >=3mm, shorten <=3mm] (4);
  \path (5)  edge  [green, skip loop =6mm, shorten >=3mm, shorten <=3mm] (7);
  \path (6)  edge  [green, skip loop =7mm, shorten >=3mm, shorten <=3mm] (10);
  \path (8)  edge  [green, skip loop =6mm, shorten >=3mm, shorten <=3mm] (9);
  \end{scope}
  \end{tikzpicture} \vspace{5mm}
  \end{center}
  	
  \item[3.] The algorithm finished, leaving the following set of tensors:
  \begin{eqnarray*}
  & & \hughf{(b^{2})}_{j_{(2,1)}}	=	\tensor*{\delta}{*^{2}_{j_{(2,1)}}} \\
  & & \hughf{(b^{3})}_{j_{(3,1)}}	=	\tensor*{\delta}{*^{3}_{j_{(3,1)}}} \\
  & & \hughf{(b^{4})}_{i_{(4,1)}i_{(4,2)}} = \tensor*{\delta}{*^{3}_{i_{(4,1)}}^{6}_{i_{(4,2)}}} + \tensor*{\delta}{*^{2}_{i_{(4,1)}}^{8}_{i_{(4,2)}}} \\
  & & \hughf{(b^{5})}_{j_{(5,1)}j_{(5,2)}j_{(5,3)}}	=	\tensor*{\delta}{*^{6}_{j_{(5,1)}}^{7}_{j_{(5,2)}}^{4}_{j_{(5,2)}}} + \tensor*{\delta}{*^{8}_{j_{(5,1)}}^{9}_{j_{(5,2)}}^{4}_{j_{(5,2)}}} + \tensor*{\delta}{*^{10}_{j_{(5,1)}}^{6}_{j_{(5,2)}}^{4}_{j_{(5,2)}}} \\
  & & \hughf{(b^{6})}_{i_{(6,1)}}	=	\tensor*{\delta}{*^{4}_{i_{(6,1)}}} \\
  & & \hughf{(b^{7})}_{i_{(7,1)}} =	\tensor*{\delta}{*^{5}_{i_{(7,1)}}}
  \end{eqnarray*}
  \end{enumerate}
  
\noindent  The above algorithm provides a set of partial permutations with entries containing the value $1$ at positions which are of use in the final proof of the necessity of connectedness having to be satisfied. However, they are still \emph{partial} permutations, and in order to discuss objects of the form $F(\mathbf{C}_{n},\mathbf{C}_{n})$ full permutations must be used. It is therefore necessary to complete the permutations. Completing permutations is non-trivial, but by virtue of a number of properties bestowed on each of the partial permutations described in Algorithm \ref{DisconnectTupleAlg}, we are indeed capable of performing such a task.
  
  \begin{lem} \label{ConAlgCompleteLemma} Every partial higher-order
    permutation $\hughf{(b^{m})}_{\mathbf{i}}$ where $m \in [2,M]$ which is an
    output tensor of Algorithm~\ref{DisconnectTupleAlg} can be
    completed to form a full higher-order permutation~$\hughf{(c^{m})}_{\mathbf{i}}$.
  \end{lem}
  \proof
    It is known from Definition~\ref{TensorExamDefns} that for any $L$ and $n$ it is always possible to
    find the tensor $\cycle(L,n,0)_{i_{1} \cdots i_{L}}$ in $\Perm(L,n)$. It is also well established that using a permutation over $[n]$ on the numbering
    used on a single index of a permutation in $\Perm(L,n)$ for any $L \in \N^{+}$ always produces another equally valid permutation in the same set.
    
  Suppose that $b_{i_{1} \cdots i_{L}} \in \PPerm(L,n)$ is one of the partial permutations generated by Algorithm \ref{DisconnectTupleAlg} after being given appropriate inputs. Then we can refer back to the tuples $\mathbf{x}^{1}, \ldots ,\mathbf{x}^{K'}$ created in the middle of the algorithm which describe the positions where its entries are equal to $1$, and also the positions in each of the tuples with an `exit marker'. From the definition of the procedure, we can be assured that if the $j^{th}$ position of the $k^{th}$ tuple is marked, then $l^{k}_{j} \neq l^{k'}_{j}$ for all $k \neq k'$. In other words, the number found in the position of a tuple's exit marker is not used again in the same position of another tuple. With
    this information, we define a set of partial functions in $\N^{+}$
    $\{\alpha_{l}:l \in [L]\}$ in the following manner:
    \begin{enumerate}
    \item Let $i = 1$, $k = 1$ and $l = 1$.
    \item If $\alpha_{m}(x^{k}_{l})$ has not yet been assigned a value
      and the tuple position in question was not marked an exit entry in Algorithm~\ref{DisconnectTupleAlg}, then we say that
      $\alpha_{m}(x^{k}_{l}) = 2^{i}$ and then increment $i$. Go to
      Step~3.
    \item If $l < T$, increment $l$ and go to Step~2; otherwise
      increment $k$, set $l$ back to $1$ and go to Step~4.
    \item If $k \leq K$, then go to Step~2; otherwise, go to Step~5.
    \item Reset $k$ and $l$.
    \item If $\alpha_{l}(x^{k}_{l})$ has not yet been assigned a value,
      then set $\alpha_{l}(x^{k}_{l}) = n - \textstyle\sum_{j \neq l}
      \alpha_{j}(x^{k}_{j})$ and go to Step~7. Otherwise increment $l$
      and repeat Step~6.
    \item If $k < K$, increment $k$ and let $l = 1$, and return to Step
      6. Otherwise, terminate the algorithm.
    \end{enumerate}
  
\noindent     This algorithm is well defined, and each partial function
    $\alpha_{l}$ is injective with all defined source and target values
    in $[n]$. As such, it is possible to restrict each of the partial
    functions to act upon $[n]$, and then to extend them to full
    $2$-permutations over (i.e. bijective endomorphisms on) $[n]$. One way in which such an extension of a partial function $\alpha_{m}$ could be formed is inductively: by assigning the lowest value in $[n]$ not already in the image of $\alpha_{m}$ to be the image of the lowest number in $[n]$ not in domain of definition, and continuing similarly with the new definition of the partial function until it becomes total.
    
    \p We now define the tensor $c_{i_{1} \cdot i_{L}}$ for $b_{i_{1} \cdots i_{L}}$ in the following manner:
      \begin{displaymath}
      c_{i_{1}\cdots i_{L}} = \left\{ 
        \begin{array}{ll}
          1 & \text{if } \textstyle\sum_{j=1}^{L}\alpha_{j}(i_{j}) \equiv 0
          \text{ mod } n \\ 
          0 & \text{otherwise} 
        \end{array}
      \right.
    \end{displaymath} 
    The tensor is created by using $2$-permutations on each of the entry positions of $\cycle(L,n,0)$, and so belongs to the set $\Perm(L,n)$. Furthermore, since $c_{x_{1} \cdots x_{T}} = 1$ for each tuple $(x_{1},\ldots,x_{T})$ where $b_{x_{1} \cdots x_{T}} = 1$ by design, it is a completion of $b_{i_{1} \cdots i_{L}}$, and as such we have satisfied the original statement as desired. \qed
  
  
  Much in the same manner as with Lemma~\ref{AcyclicLemma}, we are finally in a position to prove that the proof structures being modelled in linear combinations in the double-glued category are always proof nets.
  
  \begin{prop} \label{ConnectedLemma} Every MDNF transformation in $\GC$
    modelling a linear combination of acyclic proof structures is
    modelling a linear combination of cut-free proof nets.
  \end{prop}
  \proof
    Let $F$ be an MDNF formula as in Lemma~\ref{AcyclicLemma}, and let \\ $\tau = (\tau_{\textbf{R}} \in
    \C\left[I,|\bbrk{F}|(\mathbf{R},\mathbf{R})\right])_{\mathbf{R} \in
      \C^{N}}$ be an MDNF transformation in $\C$ modelling a linear
    combination of acyclic yet disconnected proof structures. We consider $\bbrk{F}\left(\textbf{C}_{n},\textbf{C}_{n}\right)_{Val}$
    where $n$ is the output integer given by
    Algorithm~\ref{DisconnectTupleAlg} if the linear combination described by $\tau$ is used as an input. From
    Section~\ref{SectionMDNFObjects} we know the criteria the tensor
    representations of the values of this object must satisfy.
  
  \p As seen in earlier deductions, for the integer $n$ acquired from the algorithm, we have the tensor representation
    \begin{displaymath}
      \tensor*{\tau}{*^{j_{(1, 1)}}_{i_{(1,
            1)}}^{\;\cdots\;}_{\;\cdots}^{j_{(M, 
            N_M)}}_{i_{(M,P_{M})}}}
      = \textstyle\sum_{\beta} s_{\beta}\cdot
      \tensor*{\delta}{*^{j_{\beta(1,1)}}_{i_{(1,
            1)}}^{\;\cdots\;}_{\;\cdots}^{j_{\beta(M, P_{M})}}_{i_{(M,
            P_M)}}}.
    \end{displaymath}
    Algorithm~\ref{DisconnectTupleAlg} and Lemma~\ref{ConAlgCompleteLemma} makes it possible to obtain a
    set of (full) permutations $\{c^{m}:m \in [2,M]\}$ for the coming argument.
  
    \p Take any set of axiom links described by a bijection, $\zeta$ say,
    such that $s_{\zeta} \neq 0$, and consider the tensor
    $\tensor*{\delta}{*^{j_{\zeta(1,1)}}_{i_{(1,
          1)}}^{\;\cdots\;}_{\;\cdots}^{j_{\zeta(M, P_{M})}}_{i_{(M,
          P_M)}}}$. It is possible to factorise this tensor into a
    product of Kronecker deltas, one for each component in the
    switching of $\zeta$. Multiplying the first of these smaller
    component tensors not containing the indices concerning the first
    block, $\tensor*{\delta}{*^{j_{\zeta(\theta(1),1)}}_{i_{(\theta(1),
          1)}}^{\;\cdots\;}_{\;\cdots}^{j_{\zeta(\theta(M'), P_{\theta
            (M')})}}_{i_{(\theta(M'), P_{\theta(M')})}}}$ say, with
    each of the permutations $b^{\theta(1)},\ldots,b^{\theta(M')}$
    corresponding to the blocks within the component in question, the zero scalar is produced.
    
    \begin{clmenv}
    \begin{cLm}
    Using the definitions of $\zeta$ and $\theta$ given above, 
    \begin{displaymath}
    \tensor*{\delta}{*^{j_{\zeta(\theta(1),1)}}_{i_{(\theta(1),
            1)}}^{\;\cdots\;}_{\;\cdots}^{j_{\zeta(\theta(M'),
            P_{\theta (M')})}}_{i_{(\theta(M'),
            P_{\theta(M')})}}}
      \cdot
      \textstyle\prod_{m=1}^{M'}
      \tensor*{\left(c^{\theta(m)}\right)}{*^{j_{(\theta(m),
            1)}}_{i_{(\theta(m), 1)}}^{\;\cdots\;}_{\;\cdots}^{j_{(\theta(m),
            N_{\theta(m)})}}_{i_{(\theta(m), P_{\theta(m)})}}}
      = 0
    \end{displaymath}
    \end{cLm}
    \proof
    It is easy to show that there is a partial ordering induced by every tree and every choice of node within it. Given a tree $G$ and vertex $w$, we can create an ordering $\leq_{(G,w)}$ defined as follows: if the unique path starting from a vertex $u$ and finishing at $w$ passes through the vertex $v$, then we say that $u \leq_{(G,w)} v$. The point $w$ is therefore maximal in this ordering. This is a partial order if we close this under reflexivity.
    
    \p We consider the blocks with numbers in the image of $\theta$ (that is, the blocks of the second component) observing the tree partial ordering induced by the leaf with the highest block number $\theta(q)$. With block $\theta(q)$ being a leaf, the tensor $c^{\theta(q)} \,=\, \delta_{k_{(\theta(q),1)}}^{L}$ for some constant $L$, where the $k$-index replaces either an $i$- or $j$-index depending on the polarity of the literal in the block. It should also be noted that, by design of the algorithm, $L$ is a smaller number than any of the numbers given exit markers within the component which are not in a leaf block, and greater than all other numbers given to leaves in the component. We use an inductive argument on the blocks in this order to demonstrate that, for any block $\theta(m)$ except the reference leaf, a composition of the tensors in $\{c^{\theta(m')}:\theta(m') \leq_{G} \theta(m)\}$ with all the Kronecker deltas pairs whose indices are shared with two permutations in that set reduces to a constant tensor on the index with the exit marker created at this point in Algorithm \ref{DisconnectTupleAlg} and the number with said marker. That is,
    \begin{eqnarray*}
   \textstyle\prod_{m' <_{(G,\theta(q))} \theta(m)}\tensor*{\delta}{*^{j_{\zeta(m',1)}}_{i_{(m',
            1)}}^{\;\cdots\;}_{\;\cdots}^{j_{\zeta(m',
            P_{m'})}}_{i_{(m',
            P_{m'})}}} 
      \cdot \textstyle\prod_{m'\leq_{(G,\theta(q))}\theta(m)}\tensor*{\left(c^{m'}\right)}{*^{j_{(m',
            1)}}_{i_{(m', 1)}}^{\;\cdots\;}_{\;\cdots}^{j_{(m',
            N_{m'})}}_{i_{(m', P_{\theta(m')})}}} & = & \delta_{k}^{x}
    \end{eqnarray*}
    where $k$ is the sole index not to be composed, and $x$ is the number connected to the exit marker in the tuple generated by the algorithm the appropriate linking.
    \begin{itemize}
    \item The base case occurs when block $\theta(m)$ is a leaf. The first product of Kronecker deltas is then empty, and the second only contains the permutation $c^{\theta(m)}$. Since block $\theta(m)$ is a leaf, it must be the case that there is only one literal in the block, and therefore the permutation is a $1$-permutation, i.e. a constant tensor of dimension $1$. The algorithm defines the non-zero entry to occur at the $x^{th}$ position, where $x$ is the single value given to the block in Algorithm~\ref{DisconnectTupleAlg}, as desired.
    \item If not, then the composition can be split further. Let $H(m)$ denote the neighbourhood of block~$m$ in the graph and $E \,=\, \left\{ m' \in H(\theta(m))\,:\, m' <_{(G,\theta(q))} \right\} \,\neq\, \varnothing$. We can rewrite the composition above as follows:
    
    \begin{align*}
    \prod_{m' \in E} \left(\delta_{k_{m'}}^{l_{\xi{m'}}} \prod_{m'' <_{(G,\theta(q))} m'} \tensor*{\delta}{*^{j_{\zeta(m'',1)}}_{i_{(m'',
              1)}}^{\;\cdots\;}_{\;\cdots}^{j_{\zeta(m'',
              P_{m''})}}_{i_{(m'',
              P_{m''})}}}\right. &\cdot\, \left.\textstyle\prod_{m''\leq_{(G,\theta(q))}m'}\tensor*{\left(c^{m''}\right)}{*^{j_{(m'',
                        1)}}_{i_{(m'', 1)}}^{\;\cdots\;}_{\;\cdots}^{j_{(m'',
                        N_{m''})}}_{i_{(m'', P_{m''})}}}  \right) \,\\
    &\cdot\, \tensor*{\left(c^{\theta(m)}\right)}{*^{j_{(\theta(m),
                                              1)}}_{i_{(\theta(m), 1)}}^{\;\cdots\;}_{\;\cdots}^{j_{(\theta(m),
                                              N_{\theta(m)})}}_{i_{(\theta(m), P_{\theta(m)})}}}
    \end{align*}
    
    By the induction hypothesis, the inside of the bracket can be simplified, leaving
    \begin{align*}
    \left( \prod_{m' \in E} \delta_{k_{m}}^{x_{m'}} \cdot \delta_{k_{m'}}^{l_{\xi(m')}} \right) &\cdot \tensor*{\left(c^{\theta(m)}\right)}{*^{j_{(\theta(m),1)}}_{i_{(\theta(m), 1)}}^{\;\cdots\;}_{\;\cdots}^{j_{(\theta(m),N_{\theta(m)})}}_{i_{(\theta(m), P_{\theta(m)})}}} \\
    &=	\prod_{m' \in E} \delta_{l_{\xi(m')}}^{x_{m'}} \cdot \tensor*{\left(c^{\theta(m)}\right)}{*^{j_{(\theta(m),1)}}_{i_{(\theta(m), 1)}}^{\;\cdots\;}_{\;\cdots}^{j_{(\theta(m),N_{\theta(m)})}}_{i_{(\theta(m), P_{\theta(m)})}}} \\
    &=	\tensor*{\left(c^{\theta(m)}\right)}{*^{x_{\pi_{1}(\zeta^{-1}(\theta(m),1))}}_{x_{\pi_{1}(\zeta(\theta(m),1))}}^{\;\cdots\;}_{\;\cdots}^{x_{\pi_{1}(\zeta^{-1}(\theta(m),y))}}_{i_{(\theta(m),y)}}^{\;\cdots\;}_{\;\cdots}^{x_{\pi_{1}(\zeta^{-1}(\theta(m),N_{m}))}}_{x_{\pi_{1}(\zeta(\theta(m),P_{m}))}}} \\ &\quad \mbox{ or } \tensor*{\left(c^{\theta(m)}\right)}{*^{x_{\pi_{1}(\zeta^{-1}(\theta(m),1))}}_{x_{\pi_{1}(\zeta(\theta(m),1))}}^{\;\cdots\;}_{\;\cdots}^{j_{(\theta(m),y)}}_{x_{\pi_{1}(\zeta(\theta(m),y))}}^{\;\cdots\;}_{\;\cdots}^{x_{\pi_{1}(\zeta^{-1}(\theta(m),N_{m}))}}_{x_{\pi_{1}(\zeta(\theta(m),P_{m}))}}}
    \end{align*}
    
    Since $c^{\theta(m)}$ is a full $(P_{m}+N_{m})$-permutation, it must be the case that our final tensor, with constants taking the places of all bar one of the entry positions, is a $1$-permutation as desired. Furthermore, the Kronecker deltas which have been fed into the equation are in agreement with exactly one set of tuples that is provided by the algorithm for $c^{\theta(m)}$. We therefore know that the number that the remaining free variable for the tensor must equal the final unused number from the tuple, which is the value at the exit marker. Our equation reduces to the form $\delta^{x}_{k}$ with free index $k$ and constant $x$ as desired.

    \end{itemize}
    Now we refer back to the original claim. The composition
      \begin{eqnarray*}
   & & \tensor*{\delta}{*^{j_{\zeta(\theta(1),1)}}_{i_{(\theta(1),
            1)}}^{\;\cdots\;}_{\;\cdots}^{j_{\zeta(\theta(M'),
            P_{\theta (M')})}}_{i_{(\theta(M'),
            P_{\theta(M')})}}}
  \cdot          
            \textstyle\prod_{m=1}^{M'}
      \tensor*{\left(c^{\theta(m)}\right)}{*^{j_{(\theta(m),
            1)}}_{i_{(\theta(m), 1)}}^{\;\cdots\;}_{\;\cdots}^{j_{(\theta(m),
            N_{\theta(m)})}}_{i_{(\theta(m), P_{\theta(m)})}}} \\ 
            & & \quad\quad\quad\quad = 
             \tensor*{\delta}{*^{j_{\zeta(\theta(1),1)}}_{i_{(\theta(1),
            1)}}^{\;\cdots\;}_{\;\cdots}^{j_{\zeta(\theta(M'),
            P_{\theta (M')})}}_{i_{(\theta(M'),
            P_{\theta(M')})}}}
            \cdot
                \textstyle\prod_{m \neq q}
      \tensor*{(c^{\theta(m)})}{*^{j_{(\theta(m),
            1)}}_{i_{(\theta(m), 1)}}^{\;\cdots\;}_{\;\cdots}^{j_{(\theta(m),
            N_{\theta(m)})}}_{i_{(\theta(m), P_{\theta(m)})}}} \cdot \delta_{k}^{L}
     \end{eqnarray*}
           
    reduces to the very simple $\delta_{k_{(\theta(q),1)}}^{x}\delta_{k_{(\theta(q),1)}}^{L'} = \delta_{x}^{L}$, where the $k$-index is a substitute for an $i$- or $j$-index, depending on whether the literal for the component $\theta(q)$ is positive or negative as before. By the design of the algorithm, this final value $x$ is assured of being not equal to $L$; and therefore it must be the case that $\delta_{x}^{L} = 0$ as desired. \qed
    \end{clmenv}
      
    \noindent Multiplying a zero tensor with any other tensor results in another
    zero tensor, and as such we know that
    \begin{displaymath}
      \tensor*{\delta}{*^{j_{\zeta(1,1)}}_{i_{(1,
            1)}}^{\;\cdots\;}_{\;\cdots}^{j_{\zeta(M, P_{M})}}_{i_{(M,
            P_M)}}} \cdot
      \textstyle\prod_{m=2}^{M}    \tensor*{\left(c^{m}\right)}{*^{j_{(m,
            1)}}_{i_{(m, 1)}}^{\;\cdots\;}_{\;\cdots}^{j_{(m,
            N_{m})}}_{i_{(m, P_{m})}}} =
      \tensor*{0}{*^{j_{(1,1)}}_{i_{(1,1)}}^{\;\cdots\;}_{\;\cdots}^{j_{(1, 
            N_1)}}_{i_{(1,P_{1})}}}.
    \end{displaymath}
    Since
    $\tensor*{\tau}{*^{j_{(1,1)}}_{i_{(1,1)}}^{\;\cdots\;}_{\;\cdots}^{j_{(M,
          N_M)}}_{i_{(M,P_{M})}}}$ is merely a linear combination of
    tensors like $ \tensor*{\delta}{*^{j_{\zeta(1,1)}}_{i_{(1,
          1)}}^{\;\cdots\;}_{\;\cdots}^{j_{\zeta(M, P_{M})}}_{i_{(M,
          P_M)}}}$, and $\zeta$ was arbitrary, we know that
    \begin{displaymath}
      \tensor*{\tau}{*^{j_{(1,1)}}_{i_{(1,
            1)}}^{\;\cdots\;}_{\;\cdots}^{j_{(M, N_{M})}}_{i_{(M,
            P_M)}}} \cdot
      \textstyle\prod_{m=2}^{M}    \tensor*{\left(c^{m}\right)}{*^{j_{(m,
            1)}}_{i_{(m, 1)}}^{\;\cdots\;}_{\;\cdots}^{j_{(m,
            N_{m})}}_{i_{(m, P_{m})}}} = 
      \tensor*{0}{*^{j_{(1,1)}}_{i_{(1,1)}}^{\;\cdots\;}_{\;\cdots}^{j_{(1, 
            N_1)}}_{i_{(1,P_{1})}}}.
    \end{displaymath}
    This means that
    $\tensor*{\tau}{*^{j_{(1,1)}}_{i_{(1,1)}}^{\;\cdots\;}_{\;\cdots}^{j_{(M,
          N_M)}}_{i_{(M,P_{M})}}} \notin
    \bbrk{F}\left(\textbf{B}_{n},\textbf{B}_{n}\right)_{Val}$, and so
    $\tau_{\mathbf{R}}$ does not meet the criteria to be found in $\GC\left[I,\bbrk{F}\left(\textbf{B}_{n},\textbf{B}_{n}\right)\right]$.
    The \mll\ transformation $\tau$ in $\C$ is therefore unable to be
    translated into $\GC$ either. \qed
  
  \subsection{Uniqueness} \label{SubsectionUniqueness}
  
  Non-simple linear combinations of two or more proof structures can never be proof nets themselves. Because of this, their representations must be demonstrated to have been eradicated from the categorical model by the double glueing construction before we can declare MDNF~full completeness proved.
  
  \p The sections above allow us to take as fact that every \mll~transformation in the glued category $\GC$ is not only the representation of a linear combination of proof structures, but a linear combination of proof \emph{nets}. Furthermore, it is also known that the scalars that each of the proof nets being modelled is multiplied by must sum to the semiring multiplicative unit $1$. Interestingly, the information here is sufficient to provide MDNF~full completeness results for a number of categories of the form $\GC$ for some category $\C$: namely those whose semiring of scalars have the property that if a sum $\sum_{i}s_{i}$ happens to equal $1$, then there is exactly one $x$ such that $s_{x} = 1$, and every other $s_{i} = 0$. An example of this type of category is the category of semimodules over $\N$.
  
  \p However, such grandiose claims cannot be written about the majority of compact closed categories with finite biproducts. In particular, the categories $\Rel$ and $\FDVec$, which are the examples investigated in \cite{Tan97} in detail, do not benefit from having this property. A more comprehensive proof for uniqueness is required, and this unsurprisingly takes the same combinatorial form as seen throughout this chapter. We show the intuition to the proof below.
  
  \p For the sake of the coming argument, we assume that we do not have the result from Lemma \ref{OnlyOneLemma} (we use it in a more restricted form later). Suppose we have an \mll~transformation $\C$ which can be thought of as representing the sum of the two sets of axiom links over the formula below, itself being described by a functor $F$.
  
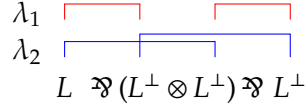
  \begin{center} \vspace{5mm}
  \begin{tikzpicture}
    [auto, node
    distance=5mm, skip loop/.style={to path={-- ++(0,#1) -| (\tikztotarget)}}] \tikzstyle{every node} = [text depth=-5pt,text height=0.5ex]
    \node (z1) { };
     \node (1) {$L$} [right of= z1]; \node (a) [right of=1] {$\invamp$} ;
    \node (2) [right of= a] {$(L^{\bot}$}; \node (b) [right of=2] {$\tens$} ;
    \node (3)  [right of= b] {$L^{\bot})$}; \node (c)  [right of=3] {$\invamp$};
    \node (4)  [right of= c] {$L^{\bot}$};
    \node (t1) [above of= z1, xshift=-5mm] {$\lambda_{2}$};
    \node (t2) [above of= t1] {$\lambda_{1}$};
  \begin{scope}
  \path   (1)  edge [blue, skip loop =6mm, shorten >=3mm, shorten <=3mm]  (3);
  \path (2)  edge  [blue, skip loop =7mm, shorten >=3mm, shorten <=3mm] (4);
  \path (1) edge [red, skip loop = 11mm, shorten >=8mm, shorten <=8mm]  (2);
  \path (3) edge  [red, skip loop = 11mm, shorten >=8mm, shorten <=8mm] (4);
  \end{scope}
  \end{tikzpicture} \vspace{5mm}
  \end{center}
 
 \noindent Both linkings successfully describe proof nets for the formula, and so we are looking at a scenario concerning a linear combination of proof nets. The tensor representations of the two linkings $\lambda_{1}$ and $\lambda_{2}$, given by the red and blue linkings respectively, using the standard indices for each literal, are $\tensor*{\bar{(\lambda_{1})}}{*^{j_{(2,1)}}_{i_{(1,1)}}^{j_{(2,2)}}_{i_{(3,1)}}} = \tensor*{\delta}{*^{j_{(2,1)}}_{i_{(1,1)}}^{j_{(2,2)}}_{i_{(3,1)}}}$ and $\tensor*{\bar{(\lambda_{2})}}{*^{j_{(2,1)}}_{i_{(1,1)}}^{j_{(2,2)}}_{i_{(3,1)}}} = \tensor*{\delta}{*^{j_{(2,2)}}_{i_{(1,1)}}^{j_{(2,1)}}_{i_{(3,1)}}}$ respectively. The description of the linear combination of proof nets being discussed is $\tensor*{\bar{\Lambda}}{*^{j_{(2,1)}}_{i_{(1,1)}}^{j_{(2,2)}}_{i_{(3,1)}}} = \tensor*{\delta}{*^{j_{(2,1)}}_{i_{(1,1)}}^{j_{(2,2)}}_{i_{(3,1)}}} + \tensor*{\delta}{*^{j_{(2,2)}}_{i_{(1,1)}}^{j_{(2,1)}}_{i_{(3,1)}}}$.
  
  \p It is simple enough to find partial permutations for both of the first two blocks in the formula of some order which produces a constant tensor when composed with $\tensor*{\bar{\lambda_{1}}}{*^{j_{(2,1)}}_{i_{(1,1)}}^{j_{(2,2)}}_{i_{(3,1)}}}$ yet creates a zero tensor when composed with $\tensor*{\bar{\lambda_{2}}}{*^{j_{(2,1)}}_{i_{(1,1)}}^{j_{(2,2)}}_{i_{(3,1)}}}$. If we let $(u^{1})_{i_{(1,1)}} = \delta^{1}_{i_{(1,1)}}$ and $(u^{2})_{j_{(2,1)}j_{(2,2)}} = \tensor*{\delta}{*^{1}_{j_{(2,1)}}^{2}_{j_{(2,2)}}}$, then
  \begin{eqnarray*}
  \tensor*{\bar{(\lambda_{1})}}{*^{j_{(2,1)}}_{i_{(1,1)}}^{j_{(2,2)}}_{i_{(3,1)}}}(u^{1})_{i_{(1,1)}}(u^{2})_{j_{(2,1)}j_{(2,2)}} & = & \tensor*{\delta}{*^{j_{(2,1)}}_{i_{(1,1)}}^{j_{(2,2)}}_{i_{(3,1)}}}\delta^{1}_{i_{(1,1)}}\tensor*{\delta}{*^{1}_{j_{(2,1)}}^{2}_{j_{(2,2)}}} \\
  																	& = & \tensor*{\delta}{*^{1}_{1}^{2}_{i_{(3,1)}}} = \delta^{2}_{i_{(3,1)}} \\
  \tensor*{\bar{(\lambda_{2})}}{*^{j_{(2,1)}}_{i_{(1,1)}}^{j_{(2,2)}}_{i_{(3,1)}}}(u^{1})_{i_{(1,1)}}(u^{2})_{j_{(2,1)}j_{(2,2)}} & = & \tensor*{\delta}{*^{j_{(2,2)}}_{i_{(1,1)}}^{j_{(2,1)}}_{i_{(3,1)}}}\delta^{1}_{i_{(1,1)}}\tensor*{\delta}{*^{1}_{j_{(2,1)}}^{2}_{j_{(2,2)}}} \\
  																	& = & \tensor*{\delta}{*^{2}_{1}^{1}_{i_{(3,1)}}} = 0_{i_{(3,1)}}
  \end{eqnarray*}

\noindent  By symmetry, it must also be possible to reverse this effect by letting $(v^{1})_{i_{(1,1)}} = \delta^{1}_{i_{(1,1)}}$ and $(v^{2})_{j_{(2,1)}j_{(2,2)}} = \tensor*{\delta}{*^{3}_{j_{(2,1)}}^{1}_{j_{(2,2)}}}$.
  \begin{eqnarray*}
  \tensor*{\bar{(\lambda_{1})}}{*^{j_{(2,1)}}_{i_{(1,1)}}^{j_{(2,2)}}_{i_{(3,1)}}}(v^{1})_{i_{(1,1)}}(v^{2})_{j_{(2,1)}j_{(2,2)}} & = & \tensor*{\delta}{*^{j_{(2,1)}}_{i_{(1,1)}}^{j_{(2,2)}}_{i_{(3,1)}}}\delta^{1}_{i_{(1,1)}}\tensor*{\delta}{*^{3}_{j_{(2,1)}}^{1}_{j_{(2,2)}}} \\
  																	& = & \tensor*{\delta}{*^{3}_{1}^{1}_{i_{(3,1)}}} = 0_{i_{(3,1)}} \\
  \tensor*{\bar{(\lambda_{2})}}{*^{j_{(2,1)}}_{i_{(1,1)}}^{j_{(2,2)}}_{i_{(3,1)}}}(v^{1})_{i_{(1,1)}}(v^{2})_{j_{(2,1)}j_{(2,2)}} & = & \tensor*{\delta}{*^{j_{(2,2)}}_{i_{(1,1)}}^{j_{(2,1)}}_{i_{(3,1)}}}\delta^{1}_{i_{(1,1)}}\tensor*{\delta}{*^{3}_{j_{(2,1)}}^{1}_{j_{(2,2)}}} \\
  																	& = & \tensor*{\delta}{*^{1}_{1}^{3}_{i_{(3,1)}}} = \delta^{3}_{i_{(3,1)}}
  \end{eqnarray*}
  
  There are two significant points in the development of these partial permutations. Firstly, letting $(u \cup v)_{\mathbf{i}} = u_{\mathbf{i}} \cdot v_{\mathbf{i}}$ (not using Einstein notation), the unions $w^{1} = u^{1} \cup v^{1}$ and $w^{2} = u^{2} \cup v^{2}$ are both partial permutations, where $u_{I} \cup v_{I} = 0$ if $u_{I} = 0 = v_{I}$ and $1$ otherwise. Secondly, the constant tensors created by the compositions $\tensor*{\bar{(\lambda_{1})}}{*^{j_{(2,1)}}_{i_{(1,1)}}^{j_{(2,2)}}_{i_{(3,1)}}}(u^{1})_{i_{(1,1)}}(u^{2})_{j_{(2,1)}j_{(2,2)}}$ and $\tensor*{\bar{(\lambda_{2})}}{*^{j_{(2,1)}}_{i_{(1,1)}}^{j_{(2,2)}}_{i_{(3,1)}}}(v^{1})_{i_{(1,1)}}(v^{2})_{j_{(2,1)}j_{(2,2)}}$ are not identical. The effect that these properties together have is that we can find a pair of partial permutations which act on the tensor sum $\tensor*{\bar{(\lambda_{1})}}{*^{j_{(2,1)}}_{i_{(1,1)}}^{j_{(2,2)}}_{i_{(3,1)}}}(v^{1})_{i_{(1,1)}}(v^{2})_{j_{(2,1)}j_{(2,2)}} + \tensor*{\bar{(\lambda_{2})}}{*^{j_{(2,1)}}_{i_{(1,1)}}^{j_{(2,2)}}_{i_{(3,1)}}}(u^{1})_{i_{(1,1)}}(u^{2})_{j_{(2,1)}j_{(2,2)}}$, and the resulting tensor is the sum of two constant tensors where the constants are different (meaning they cannot interfere with one another and cancel each other out).
  \begin{eqnarray*}
  \tensor*{\bar{\Lambda}}{*^{j_{(2,1)}}_{i_{(1,1)}}^{j_{(2,2)}}_{i_{(3,1)}}}(w^{1})_{i_{(1,1)}}(w^{2})_{j_{(2,1)}j_{(2,2)}} & = &  \tensor*{\delta}{*^{j_{(2,1)}}_{i_{(1,1)}}^{j_{(2,2)}}_{i_{(3,1)}}}(u^{1})_{i_{(1,1)}}(u^{2})_{j_{(2,1)}j_{(2,2)}} + \tensor*{\delta}{*^{j_{(2,2)}}_{i_{(1,1)}}^{j_{(2,1)}}_{i_{(3,1)}}}(v^{1})_{i_{(1,1)}}(v^{2})_{j_{(2,1)}j_{(2,2)}} \\
  & = & \delta_{i_{(3,1)}}^{2} + \delta_{i_{(3,1)}}^{3}
  \end{eqnarray*}
  There are clearly two non-zero entries in the resultant tensor, which means we can infer that the tensor $\hat{\Lambda}$ cannot belong to the set of values $F(\mathbf{A}_{n},\mathbf{A}_{n})$ for any number $n \geq 3$.
  
  \p The idea behind the coming proof can be extrapolated from this very simple example with relative ease. Given an \mll~transformation $\tau$ in $\C$ modelling a non-simple linear combination of proof nets, we can show it does not manifest itself in any form in $\GC$. An algorithm is given which takes linkings being described by $\tau$ and creates partial permutations for all bar one of the blocks in the formula such that composing them with the tensor representation of two axiom links leaves a constant tensor, minimising interference between tensor representations of each linking involved. Composing these permutations with the a tensor representation of the linear combination of all the axiom links gives a tensor containing more than one non-zero entry, and therefore the tensor $\tau_{n\I}$ is not in $F(\mathbf{A}_{n},\mathbf{A}_{n})_{Val}$ for some $n$, thus $\tau$ cannot be found in $\GC$. We start by giving the algorithm which begins this process.
  
  \begin{algo} \label{UniqueTupleAlg} Input: A non-trivial linear
    combination of MDNF proof nets for a sequent containing $M$ blocks, the $m^{th}$ of which having size $T_{m}$ for each $m$.
    \\ Output: A number $n \in \N^{+}$; tensors $c^{1},\ldots,c^{M}$
    such that $c^{m}_{i_{1} \cdots i_{T_{m}}} \in \PPerm(T_{m},n)$ for
    each $m$ except for one leaf.
  \end{algo}
  \begin{enumerate}
  \item Let $\lambda_{1}$ and $\lambda_{2}$ be two distinct sets of
    axiom links in the given linear combination of MDNF proof
    structures. We set the valencies $v_{1},\ldots,v_{M}$ such that
    $v_{m} = T_{m}$ for all $m \in [M]$, except for the last leaf,
    block $l$ say, for which $v_{l} = 0$. We let $i = 0$ and $k=1$.
  \item Choose the first block $m$ with $v_{m} = 1$.
    \begin{enumerate}
    \item If one does exist, then mark the sole literal not yet
      allocated a number as an `exit' and go to Step 3.
    \item If there are no more blocks of valency $1$, create a new
      $T_{m}$ tuple $\mathbf{x}$ for each block $m$ in the component,
      and let $x_{i} = u_{i}$ for every $i \in [T_{m}]$, where $u_{i}$
      is the label given to the $i^{th}$ literal of the block. Remove
      all the labels. If $k=1$, increment $k$, reset the valencies and
      restart Step~2; otherwise go to Step~4.
    \end{enumerate}
  
  \item Check to see whether the numbers given to each of the non-exit
    literals in the block correspond exactly to those of a tuple
    $\mathbf{t} = (t_{1},\ldots,t_{T_{m}})$ already associated with the
    block.
    \begin{enumerate}
    \item If so, then assign the final unused number in $\mathbf{t}$ to
      the exit literal.
    \item If not, increment $i$ and then assign the new value of $i$ to
      the exit literal.
    \end{enumerate}
    Assign the number to the literal with which the exit literal shares
    an axiom link in $\lambda_{k}$. Decrement the valencies of both
    block $m$ and the the block with which the exit literal of block
    $m$ is linked by an axiom link in $\lambda_{k}$. Restart Step~2.
  
  \item Declare that $n = i$, and we state that for each $m \in [M]$,
    \begin{displaymath}
      \hughf{(c^{m})}_{i_{1} \cdots i_{T_{m}}} = \left\{ 
        \begin{array}{ll}
          1 & \text{if }(i_{1},\ldots,i_{T_{m}}) \text{ is a tuple for } m \\
          0 & \text{otherwise.} 
        \end{array}
      \right. 
    \end{displaymath} 
  \end{enumerate}
  
\noindent  The algorithm above certainly terminates and produces partial permutations. By virtue of the proof structures being described by $\lambda_{1}$ and $\lambda_{2}$ being acyclic, there is always a block with valency $1$ at every stage of the algorithm for a single value of $k$ when one is needed until all the blocks bar leaf $l$ have been visited. This assures that each block is considered in Step~2 exactly twice, and therefore at most two entries are found in each tensor of a block $m \neq l$. On top of that, at the point where all bar one of the positions in a new tuple have been decided when $k=2$ it is checked whether they all correspond exactly to the first tuple created when $k=1$. If they do then the second tuple is completed to be a clone of the first; if not then a completely new number is used in the unfilled position. This ensures that there is never a situation where the two tuples differ in exactly one coordinate, and as such the tuples $c^{1},\ldots,c^{M}$ (excluding $c^{l}$) have to be partial permutations.
  
  \p The partial permutations created need to have the effect of annihilating tensor representations of all possible axiom links on a sequent other than the two that have been selected for use in Algorithm \ref{UniqueTupleAlg}. Fortunately, the algorithm does do this naturally, and in fact makes sure that the representations of $\lambda_{1}$ and $\lambda_{2}$ only make use of one non-zero entry in each of the permutations.
  
  \begin{exa} \label{UniqueExam}
  Consider the sum of the three linkings $\lambda_{1}$, $\lambda_{2}$ and $\lambda_{3}$ (given in red, blue and green respectively) in the diagram below.
  \end{exa}
  
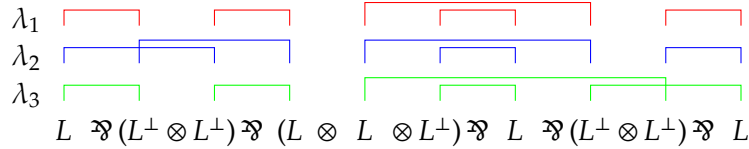
\begin{center} \vspace{2mm}
  \begin{tikzpicture}
    [auto, node
    distance=5mm, skip loop/.style={to path={-- ++(0,#1) -| (\tikztotarget)}}] \tikzstyle{every node} = [text depth=-5pt,text height=0.5ex]
   \node (z1) { }; 
      \node (1) [right of=z1] {$L$}; \node (a) [right of=1] {$\invamp$} ;
     \node (2) [right of=a] {$(L^{\bot}$}; \node (b) [right of=2] {$\tens$} ;
     \node (3)  [right of=b] {$L^{\bot})$}; \node (c)  [right of=3] {$\invamp$};
     \node (4)  [right of=c] {$(L$}; \node(d) [right of=4] {$\tens$} ;
     \node (5) [right of=d] {$L$}; \node (e) [right of=5] {$\tens$} ;
     \node (6) [right of=e] {$L^{\bot})$};  \node (f) [right of=6] {$\invamp$} ;
     \node (7)  [right of=f] {$L$}; \node (g)  [right of=7] {$\invamp$};
     \node (8) [right of=g] {$(L^{\bot}$};  \node (h) [right of=8] {$\tens$} ;
     \node (9)  [right of=h] {$L^{\bot})$}; \node (i)  [right of=9] {$\invamp$};
     \node (10)  [right of=i] {$L$};
    \node (t1) [above of= z1] {$\lambda_{3}$};
    \node (t2) [above of= t1] {$\lambda_{2}$};
    \node (t3) [above of= t2] {$\lambda_{1}$};
  \begin{scope}
  \path (1)  edge  [red, skip loop =16mm,shorten >=13mm, shorten <=13mm] (2);
  \path (3)  edge  [red, skip loop =16mm,shorten >=13mm, shorten <=13mm] (4);
  \path (5)  edge  [red, skip loop =17mm,shorten >=13mm, shorten <=13mm] (8);
  \path (6)  edge  [red, skip loop =16mm,shorten >=13mm, shorten <=13mm] (7);
  \path (9)  edge  [red, skip loop =16mm,shorten >=13mm, shorten <=13mm] (10);
  \path (1)  edge  [blue, skip loop =11mm,shorten >=8mm, shorten <=8mm] (3);
  \path (2)  edge  [blue, skip loop =12mm,shorten >=8mm, shorten <=8mm] (4);
  \path (5)  edge  [blue, skip loop =12mm,shorten >=8mm, shorten <=8mm] (8);
  \path (6)  edge  [blue, skip loop =11mm,shorten >=8mm, shorten <=8mm] (7);
  \path (9)  edge  [blue, skip loop =11mm,shorten >=8mm, shorten <=8mm] (10);
  \path (1)  edge  [green, skip loop =6mm, shorten >=3mm, shorten <=3mm] (2);
  \path (3)  edge  [green, skip loop =6mm, shorten >=3mm, shorten <=3mm] (4);
  \path (5)  edge  [green, skip loop =7mm, shorten >=3mm, shorten <=3mm] (9);
  \path (6)  edge  [green, skip loop =6mm, shorten >=3mm, shorten <=3mm] (7);
  \path (8)  edge  [green, skip loop =6mm, shorten >=3mm, shorten <=3mm] (10);
  \end{scope}
  \end{tikzpicture} \vspace{2mm}
  \end{center}
  \textit{The tensor representation of this linear combination is}
  \begin{displaymath}
   \tensor*{\delta}{*^{j_{(2,1)}}_{i_{(1,1)}}^{j_{(2,2)}}_{i_{(3,1)}}^{j_{(5,1)}}_{i_{(3,2)}}^{j_{(3,1)}}_{i_{(4,1)}}^{j_{(5,2)}}_{i_{(6,1)}}} + \tensor*{\delta}{*^{j_{(2,2)}}_{i_{(1,1)}}^{j_{(2,1)}}_{i_{(3,1)}}^{j_{(5,1)}}_{i_{(3,2)}}^{j_{(3,1)}}_{i_{(4,1)}}^{j_{(5,2)}}_{i_{(6,1)}}} + \tensor*{\delta}{*^{j_{(2,1)}}_{i_{(1,1)}}^{j_{(2,2)}}_{i_{(3,1)}}^{j_{(5,2)}}_{i_{(3,2)}}^{j_{(3,1)}}_{i_{(4,1)}}^{j_{(5,1)}}_{i_{(6,1)}}}.
   \end{displaymath}
  \textit{We apply Algorithm~\ref{UniqueTupleAlg}, choosing $\lambda_{1}$ and $\lambda_{2}$ as the two (synonymous) input linkings.}
  \vspace{5mm}
  \begin{enumerate}[label={\cW4 \& }\arabic*.]
  \item[1.] Block $6$ is the right-most block containing only one literal, so the valencies of each block are given as follows:
  \begin{displaymath} \vspace{2mm}
  v_{1} =  1, \; v_{2} =  2, \; v_{3} =  3, \; v_{4} =  1, \; v_{5} =  2, \; v_{6} =  0
  \end{displaymath} 
  \item[2 \& 3.] We deal first with $\lambda_{1}$ ($k=1$). Block $1$ has valency $1$, and no tuples have been given to it already (as is always the case when $k=1$), so the number $1$ is allocated to it.
  
  Continuing along the same lines, looking for blocks of valency $1$ from left to right at each iteration of Step~2, we obtain the following tuples.
  \begin{center} \vspace{-5mm}
  \begin{tikzpicture}
    [auto, node
    distance=5mm, skip loop/.style={to path={-- ++(0,#1) -| (\tikztotarget)}}] \tikzstyle{every node} = [text depth=-5pt,text height=0.5ex]
   \node (z1) { }; 
      \node (1) [right of=z1] {$L$}; \node (a) [right of=1] {$\invamp$} ;
     \node (2) [right of=a] {$(L^{\bot}$}; \node (b) [right of=2] {$\tens$} ;
     \node (3)  [right of=b] {$L^{\bot})$}; \node (c)  [right of=3] {$\invamp$};
     \node (4)  [right of=c] {$(L$}; \node(d) [right of=4] {$\tens$} ;
     \node (5) [right of=d] {$L$}; \node (e) [right of=5] {$\tens$} ;
     \node (6) [right of=e] {$L^{\bot})$};  \node (f) [right of=6] {$\invamp$} ;
     \node (7)  [right of=f] {$L$}; \node (g)  [right of=7] {$\invamp$};
     \node (8) [right of=g] {$(L^{\bot}$};  \node (h) [right of=8] {$\tens$} ;
     \node (9)  [right of=h] {$L^{\bot})$}; \node (i)  [right of=9] {$\invamp$};
     \node (10)  [right of=i] {$L$};
    \node (t1) [above of= z1] {$\lambda_{1}$};
    \node (t2) [above of= t1] { };
    \node (t3) [above of= t2] { };
    \node (11) [below of=1] {[1]}; 
    \node (12) [below of=2] {[1};  \node (1b) [below of=b] {,};
    \node (13) [below of=3] {2]}; 
    \node (14) [below of=4] {[2};  \node (1d) [below of=d] {,};
    \node (15) [below of=5] {4};  \node (1e) [below of=e] {,};
    \node (16) [below of=6] {3]}; 
    \node (17) [below of=7] {[3]}; 
    \node (18) [below of=8] {[4};  \node (1h) [below of=h] {,};
    \node (19) [below of=9] {5]}; 
    \node (20) [below of=10] {[5]};
  \begin{scope}
  \path (1)  edge  [red, skip loop =6mm, shorten >=3mm, shorten <=3mm] (2);
  \path (3)  edge  [red, skip loop =6mm, shorten >=3mm, shorten <=3mm] (4);
  \path (5)  edge  [red, skip loop =7mm, shorten >=3mm, shorten <=3mm] (8);
  \path (6)  edge  [red, skip loop =6mm, shorten >=3mm, shorten <=3mm] (7);
  \path (9)  edge  [red, skip loop =6mm, shorten >=3mm, shorten <=3mm] (10);
  \end{scope}
  \end{tikzpicture} \vspace{2mm}
  \end{center}
  
  \item[2 \& 3.] The same is done for $k=2$, i.e. for the linking $\lambda_{2}$, only making sure tuples differing in exactly one position from the ones created when considering $\lambda_{1}$ do not occur (which in this case only occurs at leaf blocks $1$ and $3$). We obtain the following:
  
  \begin{center} \vspace{-5mm}
  \begin{tikzpicture}
    [auto, node
    distance=5mm, skip loop/.style={to path={-- ++(0,#1) -| (\tikztotarget)}}] \tikzstyle{every node} = [text depth=-5pt,text height=0.5ex]
   \node (z1) { }; 
      \node (1) [right of=z1] {$L$}; \node (a) [right of=1] {$\invamp$} ;
     \node (2) [right of=a] {$(L^{\bot}$}; \node (b) [right of=2] {$\tens$} ;
     \node (3)  [right of=b] {$L^{\bot})$}; \node (c)  [right of=3] {$\invamp$};
     \node (4)  [right of=c] {$(L$}; \node(d) [right of=4] {$\tens$} ;
     \node (5) [right of=d] {$L$}; \node (e) [right of=5] {$\tens$} ;
     \node (6) [right of=e] {$L^{\bot})$};  \node (f) [right of=6] {$\invamp$} ;
     \node (7)  [right of=f] {$L$}; \node (g)  [right of=7] {$\invamp$};
     \node (8) [right of=g] {$(L^{\bot}$};  \node (h) [right of=8] {$\tens$} ;
     \node (9)  [right of=h] {$L^{\bot})$}; \node (i)  [right of=9] {$\invamp$};
     \node (10)  [right of=i] {$L$};
    \node (t1) [above of= z1] {$\lambda_{2}$};
    \node (t2) [above of= t1] { };
    \node (t3) [above of= t2] { };
    \node (11) [below of=1] {[1]}; 
    \node (12) [below of=2] {[6};  \node (1b) [below of=b] {,};
    \node (13) [below of=3] {1]}; 
    \node (14) [below of=4] {[6};  \node (1d) [below of=d] {,};
    \node (15) [below of=5] {7};  \node (1e) [below of=e] {,};
    \node (16) [below of=6] {3]}; 
    \node (17) [below of=7] {[3]}; 
    \node (18) [below of=8] {[7};  \node (1h) [below of=h] {,};
    \node (19) [below of=9] {8]}; 
    \node (20) [below of=10] {[8]};
  \begin{scope}
  \path (1)  edge  [blue, skip loop =6mm, shorten >=3mm, shorten <=3mm] (3);
  \path (2)  edge  [blue, skip loop =7mm, shorten >=3mm, shorten <=3mm] (4);
  \path (5)  edge  [blue, skip loop =7mm, shorten >=3mm, shorten <=3mm] (8);
  \path (6)  edge  [blue, skip loop =6mm, shorten >=3mm, shorten <=3mm] (7);
  \path (9)  edge  [blue, skip loop =6mm, shorten >=3mm, shorten <=3mm] (10);
  \end{scope}
  \end{tikzpicture} \vspace{5mm}
  \end{center}
  \item[4.] Joining the two sets of tuples together, we get
  
  
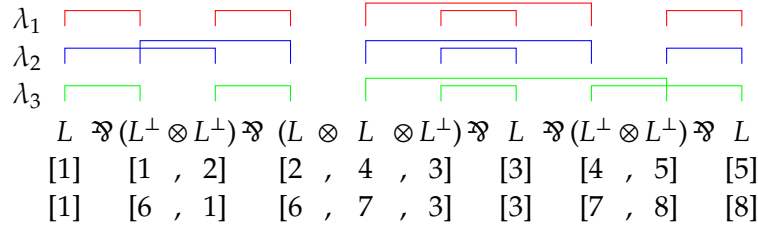
\begin{center} \vspace{5mm}
  \begin{tikzpicture}
    [auto, node
    distance=5mm, skip loop/.style={to path={-- ++(0,#1) -| (\tikztotarget)}}] \tikzstyle{every node} = [text depth=-5pt,text height=0.5ex]
   \node (z1) { }; 
      \node (1) [right of=z1] {$L$}; \node (a) [right of=1] {$\invamp$} ;
     \node (2) [right of=a] {$(L^{\bot}$}; \node (b) [right of=2] {$\tens$} ;
     \node (3)  [right of=b] {$L^{\bot})$}; \node (c)  [right of=3] {$\invamp$};
     \node (4)  [right of=c] {$(L$}; \node(d) [right of=4] {$\tens$} ;
     \node (5) [right of=d] {$L$}; \node (e) [right of=5] {$\tens$} ;
     \node (6) [right of=e] {$L^{\bot})$};  \node (f) [right of=6] {$\invamp$} ;
     \node (7)  [right of=f] {$L$}; \node (g)  [right of=7] {$\invamp$};
     \node (8) [right of=g] {$(L^{\bot}$};  \node (h) [right of=8] {$\tens$} ;
     \node (9)  [right of=h] {$L^{\bot})$}; \node (i)  [right of=9] {$\invamp$};
     \node (10)  [right of=i] {$L$};
    \node (t1) [above of= z1] {$\lambda_{3}$};
    \node (t2) [above of= t1] {$\lambda_{2}$};
    \node (t3) [above of= t2] {$\lambda_{1}$};
    \node (11) [below of=1] {[1]}; 
    \node (12) [below of=2] {[1};  \node (1b) [below of=b] {,};
    \node (13) [below of=3] {2]}; 
    \node (14) [below of=4] {[2};  \node (1d) [below of=d] {,};
    \node (15) [below of=5] {4};  \node (1e) [below of=e] {,};
    \node (16) [below of=6] {3]}; 
    \node (17) [below of=7] {[3]}; 
    \node (18) [below of=8] {[4};  \node (1h) [below of=h] {,};
    \node (19) [below of=9] {5]}; 
    \node (20) [below of=10] {[5]};
    \node (21) [below of=11] {[1]}; 
    \node (22) [below of=12] {[6};  \node (2b) [below of=1b] {,};
    \node (23) [below of=13] {1]}; 
    \node (24) [below of=14] {[6};  \node (2d) [below of=1d] {,};
    \node (25) [below of=15] {7};  \node (2e) [below of=1e] {,};
    \node (26) [below of=16] {3]}; 
    \node (27) [below of=17] {[3]}; 
    \node (28) [below of=18] {[7};  \node (2h) [below of=1h] {,};
    \node (29) [below of=19] {8]}; 
    \node (30) [below of=20] {[8]};
  \begin{scope}
  \path (1)  edge  [red, skip loop =16mm,shorten >=13mm, shorten <=13mm] (2);
  \path (3)  edge  [red, skip loop =16mm,shorten >=13mm, shorten <=13mm] (4);
  \path (5)  edge  [red, skip loop =17mm,shorten >=13mm, shorten <=13mm] (8);
  \path (6)  edge  [red, skip loop =16mm,shorten >=13mm, shorten <=13mm] (7);
  \path (9)  edge  [red, skip loop =16mm,shorten >=13mm, shorten <=13mm] (10);
  \path (1)  edge  [blue, skip loop =11mm,shorten >=8mm, shorten <=8mm] (3);
  \path (2)  edge  [blue, skip loop =12mm,shorten >=8mm, shorten <=8mm] (4);
  \path (5)  edge  [blue, skip loop =12mm,shorten >=8mm, shorten <=8mm] (8);
  \path (6)  edge  [blue, skip loop =11mm,shorten >=8mm, shorten <=8mm] (7);
  \path (9)  edge  [blue, skip loop =11mm,shorten >=8mm, shorten <=8mm] (10);
  \path (1)  edge  [green, skip loop =6mm, shorten >=3mm, shorten <=3mm] (2);
  \path (3)  edge  [green, skip loop =6mm, shorten >=3mm, shorten <=3mm] (4);
  \path (5)  edge  [green, skip loop =7mm, shorten >=3mm, shorten <=3mm] (9);
  \path (6)  edge  [green, skip loop =6mm, shorten >=3mm, shorten <=3mm] (7);
  \path (8)  edge  [green, skip loop =6mm, shorten >=3mm, shorten <=3mm] (10);
  \end{scope}
  \end{tikzpicture} \vspace{5mm}
  \end{center}
  Therefore the tensors $u^{1},\ldots,u^{5}$ can now be formed
  \begin{eqnarray*}
  \hughf{(u^{1})}_{i_{(1,1)}} & = & \delta_{i_{(1,1)}}^{1} \\
  \hughf{(u^{2})}_{j_{(2,1)}j_{(2,2)}} & = & \tensor*{\delta}{*^{1}_{j_{(2,1)}}^{2}_{j_{(2,2)}}} + \tensor*{\delta}{*^{6}_{j_{(2,1)}}^{1}_{j_{(2,2)}}} \\
  \hughf{(u^{3})}_{i_{(3,1)}i_{(3,2)}j_{(3,1)}} & = & \tensor*{\delta}{*^{2}_{i_{(3,1)}}^{4}_{i_{(3,2)}}^{3}_{j_{(3,1)}}} + \tensor*{\delta}{*^{6}_{i_{(3,1)}}^{7}_{i_{(3,2)}}^{3}_{j_{(3,1)}}} \\
  \hughf{(u^{4})}_{i_{(4,1)}} & = & \delta_{i_{(4,1)}}^{3} \\
  \hughf{(u^{5})}_{j_{(5,1)}j_{(5,2)}} & = & \tensor*{\delta}{*^{4}_{j_{(5,1)}}^{5}_{j_{(5,2)}}} + \tensor*{\delta}{*^{7}_{j_{(5,1)}}^{8}_{j_{(5,2)}}} \\
  \hughf{(u^{6})}_{i_{(6,1)}} & = & \delta_{i_{(6,1)}}^{5} + \delta_{i_{(6,1)}}^{8}
  \end{eqnarray*}
  \end{enumerate}
  
\noindent  Composing the five partial permutations with the representation of $\lambda_{1}$ ends with a constant tensor.
  \begin{eqnarray*}
  \tensor*{\bar{(\lambda_{1})}}{*^{j_{(2,1)}}_{i_{(1,1)}}^{j_{(2,2)}}_{i_{(3,1)}}^{j_{(3,1)}}_{i_{(3,2)}}^{j_{(5,1)}}_{i_{(4,1)}}^{j_{(5,2)}}_{i_{(6,1)}}} \cdot \prod_{m=1}^{5}u^{m}_{\mathbf{i}_{m}\mathbf{j}_{m}} & = & \tensor*{\delta}{*^{j_{(2,1)}}_{i_{(1,1)}}^{j_{(2,2)}}_{i_{(3,1)}}^{j_{(5,1)}}_{i_{(3,2)}}^{j_{(3,1)}}_{i_{(4,1)}}^{j_{(5,2)}}_{i_{(6,1)}}} \cdot \delta_{i_{(1,1)}}^{1} (\tensor*{\delta}{*^{1}_{j_{(2,1)}}^{2}_{j_{(2,2)}}} + \tensor*{\delta}{*^{6}_{j_{(2,1)}}^{1}_{j_{(2,2)}}}) \\
  & & (\tensor*{\delta}{*^{2}_{i_{(3,1)}}^{4}_{i_{(3,2)}}^{3}_{j_{(3,1)}}} + \tensor*{\delta}{*^{6}_{i_{(3,1)}}^{7}_{i_{(3,2)}}^{3}_{j_{(3,1)}}}) \delta_{i_{(4,1)}}^{3} (\tensor*{\delta}{*^{4}_{j_{(5,1)}}^{5}_{j_{(5,2)}}} + \tensor*{\delta}{*^{7}_{j_{(5,1)}}^{8}_{j_{(5,2)}}}) \\
  	& = & \tensor*{\delta}{*^{1}_{1}^{2}_{2}^{4}_{4}^{3}_{3}^{5}_{i_{(6,1)}}} + \tensor*{\delta}{*^{1}_{1}^{2}_{2}^{7}_{4}^{3}_{3}^{8}_{i_{(6,1)}}} + \tensor*{\delta}{*^{1}_{1}^{2}_{6}^{4}_{7}^{3}_{3}^{5}_{i_{(6,1)}}} \\ 
  	&  & +\,\, \tensor*{\delta}{*^{1}_{1}^{2}_{6}^{7}_{7}^{3}_{3}^{8}_{i_{(6,1)}}} + \tensor*{\delta}{*^{6}_{1}^{1}_{2}^{4}_{4}^{3}_{3}^{5}_{i_{(6,1)}}} + \tensor*{\delta}{*^{6}_{1}^{1}_{2}^{7}_{4}^{3}_{3}^{8}_{i_{(6,1)}}} \\
  						&   & +\,\, \tensor*{\delta}{*^{6}_{1}^{1}_{6}^{4}_{7}^{3}_{3}^{5}_{i_{(6,1)}}} + \tensor*{\delta}{*^{6}_{1}^{1}_{6}^{7}_{7}^{3}_{3}^{8}_{i_{(6,1)}}} = \delta_{i_{(6,1)}}^{5}
  \end{eqnarray*}
  With similar equations it is easily shown that $\tensor*{\bar{(\lambda_{2})}}{*^{j_{(2,1)}}_{i_{(1,1)}}^{j_{(2,2)}}_{i_{(3,1)}}^{j_{(3,1)}}_{i_{(3,2)}}^{j_{(5,1)}}_{i_{(4,1)}}^{j_{(5,2)}}_{i_{(6,1)}}} \cdot \prod_{m=1}^{5}\hughf{(u^{m})}_{\mathbf{i}_{m}\mathbf{j}_{m}} = \delta_{i_{(6,1)}}^{8}$, and that $\tensor*{\bar{(\lambda_{3})}}{*^{j_{(2,1)}}_{i_{(1,1)}}^{j_{(2,2)}}_{i_{(3,1)}}^{j_{(3,1)}}_{i_{(3,2)}}^{j_{(5,1)}}_{i_{(4,1)}}^{j_{(5,2)}}_{i_{(6,1)}}} \cdot \prod_{m=1}^{5}\hughf{(u^{m})}_{\mathbf{i}_{m}\mathbf{j}_{m}} = 0_{i_{(6,1)}}$. The components affected by the tensor representations of $\lambda_{1}$ and $\lambda_{2}$ are different from one another, meaning they do not interfere with one another. Similarly, the tensor representation for $\lambda_{3}$ and partial permutations reduce to the zero morphism, and therefore has no effect on any component. The permutations composed with the tensor representation for the sum of all the linkings is therefore found to be $\delta_{i_{(6,1)}}^{5} + \delta_{i_{(6,1)}}^{8}$, which has two non-zero entries. This proves that the linking combination cannot be modelled in $\GC$.
  
  \p It can also be seen that this lack of interference between different proof nets in the final result means that scalar multiples of the three linkings in question in the example can be summed together in a linear combination and the same concept of proof still holds. If we consider an \mll~transformation $\tau'$ in the underlying category of such a form, then composing $\tau'_{n\I}$ with the same five partial permutations for suitable $n$ results in a tensor with two non-zero entries, with the fifth and eighth positions filled by the scalars multipled to $\lambda_{1}$ and $\lambda_{2}$ repectively. The exact values within the entries are irrelevant---it only matters that two are non-zero.
  
  \begin{lem} \label{UniqueAlgSepsAndAnnisClaim}
  Let $\tau:\K \longrightarrow F$ be an MDNF~transformation in a compact closed category $\C$ with finite biproducts modelling a linear combination of at least two distinct proof nets, $\{\lambda_{a}:a \in [K]\}$ say, over the MDNF~formula modelled by $F$ containing $M$ blocks, the $l^{th}$ of which is the last leaf. Letting $\lambda_{1}$ and $\lambda_{2}$ be the two primary inputs to Algorithm \ref{UniqueTupleAlg}, producing $M-1$ partial permutations $\{\hughf{(u^{m})}_{\mathbf{i}_{(m,-)}\mathbf{j}_{(m,-)}}: m \in [M]\backslash \{l\}\}$, we find that
  \begin{eqnarray*}
  (\bar{\lambda_{1}})_{\mathbf{i}}^{\mathbf{j}} \cdot \prod_{m \neq l}\hughf{(u^{m})}_{\mathbf{i}_{(m,-)}\mathbf{j}_{(m,-)}} & = & \delta_{k}^{x} \mbox{, and} \\
  (\bar{\lambda_{2}})_{\mathbf{i}}^{\mathbf{j}} \cdot \prod_{m \neq l}\hughf{(u^{m})}_{\mathbf{i}_{(m,-)}\mathbf{j}_{(m,-)}} & = & \delta_{k}^{y}
  \end{eqnarray*}
  where $x$ and $y$ are the first and second values offered to leaf $l$ in the algorithm, and $k$ is the index $i_{(l,1)}$ or $j_{(l,1)}$, depending on the polarity of the literal denoted by leaf $l$. Furthermore, for every $a > 2$,
  \begin{displaymath}
  (\bar{\lambda_{a}})_{\mathbf{i}}^{\mathbf{j}} \cdot \prod_{m \neq l}\hughf{(u^{m})}_{\mathbf{i}_{(m,-)}\mathbf{j}_{(m,-)}} = 0_{k}
  \end{displaymath}
  \end{lem}
  \proof
  The tensor representation of a linking of a proof net is represented by a product of Kronecker deltas, where two indices are in the same delta if the literals to which they are associated are linked by an axiom link. As such, if two indices $i_{(m_{i},l_{i})}$ and $j_{(m_{j},l_{j})}$ associated with literals connected by an axiom link in $\lambda_{1}$ are forced to be given different values $p$ and $q$ say, then we know that $(\bar{\lambda_{1}})^{\mathbf{j}}_{\mathbf{i}} \cdot \delta^{p}_{i_{(m_{i},l_{i})}} \cdot \delta^{q}_{j_{(m_{j},l_{j})}} = 0^{\mathbf{j}'}_{\mathbf{i}'}$ for appropriate index sets $\mathbf{i}'$ and $\mathbf{j}'$.
  
  \p The partial permutations, containing either one or two non-zero entries, can be multiplied together and the distribution law can be used to give a sum of $2^{M'}$ products of constant tensors ($M'$ being the number of blocks given partial permutations with two non-zero entries), with each index being given a number except $k$. These $2^{M'}$ assignments of numbers are determined by the combinations of first and second tuples associated with the sets of indices for each block---Example \ref{UniqueExam} provides a good example of this.
  
  \p Of these combinations, only two are capable of being composed with a tensor representation of a set of axiom links to create a non-zero tensor: the one where all the constants are chosen to be from the top tuples of each block in the algorithm; and the one from all the bottom tuples. This is because the algorithm continuously uses new numbers when an arbitrary number must be selected (Step~3(b)), and so mixing up the top and bottom tuples leaves more than one number given to only one index, thus ensuring it cannot be part of an axiom link pairing without creating a zero tensor. An example of this argument in action can be seen in Example~\ref{UniqueExam} at the start of Step~4: we see that $(1,2)$ of block~2 and $(6,7,3)$ of block $3$ only have one value $2$ within the two of them together.
  
  \p For the tensor representation of $\lambda_{1}$, only the combination developed from the top tuples gives a non-zero tensor. The second tuples are defined so that two literals are only given the same number in their tuples if they share an axiom link in $\lambda_{2}$; and since $\lambda_{1}$ and $\lambda_{2}$ are distinct, there must be at least one pair of literals joined by an axiom link in $\lambda_{1}$ which does not exist in $\lambda_{2}$, thus leading to a zero tensor. The one index from $\bar{\lambda_{1}}$ which does not have a number assigned to it by a partial permutation is clearly $k$, which is connected to the sole literal in a block other than $l$ assigned the number $x$ in the top tuple. The composition of $(\bar{\lambda_{1}})_{\mathbf{i}}^{\mathbf{j}}$ with the contsant tensors for relating to the top tuples from the $M-1$ blocks which are not block $l$ therefore reduces to a product of Kronecker deltas of the form $\delta_{z}^{z}$ for various $z$ and $\delta_{k}^{x}$. Since $\delta_{z}^{z} = 1$ for all $z$, this becomes simply $\delta_{k}^{x}$; and so it follows that $(\bar{\lambda_{1}})_{\mathbf{i}}^{\mathbf{j}} \cdot \prod_{m \neq l}\hughf{(u^{m})}_{\mathbf{i}_{(m,-)}\mathbf{j}_{(m,-)}} = \delta_{k}^{x}$ as desired.
  
  \p The argument for the representation of $\lambda_{2}$ is fundamentally identical, replacing the bottom tuples for the top tuples throughout the above paragraph. For every other $\lambda_{a}$ in the linear combination when composed any one of the $2^{M'}-2$ combinations mixing top and bottom tuples equates to zero for the same reasons as for $\lambda_{1}$ and $\lambda_{2}$ before. Each of them is also distinct from both $\lambda_{1}$ and $\lambda_{2}$ by definition, meaning that neither the combination of all the top tuples nor that of the bottom tuples induce constant tensors which compose with $\bar{\lambda_{a}}$ to give a non-zero tensor due to the same principle as why the bottom tuples do not for $\bar{\lambda_{1}}$. \qed
  
  We are now able to give the proof for a lemma proving the simplicity of the linear combinations of proof nets capable of being modelled in $\GC$.
  
  \begin{prop} \label{UniqueLemma} Every MDNF transformation in $\GC$
    modelling a linear combination of proof nets is modelling a unique
    proof net with scalar~$1$.
  \end{prop}
  \proof
    Let $F$ be an MDNF formula, and consider
    $\bbrk{F}(\mathbf{A}_{n},\mathbf{A}_{n})$ for arbitrary $n$, where
    $\mathbf{A}_{n}$ is defined as in Section~\ref{SectionMDNFObjects}. Again,
    we use the criteria also given in Section~\ref{SectionMDNFObjects} that the
    tensor representation of an arrow $f \in
    \C[\I,|\bbrk{F}(\mathbf{A}_{n},\mathbf{A}_{n})|]$ must satisfy in
    order to be found in~$\GC$. Suppose that $\tau$ is an MDNF
    transformation in $\C$ which models a linear combination of two or
    more proof nets. Then, for $\mathbf{R} = (n\I,\ldots,n\I)$,
    $\tau_{\mathbf{R}}$ once again takes the tensorial form
    \begin{displaymath}
      \tensor*{\tau}{*^{j_{(1,1)}}_{i_{(1,1)}}^{\;\cdots\;}_{\;\cdots}^{j_{(M, 
            N_M)}}_{i_{(M,P_{M})}}}
      = \textstyle\sum_{\beta} s_{\beta}\cdot
      \tensor*{\delta}{*^{j_{\beta(1,1)}}_{i_{(1,
            1)}}^{\;\cdots\;}_{\;\cdots}^{j_{\beta(M, P_{M})}}_{i_{(M,
            P_M)}}}
    \end{displaymath}
    and we know that there are two bijections, $\zeta_{1}$ and
    $\zeta_{2}$ say, corresponding to axiom link sets $\lambda_{1}$ and
    $\lambda_{2}$ respectively, where $s_{\zeta_{1}} \neq 0 \neq
    s_{\zeta_{2}}$. We use these two sets of axiom links in
    Algorithm~\ref{UniqueTupleAlg} to produce partial permutations
    $\{c^{m}: m \in [M]\}$ for all of the blocks, and to fix the
    required size of the number $n \in \N^{+}$ to prove the lemma.
  
    Let $x_{1}$ and $x_{2}$
    be the first and second tuple values associated with the last leaf
    (block $l$) from the algorithm. Then we find that for $k \in
    \{1,2\}$
    \begin{displaymath}
      \tensor*{\delta}{*^{x_{k}}_{r_{(l,1)}}} 
      \cdot
      \tau^{\mathbf{j}}_{\mathbf{i}}\cdot\textstyle\prod_{m \neq
        l}\tensor*{(c^{m})}{*^{j_{(m, 1)}}_{i_{(m,
            1)}}^{\;\cdots\;}_{\;\cdots}^{j_{(m, N_{m})}}_{i_{(m,
            P_{m})}}} = s_{\zeta_l} \neq 0 
    \end{displaymath}
    where the index $r_{(l,1)}$ equals $i_{(l,1)}$ or $j_{(l,1)}$
    depending on whether the literal of leaf $l$ is positive or
    negative.
    
    \p We have shown that there are at least two non-zero entries in \\
    $\tau^{\mathbf{j}}_{\mathbf{i}}\cdot\textstyle\prod_{m \neq
      l}\tensor*{(c^{m})}{*^{j_{(m, 1)}}_{i_{(m,
          1)}}^{\;\cdots\;}_{\;\cdots}^{j_{(m, N_{m})}}_{i_{(m,
          P_{m})}}}$, and so $\tau^{\mathbf{j}}_{\mathbf{i}} \notin
    \bbrk{F}(\mathbf{A}_{n},\mathbf{A}_{n})_{Val}$. The arrow
    $\tau_{\mathbf{R}} \in
    \C[\I,|\bbrk{F}(\mathbf{A}_{n},\mathbf{A}_{n})|]$ does not
    therefore exist in
    $\GC[\I,\bbrk{F}(\mathbf{A}_{n},\mathbf{A}_{n})]$, and so $\tau$ is
    not an \mll\ transformation.
  
    \p From the above, we know that every \mll\ transformation in $\GC$
    models a scalar multiple of a cut-free proof net. Lemma~\ref{OnlyOneLemma} ensures that those scalar multiples with scalar not equal to $1$ are not possible, and so we have proved the result. \qed
    
    \begin{thm} \label{MDNFFC} If\/ $\C$ is a compact closed category
      with biproducts satisfying feeble full completeness, then $\GC$
      satisfies MDNF~full completeness.
    \end{thm}
    
    \subsection{Extending to \mll~Full Completeness} \label{SectionMLLExtend}
    
    The previous subsections contain results which combine to produce a full completeness result for \mll~transformations to target functors describing sequents in the multiplicative disjunctive normal form. Although a strong connection between any family of \mll~transformations and proof nets in \mll\ is advantageous, it is certainly no substitute for a `complete' \mll~full completeness theorem. One of the remarkable properties of the categories in which we are interested is that the earlier MDNF~full completeness proof can be extended using one lemma and a couple of supplementary algorithms so that it works for \emph{all} \mll~functors, giving the full completeness theorem originally proposed.
    
    \p As stated in Chapter~\ref{CatModels}, in every $*$-autonomous category $\D$ there are natural
    transformations
    \begin{displaymath}
    \begin{array}{l}
    w^{LL} = (w^{LL}_{A,B,C}: A \otimes (B \invamp C) \rightarrow (A \otimes B) \invamp C)_{A,B,C \in \D} \\
    w^{LR} = (w^{LR}_{A,B,C}: A \otimes (B \invamp C) \rightarrow (A \otimes C) \invamp B)_{A,B,C \in \D} \\
    w^{RL} = (w^{RL}_{A,B,C}: (A \invamp B) \otimes C \rightarrow B \invamp (A \otimes C))_{A,B,C \in \D} \\
    w^{RR} = (w^{RR}_{A,B,C}: (A \invamp B) \otimes C \rightarrow A \invamp (B \otimes C))_{A,B,C \in \D},
    \end{array}
    \end{displaymath}
    and these are canonically isomorphic to appropriate compositions of the associativity and symmetry isomorphisms $\alpha$ and $\sigma$ if $\D$ happens to be compact closed, making them bijections (no two distinct \mll~transformations compose with the same weak distributivity transformation to give the same result). The tensor representations of these natural isomorphisms are both described solely by Kronecker deltas as demonstrated in Section~\ref{SectionTransExtensors}. Due to the small diagrammatic argument on page~\pageref{DiNatsComposeStatement}, an \mll~transformation $\tau$ in $\C$ may be composed with a sequence of natural transformations built from the weak distributivity transformations and the associativity and symmetry isomorphisms of $\C$ to
    produce an MDNF~transformation $\tau'$. Furthermore, $\tau$ describes an \mll~transformation in $\GC$ only if $\tau'$ does.
    
    \begin{prop} \label{MDNFtoMLLProp}
    Let $\fctl{\tau}{\K}{F}$ be an \mll~transformation in $\C$, and let $\fctl{\bar{w}}{F}{G}$ be a natural transformation built from the weak distributivity, associativity and symmetry natural transformations of $\C$. Then $\tau$ does not exist \hugh{in} $\GC$ if $\bar{w} \after \tau$ does not.
    \end{prop}
    \proof
    The transformation $\bar{w}$ is in $\GC$, and so we know that if $\tau$ is dinatural in the glued category then so does $\bar{w} \after \tau$, since both are well-defined in $\GC$. The statement of the proposition is the contrapositive of this fact. \qed
    
    It is always possible to find a natural
    transformations which preserves the cyclicity of at least one
    modelled cyclic proof structure between \mll~transformations. As such it is also possible to find a composition of natural transformations which not only preserves the existence of a cyclic proof structure modelled but ensures at least one of these cycles never passes through a $\invamp$-vertex in its switching\footnote{In fact, if the cycle does not pass through a $\invamp$-vertex the cycle exists in all switchings of the proof structure}.
    
    \begin{algo} \label{MLLtoMDNFCycleAlg}
    Input: An \mll~transformation $\tau:\K \longrightarrow F$ describing a linear combination of proof structures over a common sequent with at least one structure breaking the acyclicity criterion. \\
    Output: An \mll~transformation $\tau'$ describing a linear combination of proof structures over a common sequent with at least one structure containing a switching cycle passing through no $\invamp$-vertices.
    \end{algo}
    \begin{enumerate}
    \item Select one of the proof structures associated with $\tau$ which fails the acyclicity criterion. Choose a cycle of minimum length from one of its switchings. We let $\tau_{0} = \tau$ and $F_{0} = F$, and we name the cycle $C_{0}$. Let $n = 0$, and whenever we refer to subformulae called $X$, $Y$ and $Z$, we name their lowest vertices as in a parse tree $x$, $y$ and $z$ respectively.
    
    \item Search for a position in $F_{n}$ of the form $X \otimes \left(Y \invamp Z\right)$ for some subformulae $X$, $Y$ and $Z$, the corresponding edge $\otimes-\invamp$ of which is in the cycle $C_{n}$.
    	\begin{enumerate}
    	\item If one should exist, and $C_{n}$ passes through $y$, then let
    	\begin{displaymath}
    	\tau_{n+1} = \left(1 \cdots w^{LL} \cdots 1\right) \circ \tau_{n}.
    	\end{displaymath}
    The functor $F_{n+1}$ is defined to be the target functor of $\tau_{n+1}$, which is the same as $F_{n}$ except that the subformula $X \otimes \left(Y \invamp Z\right)$ is replaced by $\left(X \otimes Y\right) \invamp Z$. The cycle $C_{n+1}$ passes through the same vertices as those in $C_{n}$ in the identical sections of $F_{n+1}$ and $F_{n}$. If $C_{n}$ contains the path $x-\otimes-\invamp-y$, then the cycle $C_{n+1}$ is completed by adding $x-\otimes-y$; otherwise, $\invamp-\otimes-y$ is used to connect the ends together. Increment $n$ and restart Step 2.
    	\item If one should exist, but $C_{n}$ passes through $z$ instead, let
    	\begin{displaymath}
    	\tau_{n+1} = \left(1 \cdots w^{LR} \cdots 1\right) \circ \tau_{n}.
    	\end{displaymath}
    	 The functor $F_{n+1}$ is defined similarly to before: $X \otimes \left(Y \invamp Z\right)$ is replaced by $\left(X \otimes Z\right) \invamp Y$. The cycle $C_{n+1}$ passes through the same vertices as those in $C_{n}$ in the identical sections of $F_{n+1}$ and $F_{n}$. If $C_{n}$ contains $x-\otimes-\invamp-z$, then the cycle $C_{n+1}$ is completed by adding the path $x-\otimes-z$; otherwise, $\invamp-\otimes-z$. Increment $n$ and restart Step 2.
    	\item If one does not exist, then move to Step 3.
    	\end{enumerate}
    	
    \item Search for a position in $F_{n}$ of the form $\left(X \invamp Y\right) \otimes Z$ for some subformulae $X$, $Y$ and $Z$, the corresponding edge $\invamp-\otimes$ of which is in the cycle $C_{n}$.
    	\begin{enumerate}
    	\item If one should exist, and $C_{n}$ passes through $x$, then let
    	\begin{displaymath}
    	\tau_{n+1} = \left(1 \cdots w^{RL} \cdots 1\right) \circ \tau_{n}.
    	\end{displaymath}
    The functor $F_{n+1}$ is defined to be the target functor of $\tau_{n+1}$, which is the same as $F_{n}$ except that the subformula $\left(X \invamp Y\right) \otimes Z$ is replaced by $Y \invamp \left(X \otimes Z\right)$. The cycle $C_{n+1}$ passes through the same vertices as those in $C_{n}$ in the identical sections of $F_{n+1}$ and $F_{n}$. If $C_{n}$ contains the path $x-\invamp-\otimes-z$, then the cycle $C_{n+1}$ is completed by adding $x-\otimes-z$; otherwise, $x-\otimes-\invamp$ is used to connect the ends together. Increment $n$ and restart Step 2.
    	\item If one should exist, but $C_{n}$ passes through $y$ instead, let 
    	\begin{displaymath}
    	\tau_{n+1} = \left(1 \cdots w^{RR} \cdots 1\right) \circ \tau_{n}.
    	\end{displaymath}
    	The functor $F_{n+1}$ is defined similarly to before: $\left(X \invamp Y\right) \otimes Z$ is replaced by $X \invamp \left(Y \otimes Z\right)$. The cycle $C_{n+1}$ passes through the same vertices as those in $C_{n}$ in the identical sections of $F_{n+1}$ and $F_{n}$. If $C_{n}$ contains $y-\otimes-\invamp-z$, then the cycle $C_{n+1}$ is completed by adding the path $y-\otimes-z$; otherwise, $\invamp-\otimes-z$. Increment $n$ and go back to Step 2.
    	\item If one does not exist, then define $\tau' = \tau_{n}$, $F' = F_{n}$ and $C' = C_{n}$
    	\end{enumerate}
    \end{enumerate}
    
\noindent     Intuitively it is simple to see why this algorithm works. Suppose, without loss of generality, that one of the scenarios expected for Step~2(a) unfolds (dual arguments can be given for each of the other four situations). We can think of the proof structure with the cycle $C_{n}$ for some $n$ taking the form of one of the left-hand diagrams in Figures~7.1 and~7.2 below.
    
    \begin{figure}[htbp] \label{DissociateFigure}
    \caption{The effect of Algorithm~\ref{MLLtoMDNFCycleAlg}}
    \label{setfig}
    \begin{center}
    \begin{tikzpicture}
          [node distance=.3cm and .5cm, box/.style={
    rectangle,minimum size=6mm, very thick, draw=#1!50!black!50,
    top color=#1!50!black!20, bottom color=#1!50!black!20}, 
    skip loop/.style={to path={-- ++(0,#1) -| (\tikztotarget)}}]
    \tikzstyle{myarrows}=[line width=3mm,draw=black,-triangle 45,postaction={draw, line width=5mm, shorten >=7mm, -}]
    \node (X1) [box=red,] {$X$};
    \node (Y1) [box=blue, right=of X1] {$Y$};
    \node (Z1) [box=green, right=of Y1] {$Z$};
    \node (x1) [below=of X1] {$x$};
    \node (y1) [below=of Y1] {$y$};
    \node (z1) [below=of Z1] {$z$};
    \node (p1) [below left=of z1, xshift=5mm] {$\invamp$};
    \node (t1) [below left=of p1, xshift=5mm] {$\tens$};
    \node (W1) [right=of Z1] {};
    \node (W2) at (t1 -| W1) {};
    \node (W0) [above= 2mm of W1] {};
    \node (W3) [below= 2mm of W2] {};
    \node (tail) [below right=of z1, xshift=5mm, yshift=2mm] {};
    \node (head) [right= 20mm of tail] {};
    \node (X2) [box=red, right= 4cm of Z1] {$X$};
    \node (Y2) [box=blue, right=of X2] {$Y$};
    \node (Z2) [box=green, right=of Y2] {$Z$};
    \node (x2) [below=of X2] {$x$};
    \node (y2) [below=of Y2] {$y$};
    \node (z2) [below=of Z2] {$z$};
    \node (t2) [below right=of x2, xshift=-5mm] {$\tens$};
    \node (p2) [below right=of t2, xshift=-5mm] {$\invamp$};
    \node (V1) [right=of Z2] {};
    \node (V2) at (p2 -| V1) {};
    \node (V0) [above= 2mm of V1] {};
    \node (V3) [below= 2mm of V2] {};
    \path (X1) edge [orange, very thick, skip loop=6mm] (Y1);
    \path (X1) edge [orange ](x1);
    \path (Y1) edge [orange] (y1);
    \path (Z1) edge  (z1);
    \path (x1) edge [orange] (t1);
    \path (y1) edge [orange] (p1);
    \path (z1) edge (p1);
    \path (p1) edge [orange] (t1);
    \path (X2) edge [orange, very thick, skip loop=6mm] (Y2);
    \path (X2) edge [orange] (x2);
    \path (Y2) edge [orange] (y2);
    \path (Z2) edge  (z2);
    \path (x2) edge [orange] (t2);
    \path (y2) edge [orange] (t2);
    \path (z2) edge  (p2);
    \path (p2) edge (t2);
    \draw [myarrows](tail)--(head);
    \end{tikzpicture}
    \end{center}
    \end{figure}
    
        \begin{figure}[htbp] \label{DissociateFigure2}
        \caption{The effect of Algorithm~\ref{MLLtoMDNFCycleAlg}}
        \label{setfig}
        \begin{center}
        \begin{tikzpicture}
              [node distance=.3cm and .5cm, box/.style={
        rectangle,minimum size=6mm, very thick, draw=#1!50!black!50,
        top color=#1!50!black!20, bottom color=#1!50!black!20}, 
        skip loop/.style={to path={-- ++(0,#1) -| (\tikztotarget)}}]
        \tikzstyle{myarrows}=[line width=3mm,draw=black,-triangle 45,postaction={draw, line width=5mm, shorten >=7mm, -}]
        \node (X1) [box=red,] {$X$};
        \node (Y1) [box=blue, right=of X1] {$Y$};
        \node (Z1) [box=green, right=of Y1] {$Z$};
        \node (x1) [below=of X1] {$x$};
        \node (y1) [below=of Y1] {$y$};
        \node (z1) [below=of Z1] {$z$};
        \node (p1) [below left=of z1, xshift=5mm] {$\invamp$};
        \node (t1) [below left=of p1, xshift=5mm] {$\tens$};
        \node (W1) [right=of Z1] {};
        \node (W2) at (t1 -| W1) {};
        \node (W0) [above= 2mm of W1] {};
        \node (W3) [below= 2mm of W2] {};
        \node (tail) [below right=of z1, xshift=5mm, yshift=2mm] {};
        \node (head) [right= 20mm of tail] {};
        \node (X2) [box=red, right= 4cm of Z1] {$X$};
        \node (Y2) [box=blue, right=of X2] {$Y$};
        \node (Z2) [box=green, right=of Y2] {$Z$};
        \node (x2) [below=of X2] {$x$};
        \node (y2) [below=of Y2] {$y$};
        \node (z2) [below=of Z2] {$z$};
        \node (t2) [below right=of x2, xshift=-5mm] {$\tens$};
        \node (p2) [below right=of t2, xshift=-5mm] {$\invamp$};
        \node (V1) [right=of Z2] {};
        \node (V2) at (p2 -| V1) {};
        \node (V0) [above= 2mm of V1] {};
        \node (V3) [below= 2mm of V2] {};
        \path (Y1) edge [orange, very thick, skip loop=6mm] (W1);
        \path (W0) edge [orange,very thick] (W3);
        \path (t1) edge [orange, very thick, skip loop=-6mm] (W2);
        \path (X1) edge (x1);
        \path (Y1) edge [orange] (y1);
        \path (Z1) edge  (z1);
        \path (x1) edge (t1);
        \path (y1) edge [orange] (p1);
        \path (z1) edge (p1);
        \path (p1) edge [orange] (t1);
        \path (Y2) edge [orange, very thick, skip loop=6mm] (V1);
        \path (V0) edge [orange,very thick] (V3);
        \path (p2) edge [orange, very thick, skip loop=-6mm] (V2);
        \path (X2) edge (x2);
        \path (Y2) edge [orange] (y2);
        \path (Z2) edge  (z2);
        \path (x2) edge (t2);
        \path (y2) edge [orange] (t2);
        \path (z2) edge  (p2);
        \path (p2) edge [orange] (t2);
        \draw [myarrows](tail)--(head);
        \end{tikzpicture}
        \end{center}
        \end{figure}
    
    \p The diagrams highlight the two possible key paths from $C_{n}$ in $F_{n}$ (which also happen to the the edges which still exist in one of the switchings causing the cycle).
    
    \begin{itemize}
    
    \item In the first situation the path is composed of the following: the switching path linking the subformulae $X$ and $Y$, which may pass through any number of other subgraphs of a switching; the edge from $x$ to the $\otimes$-vertex; the switch edge from the $\invamp$-vertex to $y$; and the edge between the $\otimes$- and $\invamp$-vertices denoting the synonymous connectives written explicitly in $X \otimes (Y \invamp Z)$. It is trivially true that there must be a path from the literals connected to the highlighted axiom link through their respective subformulae $X$ and $Y$ to $x$ and $y$.
    
    \p The natural transformation $w^{LL}$ converts that proof structure from Figure~7.1 and the cycle described within it to the one to its right. The weak distributivity transformation does not affect which literals are connected to one another via an axiom link, and also does not affect any subformulae of $F_{n}$ except $X \otimes (Y \invamp Z)$. The only change is a `shuffling' of the $\otimes$- and $\invamp$-vertices in the primary subformula being considered. The cycle itself can be thought of as having been reduced by the algorithm so it contains all the same vertices except possibly for the $\invamp$-vertex, which may have been removed.
    
    \item In the second scenario we have a slightly longer path: the switching path linking the lowest $\otimes$-vertex in the subgraph back to $Y$ at the top, which may pass through any number of axiom links; the switch edge from the $\invamp$-vertex to $y$; and the edge between the $\otimes$- and $\invamp$-vertices denoting the synonymous connectives written explicitly in $X \otimes (Y \invamp Z)$.
    
    \p The natural transformation $w^{LL}$ converts the left proof structure of Figure~7.2 to the one on its right, with the cycle highlighted in the second diagram still existing. In this situation the length of the cycle has not been reduced, but a $\otimes$-vertex has effectively been moved further up the proof structure, which creates new possibilities for scenarios such as that seen in the previous point to be found. 
    
    \end{itemize}
    
\noindent    The lowest vertices in a cycle in the switching of a proof structure must be $\otimes$-vertices, with the cycle flowing through both of its argument edges. Whenever we meet such a situation as seen in the first case, the length of the cycle is reduced by one: a $\invamp$-vertex is removed. In the second case the length of the cycle is not changed, but the alteration in the form of the cycle in effect pushes the $\invamp$-vertex further down the structure until it inevitably meets a $\otimes$-vertex of maximal depth, at which point the vertex is removed.
    
    \p Since the other four steps 2(b), 3(a) and 3(b) of Algorithm~\ref{MLLtoMDNFCycleAlg} are identical in concept to Step~2(a) they also steadily remove these vertices. Since the algorithm stops once there are no $\invamp$-vertices in the cycle, and eventually in the worst case all $\invamp$-vertices will find themselves under the $\otimes$-vertices, this principle assures termination.
    
    \p It is also always possible to produce an MDNF~transformation from an \mll\ transformation using an algorithm such as the one below.
    
    \begin{algo} \label{MDNFAlg}
    Input: An \mll~transformation $\tau:\K \longrightarrow F$ describing a linear combination of proof structures over a common sequent. \\
    Output: An MDNF~transformation $\tau'$ describing a linear combination of proof structures over a common sequent.
    
    \p An algorithm which takes an MLL transformation describing a linear combination of proof structures over a common sequent, and produces another MLL transformation representing a linear combination of proof structures, only over an MDNF~sequent. The new sequent will preserve the number of tensor and par operators within the formulae.
    \end{algo}
    \begin{enumerate}
    \item Let $\tau_{0} = \tau: \K \longrightarrow F$ and $F_{0} = F$. Let $n = 0$, and whenever we refer to subformulae called $X$, $Y$ and $Z$, we name their lowest vertices as in a parse tree $x$, $y$ and $z$ respectively.
    
    \item Search for any position in $F_{n}$ of the form $X \otimes \left(Y \invamp Z\right)$ for some subformulae $X$, $Y$ and $Z$.
    	\begin{enumerate}
    	\item If one should exist, then let $\tau_{n+1} = \left(1 \cdots w^{LL} \cdots 1\right) \circ \tau_{n}$. The functor $F_{n+1}$ is defined to be the target functor of $\tau_{n+1}$. Increment $n$ and go back to Step 2.
    	\item If one does not exist, then move to Step 3.
    	\end{enumerate}
    	
    \item Search for any position in $F_{n}$ of the form $\left(X \invamp Y\right) \otimes Z$ for some subformulae $X$, $Y$ and $Z$.
    	\begin{enumerate}
    	\item If one should exist, then let $\tau_{n+1} = \left(1 \cdots w^{RR} \cdots 1\right) \circ \tau_{n}$. The functor $F_{n+1}$ is defined to be the target functor of $\tau_{n+1}$. Increment $n$ and go back to Step 2.
    	\item If one does not exist, then let $\hat{\tau} = \tau_{n}$, $\hat{F} = F_{n}$ and $\hat{C} = C_{n}$.
    	\end{enumerate}
    \end{enumerate}
    
\noindent    Algorithm \ref{MDNFAlg} is far simpler than Algorithm~\ref{MLLtoMDNFCycleAlg}, because there is no need to consider the preservation of anything: we are merely composing an \mll~transformation with a number of natural transformations which lead to the creation of an MDNF~transformation. However, it is useful to make the following observation.
    
    \begin{obs}
    Let $\tau:\K \longrightarrow F$ be an \mll~transformation modelling a linear combination of proof structures over a sequent, at least one of which fails the acyclicity criterion. Then applying Algorithms~\ref{MLLtoMDNFCycleAlg} and~\ref{MDNFAlg} in that order produces an MDNF~transformation $\tau''$ modelling a linear combination of proof structures, with at least one being cyclic.
    \end{obs}
    
    It is now possible to prove \mll~full completeness for $\GC$.
    
    \begin{thm}\label{fcomp}
      If\/ $\C$ is a compact closed category with finite biproducts
      satisfying feeble full completeness then\/ $\GC$ satisfies \mll~full completeness.
    \end{thm}
    \proof
      Let $\tau$ be an \mll~transformation in $\C$ modelling a linear
      combination of proof structures, at least one of which is cyclic.
      Using Algorithms \ref{MLLtoMDNFCycleAlg} and \ref{MDNFAlg}, we obtain a family of arrows $\tau''$ modelling a linear combination of MDNF~proof
      structures, with at least one of these containing a switching
      cycle. By Proposition~\ref{MDNFFC}, $\tau''$ is not an
      MDNF~transformation in $\GC$; and therefore it immediately follows
     from Proposition~\ref{MDNFtoMLLProp} that $\tau$ is not either.
    
      \p If we instead assume that $\tau$ models a linear combination of acyclic yet disconnected \mll~proof structures, Algorithm \ref{MDNFAlg} produces an MDNF~transformation $\tau'$ when introduced to $\tau$, and this new transformation is also a linear combination of \hughf{acyclic, disconnected} proof structures. Proposition~\ref{MDNFFC} once again allows us to say that $\tau'$---and consequently $\tau$---does not exist in $\GC$. The same principle can be used for the case where $\tau$ is a non-simple linear combination of proof nets, and so no \mll~transformations not representing a unique proof net is found in $\GC$. \qed
    
    We can in fact improve on the theorem above greatly with near enough no effort by making a simple observation. For any choice of category $\C$, the only objects used in the lemmata in this section are in the sets $\{A_{n}:n \in \N^{+}\}$ and $\{B_{n}:n \in \N^{+}\}$; and all of these are found in the collection of objects for any orthagonality category $\GsC$ with focus $S \supseteq \mathbf{2} = \{0_{\I,\bot},\iota\}$. The values and covalues of a $\GsC$-object $Z$ also describe the set of morphisms between the category's tensor unit $\I_{S}=(\I,\{1_{\I}\},S)$ and $Z$ and between $Z$ and $\I_{S}^{\bot}$ respectively (Fact~\ref{CoValueArrowProp}). Because of this, the arguments from previous chapters and earlier in this chapter using the properties of the values and covalues may be replicated verbatim to give an \mll~full completeness theorem for all categories of this more restricted form.
    
    \begin{thm}
      If\/ $\C$ is a compact closed category with finite biproducts
      satisfying feeble full completeness then\/ $\GsC$ satisfies
      \mll~full completeness for any $S \supseteq
      \{0_{\I,\bot},\iota\}$.
    \end{thm}
    
    \begin{cor}
      For a tensor-generated compact closed category with
      biproducts\/ $\C$ the category\/ $\GC$ satisfies\/ \mll\ full completeness.
    \end{cor}
    
    Note that biproducts are necessary for this result despite the
      fact that we are not addressing the additive connectives here.
      Indeed, it is very easy to produce a (tensor-generated) compact
      closed category without biproducts which does not create a fully
      complete model under the double glueing construction (the full
      subcategory of~$\FDVec$ containing only the tensor unit $\R$ for
      example).
    
    \section{\mllmix~Full Completeness for $\GiC$} \label{SectionMLLMixFC}
    
    The theorem given in the previous section has the effect of generalising the primary results found in first four chapters of \cite{Tan97}, which concern the categories $\GRel$ and $\GFDVec$ for arbitrary field $\F$ of characteristic $0$. However, the manner in which Tan proves her theorem for $\GRel$ actually has a positive side effect that the proof above is unable to replicate. The lemmata proving acyclicity and uniqueness are not stated with an assumption of an already derived connectedness lemma; and the objects used in the proofs of the acyclicity and uniqueness are found in the collection of objects in the orthogonality subcategory $\mathbf{G}_{1}\mathbf{Rel}$ with focus the singleton $\{\iota\}$ for $\iota = \lambda_{\I^{*}} \after v_{\I,\I,\I}(\rho_{\I})$, which is defined earlier in Section~\ref{SectionTransExtensors}. The category $\mathbf{G}_{1}\Rel$ is known to be a categorical model of \mllmix, and every dinatural transformation in $\mathbf{G}_{1}\Rel$ is inherited from $\Rel$. All these facts together allow us to conclude the following:
    
    \begin{fact}\hugh{\cite{Tan97}}
    The category $\mathbf{G}_{1}\Rel$ satisfies \mllmix~full completeness\footnote{The category in question was actually named $\mathbb{S}$ in \cite{Tan97}.}.
    \end{fact}
    
    This result can in fact also be generalised, though not the same extent as that seen in Section~\ref{SectionGCFC}. In this section we demonstrate that the category $\GiC$ satisfies \mllmix~full completeness if $\C$ is, like $\Rel$, a \emph{zero-sum-free} compact closed category with biproducts.
    \hugh{\begin{defi}
    A semiring $S$ is \emph{zero-sum-free} if its additive unit $0$ is its sole element with an additive inverse. A compact closed category with finite biproducts is \emph{zero-sum-free} if it is enriched over a category $\mathbf{SMod}_{S}$ for which the $S$ is zero-sum-free.
    \end{defi}}
    
     Though the use of counterexample test objects to show certain \mll~transformations found in $\C$ cannot be seen in the glued category is still abundantly present, the flavour of the coming proof resembles the deduction of Tan more closely than the earlier proofs:
     \begin{enumerate}[label=(\autoref{SubsectionMixUnique})]
     \item[(\autoref{SubsectionMixUnique})] We start by showing that the \mll~transformations which are found in both $\C$ and $\GiC$ must model simple linear combinations of proof structures. This is done by using reducing the problem to one for \mll~functors built using no tensor functors (Proposition~\ref{MDNFMixToMLLProp}). By necessity the proof of uniqueness (Proposition~\ref{MixUniqueLemma}) takes a slightly less combinatorial shape than before and has a more algebraic feel, with its basis being simultaneous equations.
     \item[(\autoref{SubsectionMixAcyclic})] This is followed by the proof that the proof structures being modelled must also be acyclic and therefore correct with respect to the Danos-Regnier \mllmix~proof net criterion. Algorithm~\ref{MixCycleAlg} reduces the target to the most simple standard form, leaving a comparartively simple argument for Proposition~\ref{MixAcyclicLemma}. The theorem immediately follows.
     \end{enumerate}
     In spite of the small change in proof style, the index notation for values and covalues of the objects of the form $(n\I,U,X)$ that has been employed greatly in this chapter continues to be utilised ubiquitously.
    
    \subsection{Uniqueness} \label{SubsectionMixUnique}
    
    In this full subcategory of $\GC$ in which we now reside, we have restrictions on the sets of values and covalues which we may choose for our test objects to prove certain transformations from the underlying category do not become transformations in our new location. Every value must compose with every covalue to produce the morphism $\fctl{\iota}{\I}{\bot}$ and vice versa. This is remarkably restrictive, primarily because it becomes harder to find valid test objects which are self-dual and still useful, such as the objects $A_{n}$ and $B_{n}$ described in Section \ref{SectionMDNFObjects} in $\GC$. However, this is counterbalanced by the fact that categorical models of \mllmix\ are equipped with the `Mix' natural transformation $\fctl{\Mix}{-~\otimes~-}{-~\invamp~-}$, which permits great simplifications to the set of \mll~formulae which must be considered to produce the desired results. The proof of the simplicity of all linear combinations of proof structures modelled in $\GiC$ demonstrates both these changes to the playing field in the clearest manner.
    
    \p Proposition~\ref{MDNFtoMLLProp} in the previous section allowed us to simplify the entire of the \mll~full completeness proof of $\GC$ to one for MDNF~full completeness. Now using the mix natural transformation, we can produce a more extreme version of this proposition whose proof follows from an identical concept, which is possible due to $\Mix$ being an isomorphism in compact closed categories.
    \begin{prop} \label{MDNFMixToMLLProp}
    Let $\fctl{\tau}{\K}{F}$ be an \mll~transformation in $\C$, and let $\fctl{\bar{w}}{F}{G}$ be a natural transformation built from the `Mix', weak distributivity, associativity and symmetry natural transformations of $\C$. Then $\tau$ does not translate into $\GiC$ if $\bar{w} \after \tau$ does not.
    \end{prop}
    
    For the uniqueness proof the actual connectives that are in each formula being modelled become irrelevant---only the number of linkings is important. Because of this, a natural transformation eradicating all uses of the tensor product `$\otimes$', leaving a formula of the form $\tbiginvamp_{m=1}^{M}(L_{\phi(m)} \invamp L^{\bot}_{\phi(m)})$, is a reasonable suggestion for $\bar{w}$ in the above proposition.
    
    \begin{cor} \label{AllParToMLLProp}
    Every \mllmix~transformation in $\GiC$ models a single proof structure if and only if every \mllmix~transformation to a functor leading to objects of the form $\Parf_{M}(\mathbf{L},\mathbf{L}) = \tbiginvamp_{m=1}^{M}(L_{\phi(m)} \invamp L^{\bot}_{\phi(m)})$ for some $\phi:[M] \longrightarrow [N]$ does.
    \end{cor}
    
    We now follow the same procedure of choosing a single family of objects in $\GiC$, $\{S_{n}: n \in \N^{+}\}$, where $US_{n} = n\I$ for each $n$, so that the set of values of the object $\Parf_{M}(S_{n},S_{n})$ does not contain the tensor representation of an \mll~transformation modelling a non-simple linear combination of axiom links. We define each of the $S_{n}$ as follows:
    \begin{displaymath}
    S_{n} := (n\I,\{\delta_{ix}:x \in [n]\},\{1_{i}\}),
    \end{displaymath}
    where $1_{i}$ is the tensor with entries equalling $1$ for all $i$. The values and covalues of $\Parf_{M}$ can be found from this definition for any $M$ easily.
    
    \begin{lem} \label{ClaimMixObjectProperties}
    For each $M \in \N^{+}$,
    \begin{itemize}
    \item $\Parf_{M}(S_{n},S_{n})_{Val}$ \\ $\;\; = \left\{\tensor*{z}{*^{j_{1}}_{i_{1}}^{\;\cdots\;}_{\;\cdots}^{j_{M}}_{i_{M}}} : \forall (x_{1},\ldots,x_{M}) \in [n]^{M}\, \forall k \in [M]\,\, z_{\mathbf{j}}^{\mathbf{i}} \cdot \prod_{m \neq k} (\delta_{i_{m}}^{x_{m}}1_{j_{m}}) \in \Perm(2,n)\right\}$
    \item $\Parf_{M}(S_{n},S_{n})_{CoVal} = \left\{\prod_{m=1}^{M}\delta_{i_{m}}^{x_{m}}1_{j_{m}} : (x_{1},\ldots,x_{M}) \in [n]^{M}\right\}$
    \end{itemize}
    \end{lem}
    \proof
    We follow a similar argument to that seen in Lemma~\ref{ClaimTensorPowers}, which is based around induction. We start by considering the covalues. The proof that the covalues of $\Parf_{1}(S_{n},S_{n})$ is near enough trivial.
    \begin{eqnarray*}
    \Parf_{1}(S_{n},S_{n})_{CoVal} = (S_{n} \invamp S_{n}^{\bot})_{CoVal} & = & \{y_{i} \cdot v_{j} : y_{i} \in (S_{n})_{CoVal},\,\,(S_{n}^{\bot})_{CoVal}\} \\
    													&	=	&	\{\delta_{i}^{x}1_{j} : x \in [n]\}
    \end{eqnarray*}
    It now takes minimal effort to show that the result desired is true for
    \begin{displaymath}
    \Parf_{M+1}(S_{n},S_{n}) = \Parf_{M}(S_{n},S_{n}) \invamp \Parf_{1}(S_{n},S_{n})
    \end{displaymath}
    whilst assuming the claim is true for $\Parf_{K}(S_{n},S_{n})$ for all $K \leq M$.
    \begin{eqnarray*}
    \Parf_{M+1}(S_{n},S_{n})_{CoVal}	& = & (\Parf_{M}(S_{n},S_{n}) \invamp \Parf_{1}(S_{n},S_{n}))_{CoVal} \\
    														& = & \{t^{\mathbf{j}}_{\mathbf{i}} \cdot u^{j_{M+1}}_{i_{M+1}} : t^{\mathbf{j}}_{\mathbf{i}} \in \Parf_{M}(S_{n},S_{n})_{CoVal},\,\,u^{j_{M+1}}_{i_{M+1}} \in \Parf_{1}(S_{n},S_{n})_{CoVal}\} \\
    														& = & \left\{\left(\textstyle\prod_{m=1}^{M}\delta_{i_{m}}^{x_{m}}1_{j_{m}}\right) \cdot \delta_{i_{M+1}}^{x_{m}}1_{j_{m}} : (x_{1},\ldots,x_{M}) \in [n]^{M},\,\,x_{M+1} \in [n]\right\} \\
    														& = & \left\{\textstyle\prod_{m=1}^{M+1}\delta_{i_{m}}^{x_{m}}1_{j_{m}} : (x_{1},\ldots,x_{M+1}) \in [n]^{M+1}\right\}
    \end{eqnarray*}
    
    Now we deal with the values of these object amalgamations. The base case shows us that the values of $\Parf_{1}(S_{n},S_{n})$ are the permutations over $[n]$.
    \begin{eqnarray*}
    \Parf_{1}(S_{n},S_{n})_{Val} = (S_{n} \invamp S_{n}^{\bot})_{Val}	& = & \{z_{i}^{j} : \forall y_{i} \in (S_{n})_{CoVal}\, z_{i}^{j}y_{i} \in (S_{n}^{\bot})_{Val}, \\
    													&		&	\qquad \forall v_{j} \in (S_{n}^{\bot})_{CoVal} \, z_{i}^{j}v_{j} \in (S_{n}^{\bot})_{Val}\} \\
    													& = & \{z_{i}^{j} : \exists y \in [n]\, z_{i}^{j}1_{i} = \delta_{jy},\,\, \forall y \in [n] \\
    													&		&	\qquad\qquad\qquad\quad z_{i}^{j}\delta_{jy}=z_{i}^{y}=1_{i}\} \\
    													& = & \Perm(2,n)
    \end{eqnarray*}
    The inductive case follows.
    \begin{eqnarray*}
    \Parf_{M+1}(S_{n},S_{n})_{Val}	& = & (\Parf_{M}(S_{n},S_{n}) \invamp \Parf_{1}(S_{n},S_{n}))_{Val} \\
    												& = & \{\tensor*{z}{*^{\mathbf{j}}_{\mathbf{i}}^{j_{M+1}}_{i_{M+1}}} : \forall w_{\mathbf{i}}^{\mathbf{j}} \in \Parf_{M}(S_{n},S_{n})_{CoVal}\, \tensor*{z}{*^{\mathbf{j}}_{\mathbf{i}}^{j_{M+1}}_{i_{M+1}}}w_{\mathbf{i}}^{\mathbf{j}} \in \Parf_{1}(S_{n},S_{n})_{Val}, \\											
    					&  & \quad\quad \forall u_{i_{M+1}}^{j_{M+1}} \in \Parf_{1}(S_{n},S_{n})_{Val}\, \tensor*{z}{*^{\mathbf{j}}_{\mathbf{i}}^{j_{M+1}}_{i_{M+1}}}u_{i_{M+1}}^{j_{M+1}} \in \Parf_{M}(S_{n},S_{n})_{Val}\} \\
    												& = & \{\tensor*{z}{*^{\mathbf{j}}_{\mathbf{i}}^{j_{M+1}}_{i_{M+1}}} : \forall (x_{1},\ldots,x_{M}) \in [n]^{M}\, \tensor*{z}{*^{\mathbf{j}}_{\mathbf{i}}^{j_{M+1}}_{i_{M+1}}} \cdot \textstyle\prod_{m=1}^{M}\delta_{i_{m}}^{x_{m}} \in \Perm(2,n), \\		
    					&  & \quad\quad \forall x_{M+1} \in [n]\, \tensor*{z}{*^{\mathbf{j}}_{\mathbf{i}}^{j_{M+1}}_{i_{M+1}}}\delta_{i_{M+1}}^{m_{M+1}}1_{j_{M+1}} \in \Parf_{M}(S_{n},S_{n})_{Val}\} \\							
    												& = & \{\tensor*{z}{*^{\mathbf{j}}_{\mathbf{i}}^{j_{M+1}}_{i_{M+1}}} : \forall \mathbf{x} \in [n]^{M}\, \tensor*{z}{*^{\mathbf{j}}_{\mathbf{i}}^{j_{M+1}}_{i_{M+1}}}\delta_{\mathbf{i}}^{\mathbf{x}} \in \Perm(2,n), \\
    					&  & \quad\quad \forall x_{M+1} \in [n]\, \tensor*{z}{*^{\mathbf{j}}_{\mathbf{i}}^{j_{M+1}}_{i_{M+1}}}\delta_{i_{M+1}}^{m_{M+1}}1_{j_{M+1}} \in
    \{q_{\mathbf{i}}^{\mathbf{j}} : \forall \mathbf{x} \in [n]^{M}\, \forall k \in [M] \\ 
        &  & \quad\quad\quad\quad q_{\mathbf{j}}^{\mathbf{i}} \cdot \textstyle\prod_{m \neq k} (\delta_{i_{m}}^{x_{m}}1_{j_{m}}) \in \Perm(2,n)\}
    					\} \\
    												& = & \{\tensor*{z}{*^{j_{1}}_{i_{1}}^{\;\cdots\;}_{\;\cdots}^{j_{M+1}}_{i_{M+1}}} : \forall (x_{1},\ldots,x_{M+1}) \in [n]^{M+1}\, \forall k \in [M+1] \\
    												&
                                            &	\qquad\qquad\qquad
                                              z_{\mathbf{j}}^{\mathbf{i}}
                                              \cdot \textstyle\prod_{m
                                              \neq k}
                                              (\delta_{i_{m}}^{x_{m}}1_{j_{m}})
                                              \in \Perm(2,n)\}
\rlap{\hbox to 98 pt{\hfill\qEd}}
    \end{eqnarray*}\smallskip
    
\noindent    Having access to the form of the values of $\Parf_{M}(S_{n},S_{n})$ for each value of $M$ and $n$ makes it simple enough to prove uniqueness for $\GiC$ as wished.
    
    \begin{prop} \label{MixUniqueLemma}
    Every \mll~transformation in $\GiC$ for zero-sum-free $\C$ models a unique proof structure.
    \end{prop}
    \proof
    We show that transformations are as desired for the \mll~functor $\Parf_{M}$; Corollary \ref{AllParToMLLProp} forces the result to work for all \mll~functors.  At this point we notice that $Z = (\I,\{1\},\{1\})$ utilised primarily in Section \ref{SectionZeroTransProof} is actually $S_{1}$, and so is an object in $\GiC$. We can therefore use Lemma~\ref{OnlyOneLemma} to say that the sum of every scalar involved in a linear combination of proof structures modelled in the category totals $1$.
    
    \p The formula $\Parf_{1}(L)$ is equal to $L \invamp L^{\bot}$, meaning only one valid set of axiom links can be placed on its formula, namely the linking containing only one axiom link connecting the only two literals. Since every \mll~transformation in the category must model a linear combination of proof structures, this means that a transformation to $\Parf_{1}$ has to model a scalar multiple of this one axiom link, and therefore have a tensor representation $s \cdot \delta_{ij}$ for some $s \in \C[\I,\I]$. Lemma~\ref{ClaimMixObjectProperties} tells us that $\Parf_{1}(S_{n},S_{n})_{Val} = \Perm(2,n)$ for all $n \in \N^{+}$, which only contains $s \cdot \delta_{ij}$ if $s = 1_{\I}$. The conclusion of this is that non-identity scalar multiples of the axiom link are not modelled by a transformation in the glued category.
    
    \p Now we consider the more general case of when $M \geq 2$. It is possible to follow the argument through for any $n \geq M$, so for the sake of simplicity we take $\Parf_{M}(S_{M})$. Lemma~\ref{ClaimMixObjectProperties} states that the tensors in $\Parf_{M}(S_{M})_{Val}$ are those which create full permutations when any $M-1$ of the $M$ $i$-$j$-index pairs are composed with matrices of the form $\delta_{i}^{x}1_{j}$ for each $x$. This leads to many tensor equations, but due to the symmetry and addition properties noted in Section~\ref{SectionTransExtensors} we may choose to restrict the equations somewhat without loss of generality. We only consider equations for which each $\delta_{i_{k}}^{x}$ composing with the value tensors are such that $x = k$. This leaves $M$ simultaneous equations to be solved over the semiring $\C[\I,\I]$: for each $k \in [M]$ there must be a vector of values $\mathbf{y} \in [n]^{M}$ such that
    \begin{displaymath}
    \tensor*{\omega}{*^{j_{1}}_{1}^{\;\cdots\;}_{\;\cdots}^{j_{k}}_{k}^{\;\cdots\;}_{\;\cdots}^{j_{M}}_{M}} \cdot \prod_{m \neq k}1_{j_{m}} \in \delta_{j_{k}y_{k}}.
    \end{displaymath}
    Again simplifying using the ideas from Section~\ref{SectionTransExtensors}, we can look solely at the entries where the one remaining $i$-indexed position, the $k^{th}$ position, also considered at~$k$. This means that the only entries of $\omega$ which are now of interest are of the form $\sum_{p \in S_{M}}c_{p} \cdot \tensor*{\delta}{*^{p(1)}_{j_1}^{\;\cdots\;}_{\;\cdots}^{p(M)}_{j_M}}$, where each $p:[M] \longrightarrow [M]$ is a permutation and $c_{p}$ a corresponding scalar. Entries where the $j$-indices do not form a permutation of $[M]$ are therefore always zero.
    
    \p For some set of $\mathbf{y} \in [n]^{M}$,
    \begin{displaymath}
    \sum_{p}c_{p} \cdot \tensor*{\delta}{*^{p(1)}_{j_1}^{\;\cdots\;}_{\;\cdots}^{p(k)}_{j_k}^{\;\cdots\;}_{\;\cdots}^{p(M)}_{j_M}} \cdot \prod_{m \neq k}1_{j_{m}} \,=\, \delta_{j_{k}y_{k}}
    \end{displaymath}
    This leads to $M^{2}$ linear equations of consequence, all bar $M$ of which being sums which have to equal $0$. In the context of a zero-sum-free semiring, as $\C[\I,\I]$ is from our original assumption, each of the values in the sums of these $(M^{2}-M)$ equations must equal $0$.
    
    Let $f:[M] \longrightarrow [M]$ be the function which describes the vector $\mathbf{y}$, i.e. for every $k$, $f(k) = y_{k}$. Focusing on an arbitrary $k$, we find that
    \begin{align*}
    \sum_{p \in S_{M}} c_{p} \cdot \tensor*{\delta}{*^{p(1)}_{j_1}^{\;\cdots\;}_{\;\cdots}^{p(M)}_{j_M}} \cdot \prod_{m \neq 1}1_{j_{m}} &= 0 \mbox{ when $j_{k} \,\neq\, y_{k}$}.
    \end{align*}
    Therefore, by zero-sum-freeness, for every permutation $p$ whenever $j_{k} \,\neq\, y_{k}$,
    \begin{displaymath}
    c_{p} \cdot \tensor*{\delta}{*^{p(1)}_{j_1}^{\;\cdots\;}_{\;\cdots}^{p(M)}_{j_M}} \,=\, 0.
    \end{displaymath}
    
    By symmetry, we find that for all $p \,\in\, S_{M}$, $c_{p} \cdot \tensor*{\delta}{*^{p(1)}_{j_1}^{\;\cdots\;}_{\;\cdots}^{p(M)}_{j_M}} \,=\, 0$ unless $j_{k} \,=\, y_{k}$ for all $k \,\in\, [1,M]$, which forces us to conclude that
    \begin{displaymath}
    \tensor*{\omega}{*^{j_{1}}_{1}^{\;\cdots\;}_{\;\cdots}^{j_{k}}_{k}^{\;\cdots\;}_{\;\cdots}^{j_{M}}_{M}} \,=\, c \cdot \tensor*{\delta}{*^{y_1}_{j_1}^{\;\cdots\;}_{\;\cdots}^{y_M}_{j_M}}
    \end{displaymath}
    for some $c \,\in\, \C[\I,\I]$.
    
    \p By the form of $\omega$, we already know then that $f(x) \,=\, y_{k}$ must form a permutation, and that
    \begin{displaymath}
    \tensor*{\omega}{*^{j_{1}}_{i_1}^{j_{M}}_{i_M}} \,=\, c_{p} \cdot \tensor*{\delta}{*^{i_{f(1)}}_{j_1}^{\;\cdots\;}_{\;\cdots}^{i_{f(M)}}_{j_M}}.
    \end{displaymath}
    The sum of the scalars must be unity, so $c \,=\, 1$; and so the arrow which $\omega$ describes is an instantiation of an \mll~transformation which represents a unique proof structure. \qed

    The proof above can be elucidated somewhat through the use of an example.
    
    \begin{exa} \label{MixUniqueExam}
    We view the proof of Lemma~\ref{MixUniqueLemma} for the \mll~formula $\Parf_{3}(S_{3},S_{3})$ in more detail.
    \end{exa}
    We assume that $\tensor*{\omega}{*^{j_{1}}_{1}^{j_{2}}_{2}^{j_{3}}_{3}}$ is the tensor representation of a linear combination of linkings on the formula modelled by $\Parf_{3}$. Lemma~\ref{ClaimMixObjectProperties} is provides us with the three tensor equations which must be satisfied for some numbers $y_{1}$, $y_{2}$ and $y_{3}$ found in $\{1,2,3\}$.
    \begin{eqnarray*}
    \tensor*{\omega}{*^{j_{1}}_{1}^{j_{2}}_{2}^{j_{3}}_{3}} \cdot 1_{j_{2}j_{3}} & = & \delta_{j_{1}y_{1}} \\
    \tensor*{\omega}{*^{j_{1}}_{1}^{j_{2}}_{2}^{j_{3}}_{3}} \cdot 1_{j_{1}j_{3}} & = & \delta_{j_{2}y_{2}} \\
    \tensor*{\omega}{*^{j_{1}}_{1}^{j_{2}}_{2}^{j_{3}}_{3}} \cdot 1_{j_{1}j_{2}} & = & \delta_{j_{3}y_{3}}
    \end{eqnarray*}
    
\noindent     Suppose that $y_{k} = k$ for each $k$, making each constant distinct from one another and the function $f$ proposed in the proof of the earlier lemma a permutation. This creates $3^{2} = 9$ equations:
    \begin{eqnarray}
    \tensor*{\omega}{*^{1}_{1}^{2}_{2}^{3}_{3}} + \tensor*{\omega}{*^{1}_{1}^{3}_{2}^{2}_{3}}  & = & 1 \\
    \tensor*{\omega}{*^{2}_{1}^{1}_{2}^{3}_{3}} + \tensor*{\omega}{*^{2}_{1}^{3}_{2}^{1}_{3}}  & = & 0 \\
    \tensor*{\omega}{*^{3}_{1}^{1}_{2}^{2}_{3}} + \tensor*{\omega}{*^{3}_{1}^{2}_{2}^{1}_{3}}  & = & 0 \\
    \tensor*{\omega}{*^{2}_{1}^{1}_{2}^{3}_{3}} + \tensor*{\omega}{*^{3}_{1}^{1}_{2}^{2}_{3}}  & = & 0 \\
    \tensor*{\omega}{*^{1}_{1}^{2}_{2}^{3}_{3}} + \tensor*{\omega}{*^{3}_{1}^{2}_{2}^{1}_{3}}  & = & 1 \\
    \tensor*{\omega}{*^{1}_{1}^{3}_{2}^{2}_{3}} + \tensor*{\omega}{*^{2}_{1}^{3}_{2}^{1}_{3}}  & = & 0 \\
    \tensor*{\omega}{*^{2}_{1}^{3}_{2}^{1}_{3}} + \tensor*{\omega}{*^{3}_{1}^{2}_{2}^{1}_{3}}  & = & 0 \\
    \tensor*{\omega}{*^{1}_{1}^{3}_{2}^{2}_{3}} + \tensor*{\omega}{*^{3}_{1}^{1}_{2}^{2}_{3}}  & = & 0 \\
    \tensor*{\omega}{*^{1}_{1}^{2}_{2}^{3}_{3}} + \tensor*{\omega}{*^{2}_{1}^{1}_{2}^{3}_{3}}  & = & 1
    \end{eqnarray}
    Since we are in a zero-sum-free compact closed category, we know that all the values in the sums equalling $0$ must come to zero themselves. That means we can conclude from looking at equations (2), (3), (4), (6), (7) and (8) that
    \begin{displaymath}
    \tensor*{\omega}{*^{2}_{1}^{1}_{2}^{3}_{3}} = \tensor*{\omega}{*^{2}_{1}^{3}_{2}^{1}_{3}} = \tensor*{\omega}{*^{3}_{1}^{1}_{2}^{2}_{3}} = \tensor*{\omega}{*^{3}_{1}^{2}_{2}^{1}_{3}} = \tensor*{\omega}{*^{1}_{1}^{3}_{2}^{2}_{3}} = 0.
    \end{displaymath}
    Substituting these assignments into the $9$ equations leaves the six sums totalling zero as trivial, and equations (1), (5) and (9) simplify to become
    \begin{displaymath}
    \tensor*{\omega}{*^{1}_{1}^{2}_{2}^{3}_{3}} = 1
    \end{displaymath}
    Note that the choices of $y_{1}$, $y_{2}$ and $y_{3}$ coincide with the order of the three superscript index numbers of $\omega$. This is to be expected, since the $y_{k} = k$ makes it necessary for the $k^{th}$ superscript index position to be $k$ in order to for it to be in the equation for which the other $(M-1)$ $j$-indices are changed in are added together and equate to~$1$. As such, the given position is the only one which fits this criteria for all $k$.
    
    Due to $\omega$ being a linear combination of Kronecker deltas representing axiom links between positive and negative literals, the only \mll~transformation which can be said to satisfy the criteria given here is the one modelling the unique proof structure given below.
    
    \begin{center} \vspace{5mm}
        \begin{tikzpicture}
          [auto, node
          distance=5mm, skip loop/.style={to path={-- ++(0,#1) -| (\tikztotarget)}}] \tikzstyle{every node} = [text depth=-5pt,text height=0.5ex]
           \node (1) {$L$}; \node (a) [right of=1] {$\invamp$} ;
          \node (2) [right of=a] {$L^{\bot}$}; \node (b) [right of=2] {$\invamp$} ;
          \node (3)  [right of=b] {$L$}; \node (c)  [right of=3] {$\invamp$};
          \node (4)  [right of=c] {$L^{\bot}$}; \node (d) [right of=4] {$\invamp$} ;
          \node (5) [right of=d] {$L$}; \node (e) [right of=5] {$\invamp$} ;
          \node (6) [right of=e] {$L^{\bot}$};
        \begin{scope}
        \path   (1)  edge [skip loop =6mm, shorten >=3mm, shorten <=3mm]  (2);
        \path   (3)  edge [skip loop =6mm, shorten >=3mm, shorten <=3mm]  (4);
        \path   (5)  edge [skip loop =6mm, shorten >=3mm, shorten <=3mm]  (6);
        \end{scope}
        \end{tikzpicture} \vspace{5mm}
    \end{center}
    Symmetry in permutations allows us to ascertain that all the possible linkings are possible by changing the permutation that $f$ describes to each of the other $3!-1 = 5$, but no non-simple linear combinations have been shown to be possible.
    
    \p Now, we choose values of $y_{1},y_{2},y_{3}$ so that the induced function $f$ is not a permutation. Let us first consider $y_{k} = 1$ for all $k$; these induce the following linear equations:
    \begin{eqnarray}
    \tensor*{\omega}{*^{1}_{1}^{2}_{2}^{3}_{3}} + \tensor*{\omega}{*^{1}_{1}^{3}_{2}^{2}_{3}}  & = & 1 \\
    \tensor*{\omega}{*^{2}_{1}^{1}_{2}^{3}_{3}} + \tensor*{\omega}{*^{2}_{1}^{3}_{2}^{1}_{3}}  & = & 0 \\
    \tensor*{\omega}{*^{3}_{1}^{1}_{2}^{2}_{3}} + \tensor*{\omega}{*^{3}_{1}^{2}_{2}^{1}_{3}}  & = & 0 \\
    \tensor*{\omega}{*^{2}_{1}^{1}_{2}^{3}_{3}} + \tensor*{\omega}{*^{3}_{1}^{1}_{2}^{2}_{3}}  & = & 1 \\
    \tensor*{\omega}{*^{1}_{1}^{2}_{2}^{3}_{3}} + \tensor*{\omega}{*^{3}_{1}^{2}_{2}^{1}_{3}}  & = & 0 \\
    \tensor*{\omega}{*^{1}_{1}^{3}_{2}^{2}_{3}} + \tensor*{\omega}{*^{2}_{1}^{3}_{2}^{1}_{3}}  & = & 0 \\
    \tensor*{\omega}{*^{2}_{1}^{3}_{2}^{1}_{3}} + \tensor*{\omega}{*^{3}_{1}^{2}_{2}^{1}_{3}}  & = & 1 \\
    \tensor*{\omega}{*^{1}_{1}^{3}_{2}^{2}_{3}} + \tensor*{\omega}{*^{3}_{1}^{1}_{2}^{2}_{3}}  & = & 0 \\
    \tensor*{\omega}{*^{1}_{1}^{2}_{2}^{3}_{3}} + \tensor*{\omega}{*^{2}_{1}^{1}_{2}^{3}_{3}}  & = & 0
    \end{eqnarray}
    Earlier it is explained why the only position of $\tensor*{\omega}{*^{j_1}_{1}^{j_2}_{2}^{j_3}_{3}}$ that can be non-zero is the one where $\mathbf{j} = \mathbf{y}$. Bearing this in mind, we can expect no valuations of the entries of interest satisfy the above equations. Hypothetically, only $\tensor*{\omega}{*^{1}_{1}^{1}_{2}^{1}_{3}}$ would be allowed to be non-zero; but by the constraints set upon it it is $0$ by definition. By similar arguments, no tensor representation of an \mll~transformation $\omega$ satisfies the equations when two or more of $y_{1},y_{2},y_{3}$ are equal.
    
    \p For a tensor representation of an \mll~transformation to be in the values, it must satisfy the $9$ equations for at least one of the combinations of values $y_{1}$, $y_{2}$ and $y_{3}$ can take. From the evidence above we know that such an $\tensor*{\omega}{*^{j_1}_{i_1}^{j_2}_{i_2}^{j_3}_{i_3}}$ can only do that if only one tuple $(j_{1},j_{2},j_{3})$ provides a non-zero value for $\tensor*{\omega}{*^{j_1}_{1}^{j_2}_{2}^{j_3}_{3}}$, and that position must equal zero. The only conclusion is, therefore, that $\tensor*{\omega}{*^{j_1}_{1}^{j_2}_{2}^{j_3}_{3}} = \tensor*{\omega}{*^{j_p(1)}_{i_1}^{j_p(2)}_{i_2}^{j_p(3)}_{i_3}}$ for some permutation $p$ over $3$. That is, no non-simple linear combinations of proof structures are modelled by a tensor in $\Parf_{3}(S_{3})$.
    
    \subsection{Acyclicity} \label{SubsectionMixAcyclic}
    
    Now that is has been established that $\GiC$ only contains \mll~transformations modelling unique proof structures, we can now attempt to show that those proof structures modelled are also acyclic. Once this has been shown, we will have proved \mllmix~full completeness: the connectedness condition of the Danos-Regnier criteria is rendered redundant once the `Mix' rule has been installed in the logic.
    
    \p Simplifying the problem down to dealing solely with the \mll~transformations $\Parf_{M}$ for different values of $M$ is not an option here. If we wish to prove that a cyclic structure cannot be modelled in the glued category, we need to make sure that any natural transformations do not create a transformation modelling an acyclic structure. As such, at least one of the $\otimes$-vertices must be preserved. However, ignoring the existence of $\Mix$ would be a waste of the available attributes of the category.
    
    \p The methods utilised in Section \ref{SectionGCFC} to allow us to only consider the easier-to-handle MDNF~transformations are once again of use here. There is an algorithm given below which takes MDNF~transformations modelling unique cyclic proof structures and produces other MDNF~transformations modelling a formula containing the minimum number of \linebreak $\otimes$-vertices possible whilst preserving exactly one cycle.
    
    \begin{algo} \label{MixCycleAlg}
    Input: An MDNF~transformation $\tau:\K \longrightarrow F$ describing a unique proof structure, with $M$ blocks say, breaking the acyclicity criterion. \\
    Output: An MDNF~transformation $\tau':\K \longrightarrow F'$ built by composing $\tau$ with natural transformations which describes a unique proof structure still breaking the acyclicity criterion with the minimum number of $\otimes$-vertices within it.
    \end{algo}
    \begin{enumerate}
    \item Choose a cycle from the proof structure described by $\tau$ passing through the minimum number of axiom links. We let $n = 0$, $l = 1$, $m = 1$ $\tau_{0} = \tau$, and we name the cycle $C$ (referring only to the literals). For every $n$, $F_{n}$ is the target functor of $\tau_{n}$
    \item Search for the block furthest to the left in the formula modelled with literals within $C$, calling it block $m_{1}$. Let $\tilde{\sigma}$ be a composition of the associativity and symmetry isomorphisms for the monoidal bifunctor $-\otimes-$ which modifies the block so it takes the form $((X_{1} \otimes Y_{1}) \otimes Z)$, where $X_{1}$ and $Y_{1}$ are the positions signifying the literals in $C$ and $Z$ is a tensor product containing the rest of the literals. Define $\tau_{n+1} = \binvamp_{m=1}^{m_{1}-1}1_{B_{m}} \invamp \tilde{\sigma} \invamp \binvamp_{m=m_{1}+1}^{M}1_{B_{m}}$, where $B_{m}$ is the $m^{th}$ block. If $X_{1}$ and $Y_{1}$ are connected via an axiom link go to Step~4; otherwise go to Step~3.
    \item Increment $l$. Search for the block (which we name block $m_{l}$) which contains the literal, $X_{l}$, connected to literal $Y_{l-1}$ via an axiom link. Let $\tilde{\sigma}$ be a composition of associativity and symmetry isomorphisms which modifies the block so it takes the form $((X_{l} \otimes Y_{l}) \otimes Z)$, where $Y_{l}$ is the other literal position in the block which is in $C$. Define $\tau_{n+1} = \binvamp_{m=1}^{m_{l}-1}1_{B_{m}} \invamp \tilde{\sigma} \invamp \binvamp_{m=m_{l}+1}^{M}1_{B_{m}}$. If $Y_{l}$ is connected via an axiom link to $X_{1}$ then go to Step~4; otherwise repeat Step~3.
    \item Consider block $m$, $B_{m} = L_{1,m} \otimes L_{2,m} \otimes L_{m}$ where $L_{1,m}$ and $L_{2,m}$ are the functors representing the first and second literal positions. We set the natural transformation $\MIX^{F}$ the maximal composition of `mix' transformations on an \mll~functor $F$ which eradicates all instances of $\otimes$ and replaces them with $\invamp$ whilst maintaining the order and bracketing of the inputs of the functor.
    	\begin{enumerate}
    	\item If it does not contain a literal in $C$, then let $\mu_{m} = \MIX^{B_{m}}$.
    	\item If it does, then let $\mu_{m} = \Mix \after ((1_{\GiC} \otimes 1_{\GiC}) \otimes \MIX^{L_{m}})$
    	\end{enumerate}
    If $m < M$, then increment the number, and repeat Step~4. Otherwise let \linebreak $\tau_{n+1} = \binvamp_{m=1}^{M}\mu_{m} \after \tau_{n}$, increment $n$ and go to Step~5.
    \item Let $\tilde{\beta}$ be a natural isomorphism built using the associativity and symmetry isomorphisms corresponding to the bifunctor $-\invamp-$ so that all the blocks in the proof structure described by $\tau_{n}$ containing only one literal are to the left of all those containing two literals, and the two-literal blocks maintain the same order with respect to each other as found in $\tau_{n}$. We let $\tau' = \tilde{\beta} \after \tau_{n}$, and terminate the algorithm.
    \end{enumerate}
    
\noindent    The algorithm above certainly terminates---after all, each step changing an \mll~transformation can only deal with a finite set of finite blocks. Equally, the output dinatural transformation does indeed describe a cyclic proof structure: in every switching there is a cycle $X_{1}-\otimes-Y_{1}-\cdots-X_{l}-\otimes-Y_{l}$, and the $\otimes$-vertices in the cycle are the only ones which appear in the structure. Together with Algorithms \ref{MLLtoMDNFCycleAlg} and \ref{MDNFAlg}, we can find natural transformations from $\C$ which continue to exist in $\GiC$ which take any \mll~transformation describing a unique cyclic proof structure in $\C$ and create another one which models a proof structure with shape much like the proof structure given below.
    
    \begin{center} \label{MixCycleFigure} \vspace{5mm}
    \begin{tikzpicture}
      [auto, node
      distance=5mm, skip loop/.style={to path={-- ++(0,#1) -| (\tikztotarget)}}] \tikzstyle{every node} = [text depth=-5pt,text height=0.5ex]
       \node (1) {$L$}; \node (a) [right of=1] {$\invamp$} ;
      \node (2) [right of=a] {$L^{\bot}$}; \node (b) [right of=2] {$\invamp$} ;
      \node (3)  [right of=b] {$\cdots$}; \node (c)  [right of=3] {$\invamp$};
      \node (4)  [right of=c] {$L$}; \node (d) [right of=4] {$\invamp$} ;
      \node (5) [right of=d] {$L^{\bot}$}; \node (e) [right of=5] {$\invamp$} ;
      \node (6) [right of=e] {$(L$}; \node (f) [right of=6] {$\tens$} ;
      \node (7) [right of=f] {$L^{\bot})$}; \node (g) [right of=7] {$\invamp$} ;
      \node (8)  [right of=g] {$\cdots$}; \node (h)  [right of=8] {$\invamp$};
      \node (9)  [right of=h] {$(L$}; \node (i) [right of=9] {$\tens$} ;
      \node (10) [right of=i] {$L^{\bot})$};
    \begin{scope}
    \path   (1)  edge [skip loop =6mm, shorten >=3mm, shorten <=3mm]  (2);
    \path   (4)  edge [skip loop =6mm, shorten >=3mm, shorten <=3mm]  (5);
    \path   (6)  edge [skip loop =7mm, shorten >=3mm, shorten <=3mm]  (10);
    \path   (7)  edge [dashed, skip loop =5mm, shorten >=3mm, shorten <=3mm]  (9);
    \end{scope}
    \end{tikzpicture} \vspace{5mm}
    \end{center}
    
\noindent    This is a very regular form of \mll~formula being modelled. The formulae can be separated into two distinct parts. If we instantiate the functor with tuples containing the same self-dual object built over a $\C$-object $n\I$ for some $n \in \N^{+}$ in every entry, $D$ say, we can say the tensor representations of the values and covalues in the resulting object take the form $F(\mathbf{D},\mathbf{D}) = \Gamma_{g}(D) \invamp \Delta_{d}(D)$ for the MDNF~functors $\Gamma_{g}$ and $\Delta_{d}$ for some $g \in \N$, $d \in \N^{+}$, where
    \begin{eqnarray*}
    \Gamma_{g}(D,D) = D^{\invamp 2g} & & \Delta_{d}(D,D) = (D^{\otimes 2})^{\invamp d}
    \end{eqnarray*}
    
    As has been the standard throughout this chapter, a single test object is required for all inputs of the total resulting \mll~transformation to show that acyclic proof structures are not modelled in $\GiC$. An object which is sufficient for the task whilst remaining in the confines of $\GiC$ is
    \begin{displaymath}
    D := (2\I,\{\delta_{ix}: x \in \{1\}\},\{\delta_{ix}: x \in \{1\}\}).
    \end{displaymath}
    
    The values and covalues of objects $F(\mathbf{D},\mathbf{D}) = \Gamma_{g}(D,D) \invamp \Delta_{d}(D,D)$, where the $N$-dimensional tuple is $\mathbf{D} = (D,\ldots,D) \in (\GiC)^{N}$ and $F$ is the target functor of an MDNF~transformation which is an output of Algorithm~\ref{MixCycleAlg} can be found easily by using inductive methods, as seen in the claim below.
    
    \begin{lem} \label{GammaDeltaClaim}
    For every $n \geq 2$, $\Gamma_{n}(D,D) = ((2\I)^{\invamp 2n},\Xi_{2n},\{\tensor*{\delta}{*^{1}_{i_{1}}^{\;\cdots\;}_{\;\cdots}^{1}_{i_{n}}^{1}_{j_{1}}^{\;\cdots\;}_{\;\cdots}^{1}_{j_{n}}}\})$, where
    \begin{displaymath}
    \Xi_{n} = \{w_{i_{1} \cdots i_{n}} : w_{1 \cdots 1} = 1;\, \forall \mathbf{x} \mbox{ containing $(n-1)$ $1$-entries}, \,w_{\mathbf{x}} = 0\}.
    \end{displaymath}
    Similarly, $\Delta_{n}(D,D) = (((2\I)^{\otimes 2})^{\invamp n},\{z_{\mathbf{i}}^{\mathbf{j}}: \forall k \bforall_{m \neq k} z_{\mathbf{i}}^{\mathbf{j}} w_{i_{m}j_{m}} \in \Xi_{2},\, z_{\mathbf{i}}^{\mathbf{j}} \cdot \prod_{m \neq k}w_{i_{m}j_{m}} = \tensor*{\delta}{*^{1}_{i_{1}}^{1}_{j_{2}}}\},\prod_{m=1}^{n}\Xi_{2})$.
    \end{lem}
    \proof
    The object $\Gamma_{n}(D,D) = D^{\invamp 2n}$, and so we show that $D^{\invamp n}$ has the form of $\Gamma_{n}(D)$ is as seen above, only with $2n$ replaced by $n$, starting with $n = 2$.
    \begin{eqnarray*}
    D \invamp D	& = & (2\I \invamp 2\I,\{w_{ij}: w_{1j} = \delta_{j1},\,w_{i1} = \delta_{i1}\},\{u_{i}v_{j}:u_{i},v_{j} \in D_{CoVal}\}) \\
    						& = & ((2\I)^{\invamp 2},\Xi_{2},\{\tensor*{\delta}{*^{1}_{i}^{1}_{j}}\})
    \end{eqnarray*}
    The inductive case is proved as follows:
    \begin{eqnarray*}
    D^{\invamp (n+1)}	& = & D^{\invamp n} \invamp D \\
    								& = & ((2\I)^{\invamp n} \invamp (2\I),\{w_{i_{1} \cdots i_{n} j}:w_{\mathbf{i}1} \in \Xi_{n},\,w_{1 \cdots 1 j} = \delta_{j1}\},\{\tensor*{\delta}{*^{1}_{i_{1}}^{\;\cdots\;}_{\;\cdots}^{1}_{i_{n}}} \cdot \delta_{j}^{1}\}) \\
    								& = & ((2\I)^{\invamp n+1},\Xi_{n+1},\{\tensor*{\delta}{*^{1}_{i_{1}}^{\;\cdots\;}_{\;\cdots}^{1}_{i_{n}}^{1}_{j}}\})
    \end{eqnarray*}
    
\enlargethispage{\baselineskip}
\noindent    The result is exactly as desired, and so the claim concerning $\Gamma_{n}(D,D)$ is solved by corollary. We now move to demonstrate $\Delta_{n}(D,D)$ takes the form wanted. We can conclude the following due to the functor $-\otimes-$ being the de Morgan dual of the functor $-\invamp-$:
    \begin{displaymath}
    \Delta_{1}(D,D) = (2\I \otimes 2\I,\{\tensor*{\delta}{*^{1}_{i}^{1}_{j}}\},\Xi_{2})
    \end{displaymath}
    The final result immediately follows. \qed
    
    Proposition~\ref{MDNFMixToMLLProp}, in collaboration with Algorithms~\ref{MLLtoMDNFCycleAlg} and \ref{MixCycleAlg}, means the following statement is true: if we can show that \mll~transformations in the form seen in the above figure are not capable of existing in $\GiC$, then no cyclic proof structures may be modelled in the category $\C$ under the orthogonalised glueing.
    
    \begin{prop} \label{MixAcyclicLemma}
    Every \mll~transformation in $\GiC$ modelling a unique proof structure models an acyclic proof structure.
    \end{prop}
    \proof
    Due to the earlier discussion it is only necessary to consider \mll~transformations of the shape seen in Figure~\ref{MixCycleFigure} due to Algorithms~\ref{MLLtoMDNFCycleAlg} and \ref{MixCycleAlg} and Proposition~\ref{MDNFMixToMLLProp}. Suppose that $F = \Gamma \invamp \Delta$ be such a functor, with $\Gamma$ containing all the one-literal blocks and $\Delta$ all the two-literal blocks. We use the test object $D = (2\I,\{\delta_{ix}\},\{\delta_{ix}\})$, meaning that we are able to use the calculations given in Lemma~\ref{GammaDeltaClaim} to find the values of $F(\mathbf{D},\mathbf{D}) = \Gamma_{g}(\mathbf{D},\mathbf{D}) \invamp \Delta_{d}(\mathbf{D},\mathbf{D})$ for some $g\in \N$, $d \in \N^{+}$.
    
    \p Suppose that $\Gamma$ is a functor describing a subformula containing no literals, making $F(\mathbf{D},\mathbf{D}) = \Delta_{d}(D,D)$ for a positive integer $d$. The values of this object are therefore exactly as seen in Lemma~\ref{GammaDeltaClaim}. To simplify matters, instead of making $i$- and $j$-indices correspond to positive and negative literal positions respectively as earlier, the indices $i_{m}$ and $j_{m}$ now relate to the first and second literals in the $m^{th}$ block in $\Delta_{d}(D)$ for all $m \in [d]$. We can therefore assume that the proof structure being discussed, which we say has linking $\lambda$, has tensor representation $\tensor*{\bar{\lambda}}{*^{j_{1}}_{i_{1}}^{\;\cdots\;}_{\;\cdots}^{j_{d}}_{i_{d}}} = \tensor*{\delta}{*^{j_{d}}_{i_{1}}^{j_{1}}_{i_{2}}^{\;\cdots\;}_{\;\cdots}^{j_{d-1}}_{i_{d}}}$.
    
    \p For $d = 1$, the situation is clear: $\Delta_{1}(D,D)_{Val} = \{\tensor*{\delta}{*^{1}_{i_{1}}^{1}_{j_{1}}}\}$, which clearly does not contain $\delta_{i_{1}j_{1}}$ as required for the cyclic proof structure to be modelled. For larger $d$, we need to use covalues from $D \otimes D$.
    \begin{displaymath}
    (D \otimes D^{\bot})_{CoVal} = \Xi_{2} = \{w_{ij}:w_{11} = 1,\,w_{12} = 0 = w_{21}\}
    \end{displaymath}
    For every block $m \in [2,d]$, we use $\delta_{i_{m}}^{j_{m}}$, which indeed does belong to the set of covalues. We find that
    \begin{eqnarray*}
    \tensor*{\bar{\lambda}}{*^{j_{1}}_{i_{1}}^{\;\cdots\;}_{\;\cdots}^{j_{d}}_{i_{d}}} \cdot \prod_{m=2}^{d}\delta_{i_{m}}^{j_{m}} & = & \tensor*{\delta}{*^{j_{d}}_{i_{1}}^{j_{1}}_{i_{2}}^{\;\cdots\;}_{\;\cdots}^{j_{d-1}}_{i_{d}}} \cdot \tensor*{\delta}{*^{j_{2}}_{i_{2}}^{\;\cdots\;}_{\;\cdots}^{j_{d}}_{i_{d}}} \\
    				& = & \delta^{j_{1}}_{i_{1}}
    \end{eqnarray*}
    The tensor $\delta^{j_{1}}_{i_{1}}$ does not equal $\tensor*{\delta}{*^{1}_{1}^{j_{1}}_{i_{1}}}$, and so it does not belong to the set of values for $D \otimes D$. As such, $\bar{\lambda}$ does not satisfy the criteria expected of all elements of the set of values for the object $\Delta_{d}(D,D)$. Because of this, the MDNF~transformation to $F$ does not lift to $\GiC$.
    
    \p We extend this result so that it works for an non-empty $\Gamma$. In this scenario we expect $F(\mathbf{D},\mathbf{D}) = \Gamma_{g}(D,D) \invamp \Delta_{d}(D,D)$ for positive $g$ and $d$, and by standard tensor calculations we note that, when $\mathbf{k}$ and $\mathbf{l}$ act as superindices for the positive and negative literals respectively,
    \begin{eqnarray*}
    F(\mathbf{D},\mathbf{D})_{Val} & = & \{\tensor*{z}{^{\,}_{\mathbf{k}}^{\mathbf{j}}_{\mathbf{i}}}: \forall \gamma_{\mathbf{k}} \in \Gamma_{g}(D,D)_{Coval}\,\tensor*{z}{^{\mathbf{l}}_{\mathbf{k}}^{\mathbf{j}}_{\mathbf{i}}}\gamma_{\mathbf{k}} \in \Delta_{d}(D,D), \\
    & & \quad\quad\quad \forall w^{\mathbf{j}}_{\mathbf{i}} \in \Delta_{d}(D,D)_{CoVal}\,\tensor*{z}{^{\mathbf{l}}_{\mathbf{k}}^{\mathbf{j}}_{\mathbf{i}}}w^{\mathbf{j}}_{\mathbf{i}} \in \Gamma_{g}(D,D)_{Val}\}
    \end{eqnarray*}
    
    Lemma~\ref{GammaDeltaClaim} tells us that there is only covalue for $\Gamma_{g}(D,D)$, namely $\tensor*{\delta}{*^{1}_{k_{1}}^{\;\cdots\;}_{\;\cdots}^{1}_{l_{g}}}$. The tensor representation of the transformation is
    \begin{displaymath}
    \tensor*{\bar{\lambda}}{*^{\mathbf{l}}_{\mathbf{k}}^{\mathbf{j}}_{\mathbf{i}}} = \tensor*{\delta}{*^{l_{1}}_{k_{1}}^{\;\cdots\;}_{\;\cdots}^{l_{g}}_{k_{g}}^{j_{d}}_{i_{1}}^{j_{1}}_{i_{2}}^{\;\cdots\;}_{\;\cdots}^{j_{d-1}}_{i_{d}}}
    \end{displaymath}
    Composing the two tensors presented above together, we find
    \begin{eqnarray*}
    \tensor*{\bar{\lambda}}{*^{\mathbf{l}}_{\mathbf{k}}^{\mathbf{j}}_{\mathbf{i}}} \cdot \tensor*{\delta}{*^{1}_{k_{1}}^{\;\cdots\;}_{\;\cdots}^{1}_{l_{g}}} & = & \tensor*{\delta}{*^{l_{1}}_{k_{1}}^{\;\cdots\;}_{\;\cdots}^{l_{g}}_{k_{g}}^{j_{d}}_{i_{1}}^{j_{1}}_{i_{2}}^{\;\cdots\;}_{\;\cdots}^{j_{d-1}}_{i_{d}}} \cdot \prod_{m=1}^{g}\tensor*{\delta}{*^{1}_{k_{m}}^{1}_{l_{m}}} \\
    	& = & \tensor*{\delta}{*^{1}_{1}^{\;\cdots\;}_{\;\cdots}^{1}_{1}^{j_{d}}_{i_{1}}^{j_{1}}_{i_{2}}^{\;\cdots\;}_{\;\cdots}^{j_{d-1}}_{i_{d}}} \\
    	& = & \tensor*{\delta}{*^{j_{d}}_{i_{1}}^{j_{1}}_{i_{2}}^{\;\cdots\;}_{\;\cdots}^{j_{d-1}}_{i_{d}}}
    \end{eqnarray*}
    
    We know from the base case when $g = 0$ that the tensor to which the composition reduces is not found in the set of values for $\Delta_{d}(D,D)$, irrespective of the size of $d$. We therefore conclude that $\bar{\lambda}$ is not found in the values of $F(\mathbf{D},\mathbf{D})$; and so its corresponding MDNF~transformation from $\C$ does not lift into $\GiC$. \qed
    
    \begin{thm}\label{mixfcomp}
    Let $\C$ be a zero-sum-free compact closed category with finite biproducts satisfying feeble full completeness. Then the category $\GiC$ satisfies \mllmix~full completeness.
    \end{thm}
    
    \subsection{The Necessity of Zero-Sum-Freeness}
    
    The satisfaction of full completeness for the logic \mllmix\ by degenerate categorical models under the orthogonalised glueing construction with focus $\{\iota\}$ is quite a strong result, but the form of the scalars required for the proof to operate is an unfortunate stumbling block. The proof given earlier in this section is designed to show that a compact closed category with finite biproducts satisfying feeble full completeness creates a full complete model of \mllmix\ when the `$\mathbf{G}_{1}$'-glueing is used if the homset $\C[\I,\I]$ acts as a zero-sum-free semiring.
    
    \p We now make clear the insurmountable hurdle which stops Tan's method from being further generalised. Lemma~\ref{MixUniqueLemma} is insufficient for compact closed categories enriched over semimodules over semirings containing even a single additive inverse. This is confirmed by the algorithm below.

    \begin{exa}
    We view the limitations of the proof of Lemma~\ref{MixUniqueLemma} by viewing the \mll~formula $\Parf_{3}(S_{3},S_{3})$ in more detail.
    \end{exa}
    
    Suppose that $\C = \mathbf{FDVec}_{\R}$, and let $\lambda$ be the linear combination of the following sets of axiom links.
    \begin{center} \vspace{1mm}
    \begin{tikzpicture}
      [auto, node
      distance=5mm, skip loop/.style={to path={-- ++(0,#1) -| (\tikztotarget)}}] \tikzstyle{every node} = [text depth=-5pt,text height=0.5ex]
     \node (z1) { }; \node (z2) [right of=z1] {}; 
        \node (1) [right of=z2] {$L$}; \node (a) [right of=1] {$\invamp$} ;
       \node (2) [right of=a] {$L^{\bot}$}; \node (b) [right of=2] {$\invamp$} ;
       \node (3)  [right of=b] {$L$}; \node (c)  [right of=3] {$\invamp$};
       \node (4)  [right of=c] {$L^{\bot}$}; \node(d) [right of=4] {$\invamp$} ;
       \node (5) [right of=d] {$L$}; \node (e) [right of=5] {$\invamp$} ;
       \node (6) [right of=e] {$L^{\bot}$}; 
      \node (t1) [above of= z1] {$-1$};
      \node (t2) [above of= t1] {$-1$};
      \node (t3) [above of= t2] {$2$};
      \node (s1) [below of= z1] {$-1$};
      \node (s2) [below of= s1] {$1$};
      \node (s3) [below of= s2] {$1$};
      \node (T1) [above of= z2] {$\times$};
      \node (T2) [above of= T1] {$\times$};
      \node (T3) [above of= T2] {$\times$};
      \node (S1) [below of= z2] {$\times$};
      \node (S2) [below of= S1] {$\times$};
      \node (S3) [below of= S2] {$\times$};
    \begin{scope}
    \path (1)  edge  [red, skip loop =16mm,shorten >=13mm, shorten <=13mm] (2);
    \path (3)  edge  [red, skip loop =16mm,shorten >=13mm, shorten <=13mm] (4);
    \path (5)  edge  [red, skip loop =16mm,shorten >=13mm, shorten <=13mm] (6);
    \path (1)  edge  [blue, skip loop =11mm,shorten >=8mm, shorten <=8mm] (2);
    \path (3)  edge  [blue, skip loop =12mm,shorten >=8mm, shorten <=8mm] (6);
    \path (4)  edge  [blue, skip loop =11mm,shorten >=8mm, shorten <=8mm] (5);
    \path (1)  edge  [green, skip loop =7mm, shorten >=3mm, shorten <=3mm] (6);
    \path (2)  edge  [green, skip loop =6mm, shorten >=3mm, shorten <=3mm] (5);
    \path (3)  edge  [green, skip loop =6mm, shorten >=3mm, shorten <=3mm] (4);
    \path (1)  edge  [magenta, skip loop =-7mm, shorten >=3mm, shorten <=3mm] (4);
    \path (2)  edge  [magenta, skip loop =-6mm, shorten >=3mm, shorten <=3mm] (3);
    \path (5)  edge  [magenta, skip loop =-5mm, shorten >=3mm, shorten <=3mm] (6);
    \path (1)  edge  [orange, skip loop =-12mm,shorten >=8mm, shorten <=8mm] (4);
    \path (3)  edge  [orange, skip loop =-10mm,shorten >=8mm, shorten <=8mm] (6);
    \path (5)  edge  [orange, skip loop =-11mm,shorten >=8mm, shorten <=8mm] (2);
    \path (1)  edge  [gray, skip loop =-16mm,shorten >=13mm, shorten <=13mm] (6);
    \path (3)  edge  [gray, skip loop =-15mm,shorten >=13mm, shorten <=13mm] (2);
    \path (5)  edge  [gray, skip loop =-15mm,shorten >=13mm, shorten <=13mm] (4);
    \end{scope}
    \end{tikzpicture} \vspace{1mm}
    \end{center}
    
    Letting $\mathcal{S}_{n}$ and $\mathcal{A}_{n}$ denoting the symmetric and alternating groups of $n$ elements respectively, the tensor representation of the \mll~transformation is written \vspace{2mm}
    \begin{displaymath} \tensor*{\bar{\lambda}}{*^{j_{1}}_{i_{1}}^{j_{2}}_{i_{2}}^{j_{3}}_{i_{3}}} = \sum_{p \in S_{3}}s_{p}\tensor*{\delta}{*^{j_{p(1)}}_{i_{1}}^{p(2)}_{i_{2}}^{p(3)}_{i_{3}}}\,,
    \end{displaymath}
   where the scalars in the set $\{s_{p}:p \in \mathcal{S}_{3}\}$ are as follows: \vspace{2mm}
    \begin{displaymath}
    s_{p} := \left\{\begin{array}{ll}
    				2 & \text{if } p = (1\,2\,3) \\
    				1 & \text{if } p \in \mathcal{A}_{3} \backslash (1\,2\,3) \\ 
            -1 & \text{otherwise}        
    \end{array}
    \right. .
    \end{displaymath} \vspace{2mm}
    Note that this means the following essential entries have the following values within them:
    \begin{itemize}
    \item $\tensor*{\bar{\lambda}}{*^{1}_{1}^{2}_{2}^{3}_{3}} = 2$;
    \item $\tensor*{\bar{\lambda}}{*^{2}_{1}^{3}_{2}^{1}_{3}} = \tensor*{\bar{\lambda}}{*^{3}_{1}^{1}_{2}^{2}_{3}} = 1$;
    \item $\tensor*{\bar{\lambda}}{*^{1}_{1}^{3}_{2}^{2}_{3}} = \tensor*{\bar{\lambda}}{*^{3}_{1}^{2}_{2}^{1}_{3}} = \tensor*{\bar{\lambda}}{*^{2}_{1}^{1}_{2}^{3}_{3}} = 0$.
    \end{itemize}
    \vspace{2mm}
    
    We remind ourselves that there are three tensor equations which must be satisfied for some $y_{1}$, $y_{2}$ and $y_{3}$ found in $\{1,2,3\}$:
    \vspace{1mm}
    \begin{eqnarray*}
    \tensor*{\bar{\lambda}}{*^{j_{1}}_{1}^{j_{2}}_{2}^{j_{3}}_{3}} \cdot 1_{j_{2}j_{3}} & = & \delta_{j_{1}y_{1}} \\
    \tensor*{\bar{\lambda}}{*^{j_{1}}_{1}^{j_{2}}_{2}^{j_{3}}_{3}} \cdot 1_{j_{1}j_{3}} & = & \delta_{j_{2}y_{2}} \\
    \tensor*{\bar{\lambda}}{*^{j_{1}}_{1}^{j_{2}}_{2}^{j_{3}}_{3}} \cdot 1_{j_{1}j_{2}} & = & \delta_{j_{3}y_{3}}
    \end{eqnarray*}
  Taking the non-trivial situation from Example~\ref{MixUniqueExam}, where $y_{k} = k$ for each $k$, the significant equations may be found.
    \begin{eqnarray}
    \tensor*{\bar{\lambda}}{*^{1}_{1}^{2}_{2}^{3}_{3}} + \tensor*{\bar{\lambda}}{*^{1}_{1}^{3}_{2}^{2}_{3}}  & = & 1 \\
    \tensor*{\bar{\lambda}}{*^{2}_{1}^{1}_{2}^{3}_{3}} + \tensor*{\bar{\lambda}}{*^{2}_{1}^{3}_{2}^{1}_{3}}  & = & 0 \\
    \tensor*{\bar{\lambda}}{*^{3}_{1}^{1}_{2}^{2}_{3}} + \tensor*{\bar{\lambda}}{*^{3}_{1}^{2}_{2}^{1}_{3}}  & = & 0 \\
    \tensor*{\bar{\lambda}}{*^{2}_{1}^{1}_{2}^{3}_{3}} + \tensor*{\bar{\lambda}}{*^{3}_{1}^{1}_{2}^{2}_{3}}  & = & 0 \\
    \tensor*{\bar{\lambda}}{*^{1}_{1}^{2}_{2}^{3}_{3}} + \tensor*{\bar{\lambda}}{*^{3}_{1}^{2}_{2}^{1}_{3}}  & = & 1 \\
    \tensor*{\bar{\lambda}}{*^{1}_{1}^{3}_{2}^{2}_{3}} + \tensor*{\bar{\lambda}}{*^{2}_{1}^{3}_{2}^{1}_{3}}  & = & 0 \\
    \tensor*{\bar{\lambda}}{*^{2}_{1}^{3}_{2}^{1}_{3}} + \tensor*{\bar{\lambda}}{*^{3}_{1}^{2}_{2}^{1}_{3}}  & = & 0 \\
    \tensor*{\bar{\lambda}}{*^{1}_{1}^{3}_{2}^{2}_{3}} + \tensor*{\bar{\lambda}}{*^{3}_{1}^{1}_{2}^{2}_{3}}  & = & 0 \\
    \tensor*{\bar{\lambda}}{*^{1}_{1}^{2}_{2}^{3}_{3}} + \tensor*{\bar{\lambda}}{*^{2}_{1}^{1}_{2}^{3}_{3}}  & = & 1
    \end{eqnarray}
    It is now seen that these equations are consistent with the entries of $\bar{\lambda}$ from earlier, and so it is deduced that $\tensor*{\bar{\lambda}}{*^{j_{1}}_{i_{1}}^{j_{2}}_{i_{2}}^{j_{3}}_{i_{3}}} \in \Parf_{3}(S_{3},S_{3})_{Val}$. As stated in Lemma~\ref{MixUniqueLemma}, other simultaneous equations using other entry positions in the tensor $\bar{\lambda}$ reduce to linear combinations of those found above using the rules from Section~\ref{SectionTransExtensors}. Furthermore, other choices of $n$ for the test object $S_{n}$ do not provide any more information when it comes to producing equations which contradict the existence of a tensor representation of $\bar{\lambda}$ in the set of values of $\Parf_{3}(S_{n},S_{n})$. It follows that the proof, or any obvious minor alterations to it, is unable to be used to disprove that $\bar{\lambda}$ describes a transformation in the double-glued category.
    
    \p It comes as no surprise that there is more than one solution once negatives are added, especially when the values in the entries are in the genuine ring, once it is realised that there are $6$ variables and the maximal linearly independent set of equations has $5$ elements. Having fewer linear independent equations than variables within them means that the kernel of the equations has positive dimension, and so a myriad of solutions may be found.
    \begin{eqnarray*}
    (\tensor*{\bar{\lambda}}{*^{3}_{1}^{1}_{2}^{2}_{3}} + \tensor*{\bar{\lambda}}{*^{3}_{1}^{2}_{2}^{1}_{3}}) - (\tensor*{\bar{\lambda}}{*^{2}_{1}^{1}_{2}^{3}_{3}} + \tensor*{\bar{\lambda}}{*^{3}_{1}^{1}_{2}^{2}_{3}}) + (\tensor*{\bar{\lambda}}{*^{1}_{1}^{2}_{2}^{3}_{3}} + \tensor*{\bar{\lambda}}{*^{2}_{1}^{1}_{2}^{3}_{3}}) - (\tensor*{\bar{\lambda}}{*^{1}_{1}^{2}_{2}^{3}_{3}} + \tensor*{\bar{\lambda}}{*^{3}_{1}^{2}_{2}^{1}_{3}}) & = & 0 \\
    (\tensor*{\bar{\lambda}}{*^{1}_{1}^{2}_{2}^{3}_{3}} + \tensor*{\bar{\lambda}}{*^{1}_{1}^{3}_{2}^{2}_{3}}) - (\tensor*{\bar{\lambda}}{*^{2}_{1}^{1}_{2}^{3}_{3}} + \tensor*{\bar{\lambda}}{*^{3}_{1}^{1}_{2}^{2}_{3}}) + (\tensor*{\bar{\lambda}}{*^{3}_{1}^{1}_{2}^{2}_{3}} + \tensor*{\bar{\lambda}}{*^{3}_{1}^{2}_{2}^{1}_{3}}) - (\tensor*{\bar{\lambda}}{*^{1}_{1}^{3}_{2}^{2}_{3}} + \tensor*{\bar{\lambda}}{*^{3}_{1}^{1}_{2}^{2}_{3}}) & = & 0 \\
    (\tensor*{\bar{\lambda}}{*^{2}_{1}^{1}_{2}^{3}_{3}} + \tensor*{\bar{\lambda}}{*^{2}_{1}^{3}_{2}^{1}_{3}}) - (\tensor*{\bar{\lambda}}{*^{2}_{1}^{1}_{2}^{3}_{3}} + \tensor*{\bar{\lambda}}{*^{3}_{1}^{1}_{2}^{2}_{3}}) + (\tensor*{\bar{\lambda}}{*^{3}_{1}^{1}_{2}^{2}_{3}} + \tensor*{\bar{\lambda}}{*^{3}_{1}^{2}_{2}^{1}_{3}}) - (\tensor*{\bar{\lambda}}{*^{2}_{1}^{3}_{2}^{1}_{3}} + \tensor*{\bar{\lambda}}{*^{3}_{1}^{2}_{2}^{1}_{3}}) & = & 0 \\
    \left. \begin{array}{r}
    (\tensor*{\bar{\lambda}}{*^{1}_{1}^{2}_{2}^{3}_{3}} + \tensor*{\bar{\lambda}}{*^{1}_{1}^{3}_{2}^{2}_{3}}) + (\tensor*{\bar{\lambda}}{*^{2}_{1}^{1}_{2}^{3}_{3}} + \tensor*{\bar{\lambda}}{*^{2}_{1}^{3}_{2}^{1}_{3}}) + (\tensor*{\bar{\lambda}}{*^{3}_{1}^{1}_{2}^{2}_{3}} + \tensor*{\bar{\lambda}}{*^{3}_{1}^{2}_{2}^{1}_{3}}) \\ - (\tensor*{\bar{\lambda}}{*^{2}_{1}^{1}_{2}^{3}_{3}} + \tensor*{\bar{\lambda}}{*^{3}_{1}^{1}_{2}^{2}_{3}}) - (\tensor*{\bar{\lambda}}{*^{1}_{1}^{2}_{2}^{3}_{3}} + \tensor*{\bar{\lambda}}{*^{3}_{1}^{2}_{2}^{1}_{3}}) - (\tensor*{\bar{\lambda}}{*^{1}_{1}^{3}_{2}^{2}_{3}} + \tensor*{\bar{\lambda}}{*^{2}_{1}^{3}_{2}^{1}_{3}})
    \end{array} \right\}
     & = & 0
    \end{eqnarray*}
    The problem becomes more apparent for formulae $\Parf_{n}(S_{n},S_{n})$ the larger $n$ becomes. As stated in the proof of Lemma~\ref{MixUniqueLemma}, the number of variable positions uniquely describing an \mll~transformation's tensor representation is $n!$, whilst the number of equations which may be considered linearly independent from one another even in the absence of negative elements is $n^{2}$; and clearly $n! > n^2$ for every $n \geq 4$.
    
    \p Of course this does not prove that non-zero-sum-free compact closed categories are incapable of being the basis of fully complete \mllmix\ models. After all, only one test object has been considered; \emph{all} objects in the focused glued category must be shown to still contain the morphisms associated to a rogue \mll~transformation from $\C$ to show that it remains a one in the glued category. Fortunately, it is possible to extend the principle from the above example to a result on the level of \mll~transformations for all the non-zero-sum-free categories.
    
    \begin{prop} \label{NoMLLMixFCforBadGiCLemma}
    Let $\C$ be a compact closed category with finite biproducts satisfying feeble full completeness. Then $\GiC$ does not satisfy \mllmix~full completeness if $\C$ is not zero-sum-free.
    \end{prop}
    \proof
    Suppose $\C$ is such a compact closed category, meaning that there exists at least one arrow $s \in \C[\I,\I]$ with an additive inverse $(-s)$. We assume without loss of generality that we are in a strict compact closed category, suppressing the usage of the isomorphism $\iota: \I \longrightarrow \I^{*}$. However, at times when associativity and unit isomorphisms are silently being used are noted for the sake of clarity, either in the equation or as a side comment. Consider the \mll~transformation $\bar{\lambda}:\K \longrightarrow - \invamp - \invamp - \invamp (-)^{\bot} \invamp (-)^{\bot} \invamp (-)^{\bot}$ found in $\C$ modelling the following linear combination of proof structures:
    
    \begin{center} \vspace{5mm}
    \begin{tikzpicture}
      [auto, node
      distance=5mm, skip loop/.style={to path={-- ++(0,#1) -| (\tikztotarget)}}] \tikzstyle{every node} = [text depth=-5pt,text height=0.5ex] \tikzstyle{every node} = [text depth=-5pt,text height=0.5ex]
     \node (z1) { }; \node (z2) [right of=z1] {}; 
        \node (1) [right of=z2] {$L$}; \node (a) [right of=1] {$\invamp$} ;
       \node (2) [right of=a] {$L^{\bot}$}; \node (b) [right of=2] {$\invamp$} ;
       \node (3)  [right of=b] {$L$}; \node (c)  [right of=3] {$\invamp$};
       \node (4)  [right of=c] {$L^{\bot}$}; \node(d) [right of=4] {$\invamp$} ;
       \node (5) [right of=d] {$L$}; \node (e) [right of=5] {$\invamp$} ;
       \node (6) [right of=e] {$L^{\bot}$}; 
      \node (t1) [above of= z1] {$-s$};
      \node (t2) [above of= t1] {$-s$};
      \node (t3) [above of= t2, xshift=-3mm] {$(1\,+\,s)$};
      \node (s1) [below of= z1] {$-s$};
      \node (s2) [below of= s1] {$s$};
      \node (s3) [below of= s2] {$s$};
      \node (T1) [above of= z2] {$\times$};
      \node (T2) [above of= T1] {$\times$};
      \node (T3) [above of= T2] {$\times$};
      \node (S1) [below of= z2] {$\times$};
      \node (S2) [below of= S1] {$\times$};
      \node (S3) [below of= S2] {$\times$};
    \begin{scope}
    \path (1)  edge  [red, skip loop =16mm,shorten >=13mm, shorten <=13mm] (2);
    \path (3)  edge  [red, skip loop =16mm,shorten >=13mm, shorten <=13mm] (4);
    \path (5)  edge  [red, skip loop =16mm,shorten >=13mm, shorten <=13mm] (6);
    \path (1)  edge  [blue, skip loop =11mm,shorten >=8mm, shorten <=8mm] (2);
    \path (3)  edge  [blue, skip loop =12mm,shorten >=8mm, shorten <=8mm] (6);
    \path (4)  edge  [blue, skip loop =11mm,shorten >=8mm, shorten <=8mm] (5);
    \path (1)  edge  [green, skip loop =7mm, shorten >=3mm, shorten <=3mm] (6);
    \path (2)  edge  [green, skip loop =6mm, shorten >=3mm, shorten <=3mm] (5);
    \path (3)  edge  [green, skip loop =5mm, shorten >=3mm, shorten <=3mm] (4);
    \path (1)  edge  [magenta, skip loop =-7mm, shorten >=3mm, shorten <=3mm] (4);
    \path (2)  edge  [magenta, skip loop =-6mm, shorten >=3mm, shorten <=3mm] (3);
    \path (5)  edge  [magenta, skip loop =-5mm, shorten >=3mm, shorten <=3mm] (6);
    \path (1)  edge  [orange, skip loop =-12mm,shorten >=8mm, shorten <=8mm] (4);
    \path (3)  edge  [orange, skip loop =-10mm,shorten >=8mm, shorten <=8mm] (6);
    \path (5)  edge  [orange, skip loop =-11mm,shorten >=8mm, shorten <=8mm] (2);
    \path (1)  edge  [gray, skip loop =-17mm,shorten >=13mm, shorten <=13mm] (6);
    \path (3)  edge  [gray, skip loop =-16mm,shorten >=13mm, shorten <=13mm] (2);
    \path (5)  edge  [gray, skip loop =-16mm,shorten >=13mm, shorten <=13mm] (4);
    \end{scope}
    \end{tikzpicture} \vspace{5mm}
    \end{center}
    Up to isomorphism the transformation can be written
    \begin{eqnarray*}
    & & (1+s)\cdot(\tensor*{\tilde{\sigma}}{*^{1}_{1}^{2}_{2}^{3}_{3}} \after (\eta \otimes \eta \otimes \eta)) + (-s)\cdot(\tensor*{\tilde{\sigma}}{*^{1}_{1}^{3}_{2}^{2}_{3}} \after (\eta \otimes \eta \otimes \eta)) + (-s)\cdot(\tensor*{\tilde{\sigma}}{*^{3}_{1}^{2}_{2}^{1}_{3}} \after (\eta \otimes \eta \otimes \eta)) \\
    & & \,\,\, + \,(-s)\cdot(\tensor*{\tilde{\sigma}}{*^{2}_{1}^{1}_{2}^{3}_{3}} \after (\eta \otimes \eta \otimes \eta)) + s\cdot(\tensor*{\tilde{\sigma}}{*^{2}_{1}^{3}_{2}^{1}_{3}} \after (\eta \otimes \eta \otimes \eta)) + s\cdot(\tensor*{\tilde{\sigma}}{*^{3}_{1}^{1}_{2}^{2}_{3}} \after (\eta \otimes \eta \otimes \eta))
    \end{eqnarray*}
    where the natural transformation $\tensor*{\tilde{\sigma}}{*^{p(1)}_{1}^{p(2)}_{2}^{p(3)}_{3}}$ for given permutation $p \in \Perm(2,3)$ is the composition of associativity and symmetry natural transformations expected of symmetric monoidal categories containing arrows with source and target in the form below.
    \begin{displaymath}
    (\tensor*{\tilde{\sigma}}{*^{p(1)}_{1}^{p(2)}_{2}^{p(3)}_{3}})_{A_{1},A_{2},A_{3},B_{1},B_{2},B_{3}}: A_1 \otimes B_1 \otimes A_2 \otimes B_2 \otimes A_3 \otimes B_3 \longrightarrow A_1 \otimes A_2 \otimes A_3 \otimes B_{p(1)} \otimes B_{p(2)} \otimes B_{p(3)}
    \end{displaymath}
    For short, we can write $\bar{\lambda} \,=\, \sum_{p \in S_{3}}s_{p} \cdot \tensor*{\tilde{\sigma}}{*^{p(1)}_{1}^{p(2)}_{2}^{p(3)}_{3}}$, where
    \begin{displaymath}
    s_{p} := \left\{\begin{array}{ll}
    				s+1 & \text{if } p = (1\,2\,3) \\
    				s & \text{if } p \in \mathcal{A}_{3} \backslash (1\,2\,3) \\ 
            -s & \text{otherwise}        
    \end{array}
    \right. .
    \end{displaymath}
    
    Let $A = (R,U,X)$ be an arbitrary object in $\GiC$, meaning that $x \after u = \iota = 1_{\I}$ for every $u \in U$ and $x \in X$. We define the following notation:
    \begin{itemize}
    \item $\mathbf{z}^{(l)}$ is the tuple $\mathbf{z}$ missing the $l^{th}$ entry;
    \item $\left\langle \mathbf{g}^{(j)},f\right\rangle_{j} \,=\, f \after \left( \bigotimes_{i<j}g_{i} \otimes 1 \otimes \bigotimes_{j<i\leq n}g_{i}\right) \after \Lambda^{n}$, where $\Lambda: \I \longrightarrow \I^{\otimes n}$ is the natural composition of the unit isomorphism.
    \end{itemize}
    
\noindent    We see that,
    \begin{eqnarray*}
    \lefteqn{\Parf_{3}(A,A)_{Val} = } \\
    & & \{f \in \C[\I,\left|\Parf_{3}(A,A)\right|]: \bforall_{k=1}^{3} u_{k}^{*} \in U\,\forall x_{k} \in X,\, \forall l\,\,\left\langle ((\mathbf{x}^{(l)},\mathbf{u}^{*})),f\right\rangle_{l} \in U, \\
    & & \quad\quad\quad\quad \left\langle (\mathbf{x},(\mathbf{u}^{*})^{(l)}),f\right\rangle_{l+3} \in X^{*}\}.
    \end{eqnarray*}
    
    \p\noindent If either $U$ or $X = \varnothing$ then it is trivially true that $\bar{\lambda}_{A}$ is in the set of values. If both sets are non-empty, let $u_{1}^{*},u_{2}^{*},u_{3}^{*} \in U^{*}$ and $x_{2},x_{3} \in X$ be arbitrary, not necessarily distinct morphisms. Noting (and suppressing usage of the isomorphism $\iota$) that $(x \otimes 1_{R^{*}}) \after \eta_{R} = \lambda_{R^{*}}^{-1} \after x^{*}$ and $(1_{R} \otimes u^{*}) \after \eta_{R} = \rho_{R}^{-1} \after u$ for all suitable arrows $u$ and $x$ in $\C$ due to dinaturality, and that $x_{i} \after u_{j} \,=\, \iota \,=\, 1_{\I}$ for any choice of $i,j$, the composition $\left\langle (x_{2},x_{3},u_{1}^{*},u_{2}^{*},u_{3}^{*}),\tensor*{\tilde{\sigma}}{*^{p(1)}_{1}^{p(2)}_{2}^{p(3)}_{3}} \after (\eta \otimes \eta \otimes \eta)\right\rangle_{1}$ can be found for each permutation $p$. 
    \begin{align*}
     &\, (1 \otimes x_{2} \otimes x_{3} \otimes u_{1}^{*} \otimes u_{2}^{*} \otimes u_{3}^{*}) \after \tensor*{\tilde{\sigma}}{*^{p(1)}_{1}^{p(2)}_{2}^{p(3)}_{3}} \after (\eta \otimes \eta \otimes \eta) \\
     &\qquad =\, \tensor*{\tilde{\sigma}}{*^{p(1)}_{1}^{p(2)}_{2}^{p(3)}_{3}} \after (1 \otimes u_{p^{-1}(1)}^{*} \otimes x_{2} \otimes u_{p^{-1}(2)}^{*} \otimes x_{3} \otimes u_{p^{-1}(3)}^{*}) \after (\eta \otimes \eta \otimes \eta) \\
     &\qquad =\, \tensor*{\tilde{\sigma}}{*^{p(1)}_{1}^{p(2)}_{2}^{p(3)}_{3}} \after (((1 \otimes u_{p^{-1}(1)}^{*}) \after \eta) \otimes ((1 \otimes u_{p^{-1}(1)}^{*}) \after \eta) \otimes ((1 \otimes u_{p^{-1}(1)}^{*}) \after \eta)) \\
     &\qquad =\, \tensor*{\tilde{\sigma}}{*^{p(1)}_{1}^{p(2)}_{2}^{p(3)}_{3}} \after ((\lambda_{\I^{*}}^{-1} \after u_{p^{-1}(1)}) \otimes \lambda_{\I^{*}}^{-1} \otimes \lambda_{\I^{*}}^{-1})
     \,=\, u_{p^{-1}(1)} \,\in\, U
    \end{align*}
    By symmetry and a similar argument we also find
    \begin{align*}
    &\, \left\langle ((\mathbf{x}^{(l)},\mathbf{u}^{*})),\tensor*{\tilde{\sigma}}{*^{p(1)}_{1}^{p(2)}_{2}^{p(3)}_{3}} \after (\eta \otimes \eta \otimes \eta)\right\rangle_{l} \, = \, u_{p^{-1}(l)} \,\in\, U \\
    &\, \left\langle (\mathbf{x},(\mathbf{u}^{*})^{(l)}),\tensor*{\tilde{\sigma}}{*^{p(1)}_{1}^{p(2)}_{2}^{p(3)}_{3}} \after (\eta \otimes \eta \otimes \eta)\right\rangle_{l+3} \, = \, x_{p(l)}^{*} \,\in\, X^{*}
    \end{align*}

\noindent    Finding the compositions $\left\langle ((\mathbf{x}^{(l)},\mathbf{u}^{*})),\tensor*{\tilde{\sigma}}{*^{p(1)}_{1}^{p(2)}_{2}^{p(3)}_{3}} \after (\eta \otimes \eta \otimes \eta)\right\rangle_{l}$ and \\ $\left\langle (\mathbf{x},(\mathbf{u}^{*})^{(l)}),\tensor*{\tilde{\sigma}}{*^{p(1)}_{1}^{p(2)}_{2}^{p(3)}_{3}} \after (\eta \otimes \eta \otimes \eta)\right\rangle_{l+3}$ now becomes a simple case of taking the correct linear combination of answers from the above calculations.
    \begin{align*}
    \left\langle (\mathbf{x}^{(l)},\mathbf{u}^{*}),\bar{\lambda}_{R}\right\rangle_{l} &= \left\langle ((\mathbf{x}^{(l)},\mathbf{u}^{*})),\sum_{p \in S_{3}}\,s_{p} \cdot (\tensor*{\tilde{\sigma}}{*^{p(1)}_{1}^{p(2)}_{2}^{p(3)}_{3}} \after (\eta \otimes \eta \otimes \eta))\right\rangle_{l} \\
    	&=	\sum_{p \in S_{3}}\,s_{p} \cdot \left\langle ((\mathbf{x}^{(l)},\mathbf{u}^{*})),\bar{\lambda}_{R}\right\rangle_{l}  \\
    	&= \left\langle ((\mathbf{x}^{(l)},\mathbf{u}^{*})),\tensor*{\tilde{\sigma}}{*^{p(1)}_{1}^{p(2)}_{2}^{p(3)}_{3}} \after (\eta \otimes \eta \otimes \eta)\right\rangle_{l} \\
    	&= 	\sum_{p \in S_{3}}\,s_{p} \cdot u_{p^{-1}(l)}
    \end{align*}
    This final sum is a linear combination of the arrows $u_{1},u_{2},u_{3}$. One finds by inspection that the scalars added together in this sum for $u_{k}$ for some $k$ are $s_{(i_{1} i_{2} i_{3})}$ and $s_{(j_{1} j_{2} j_{3})}$, the two scalars for which $i_{k} \,=\, l \,=\, j_{k}$. Since $(i_{1} i_{2} i_{3})$ and $(j_{1} j_{2} j_{3})$ are distinct but they share a value in exactly one position in this situation, it must be the case that one is an even permutation (that is, it belongs to $A_{3}$), and one is odd (and therefore is not a member of $A_{3}$). For the sake of convenience, let the former be the even permutation. If $i_{k} \,\neq\, k \,\neq\, j_{k}$, then we know that neither of the two permutations are $(1 \, 2 \, 3)$, and so we deduce that \linebreak $s_{(i_{1} i_{2} i_{3})} \,+\, s_{(j_{1} j_{2} j_{3})} \,=\, s \,+\, (-s) \,=\, 0$. If however, $i_{k} \,=\, k \,=\, j_{k}$, then one of them does equal $(1 \, 2 \, 3)$, meaning $s_{(1 \, 2 \, 3)} \,+\, s_{(j_{1} j_{2} j_{3})} \,=\, (1 + s) \,+\, (-s) \,=\, 1 \,+\, (s \,+\, (-s)) \,=\, 1$. We therefore conclude that
    \begin{align*}
    \left\langle (\mathbf{x}^{(l)},\mathbf{u}^{*}),\bar{\lambda}_{R}\right\rangle_{l} &= 	u_{l} \,\in\, U
    \end{align*}
    By a dual argument, we can find that 
    \begin{align*}
    \left\langle (\mathbf{x},(\mathbf{u}^{*})^{(l)}),\bar{\lambda}_{R}\right\rangle_{l+3} &= 	x_{l}^{*} \,\in\, X^{*}
    \end{align*}
    
\noindent    The arrows $u_{1},u_{2},u_{3} \,\in\,U$ and $x_{1},x_{2},x_{3} \,\in\,X$ are arbitrary throughout the above argument, meaning that any triple of values and covalues of $A \,=\, (R,U,X)$ could be chosen. Thus $\bar{\lambda}_{R} \in (A \invamp A \invamp A \invamp A^{\bot} \invamp A^{\bot} \invamp A^{\bot})_{Val}$. Furthermore, we may say that this for any choice of $A$, since $A$ is considered arbitrary in this proof as well. Naturally this means that the $\C$-arrows in the collection $\bar{\lambda}=\{\bar{\lambda}_{R}: R \in \C\}$ are found in all the required homsets in $\GiC$, and therefore the \mll~transformation $\bar{\lambda}$ is also in $\GiC$. The linear combination of proof structures modelled by this transformation is non-simple, and so the category does not satisfy \mllmix~full completeness. \qed

    \begin{cor}
    Let $\C$ be a compact closed category with finite biproducts satisfying feeble full completeness. The category $\GiC$ satisfies \mllmix~full completeness if and only if $\C$ is zero-sum-free.
    \end{cor}
    
\section{Conclusions}

In this paper we show that there are two simple, elegant methods
of producing categorical models for both \mll\ or~\mllmix. The
Hyland-Tan double glueing construction is seemingly the perfect
semantic description of the deductive system when applied to
tensor-generated compact closed categories with finite biproducts. 
Certainly the Danos-Regnier description of proof nets has a strongly
combinatorial flavour, and this paper shows that the same
combinatorial restrictions are imposed on the
categorical model by the double glueing construction.

\p It is notable that the tensor representations of \mll\
transformations are precisely the \emph{isotropic tensors of even
power} over the semiring of scalars. Viewed as vectors in Euclidean space, this means that such representations are exactly those which are invariant under basis change. It is therefore reasonable to suggest that the more combinatorial proofs given in this paper may be replaceable by geometric arguments. Such a methodology could unveil a different perspective on the categorical models and \mll\ itself.

\p An obvious possibility for future work is applying these techniques
to the (unitless) multiplicative additive fragment of linear logic,
\mall. Since we start from compact closed categories with finite
biproducts we know that our models, after double glueing, have both
finite sums and products~\cite{HS03} and so are models of \mall.

\p However, it is known that the Hyland-Tan construction alone cannot
produce fully complete models of \mall~\cite{Ste13}: none of the
resulting categories satisfy Joyal's softness property on the
dinatural level as is required. Categories which accurately model
\mall~proof nets are in short supply: the only example whose full proof has been published is $\G\HypCoh$ --- the category of hypercoherence spaces under the `$\mathbf{G}$'-construction --- found in~\cite{BHS05}.

\p We note that $\HypCoh$ is another example of a double-glued
category, but with a tight orthogonality upon it. \hughf{We are writing an article giving the appropriate generalisation of the definition of orthogonality found in~\cite{HS03} in order to allow the essence of this construction to be extracted. It is hoped that this will provide a starting point for finding a full completeness result for \mall\ which is as general as ours for~\mll.} One might note in
passing that the result of Blute, Hamano and Scott is based upon the
\mall~proof net criteria of Girard~\cite{Gir96} based around monomial weights, whereas there is currently no comparable result for the Hughes-van Glabbeek proof net criteria~\cite{HvG05}. We hope that our methods are suitable to address that situation.

\section*{Acknowledgements}
  \hughe{The authors would like to thank the anonymous reviewers of this paper for their constructive criticism of our originally submitted draft. Thanks also to Luke Ong and Harold Simmons, examiners of the thesis from which this article was originally drawn, for their suggestions. The commutative diagrams were drawn using the diagrams package of Paul Taylor.
  
  \noindent The second author was partially supported by the ANR projects LOGOI (10-BLAN-0213-02) and COQUAS (12-JS02-006-01), and an EPSRC studentship.}


\bibliographystyle{alpha}

\end{document}